\documentclass[aps, superscriptaddress, prb, twocolumn]{revtex4-2}
\usepackage{enumitem}
\usepackage{physics}
\usepackage{xcolor}
\usepackage{amsbsy}
\usepackage{amsfonts}
\usepackage{amsmath}
\usepackage{amstext}
\usepackage{amsthm}
\usepackage{amssymb}
\usepackage{graphicx}
\usepackage{epsfig}
\usepackage{float}
\usepackage{epstopdf}
\usepackage{bm}
\usepackage{bbm}
\usepackage{color}
\definecolor{prlblue}{rgb}{0.18,0.19,0.57}
\usepackage{multirow}
\usepackage{ifthen}
\usepackage{tikz}
\usepackage{multirow}
\usepackage{braket}
\usepackage{comment}
\usetikzlibrary{patterns.meta} 
\usetikzlibrary{calc}
\newcommand{\anb}[1]{{\color{blue} #1}}

\newcommand{\bzt}{\mathbb{Z}_2}
\newcommand{\bbz}{\mathbb{Z}}

\newcommand{\placeholder}{\text{1D $\bbz_4^f$ disorder-intrinsic state}}
\newcommand{\placeholders}{\text{1D $\bbz_4^f$ disorder-intrinsic states}}
\newcommand{\bubbleplaceholder}{\text{1D $\bbz_4^f$ disorder-intrinsic bubble}} 
\newcommand{\placeholdermodes}{\text{1D $\bbz_4^f$ disorder-intrinsic edge modes}}

\usepackage[colorlinks]{hyperref}
\usepackage{cleveref}
\makeatletter
\def\l@subsubsection#1#2{}
\makeatother

\begin{document}
\title{Classification of Average Crystalline Topological Superconductors through a Generalized Real-Space Construction}
\author{Sarvesh Srinivasan}
\email{sbs5627@psu.edu}
\affiliation{Department of Physics, The Pennsylvania State University, University Park, Pennsylvania 16802, USA}
\author{Jian-Hao Zhang}
\email{Sergio.Zhang@colorado.edu}
\affiliation{Department of Physics, The Pennsylvania State University, University Park, Pennsylvania 16802, USA}
\affiliation{Department of Physics and Center for Theory of Quantum Matter, University of Colorado Boulder, Boulder, CO 80309, USA}
\author{Yang Qi}
\email{qiyang@fudan.edu.cn}
\affiliation{State Key Laboratory of Surface Physics, Fudan University, Shanghai 200433, China}
\affiliation{Center for Field Theory and Particle Physics, Department of Physics, Fudan University, Shanghai 200433, China}
\affiliation{Collaborative Innovation Center of Advanced Microstructures, Nanjing 210093, China}
\author{Zhen Bi}
\email{zjb5184@psu.edu}
\affiliation{Department of Physics, The Pennsylvania State University, University Park, Pennsylvania 16802, USA}

\begin{abstract}
We investigate a novel class of topological superconducting phases protected by exact fermion-parity symmetry and average crystalline symmetries. These phases belong to the broader class of average crystalline symmetry-protected topological (ACSPT) states and include numerous examples of intrinsic ACSPTs -- topological phases that arise only in the presence of disorder or decoherence. Unlike conventional symmetry-protected topological (SPT) phases, which require exact symmetry protection, average SPT (ASPT) phases remain robust as long as the symmetry is restored on average across disorder realizations or mixed-state ensembles. To classify these phases, we extend the real-space block state construction framework to account for average crystalline symmetries. In this generalized setting, lower-dimensional cells are decorated with ASPT phases, and the obstruction-free conditions are reformulated to incorporate the constraints imposed by average symmetry at block intersections. This provides a physically transparent and systematic method for classifying ASPTs with spatial symmetries that are only preserved statistically. We further validate our classification using a generalized spectral sequence analysis, which serves as an independent consistency check. Our results demonstrate that many crystalline topological superconductors remain well defined under realistic imperfections, and they uncover a rich landscape of intrinsically average-symmetry-protected phases that have no analog in clean systems.

\end{abstract}

\maketitle
\tableofcontents

\section{Introduction}
Symmetry-protected topological (SPT) phases have emerged as a vibrant and rapidly advancing frontier in modern condensed matter physics, offering profound insights into how global symmetries stabilize unique topological states of matter\cite{chen2013cohomology,Chen2012science,Vishwanath2013,Wang2014interactingfermion,senthil2015symmetry,gu2014sptinteractingfermion,else2014classifying,Chen2014DDW,Bi2015NLSM}. These remarkable phases, exemplified by celebrated systems like topological insulators and superconductors\cite{qi2011topological,hasan2010colloquium}, are characterized by a gapped bulk and robust boundary states protected by specific global symmetries in the system. The boundary states, however, are not invulnerable -- once the protecting symmetry in the system is completely broken, the boundary states can typically be gapped out, rendering the system topologically trivial. In recent years, the notion of topological insulators has been extended to systems with crystalline symmetries\cite{fu2011topological,Hsieh2012TCI,ando2015topological}. This generalization has opened up a vast and realistic playground for the discovery of new topological phases in crystalline materials, offering exciting opportunities for both theoretical exploration\cite{Isobe2015,Zou2018bulk,Kruthoff2017,Po2017indicator,Khalaf2018surface,Huang2018surface,Wang2018HOTSC,Shapourian2018,Schindler2018,Shiozaki2022,Shiozaki2023,cheng2022rotation,Song2020beyond,Rasmussen2020higherorder,rasmussen2018intrinsicallyinteractingtopologicalcrystalline,Ono2021magnetic,Zhang2023interactinghigherorder,Manjunath2021,Zhang2022wire,MayMann2022,Benalcazar2017,Wladimir2017,Langbehn2017,Song2017rotation,Khalaf2018,Trifunovic2019,Geier2018, zhang2023anomalous} and experimental realization\cite{tanaka2012experimental,dziawa2012topological,okada2013observation,Zeljkovic2014,Schindler2018exp,Choi2020}. However, imperfections and disorders are ubiquitous in real-world materials, inevitably breaking these symmetries to varying degrees. This raises a critical question: can the topological phases in such systems remain robust despite the presence of symmetry-breaking disorder? Addressing this question not only has profound implications for the fundamental understanding of topological phases but also holds the key to developing practical applications in quantum materials and devices. 

Surprisingly, recent theoretical studies have revealed that, contrary to intuitive expectations, the stability of SPT phases does not always require exact symmetries. In some cases, a symmetry that is locally broken by disorders but statistically preserved across the entire ensemble of disorder realizations -- referred to as an average symmetry -- can still provide protection to certain SPT phases. This new class of topological phases is termed average symmetry-protected topological (ASPT) phases\cite{deGroot2022symmetryprotected,Ma2022ASPT1,Ma2025ASPT2}. Recently, systematic studies of ASPTs with global symmetries have garnered significant attention within the condensed matter community\cite{Lee:2022hog, Zhang:2022jul, Ma:2024kma, Guo2025lpdo, xue2024tensornetworkformulationsymmetry, Lessa2025multientanglement,Sun2025holography, Zhang2023fractonic, Guo2025swssbspt}. For crystalline SPTs, a few notable examples in the literature have demonstrated how average symmetries can protect surface states, suggesting that average crystalline symmetry-protected topological (ACSPT) phases could indeed exist\cite{Chen2025intrinsic, su2025transitions}. However, a comprehensive and systematic understanding of ACSPTs remains elusive. This paper aims to address this gap by exploring the role of average symmetries in crystalline systems and their potential to reveal new topological phases.

ASPT is also highly relevant in the context of topological phases in open quantum systems with decoherence\cite{Albert2014,Albert2018,Lieu2020}. Recent advances in quantum simulator platforms have introduced a novel approach to realizing topological phases of matter without the need for cooling a system with a topological Hamiltonian. Instead, many topological phases can be prepared through a combination of unitary circuits and measurements. This innovative method has proven to be highly productive, enabling the realization of numerous intriguing topological phases not observed in solid-state materials\cite{Satzinger2021,Semeghini2021,google2023nonabelian,Iqbal2024nonabelian}. However, in such setups, decoherence induced by system-environment coupling is an unavoidable feature that can locally break the exact symmetry of the system. In these cases, an average symmetry can emerge if the symmetry is preserved across the entire ensemble of decoherence processes. Thus, exploring ASPT in this context\cite{deGroot2022symmetryprotected,Ma2022ASPT1,Ma2025ASPT2,Lee:2022hog, Zhang:2022jul, Ma:2024kma, Guo2025lpdo, xue2024tensornetworkformulationsymmetry, Lessa2025multientanglement,Sun2025holography} is both natural and essential for advancing our understanding of topological physics in these quantum platforms.

For crystalline symmetry--protected topological (CSPT) phases in clean, pure-state systems, the \emph{block-state construction} -- also known as the \emph{topological crystal} approach -- serves as a central organizing framework\cite{thorngren2018gauging,else2019defectnetworks,song2019topological,song2017topological,huang2017building,Song2020realspace,zhang2020point2d,zhang2022crystalspt2d,zhang2022crystalspt3d}. This method partitions real space into symmetry-fixed cells -- 3D, 2D, 1D, and 0D regions denoted \(\lambda, \sigma, \tau, \mu\), respectively -- each of which is left pointwise invariant by a corresponding little group of the full crystalline symmetry. These little groups act as effective onsite symmetries within each cell, allowing one to decorate each \(p\)-cell with a \(p\)-dimensional SPT phase protected by that symmetry (e.g., a Majorana chain on a 1D rotation axis or a \(p+ip\) superconductor on a 2D mirror plane). Neighboring cells are connected via lower-dimensional junctions designed to cancel anomalies, and configurations that differ only by the nucleation and annihilation of closed SPT ``bubbles'' are identified as topologically equivalent. Following this real-space construction yields a classification of CSPT phases that agrees with results obtained independently via the \emph{crystalline equivalence principle}~\cite{thorngren2018gauging}. The block-state construction directly addresses the structure of many-body wavefunctions consistent with crystalline symmetry, making it both applicable to and especially useful for strongly interacting phases.

We note that, in the block-state construction, the cells should be understood as coarse-grained unit cells that are large compared to both the microscopic lattice constant and the correlation length of the SPT decorations (typically on the same order as the lattice constant). This scale separation ensures that the placement of lower-dimensional SPTs on the cells is well-defined and meaningful. In a gapped topological phase, universal features such as topological invariants are expected to be insensitive to microscopic details. As a result, classifying phases using blocks larger than the lattice constant should faithfully capture the same topological data. For point-group symmetric systems, a rigorous justification for this coarse-grained approach has been provided in Ref.~\cite{song2017topological,huang2017building}, where it is shown that a finite-depth symmetric unitary circuit can map a microscopic lattice model to a block-decomposed structure where the block is as large as the system size, confirming the validity of using macroscopic cells. When translation symmetry is present, large-cell constructions remain applicable under the physical assumption of a smooth state~\cite{thorngren2018gauging,huang2017building}. From a group-theoretic viewpoint, the space group associated with the enlarged (super)cell is isomorphic to the original lattice’s symmetry group, so one expects the topological classifications to coincide. Consequently, the classification of crystalline topological phases based on enlarged cells accurately reflects that of the underlying microscopic lattice.

In this work, we extend the concept of average symmetry-protected topological phases to crystalline settings, thereby generalizing crystalline SPTs to systems with average crystalline symmetry. We consider fermionic systems defined on a lattice where fermion parity remains an exact symmetry, while other crystalline symmetries -- such as point group operations -- may be broken locally due to microscopic disorder. Examples include random variations in chemical potential or pairing strength on the lattice scale, which can destroy local symmetry while preserving it statistically over larger regions. Disordered superconductors provide a natural setting for such phenomena. We also address scenarios where coupling with the environment, such as phonons or photons, leads to local decoherence that disrupts symmetry locally but retains average on a large scale.

Our framework relies on the key assumption that despite the presence of disorder or decoherence, the lattice structure remains intact and preserves crystalline symmetry on average at scales much larger than the microscopic lattice scale. Following the standard methodology of crystalline SPT construction, we partition space into cells that are significantly larger than the disorder correlation length. This scale separation ensures that microscopic fluctuations average out within each cell, allowing one to assign meaningful, average symmetry properties to each region. These coarse-grained cells can then be decorated with symmetry-protected topological states consistent with their average symmetries. In this way, the block-state construction remains applicable even in disordered or decohered systems, capturing the global topological features of these phases.

A highlight of our results is the identification of numerous cases of \emph{intrinsic} average crystalline SPT -- symmetry-protected topological phases that are enabled specifically by disorder or decoherence and have no direct counterparts in clean, pure-state systems~\cite{Ma2025ASPT2} (we will review this notion in detail later). Such exotic insulating topological phases have only recently begun to attract attention, with only a few explicit examples discussed in the literature~\cite{Ma2025ASPT2,Guo2025lpdo, Sun2025holography, Guo2025lpdo}. Our work significantly expands this landscape, providing a rich collection of examples that pave the way for further exploration of these intrinsic phases. To that end, we present a detailed comparison between the clean/pure-state classification of crystalline topological superconductors~\cite{zhang2022crystalspt2d, Heinrich2018Z2Anomaly} and the classification in the presence of disorder or decoherence. Tables.~\ref{table:2dresults_clean_vs_disordered_spinless} and~\ref{table:3dresults_clean_vs_disordered_spinless} summarize our findings for both spinless and spin-1/2 systems, clearly illustrating the emergence of a substantial number of intrinsic ACSPT phases.

    \begin{table*}[!htbp]
        \begin{center}
            \begin{tabular}{|c|c||c|c|c||c|c|c|}
                \hline 
            \multicolumn{8}{|c|}{2D ACSPTs}\\ \hline
            &  & \multicolumn{3}{|c||}{Spinless} & \multicolumn{3}{|c|}{Spin-1/2} \\ \hline
            & $G_b$ & Clean & Decoherence & Disorder & Clean & Decoherence & Disorder \\
            \hline \hline 
         1 & $p1$ & $\bzt^3$ & $\bzt^3$ & $\bzt^2$ & $\bzt^3$ & $\bzt^3$ & $\bzt^2$  \\ \hline
         2 & $p2$ & $\bzt^4$ & $\bzt^3$ & $\anb{\bzt^3}$ & $\bbz_4\times\bbz_8^3$ & $\anb{\bzt\times\bbz_4^3}$&$\bzt^3$  \\ \hline
         3 & $pm$ & $\bzt^6$ & $\bzt^5$& $\bzt^2\times\anb{\bzt}$ & $\bbz_4\times\bbz_8$ & $\anb{\bzt^3\times\bbz_4}$ &$\bzt\times\anb{\bzt^2}$  \\ \hline
         4 & $pg$ & $\bzt^3$ & $\bzt^3$&$\bzt^2$ & $\bzt^3$&$\bzt^3$ & $\bzt^2$  \\ \hline
         5 & $cm$ & $\bzt^4$ & $\bzt^3$& $\bzt^2$ & $\bzt\times\bbz_4$ & $\bzt\times\anb{\bzt^2}$&$\bzt\times\anb{\bzt}$  \\ \hline
         6 & $pmm$ & $\bzt^8$ & $\bzt^4\times\anb{\bzt^3}$&$\anb{\bzt^4}$ & $\bzt^8$ &$\anb{\bzt^8}$ &$\anb{\bzt^4}$  \\ \hline
         7 & $pmg$ & $\bzt^5$ & $\bzt^4$ &$\bzt\times\anb{\bzt^2}$ & $\bbz_4\times\bbz_8^2$ & $\anb{\bzt^2\times\bbz_4^2}$&$\bzt^2\times\anb{\bzt}$  \\ \hline
         8 & $pgg$ & $\bzt^3$ & $\bzt^2$& $\bzt\times\anb{\bzt}$ & $\bzt\times\bbz_4\times\bbz_8$ & $\bzt\times\anb{\bzt\times\bbz_4}$ & $\bzt^2$  \\ \hline
         9 & $cmm$ & $\bzt^5$ & $\bzt^2\times\anb{\bzt^2}$&$\anb{\bzt^3}$ & $\bzt^4\times\bbz_8$ & $\anb{\bzt^4\times\bbz_4}$ &$\bzt\times\anb{\bzt^2}$  \\ \hline
         10 & $p4$ & $\bzt^3\times\bbz_4$ & $\bzt^2$&$\anb{\bzt^2}$ & $\bzt\times\bbz_8^3$ & $\bzt\times\anb{\bzt^2\times\bbz_4}$&$\bzt^2$  \\ \hline
         11 & $p4m$ & $\bzt^7$ & $\bzt^2\times\anb{\bzt^3}$ &$\anb{\bzt^3}$ & $\bzt^6$ &$\anb{\bzt^6}$ & $\anb{\bzt^3}$  \\ \hline
         12 & $p4g$ & $\bzt^4$ & $\bzt^2$ & $\anb{\bzt^2}$ & $\bzt^3\times\bbz_8$ & $\bzt\times\anb{\bzt^3}$ & $\bzt\times\anb{\bzt}$  \\ \hline
         13 & $p3$ & $\bzt\times\bbz_3^3$ &  $\bzt$&$\bbz_1$ & $\bzt\times\bbz_3^3$& $\anb{\bzt}$ & $\bbz_1$  \\ \hline
         14 & $p3m1$ & $\bzt^3$ & $\bzt^2$ & $\bzt$ & $\bbz_4$ & $\anb{\bzt^2}$ &$\anb{\bzt}$  \\ \hline
         15 & $p31m$ & $\bzt^3\times\bbz_3$ & $\bzt^2$ &$\bzt$ & $\bbz_3\times\bbz_4$ & $\anb{\bzt^2}$ &$\anb{\bzt}$  \\ \hline
         16 & $p6$ & $\bzt^2\times\bbz_3^2$ & $\bzt$ & $\anb{\bzt}$ & $\bbz_3\times\bbz_8\times\bbz_{12}$ & $\anb{\bzt\times\bbz_4}$ & $\bzt$  \\ \hline
         17 & $p6m$ & $\bzt^4$ & $\bzt^2\times\anb{\bzt}$ & $\anb{\bzt^2}$ & $\bzt^4$ & $\anb{\bzt^4}$ &$\anb{\bzt^2}$  \\ \hline
         \end{tabular}
            \caption{Classification data for disordered 2D space group ASPTs . Data in \anb{blue} indicates presence of intrinsic ASPTs. Details for each case can be found in Appendix \ref{app:2d_class_details}.}
            \label{table:2dresults_clean_vs_disordered_spinless}
            \end{center}
        \end{table*}

\begin{table*}[!htbp]
    \begin{center}
        \begin{tabular}{|c|c||c|c|c||c|c|c|}
            \hline 
        \multicolumn{8}{|c|}{3D ACSPTs}\\ \hline
        &  & \multicolumn{3}{|c||}{Spinless} & \multicolumn{3}{|c|}{Spin-1/2} \\ \hline
        & $G_b$ & Clean & Decoherence & Disorder & Clean & Decoherence & Disorder \\
        \hline \hline 
     1 & $C_1$ & $\bbz_1$ & $\bbz_1$ & $\bbz_1$ & $\bbz_1$ & $\bbz_1$ & $\bbz_1$ \\ \hline
     2 & $C_i$ & $\bbz_1$ & $\bbz_1$ & $\bbz_1$ & $\bbz_1$ & $\bbz_1$ & $\bbz_1$  \\ \hline
     3 & $C_2$ & $\bbz_1$ & $\bbz_1$ & $\bbz_1$ & $\bbz_1$ & $\anb{\bzt}$ & $\anb{\bzt}$  \\ \hline
     4 & $C_{1h}$ & $\bbz_{16}$ & $\bbz_8$ & $\bbz_4$ & $\bbz_1$ & $\bbz_1$ & $\anb{\bzt}$  \\ \hline
     5 & $C_{2h}$ & $\bbz_8$ & $\bbz_4\times\anb{\bzt}$ &$\bzt\times\anb{\bzt^2}$ & $\bbz_1$ & $\anb{\bzt^3}$ &$\anb{\bzt^3}$  \\ \hline
     6 & $D_2=V$ & $\bbz_1$ & $\anb{\bzt}$&$\anb{\bzt^2}$ & $\bzt^2$ & $\anb{\bzt^4}$ &$\anb{\bzt^3}$  \\ \hline
     7 & $C_{2v}$ & $\bzt^3$ & $\bzt\times\anb{\bzt^2}$ &$\bzt\times\anb{\bzt^2}$ & $\bbz_1$ & $\anb{\bzt^2}$ &$\anb{\bzt^3}$  \\ \hline
     8 & $D_{2h}=V_h$ & $\bzt^5$ & $\bzt\times\anb{\bzt^6}$ & $\anb{\bzt^6}$ & $\bzt^3$ & $\anb{\bzt^7}$ &$\anb{\bzt^6}$  \\ \hline
     9 & $C_4$ & $\bbz_1$ & $\bbz_1$ & $\bbz_1$ & $\bbz_1$ & $\anb{\bzt}$ &$\anb{\bzt}$  \\ \hline
     10 & $S_4$ & $\bzt^2$ & $\bzt$ & $\anb{\bzt}$ & $\bzt^2$ & $\bzt\times\anb{\bzt}$ & $\bzt\times\anb{\bzt}$  \\ \hline
     11 & $C_{4h}$ & $\bbz_8\times\bzt$ & $\bbz_4\times\anb{\bzt}$ &$\bzt\times\anb{\bzt^2}$ & $\bzt$ & $\anb{\bzt^3}$ &$\anb{\bzt^3}$  \\ \hline
     12 & $D_4$ & $\bzt$ & $\bzt$ &$\anb{\bzt^2}$ & $\bzt^2$ & $\anb{\bzt^4}$ &$\anb{\bzt^3}$  \\ \hline
     13 & $C_{4v}$ & $\bzt^4$ & $\bzt\times\anb{\bzt^2}$ &$\bzt\times\anb{\bzt^2}$ & $\bbz_1$ & $\anb{\bzt^2}$ & $\anb{\bzt^3}$  \\ \hline
     14 & $D_{2d}=V_d$ & $\bzt^3$ & $\anb{\bzt\times\bbz_4}$ &$\anb{\bzt^3}$ & $\bzt$ & $\anb{\bzt^3}$ &$\anb{\bzt^3}$  \\ \hline
     15 & $D_{4h}$ & $\bzt^6$ & $\bzt\times\anb{\bzt^6}$ &$\anb{\bzt^6}$ & $\bzt^3$ & $\anb{\bzt^7}$ &$\anb{\bzt^6}$  \\ \hline
     16 & $C_3$ & $\bbz_1$ & $\bbz_1$ & $\bbz_1$ & $\bbz_1$ & $\bbz_1$ & $\bbz_1$  \\ \hline
     17 & $S_6$ & $\bbz_1$ & $\bbz_1$ & $\anb{\bzt}$ & $\bbz_1$ & $\bbz_1$ &$\anb{\bzt}$  \\ \hline
     18 & $D_3$ & $\bbz_1$ & $\bbz_1$ & $\bbz_1$ & $\bbz_1$ & $\anb{\bzt}$ &$\anb{\bzt}$  \\ \hline
     19 & $C_{3v}$ & $\bbz_{16}$ & $\bbz_8$ &$\bbz_4$ & $\bbz_1$ & $\bbz_1$ & $\anb{\bzt}$  \\ \hline
     20 & $D_{3d}$ & $\bzt^2$ & $\anb{\bzt\times\bbz_4}$ & $\anb{\bzt^3}$ & $\bbz_1$ & $\anb{\bzt^2}$ & $\anb{\bzt^3}$  \\ \hline
     21 & $C_6$ & $\bbz_1$ & $\bbz_1$ &$\bbz_1$ & $\bbz_1$ & $\anb{\bzt}$ & $\anb{\bzt}$  \\ \hline
     22 & $C_{3h}$ & $\bbz_8$ & $\bbz_4$ & $\bzt$ & $\bbz_1$ & $\bbz_1$ & $\anb{\bzt}$  \\ \hline
     23 & $C_{6h}$ & $\bbz_8$ & $\bzt\times\bbz_4$ & $\bzt\times\anb{\bzt^2}$ & $\bbz_1$ & $\anb{\bzt^2}$ & $\anb{\bzt^3}$  \\ \hline
     24 & $D_6$ & $\bbz_1$ & $\anb{\bzt}$ & $\bzt^2$ & $\bzt^2$ & $\anb{\bzt^4}$ & $\anb{\bzt^3}$  \\ \hline
     25 & $C_{6v}$ & $\bzt^3$ & $\bzt\times\anb{\bzt^2}$ & $\bzt\times\anb{\bzt^2}$ & $\bbz_1$ & $\anb{\bzt^2}$ & $\anb{\bzt^3}$  \\ \hline
     26 & $D_{3h}$ & $\bzt^3$ & $\bzt\times\anb{\bzt^2}$ & $\bzt\times\anb{\bzt^2}$ & $\bbz_1$ & $\anb{\bzt^2}$ & $\anb{\bzt^3}$  \\ \hline
     27 & $D_{6h}$ & $\bzt^5$ & $\bzt\times\anb{\bzt^6}$ & $\anb{\bzt^6}$ & $\bzt^3$ & $\anb{\bzt^7}$ & $\anb{\bzt^6}$  \\ \hline
     28 & $T$ & $\bbz_1$ & $\anb{\bzt}$ & $\bbz_1$ & $\bbz_1$ & $\anb{\bzt^2}$ & $\anb{\bzt}$  \\ \hline
     29 & $T_h$ & $\bzt^3$ & $\bzt\times\anb{\bzt^2}$ & $\anb{\bzt^2}$ & $\bzt$ & $\anb{\bzt^3}$ & $\anb{\bzt^2}$  \\ \hline
     30 & $T_d$ & $\bzt^3$ & $\anb{\bzt\times\bbz_4}$ & $\anb{\bzt^2}$ & $\bbz_1$ & $\anb{\bzt^2}$ & $\anb{\bzt^2}$  \\ \hline
     31 & $O$ & $\bzt$ & $\anb{\bzt}$ & $\anb{\bzt}$ & $\bzt$ & $\anb{\bzt^3}$ & $\anb{\bzt^2}$  \\ \hline
     32 & $O_h$ & $\bzt^5$ & $\bzt^2\times\anb{\bzt^3}$ & $\anb{\bzt^4}$ & $\bzt^2$ & $\anb{\bzt^5}$ & $\anb{\bzt^4}$  \\ \hline
    \end{tabular}
        \caption{Comparison of classification data for the 32 point groups under disorder. Data in \anb{blue} indicates intrinsic ASPTs . Details for each case can be found in Sec. \ref{sec:3d_class_details}.}
        \label{table:3dresults_clean_vs_disordered_spinless}
        \end{center}
    \end{table*}

We organize the rest of our paper as follows. In Sec. \ref{sec:generalizedblockstate}, we begin by reviewing the concept of average SPTs for onsite symmetries. We then present three representative examples of average crystalline SPTs in two and three dimensions. These examples illustrate in detail how the cell decomposition, block decoration, obstruction-free conditions, and bubble equivalence relations operate in the average symmetry settings, providing a concrete guideline for our more systematic constructions later in the paper. In Sec. \ref{sec:element of construction}, we develop a systematic description of decorated states in dimensions up to three and explain how to determine the corresponding obstruction-free conditions and bubble equivalence relations with average symmetries. These tools form the basis for our classification of ACSPTs in both 2D and 3D. We also highlight the subtle distinctions among various scenarios -- including decohered spinless systems, decohered spin-1/2 systems, disordered spinless systems, and disordered spin-1/2 systems -- and introduce a mathematical classification scheme based on the spectral sequence method. This method not only facilitates the computation of the classification but also provides a means to compare our physical construction results with rigorous mathematical predictions. In Sec. \ref{sec:results}, we present the complete classification results for two- and three-dimensional systems across all point group symmetries. Finally, we conclude in Sec. \ref{sec:conclusion}. Additional details for each case of the classification are provided in the appendix.

\section{Generalized block state construction for ACSPT -- Examples}
\label{sec:generalizedblockstate}

In this section, we present our generalized block-state construction for the decohered case using a few examples. We begin with a brief review of ASPTs with onsite symmetries, particularly on the concept of intrinsic ASPT states. Following this, we transition to the crystalline case, outlining the procedure for generalized block-state construction and demonstrating the entire process using the 2D $pmm$, 2D $p2$, and 3D $C_{2v}$ groups as detailed examples. Each of these cases exhibits interesting features that are unique to average crystalline SPTs.

\subsection{Review of ASPTs with on-site symmetries}
\label{sec:review-aspt-onsite}
We first review the classification of ASPTs for onsite symmetries in open quantum systems with decoherence. The cases for disorders can be done in a similar fashion with some care and we will comment on that in the end. The general theory of ASPT classification with onsite symmetry is developed in Ref.~\cite{Ma2025ASPT2}. Here we only state the essential results and physical intuitions. Consider a $d$-dimensional open quantum system with an exact symmetry $G$ and an average symmetry $A$. An exact symmetry dictates that the density matrix describing the state $\rho$ remains invariant under the action of a symmetry operator $U(g)$ solely on one side, as expressed by $U(g)\rho=e^{i\theta}\rho$. An average symmetry requires $\rho$ to be not invariant under the one-side action of the symmetry but only remains invariant under the simultaneous actions of the symmetry operators on both sides, namely $U(a)\rho U(a)^\dag=\rho$. The average symmetry can be interpreted as a statistical symmetry of the quantum trajectories the system goes through. 

Consider a bosonic system, with the total symmetry group being a trivial extension of the two groups, namely $\mathcal{G}=A\times G$. The classification of decohered bosonic ASPTs in $d$ spatial dimensions is described by a modified Kunneth formula of cohomology groups\cite{Chen2014DDW,Ma2022ASPT1,Ma2025ASPT2,Ma:2024kma}:
\begin{equation}\label{eq:Kunneth_dec}
     \bigoplus_{p=1}^{d+1} \mathcal{H}^{d+1-p}\left(A,h^{p}\left(G\right)\right).
\end{equation}
Here, \( h^p(G) \) denotes the classification of invertible phases in \( p-1 \) spatial dimensions with symmetry \( G \)\footnote{For the decohered case, we need to exclude invertible states that require no symmetry protection, such as the \( E_8 \) state in 2D.}.
This classification is obtained by a generalized version of the decorated domain wall method for mixed states. For more details, we refer to Ref.~\cite{Ma2025ASPT2}. Physically, the term $\mathcal{H}^{d+1-p}(A, h^p(G))$ corresponds to ASPT states which can be constructed by decorating invertible states of $G$ symmetries on $p-1$- dimensional $A$-symmetry defects in space\cite{Chen2014DDW}. This classification is closely related to the case of pure-state classification of SPT states with $G\times A$ symmetry\cite{Chen2014DDW}, with a notable difference: the absence of the $p=0$ layer. The $p=0$ layer consists of pure-state SPTs that are solely protected by the $A$ symmetry, the symmetry that becomes average in the mixed-state cases. This means that these pure-state SPTs become trivial once decoherence breaks the $A$ symmetry down to average. The physical reason is that the only nontrivial feature of this class of SPT is the Berry phase associated with the $A$ symmetry defect configurations in the quantum wavefunction. When decohered, the system loses the quantum coherence of the wavefunction, rendering these Berry phases irrelevant. On the other hand, all other layers of the Künneth formula involve nontrivial decorations of SPT phases protected by exact symmetries, which remain robust even in the presence of decoherence. We refer to these ASPTs as \emph{extrinsic} SPTs, in contrast with the \emph{intrinsic} ASPTs that will be the focus of the next section.

For bosonic systems with a product group structure between the exact and average symmetry groups, all ASPTs can be understood as descendants of pure-state SPT phases -- there are no intrinsic ASPTs in this case. This is because all decorated domain wall configurations allowed by the Künneth formula are obstruction-free in the pure-state setting, meaning that no inconsistency arises when symmetry defects, carrying SPT decorations, are moved or fused within a quantum superposition. As a result, every such ASPT admits a pure-state counterpart.

In contrast, when the total symmetry group is a nontrivial extension of the exact and average symmetry groups, these obstruction-free conditions can fail, and certain pure-state SPT phases become forbidden. In such cases, intrinsic ASPTs for bosons can arise when part of the symmetry is reduced to an average symmetry. The general structure of these intrinsic ASPTs requires the tools of the spectral sequence formalism. For a precise characterization of when intrinsic ASPTs appear in bosonic systems, we refer the reader to Ref.~\cite{Ma2025ASPT2}.

Interestingly, in fermionic systems, obstructions can arise even when the total symmetry group is a direct product of fermion parity and a bosonic symmetry group~\cite{QingRuiWangPRX2020}. As a result, intrinsic fermionic ASPTs may exist even in this seemingly simpler group structure. The emergence and mechanisms of such intrinsic fermionic ASPTs will be the focus of the following discussion.

\subsubsection{Fermionic APSTs and intrinsic ASPTs}\label{subsec:intrinsic_aspts}

We now consider systems of fermions. Our discussion begins with the classification of fermionic SPT phases\cite{gu2014sptinteractingfermion,QingRuiWangPRX2020,kapustin2015fermionic} in the pure-state setting, followed by generalizations to average symmetry and mixed-state scenarios.

Consider systems with fermion parity symmetry \( G = \mathbb{Z}_2^f \), along with a bosonic onsite symmetry denoted by \( G_b \). Physically, this setup corresponds to systems of superconductors. The relationship between the bosonic symmetry \( G_b \) and fermion parity \( \mathbb{Z}_2^f \) plays a crucial role. If \( G_b \) and \( \mathbb{Z}_2^f \) form a trivial extension -- i.e., the total symmetry group is \( G_f = G_b \times \mathbb{Z}_2^f \) -- this is commonly referred to as the \emph{spinless} fermion case. In contrast, if fermion parity nontrivially centrally extends \( G_b \), the total symmetry group takes the form \( G_f = G_b \times_{\omega_2} \mathbb{Z}_2^f \), where the extension is specified by a nontrivial class \( \omega_2 \in \mathcal{H}^2(G_b, \mathbb{Z}_2^f) = \mathbb{Z}_2 \). This setting is referred to as the \emph{spin-1/2} case.

The decorated domain wall construction can be naturally extended to fermionic systems. In this framework, possible decoration patterns are again classified by the Künneth formula:
\begin{equation}
    \bigoplus_{p=0}^{d+1} \mathcal{H}^{d-p+1}\left(G_b, h^{p}(\mathbb{Z}_2^f)\right).
    \label{eq:kunneth2}
\end{equation}
The term with \( p = d+1 \) corresponds to phases protected solely by fermion parity, while the term with \( p = 0 \) captures SPT phases protected only by the bosonic symmetry \( G_b \). The remaining terms admit a natural interpretation as domain-wall decorations: \( p \)-dimensional \( G_b \)-defects are decorated with invertible fermionic phases protected by \( \mathbb{Z}_2^f \). Explicitly, for spacetime dimensions up to \( 2+1 \), the nontrivial invertible fermionic phases with \( \mathbb{Z}_2^f \) symmetry are:
\begin{enumerate}
    \item \( h^1(\mathbb{Z}_2^f) = \mathbb{Z}_2 \): decoration with odd fermion parity states on 0-dimensional (point) defects (often referred to as complex fermion decorations)
    \item \( h^2(\mathbb{Z}_2^f) = \mathbb{Z}_2 \): decoration with Majorana chains on 1-dimensional (line) defects,
    \item \( h^3(\mathbb{Z}_2^f) = \mathbb{Z} \): decoration with \( p \pm ip \) superconductors on 2-dimensional (planar) defects.
\end{enumerate}

Let us denote the terms in Eq.~\eqref{eq:kunneth2}, corresponding to the domain-wall decoration data, as \( (n_0, n_1, \ldots, n_d, \nu_{d+1}) \), referred to as the different \emph{layers}:
\begin{align}
    n_0 &\in h^{d+1}(\mathbb{Z}_2^f), \\
    n_p &\in \mathcal{H}^p\left(G_b, h^{d+1-p}(\mathbb{Z}_2^f) \right), \quad p = 1, \ldots, d, \\
    \nu_{d+1} &\in \mathcal{H}^{d+1}\left(G_b, U_T(1)\right),
\end{align}
where \( \nu_{d+1} \) represents the bosonic SPT classification for the symmetry \( G_b \).

We now turn to the obstruction functions relevant for fermionic SPTs in the pure-state setting. Not all configurations specified by the Künneth formula correspond to valid SPT phases; certain consistency conditions must be satisfied to ensure that the decorated domain-wall structure yields a globally consistent, short-range entangled quantum state\cite{wang2021domainwalldecorationsanomalies,kapustin2014symmetryprotectedtopologicalphases,gaiotto2019symmetry,xiong2018minimalist}. These consistency conditions, or \emph{obstruction functions}, can also be organized layer by layer and are symbolically written as \( (d n_1, \ldots, d n_d, d \nu_{d+1}) \). Each obstruction function is determined by the lower-layer data \( (n_0, \ldots, \nu_{d+1}) \), with explicit forms depending on the spatial dimension \( d \).

In low dimensions, the obstruction functions are explicitly known~\cite{qingrui2018towards,QingRuiWangPRX2020}. For instance, in \( d = 1 \), they take the form:
\begin{align}
    d n_1 &= \omega_2 \cup n_0, \label{eq:obs_1d_n1} \\
    d \nu_2 &= (-1)^{\omega_2 \cup n_1}. \label{eq:obs_1d_nu2}
\end{align}
A decoration pattern is considered obstruction-free if all obstruction functions vanish, i.e., \( d n_1 = 0 \) and \( d \nu_2 = 1 \). A complete list of the obstruction functions up to \( 3+1 \)-dimensions can be found in Ref.~\cite{QingRuiWangPRX2020}. A notable feature is that, even in the spinless fermion case (i.e., when the symmetry group is a trivial extension), nontrivial obstructions can still arise -- unlike in bosonic systems, where such obstructions are absent for product group structures. All decoration patterns that pass the obstruction-free conditions represent candidate SPT phases. However, these patterns do not necessarily correspond to distinct physical phases. An additional step -- identifying and quotienting by coboundary equivalence -- is required to obtain the final classification of SPTs. For now, we do not elaborate on the details of this step, as the obstruction analysis alone provides sufficient information for addressing the average symmetry case. The crystalline analog of coboundary equivalence will be discussed in detail in later sections.

We now turn to the case of decohered fermionic ASPTs where the bosonic symmetry $G_b$ becomes average. To accommodate the physical setting of decohered systems, two key modifications must be made to the algebraic structure outlined above. First, in the decoration data, the \( p = 0 \) layer must be removed. This reflects the fact that any SPTs protected solely by the bosonic symmetry \( G_b \) become trivial once \( G_b \) is treated as an average symmetry. Such states rely on a \( G_b \) anomaly that is no longer meaningful under decoherence. Second, the final layer of the obstruction functions -- for example, Eq.~\ref{eq:obs_1d_nu2}, which captures inconsistencies in Berry phases arising from moving and fusing symmetry defects -- is no longer applicable. In decohered systems, Berry phases are not well-defined, and such obstructions are lifted. The second modification allows for the emergence of intrinsic ASPTs: decoration patterns that were obstructed in the pure-state setting may now become admissible in the mixed-state context.

To see this in action, consider a 1D example of $G_b = \bzt$. The nontrivial decorations are 1) 1D Majorana chain decoration, given by $n_0 \in h^2(\mathbb{Z}_2^f) = \mathbb{Z}_2$; and 2) complex fermion decoration on 0D $\bzt$-domain walls, given by $n_1 \in \mathcal{H}^2(\bzt,\bzt^f) = \bzt$. As $n_0,n_1,\omega_2 \in \bzt$, we can describe them in terms of binary representation, where trivial/nontrivial elements are expressed by $0/1$ and the cup product is implemented as binary multiplication. 
In the spinless case, namely $G_f=\bzt\times\bzt^f$, according to Eq. \ref{eq:obs_1d_n1} and \ref{eq:obs_1d_nu2}, there is no obstruction as $\omega_2 = 0 \pmod 2$. Therefore, in the spinless case, there are two nontrivial fermionic SPT phases corresponding to these distinct decoration patterns, both of which remain stable in the mixed-state setting. 

For the spin-1/2 case, where \( G_f = \mathbb{Z}_4^f \) and the extension class is \( \omega_2 = 1 \), the obstruction functions impose strict constraints in the pure-state setting. Specifically, the condition $dn_1 = 1 \cdot n_0 = 0$ forbids the Majorana chain decoration, and the condition $d\nu_2 = (-1)^{1 \cdot n_1} = 1$
forbids any nontrivial \( n_1 \) decoration. As a result, no nontrivial fermionic SPT phases exist in the pure-state setting for this symmetry class. However, when the \( \mathbb{Z}_2 \) symmetry is treated as an average symmetry due to decoherence, the \( d\nu_2 \) obstruction is no longer required, and the restriction on \( n_1 \) is lifted. This opens the door to an intrinsic \( \mathbb{Z}_4^f \) ASPT phase. The phase is physically characterized by the decoration of odd fermion parity states on domain walls of the average \( \mathbb{Z}_2 \) symmetry. An explicit model realizing this phase is provided in Ref.~\cite{Ma2025ASPT2}.

In the case of crystalline symmetries, the decorated domain wall picture of the on-site symmetry is generalized to the so-called block-state decoration for the crystalline SPTs, which we will detail next. We further extend the block-state construction method to include cases of average crystalline symmetry, enabling the classification of ACSPTs, including intrinsic ACSPTs. 

\subsection{Average Crystalline SPTs: Assumptions}

Now let us consider crystalline topological phases. To allow for general local interactions, a band-theoretic description is no longer adequate. Instead, the appropriate framework for classifying crystalline topological phases is a real-space approach -- namely, the block-state construction. This method has been successfully employed to classify crystalline topological phases, as demonstrated in many previous works~\cite{else2019defectnetworks,song2019topological,song2017topological,huang2017building,Song2020realspace,zhang2020point2d,zhang2022crystalspt2d,zhang2022crystalspt3d}. The block-state construction involves four key steps: cell decomposition, block decoration, obstruction-free conditions, and trivialization-free conditions. In classifying ACSPT phases, the same general structure follows, though important generalizations are required to accommodate the average symmetry setting. In the following section, we illustrate the classification procedure and compare the pure-state and mixed-state schemes through concrete examples based on open systems with decoherence. While specific details may vary across different physical contexts, the framework presented here outlines a general strategy for constructing all ACSPT phases. Variations and additional subtleties for disordered systems will be discussed in detail in Sec.~\ref{sec:element of construction}.

Our approach rests on the assumption that even in the presence of disorder or decoherence -- such as chemical potential fluctuations that locally break lattice symmetries, or interactions with phonons or photons that induce decoherence -- the underlying lattice structure remains intact. With that, we perform a cell decomposition where the size of each cell is much larger than the characteristic length scale of disorder or decoherence, which is typically on the order of the lattice spacing. This separation of scales ensures that microscopic fluctuations are effectively averaged out within each cell, allowing us to endow each cell with well-defined average symmetry properties. As a result, even though locally crystalline symmetries may be broken, the system’s topological features can still be faithfully captured by decorating these coarse-grained blocks with average symmetry-protected topological states. Furthermore, when the average symmetry relates different parts of a block -- or different blocks -- the decoration must respect this symmetry action, as otherwise the pattern would explicitly break the crystalline symmetry on macroscopic scales, contradicting the assumption of an average symmetry.

Under these assumptions, a generalized version of the block-state construction remains applicable in the presence of disorder or decoherence. As long as the blocks are sufficiently large compared to the characteristic length of the symmetry-breaking disorder or decoherence, this construction is expected to capture the universal topological features of the phase. 

\subsection{2D ACSPT with Average $pmm$ Group}

First, we consider a 2D fermionic crystalline system with $pmm$ space group, with the decoherence that renders the spatial symmetries average, leaving only the fermion parity exact. 

\subsubsection{Cell Decomposition}
The first step of this method is to perform a cell decomposition, whereby regions of the unit cell are partitioned according to the action of the crystalline symmetry \(G\) (with the total fermionic group being \(G^f = G \times_{\omega^f_2} \mathbb{Z}_2^f\)). For each space group, there is a mathematically well-defined procedure to decompose a \(d\)-dimensional Euclidean space into a union of \(p\)-dimensional cells, where \(p = 0, 1, 2, \dots, d\)\cite{huang2017building}.

\begin{figure}[!htbp]
    \centering
    \includegraphics[width=0.8\linewidth]{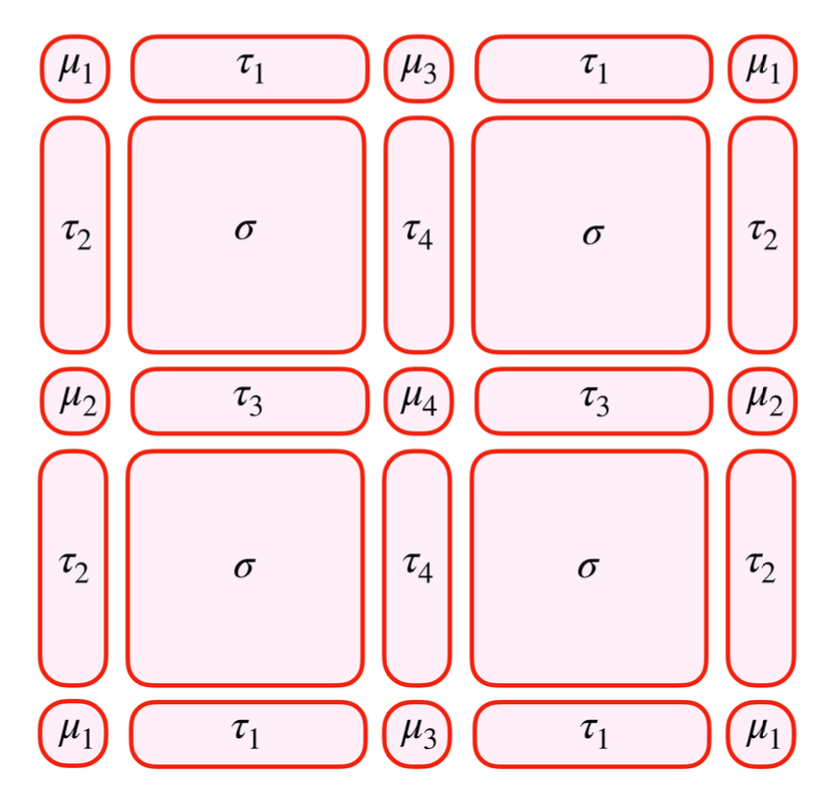}
    \caption{$pmm$ cell decomposition}\label{fig:pmm}
\end{figure}

Consider the \textit{pmm} wallpaper group where the crystalline symmetry is generated by two perpendicular mirror lines and lattice translation. The cell decomposition is illustrated in Fig. \ref{fig:pmm}. Essentially, the cell decomposition divides the unit cell into subregions based on their distinct symmetry actions. In the \(pmm\) case, the decomposition is as follows: 
\begin{itemize}
    \item \textbf{2D Blocks:} These are regions that are connected by reflection yet are not invariant under any symmetry. These 2D blocks are labeled by \(\sigma\).
    \item \textbf{1D Blocks:} Excluding the 2D blocks, what remains are 1D blocks that lie along the mirror lines. The reflection symmetry acts on these 1D blocks as an onsite \(\mathbb{Z}_2\) symmetry. Some 1D blocks are symmetry-equivalent, and there are a total of four independent (i.e., not connected by symmetry) 1D blocks, which are labeled as \(\tau_i\) for \(i = 1, 2, 3, 4\).
    \item  \textbf{0D Blocks:} In addition to the 2D and 1D blocks, there is a 0D block at the intersection points of the mirror lines. Here, the induced onsite symmetry is \(\mathbb{Z}_2 \times \mathbb{Z}_2 = \mathbb{Z}_2^2\). Due to symmetry equivalences, there are four distinct sets of 0D blocks, labeled \(\mu_i\) for \(i = 1, 2, 3, 4\).
\end{itemize}
We will use the notation \( \mu, \tau, \sigma, \lambda \) to denote 0D, 1D, 2D, and 3D cells, respectively, throughout the paper. The decomposition is conceptually straightforward. The cell decomposition for general point groups is well established in the literature\cite{song2019topological,hatcher2002algebraic}, and we will make use of these results without re-deriving them in this work.

\subsubsection{Generalized Block State Decoration}
Once we have the cell decomposition, pure-state crystalline SPT phases can be constructed systematically by decorating each lower‑dimensional cell with SPT states protected purely by on‑site symmetries. The same decorating approach carries over to average crystalline cases: here, each lower‑dimensional cell is decorated with an ASPT state protected by an average on‑site bosonic symmetry together with exact fermion parity, which we call generalized block state decoration.

We first show that the only nontrivial ASPT decoration on the 0-dimensional \(\mu\) blocks is an odd number of (complex) fermions. 0D SPT decorations are intuitive as they are classified by the linear representation of the individual symmetry groups. This is codified by the Kunneth formula for onsite SPT with d=0 and onsite symmetry $G\times\bzt^f$:
\begin{align}
    n_0 &\in h^1(\bzt^f) = \bzt \\
    \nu_1 &\in \mathcal{H}^1(G, U_T(1))
\end{align}
The first layer describes odd fermion parity decoration on $\mu$ blocks. Since any number of odd fermions are equivalent, we will often refer to this state as the complex fermion decoration. The second layer is the bosonic 0D $G$-SPT. As this is a bosonic SPT, or equivalently not protected by fermion parity, this state trivializes in our decoherence model. This justifies our initial claim. 

It is worthwhile to note here that as 0D blocks have no boundary, the SPT decoration for any setting (pure state or mixed state) is naturally obstruction-free. However, as they are surrounded by 1D and 2D blocks, they may be trivialized by bubble equivalence on these higher-dimensional blocks, as we will see in later examples. 

Consider a block \(\tau\) that is adjacent to lower-dimensional blocks \(\{\mu_{\tau} \in \partial \tau\}\) and higher-dimensional blocks \(\{\sigma_\tau \in \partial \tau\}\). When the \(\tau\) block is regarded as an isolated system, the corresponding subgroup \(G^f_\tau \subset G^f\) acts effectively onsite. Therefore, we can decorate \(\tau\) with \(G^f_\tau\)-SPTs, which we refer to as block states. (The superscript \(f\) indicates that the spatial symmetry is extended by fermion parity.) In the average cases, we allow the decoration of lower-dimensional ASPT states on the blocks. Therefore, the decoration pattern can be richer than in the clean case, as intrinsic ASPTs may be decorated on the lower-dimensional blocks. This provides one way for the emergence of intrinsic crystalline ASPTs. This situation does not occur in the $pmm$ example we consider here, but we will discuss examples with such decorations later for spin-1/2 systems in Sec. \ref{subsec:elems_dec_spinful_aspt}. 

However, not all decorations yield a consistent SPT. These decorations may leave behind edge modes on the lower-dimensional blocks \(\mu_\tau\) that reside in the bulk of the crystalline system. To achieve a gapped bulk -- a necessary condition for an SPT -- these edge modes must be gapped out in a manner that respects the symmetry of \(\mu_\tau\), which generally is a higher symmetry \(G^f_\mu\) compared to that of \(\tau\). This physical requirement is known as the \textit{obstruction-free condition}. We will discuss the generalization of the obstruction-free condition for the average cases in the next section.

\subsubsection{Generalized Obstruction-Free Condition}\label{subsubsec:pmm_obs}

Since the notion of an energy gap does not directly apply in the presence of decoherence or disorder, an alternative perspective on the obstruction‐free condition is required for the average cases. In the pure-state or clean limit of crystalline SPT phases, the obstruction-free condition for a given decoration is equivalent to requiring that the edge states of the decorated SPTs that meet at an intersection be anomaly-free with respect to the intersection’s onsite symmetry; a symmetric gap is achievable only in the absence of such an anomaly. Equivalently, the bulk corresponding to these edge degrees of freedom (to be defined precisely shortly) must form a trivial SPT. It is this concept of anomaly-free edges and the corresponding trivial bulk SPT that can be generalized to mixed-state settings.

\begin{figure}[!htbp]
    \centering
    \includegraphics[width=\linewidth]{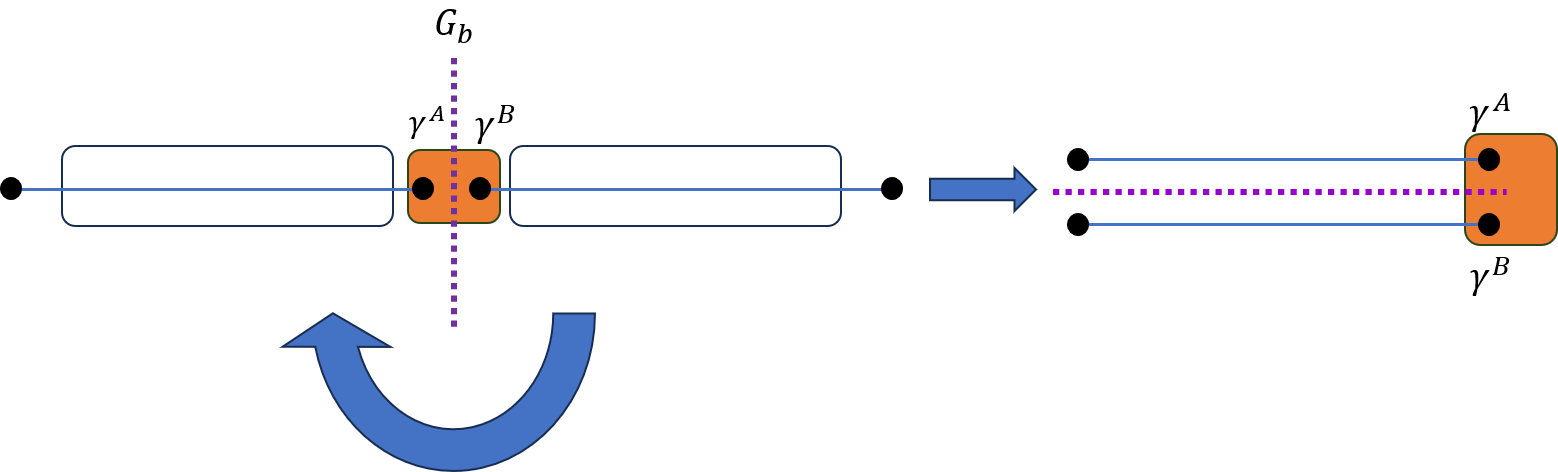}
    \caption{Demonstration of folded SPT interpretation of obstruction condition. Majorana chain decoration on the 1D blocks leads to Majorana modes on the 0D block with reflection symmetry $G_b$. The resultant state after folding is a doubled Majorana chain protected by onsite $G_b$ symmetry.}
    \label{fig:folded_Majorana}
\end{figure}

Let us first consider the clean case and see how the anomaly-free check works. Each block state decoration contributes an anomalous edge mode due to the properties of the decorated SPT. At the intersection of the blocks, the anomalies arising from these edge modes must cancel in order to achieve a symmetric gapped state. A useful method for checking this cancellation is the folding trick. One can imagine that all blocks surrounding a given intersection \(\mu\) are “folded” together, as illustrated in Fig. \ref{fig:folded_Majorana}. The resulting compound system then exhibits an onsite symmetry \(G^f_\mu\) and is referred to as a folded \(G^f_\mu\)-SPT. If this folded SPT is nontrivial, this implies the edge at \(\mu\) carries a nontrivial anomaly, meaning that the edge cannot be in a gapped symmetric state. Thus, for an obstruction-free state, the folded SPT must be trivial. The straightforward generalization of this obstruction-free condition to the average cases is the analogous requirement that the folded system must be a trivial ASPT.

This framework provides another path to construct or identify intrinsic ASPTs. In the clean case, suppose a given decoration is obstructed -- that is, it leads to a nontrivial folded SPT. In the averaged case, where spatial symmetry is completely averaged, any SPT that is purely protected by spatial symmetries is trivialized. Therefore, if the folded SPT in the clean limit is a pure bosonic \(G_\mu\)-SPT (i.e., it does not carry a mixed anomaly between \(G_\mu\) and \(\mathbb{Z}_2^f\)), then the corresponding block state decoration becomes obstruction-free once decoherence or disorder is introduced. This scenario gives rise to an intrinsic ASPT phase, as the obstruction-free condition in the averaged case is not as stringent as the clean case.

We will illustrate this phenomenon in the decohered systems of spinless fermions with \(pmm\) symmetry. First, consider the possible SPT decorations on the 1D blocks. Each 1D (\(\tau\)) block carries an onsite \(\mathbb{Z}_2\) symmetry, so we can consider decorating the block with the 1D fermionic SPT with onsite \(\mathbb{Z}_2\) symmetry. In the spinless case, this 1D fSPT is equivalent to two copies of Kitaev chains, where the \(\mathbb{Z}_2\) symmetry exchanges the two copies. At the edge, two Majorana modes \(\gamma^A\) and \(\gamma^B\) appear. In the absence of the onsite \(\mathbb{Z}_2\) symmetry, one might gap these edge modes using the fermion mass term
\[
m^{AB} = i\gamma^A\gamma^B.
\]
However, this term violates the onsite \(\mathbb{Z}_2\) symmetry -- implemented as \(G_b: \gamma^A \leftrightarrow \gamma^B\) -- since
\[
G_b: i\gamma^A\gamma^B \rightarrow i\gamma^B\gamma^A = -i\gamma^A\gamma^B.
\]
In fact, the mass term \(m^{AB}\) represents the fermionic parity at the edge, and we have shown that \([G_b, m^{AB}] \neq 0\). This is a signature that the edge carries a mixed anomaly between \(\mathbb{Z}_2^f\) and \(G_b\), precluding a symmetric gapping under decoherence. Moreover, two copies of this state -- with edge modes \(\gamma^{A,B}_{1,2}\) (shown in Fig. \ref{fig:folded_Majorana}) -- can be trivialized by the mass term
\[
m^{AB}_{12} = i\gamma^A_1\gamma^A_2 + i\gamma^B_1\gamma^B_2,
\]
which implies that the SPT on this 1D block is \(\mathbb{Z}_2\)-classified (consistent with \(\mathcal{H}^1(\mathbb{Z}_2, \mathbb{Z}_2^f) = \mathbb{Z}_2\)). It turns out that this 1D fermionic SPT survives under decoherence that breaks the $G_b$ symmetry down to average. So it is a valid block decoration in the average case as well. There is no other 1-D ASPT states that can be decorated on the 1-D blocks. 

\begin{figure}[!htbp]
    \centering
    \includegraphics[width=\linewidth]{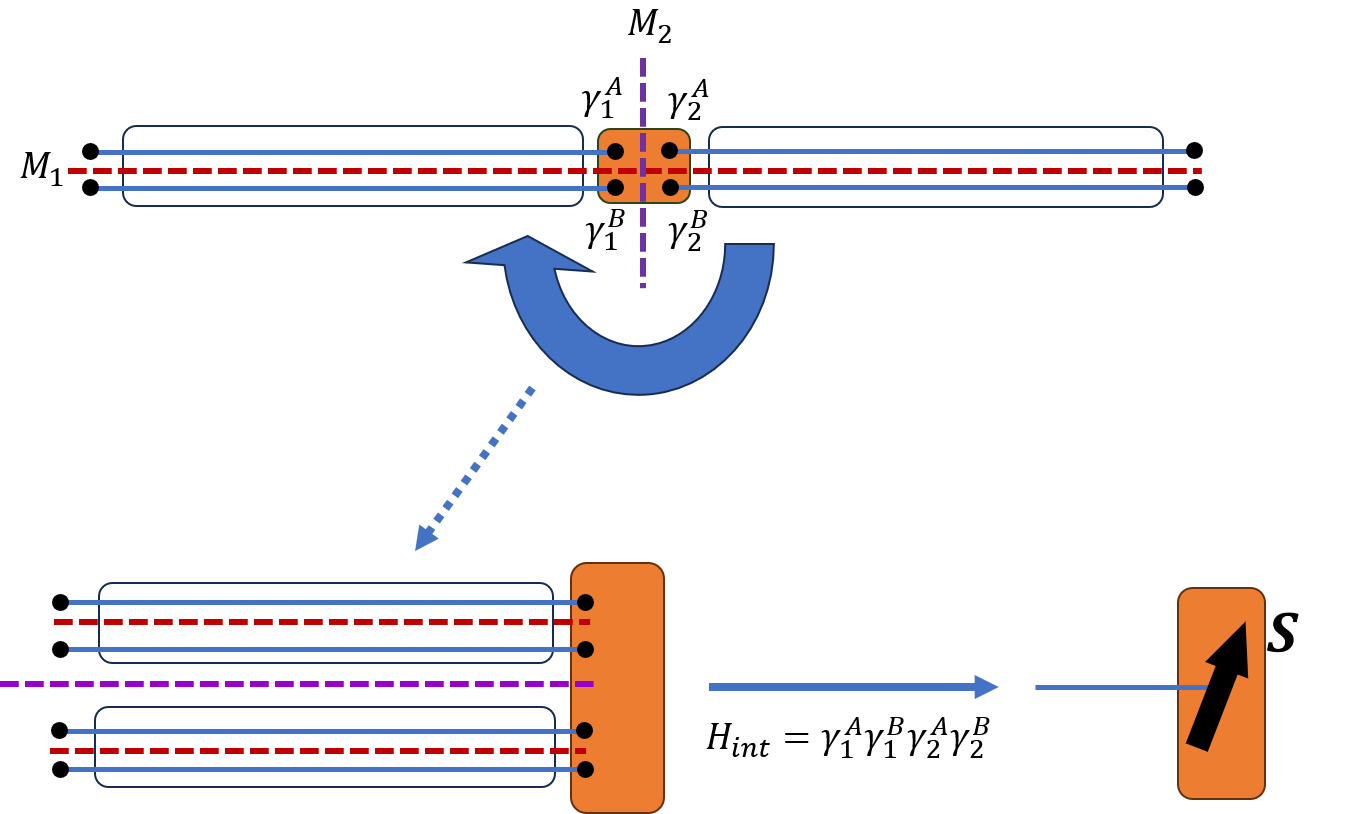}
    \caption{Decoration of $\bzt$ fSPT leads to spin-1/2 degree of freedom at the 0D block.}
    \label{fig:folded_z2_fspt}
\end{figure}

Now, consider decorating one set of 1D blocks, say \(\tau_1\), with this 1D \(\mathbb{Z}_2\)-fSPT. Since the 0D block \(\mu_1\) is shared by two \(\tau\) blocks, this decoration pattern results in four Majorana edge modes, labeled \(\gamma^{A,B}_{1,2}\) (where the subscripts label the two copies while the superscript distinguishes the internal modes of the $\bzt$ fSPT), at the intersection point located on \(\mu_1\). There are two onsite \(\mathbb{Z}_2\) symmetries on the 0D block, generated by the two mirror reflections denoted \(M_1\) and \(M_2\) (see Fig. \ref{fig:folded_z2_fspt}). The two symmetries act as follows: \(M_1: \gamma^A \leftrightarrow \gamma^B\) is the same onsite symmetry present in \(\tau_1\), while the additional symmetry at $\mu_1$ exchanges the $\bzt$ fSPT decorations \(M_2: \gamma^1 \leftrightarrow \gamma^2\). The mass term \(m^{AB}_{12}\) violates the \(M_2\) symmetry and therefore cannot gap out the system. The only viable interaction term is the fermion parity at the edge,
\[
P_{f,\mu_1} = -\gamma^A_1\gamma^B_1\gamma^A_2\gamma^B_2,
\]
which commutes with both \(M_1\) and \(M_2\). The ground state subspace of this interaction is an even fermion parity sector of these Majorana modes; however, a twofold degeneracy remains. As a result, the system is obstructed when the symmetries are exact. The twofold degenerate states form a projective representation under the two \(\mathbb{Z}_2\) symmetries arising from spatial reflections, which can be viewed effectively as a spin-\(\tfrac{1}{2}\) degree of freedom. Equivalently, the folded 1-D system (see Fig. \ref{fig:folded_z2_fspt}) is a nontrivial SPT under the two onsite \(\mathbb{Z}_2\) symmetries.

When the spatial symmetries are averaged, however, this decoration becomes obstruction-free. We provide arguments to support this. First, note that the edge modes are protected solely by bosonic symmetries, since the fermion parity at the edge commutes with both. When these bosonic symmetries are averaged due to disorder or decoherence, the degeneracy is no longer protected and the anomaly is trivialized. This reasoning can be demonstrated more explicitly by the double space formalism. In the decohered case, one can analyze the system in the Choi–Jamiołkowski double space\cite{Ma:2024kma}. After projection into the even fermion parity sector, both the ket and bra spaces host a spin-\(\tfrac{1}{2}\) degree of freedom (denoted as $\mathbf{S_L}$ and $\mathbf{S_R}$ respectively). The two spin-1/2's in the double state can be gapped out by a Heisenberg interaction \(\mathbf{S_L}\cdot\mathbf{S_R}\), resulting in a spin singlet. Importantly, this interaction is invariant under the exchange of the bra and ket spaces and the averaged symmetries, which act identically on both spin-\(\tfrac{1}{2}\) degrees of freedom. A gapped edge in the doubled state implies that the folded ASPT is trivial, rendering the decoration obstruction-free under decoherence. A similar analysis applies to the \(\mu_3\) block. Thus, although the decoration is obstructed when the symmetries are exact, it becomes obstruction-free under decoherence, leading to an intrinsic ASPT decoration.

\begin{figure}[!htbp]
    \centering
    \includegraphics[width=\linewidth]{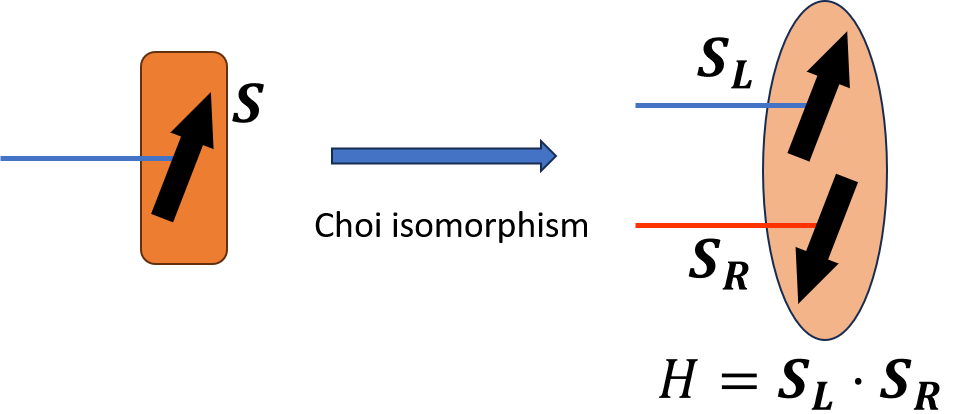}
    \caption{In the doubled state, $\bzt$ fSPT decoration has two spin-1/2 modes which can be symmetrically gapped out into a singlet.}
    \label{fig:spin_doubled_state}
\end{figure}

It is worth noting that some decorations remain obstructed even under decoherence. For instance, consider the alternative decoration on \(\tau\) blocks using a Kitaev chain. Decorating \(\tau_1\) with a Kitaev chain leaves two Majorana modes on \(\mu_1\), and the corresponding fermion parity
\[
P_{f,\mu_1} = i\gamma_1\gamma_2
\]
fails to be symmetric under \(M_2: \gamma_1 \leftrightarrow \gamma_2\). This implies that the folded SPT in this case is the 1D \(\mathbb{Z}_2\)-fSPT protected by \(M_2\); since the \(\mathbb{Z}_2\) anomaly is not bosonic, it remains obstructed under decoherence. A similar verification shows that any stacking of Kitaev chain decorations on the \(\tau\) blocks is likewise obstructed. Therefore, the 1D obstruction-free block states of the \(pmm\) group arise solely from the intrinsic ASPT decorations on the \(\tau\) blocks, yielding a naive classification of \(\mathbb{Z}_2^4\) (given the four inequivalent \(\tau\) blocks).

\subsubsection{Trivialization Condition}\label{subsec:pmm_triv}

In fact, not all obstruction-free decorations lead to nontrivial ASPTs. Since the \(\tau\) blocks lie in the bulk, one must also check whether the decorated block states on \(\tau\) might be trivialized by the so-called bubble equivalence from the higher-dimensional blocks \(\sigma\). Suppose the block \(\sigma\) is \(d_\sigma\)-dimensional; one then considers a bubble of $d_\sigma-1$-dimensional $G_\sigma$-SPT that expands from a point within \(\sigma\) to its boundary, resulting in a \(G_\sigma\)-SPT attached to the boundary of $\sigma$ block. Such a process cannot change the topological classification of the state, because the bubble can be adiabatically shrunk to vacuum. 

\begin{figure}[!htbp]
    \centering
    \includegraphics[width=\linewidth]{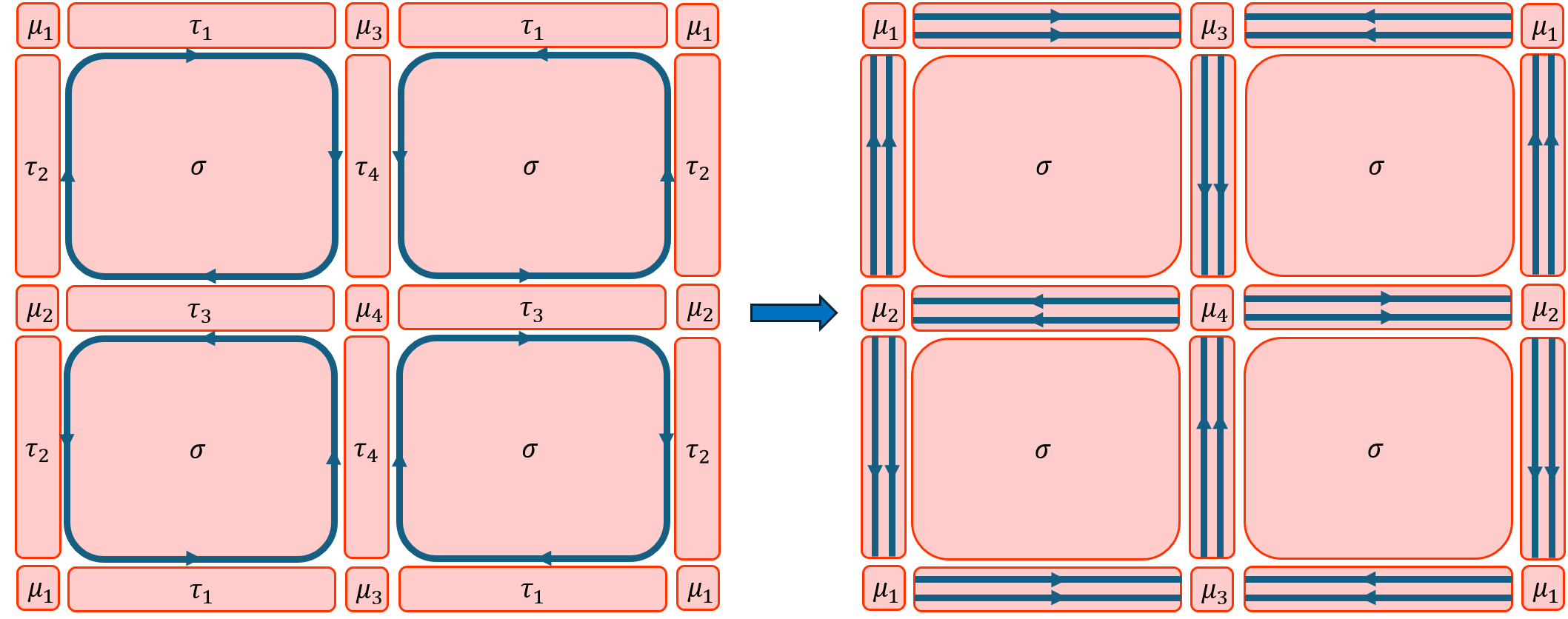}
    \caption{Majorana bubble on $\sigma$ blocks transforms to \(\mathbb{Z}_2\)-fSPT decoration on all 1D blocks}\label{fig:2d_bubble_pmm}
\end{figure}

To illustrate this in the \(pmm\) example, note that the \(\sigma\) blocks only carry a \(\mathbb{Z}_2^f\) symmetry. In this region, one may introduce a Majorana bubble -- i.e. a Majorana chain with anti-periodic boundary conditions (to ensure even fermion parity when adiabatically deformed to a point). To respect the spatial symmetry, every \(\sigma\) block must host the same bubble. As shown in Fig. \ref{fig:2d_bubble_pmm}, enlarging this Majorana bubble to the boundary of the \(\sigma\) regions produces two Majorana chains on each 1D block, with the mirror symmetries interchanging the two chains. This precisely corresponds to a decoration by a 1D \(\mathbb{Z}_2\)-fSPT on all \(\tau\) blocks. Therefore, although this decoration pattern is obstruction-free, it represents a trivial phase. Equivalently, one can begin with a 1D \(\mathbb{Z}_2\)-fSPT decoration on three of the blocks (say, \(\tau_{1,2,3}\)), and apply the Majorana bubble equivalence; this process induces two copies of the 1D \(\mathbb{Z}_2\)-fSPT on \(\tau_{1,2,3,4}\), thereby trivializing the first three blocks and leaving only a nontrivial decoration on \(\tau_4\). Consequently, the initial classification data for 1D block decorations, which was \(\mathbb{Z}_2^4\), is now reduced to \(\mathbb{Z}_2^3\).

In the above, we saw an example where a \( d \)-dimensional bubble trivializes a decoration in \( (d{-}1) \)-dimensions. This is a common scenario we will encounter. However, bubble equivalence can also potentially trivialize decorations in even lower dimensions. For instance, the same bubble states considered earlier on the boundary of \( \sigma \) blocks may, in principle, trivialize a zero-dimensional state decorated at a corner $\mu$. Although this particular trivialization does not occur in the \( pmm \) case -- for reasons we will elaborate on below -- in Sec.~\ref{subsec:p2_triv}, we describe the mechanism by which a Majorana bubble on a 2D block can trivialize 0D complex fermion decorations in the \( p2 \) case.

\begin{figure}[!htbp]
    \centering
    \includegraphics[width=0.6\linewidth]{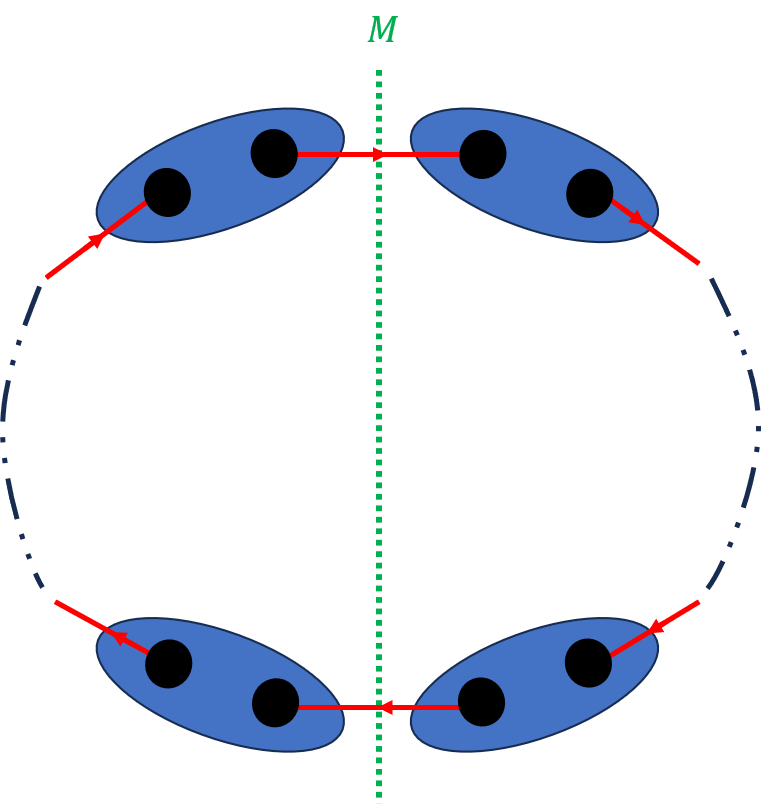}
    \caption{A Majorana bubble construction to trivialize odd fermion parity with a reflection line $M$. Since the directional links cross through $M$, the bubble state is not symmetric and hence the trivialization is incompatible with reflection.}
    \label{fig:bubble_reflection}
\end{figure}

The 0D decoration in the \( pmm \) lattice cannot be trivialized by the 2D Majorana bubble decoration due to the presence of reflection symmetry. The general reasoning is as follows: consider a 0D block \( \mu \) surrounded by a set of 2D \( \sigma \) blocks with Majorana bubbles. To trivialize the complex fermion parity decoration on the 0D block, one would need to adiabatically and symmetrically deform the Majorana bubbles into a Majorana loop with odd fermion parity encircling \( \mu \), such that shrinking this loop to the point \( \mu \) would flip its fermion parity. This process requires the Majorana loop to have periodic boundary conditions (PBC)~\cite{zhang2022crystalspt2d}, rather than anti-periodic ones. However, in the presence of reflection symmetry, at least one of the links forming the loop must cross a reflection axis (as illustrated in Fig.~\ref{fig:bubble_reflection}). The reflection operation flips the orientation of any link crossing it, thereby rendering the PBC Majorana chain asymmetric. 
As a result, it is not possible to symmetrically deform the Majorana bubbles into a PBC Majorana loop around \( \mu \), and the trivialization fails. In contrast, in cases where no reflection line interferes with the construction of a symmetric Majorana PBC loop, such trivialization of the 0D state is indeed possible, as discussed in Sec.~\ref{subsec:p2_triv}.

Just as with the two-dimensional blocks, we must also consider possible bubble equivalence on the one-dimensional blocks. The only possible 1D bubble is a fermionic insulator. Consider a set of 1D blocks, say $\tau_1$ in a trivial insulating state. On each $\tau_1$ block, one can then insert a pair of fermions and the total fermion parity is still preserved. The pair of fermions is assumed to form a trivial representation w.r.t. any onsite bosonic symmetry on the blocks. Now one can push the fermions to either end of the block, that is, we enlarge this bubble state to the boundary. The bulk is once again trivial but each boundary has one additional fermion parity odd state. The spatial symmetry, albeit weak, demands that all $\tau_1$ blocks are in the same bubble state. Now one must check that this bubble equivalence procedure does not trivialize odd fermion parity decoration on the 0D $\mu$ blocks.

In the $pmm$ case, this does not trivialize any 0D decorations, as every $\mu$ block is surrounded by an even number of each set of $\tau$ blocks. However, this becomes relevant in cases with odd rotation/dihedral symmetry, where a 0D block $\mu$ would be surrounded by an odd number of a set of 1D blocks each contributing one fermion, thus trivializing odd fermion parity decoration on $\mu$.  

The obstruction-free, trivialization-free block state decorations constitute the nontrivial ASPT states. The final step in obtaining the complete classification data is to determine whether nontrivial relations exist between different block state decorations, also known as the stacking relations of these phases. For example, decorating a 1D block with two copies of a Majorana chain (which is \(\mathbb{Z}_2\)-classified) might seem trivial, but the resulting edge modes can lead to a nontrivial decoration on 0D states. We do not address this here as all the decorations of $pmm$ stack trivially. Examples of nontrivial stacking relations will be demonstrated later in (Sec.~\ref{subsec:elems_dec_spinful_stack}).

\subsection{2D ACSPT with Average $p2$ Group}\label{subsec:eg_2}
Let us now look at the $p2$ lattice which is a parallelogrammatic unit cell with $C_2$ (2-fold) point group symmetry. While we find no examples of intrinsic ASPTs in this case, the motivation of elaborating on this construction lies in the simplicity of the point group symmetry as well as seeing the more subtle aspects of trivialization by bubble decoration. 

\subsubsection{Cell Decomposition}

\begin{figure}[!htbp]
    \centering
    \includegraphics[width=0.8\linewidth]{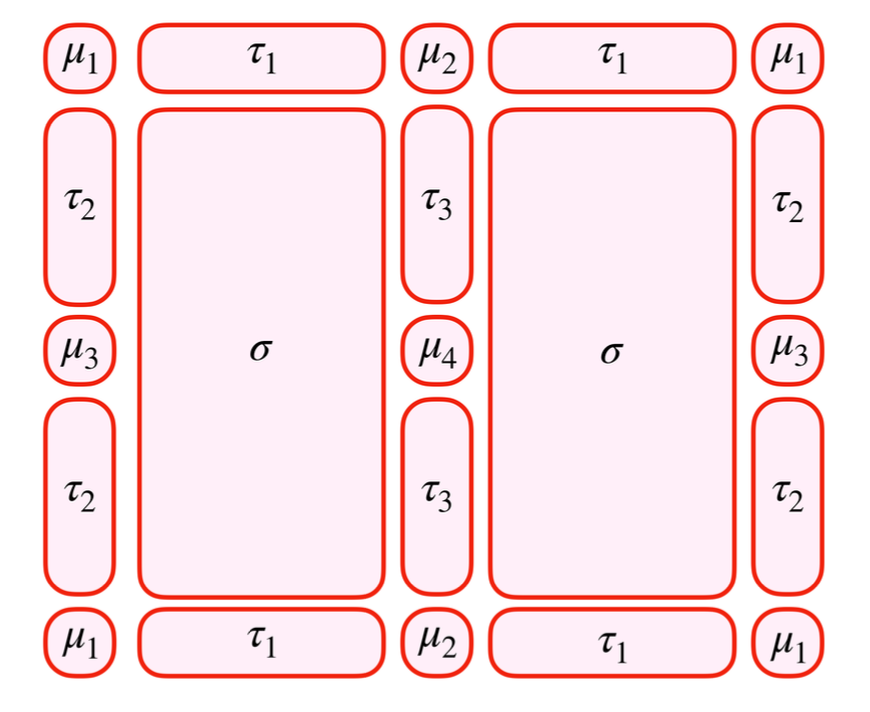}
    \caption{The $p2$ lattice}\label{fig:p2}
\end{figure}

The complete cell decomposition is shown in Fig.~\ref{fig:p2}, which is derived as follows.
We start with a parallelogrammatic lattice with a 0D block placed at the center invariant to 2-fold rotation ($\mu_4$ in the figure). By lattice translation, the opposite edges are identified with each other, as are the corners. However, the 2-fold rotation also identifies opposite corners ($\mu_1$) and edge centers ($\mu_2$ and $\mu_3$). Hence the point group also acts invariantly on these 0D blocks. The gaps on the edge are subsequently filled with 1D blocks ($\tau_1$ and $\tau_2$) while noting that these are not invariant upon action of the symmetry.
The interior of the unit cell is then split into 2 2D blocks $\sigma$ mapped to each other by rotation. Finally the remaining space is filled with a (non-invariant) set of 1D blocks ($\tau_3$), connecting the unit center cell to one set of edge centers (the choice between $\mu_2$ and $\mu_3$ is arbitrary here).

\subsubsection{Block state decoration and obstruction-free condition}\label{subsubsec:p2_deco_and_obs}

Since the 1D blocks have no symmetry, the only nontrivial decoration is a Majorana chain. However, similar to the $pmm$ case, we will find that these decorations (and their combinations) are obstructed even in the decohered case. First, we consider a Majorana chain decoration on $\tau_3$. This leaves two Majorana modes on $\mu_4$, which we denote as $\gamma_A$ and $\gamma_B$ respectively. The rotation symmetry exchanges these two Majorana modes, and the only possible mass term $m_{AB} = i\gamma_A\gamma_B$ is not symmetric under this condition ($m_{AB}\rightarrow -m_{AB}$). However, note that this mass term is precisely the fermion parity operator at the 0D center. The fermion parity and rotation being anticommutative indicates that the decoration leads to a mixed anomaly of the two symmetries at $\mu_4$. When the crystal is subject to decoherence which averages only spatial symmetry, the mixed anomaly with the exact fermion parity symmetry cannot be trivialized. Hence this decoration remains obstructed.

To further test this point, we can consider the doubled state, where each Majorana mode has a conjugate pair $\tilde{\gamma}_{A,B}$ with the addition of the conjugate symmetry $\mathcal{C}:\gamma \leftrightarrow \tilde{\gamma}$. This leads to one symmetric interaction term $\gamma_{A}\tilde{\gamma}_{A}\gamma_B\tilde{\gamma}_B$, which reduces the four-fold degeneracy to two-fold. This degeneracy cannot be removed further symmetrically. This confirms our earlier arguement.

Now that we have considered decoration of one particular set of blocks, we can easily extend this to see why all combinations of such configurations are obstructed. First, since the only 1D block that has an edge at $\mu_4$ is $\tau_3$ this anomaly cannot be canceled out by decoration on any other blocks. Hence, decoration on $\tau_3$ is completely forbidden, irrespective of decorations on any other 1D blocks. Similarly, Majorana chain decoration on $\tau_2$ is forbidden by the mixed anomaly on $\mu_3$. The only remaining possibility is Majorana chain decoration on $\tau_1$ which leaves a mixed anomaly on $\mu_1$ and $\mu_2$. Since $\mu_1$ ($\mu_2$) is also surrounded by $\tau_2$ ($\tau_3$) blocks, a Majorana chain decoration on all blocks would cancel out the mixed anomaly on $\mu_1$ and $\mu_2$, but this is obstructed by the subsequent mixed anomaly on $\mu_3$ and $\mu_4$ as discussed earlier. Hence there are no obstruction-free 1D block state decorations.

The nontrivial 0D block states are odd fermion parity on any of $\mu_{1,2,3,4}$. Note that since this decoration does not depend on the spatial symmetry, it will be a common occurrence in all wallpaper groups with 0D blocks in the cell decomposition. The classification is simply determined by the number of distinct $\mu$ blocks, in this case being $\bzt^4$ with group elements $\{(n_1,n_2,n_3,n_4);n_i=\pm 1\}$. Since 0D blocks have no boundary, they are naturally obstruction-free but the classification may be reduced by bubble equivalence as we will describe below.

\subsubsection{Trivialization by bubble decoration}\label{subsec:p2_triv}
Since there are no nontrivial 1D decorations, we only need to consider trivializations of the 0D states. The first possibility is the fermionic 1D bubble on the $\tau$ blocks. We recall from Sec.~\ref{subsec:pmm_triv} that this bubble essentially creates two complex fermions in the bulk of a 1D block and then moves them to its edge, thereby changing the fermion parity at the edge. Such a bubble operation must be consistent with the \(C_2\) symmetry. As illustrated in Fig. \ref{fig:p2}, each \(\mu\) block is surrounded by an even number of 1D blocks from each set. Consequently, any 1D bubble adds an even number of fermions to the \(\mu\) blocks, leaving the fermion parity at the 0D block unchanged.

\begin{figure}[!htbp]
    \centering
    \includegraphics[width=0.9\linewidth]{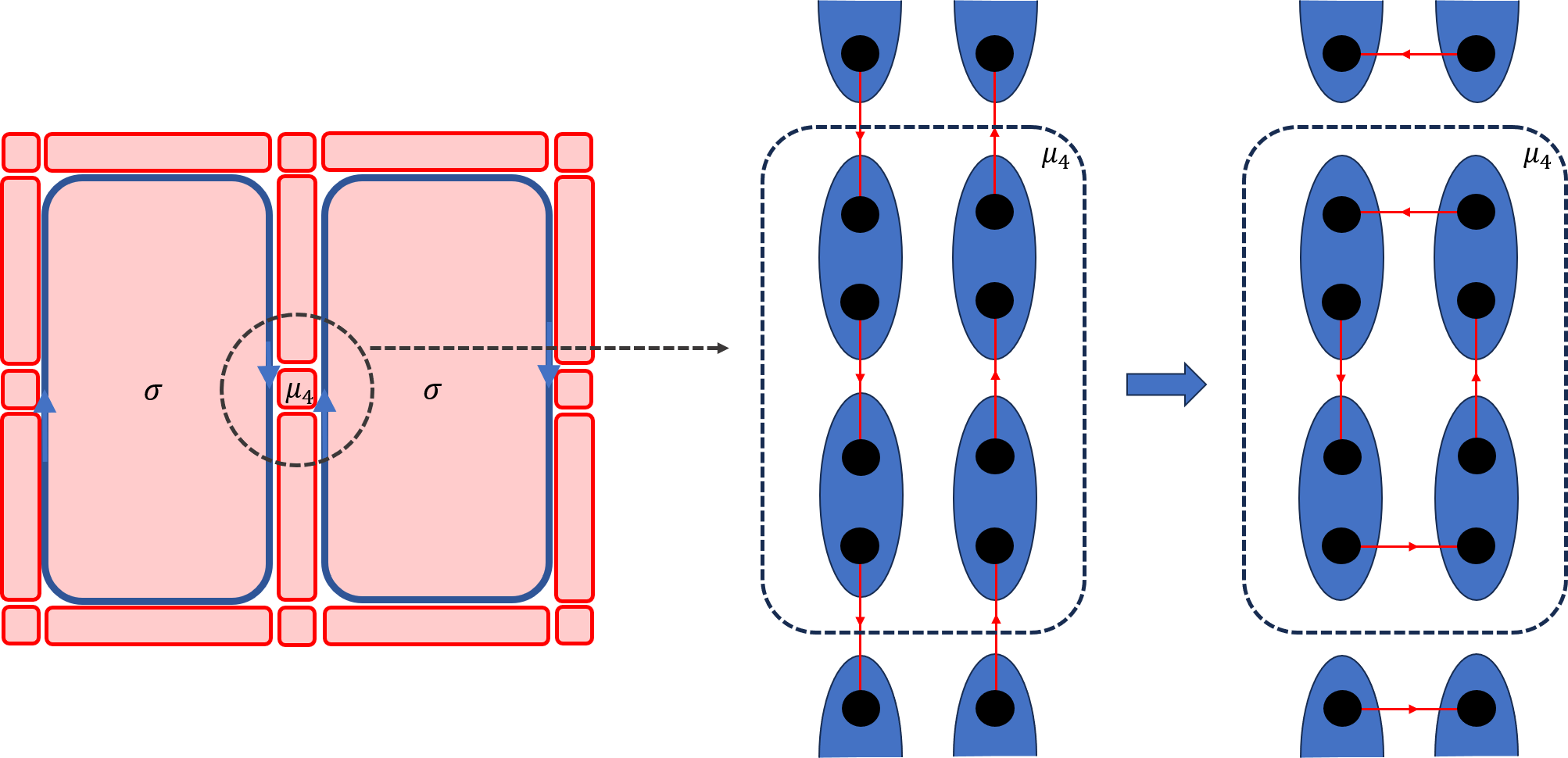}
    \caption{2D Majorana Bubble equivalence in the $p2$ lattice}
    \label{fig:bubble_2d_p2}
\end{figure}

Next, we consider the Majorana bubble decoration on the 2D blocks. We will show that by deforming the Majorana bubbles, one can change the fermion parity on the 0D blocks, thereby reducing the classification. As described in the \(pmm\) case, a Majorana bubble can be viewed as a Majorana chain along the 1D boundary of the \(\sigma\) blocks, with anti-periodic boundary conditions ensuring that the fermion parity of the chain is trivial. Near \(\mu_4\), the bubble state appears as two Majorana chains (see Fig.~\ref{fig:bubble_2d_p2}) with links oriented in opposite directions to respect the rotational symmetry. For simplicity, we assume that the \(\mu_4\) block encloses two fermions from each chain (this choice can be generalized). In the figure, an arrow from \(\gamma^A\) to \(\gamma^B\) represents the stabilizer \(i\gamma^A\gamma^B = 1\).

\begin{figure}[!htbp]
    \centering
    \includegraphics[width=0.9\linewidth]{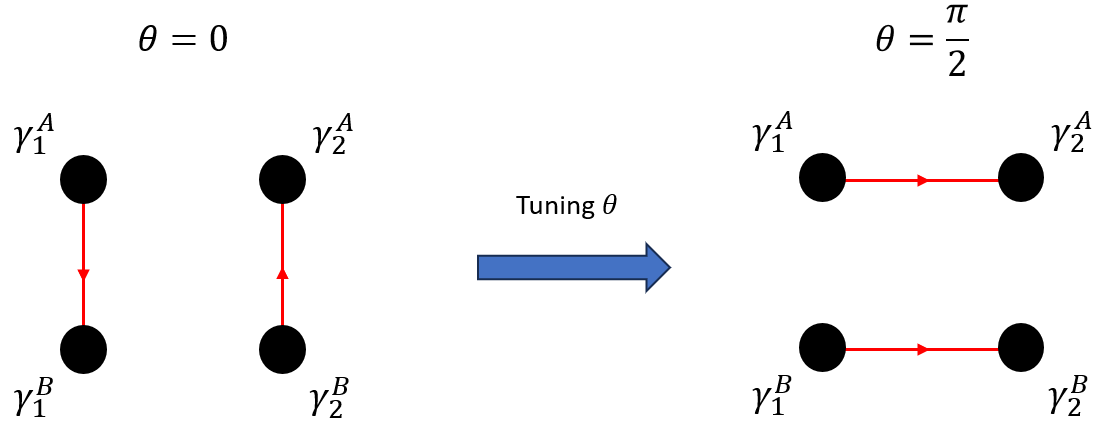}
    \caption{Adiabatic transformation of Majorana links}
    \label{fig:Htheta_adiabatic}
\end{figure}

Now, consider a local Hamiltonian of the following form
\begin{equation}\label{eq:Hadiabatic_theta}
    H = - \cos(\theta) \Bigl(i\gamma_1^A\gamma_1^B + i\gamma_2^B\gamma_2^A\Bigr) - \sin(\theta) \Bigl(i\gamma_1^A\gamma_2^A + i\gamma_1^B\gamma_2^B\Bigr).
\end{equation}
It is straightforward to show that the energy spectrum of this Hamiltonian is independent of \(\theta\). Thus, by tuning \(\theta\), we perform an adiabatic transformation on two nearby links\cite{cheng2022rotation}, namely:
\begin{align}
    \theta = 0 &\rightarrow \theta = \pi/2, \nonumber\\
    \Rightarrow\quad \Bigl(i\gamma_1^A\gamma_1^B,\, i\gamma_2^B\gamma_2^A\Bigr) &\rightarrow \Bigl(i\gamma_1^A\gamma_1^B,\, i\gamma_2^B\gamma_2^A\Bigr).
\end{align}
This transformation is diagrammatically demonstrated in Fig.~\ref{fig:Htheta_adiabatic}. Applying this transformation to the links near \(\mu_4\), while respecting the constraint of rotational symmetry, results in a small Majorana chain with periodic boundary conditions surrounding \(\mu_4\), which indicates an odd fermion parity decoration at \(\mu_4\).
By a similar argument, the bubble simultaneously changes the fermion parity at all \(\mu\) blocks. Thus, the state with nontrivial fermion parity on all the 0D blocks, \((-,-,-,-)\), is identified with the state \((+,+,+,+)\). More generally, each decoration \((n_1,n_2,n_3,n_4)\) is identified with its partner state \((-n_1,-n_2,-n_3,-n_4)\). This Majorana bubble equivalence therefore reduces the classification from \(\mathbb{Z}_2^4\) to \(\mathbb{Z}_2^3\). Hence, the final classification of obstruction-free, trivialization-free states for 2D ACSPTs in the \(p2\) group is \(\mathbb{Z}_2^3\). 

This example illustrates that the bubble equivalence relation can have effects across different dimensions, not solely limited to the immediately lower-dimensional blocks.

\subsection{3D ACSPT with Average $C_{2v}$ Group}
\label{subsec:3dC2v}
We now consider an example of decohered ACSPT with a 3D point group $C_{2v}$. This example is meant to demonstrate that in comparison to the 2D wallpaper groups, 3D crystalline point groups admit a richer set of onsite symmetries and block state decorations. While the increase in dimensions adds a layer of complexity in calculating the obstruction and trivialization conditions, we choose a simple example here which features a new set of intrinsic ACSPTs. The simplicity is apparent in its analogy to the $pmm$ group: the $C_{2v}$ group is generated by two perpendicular reflection planes sharing a common line of intersection, replacing the reflection lines and common point in the former.

\subsubsection{Cell Decomposition} 
\begin{figure}[!htbp]
    \centering
    \includegraphics[width=0.9\linewidth]{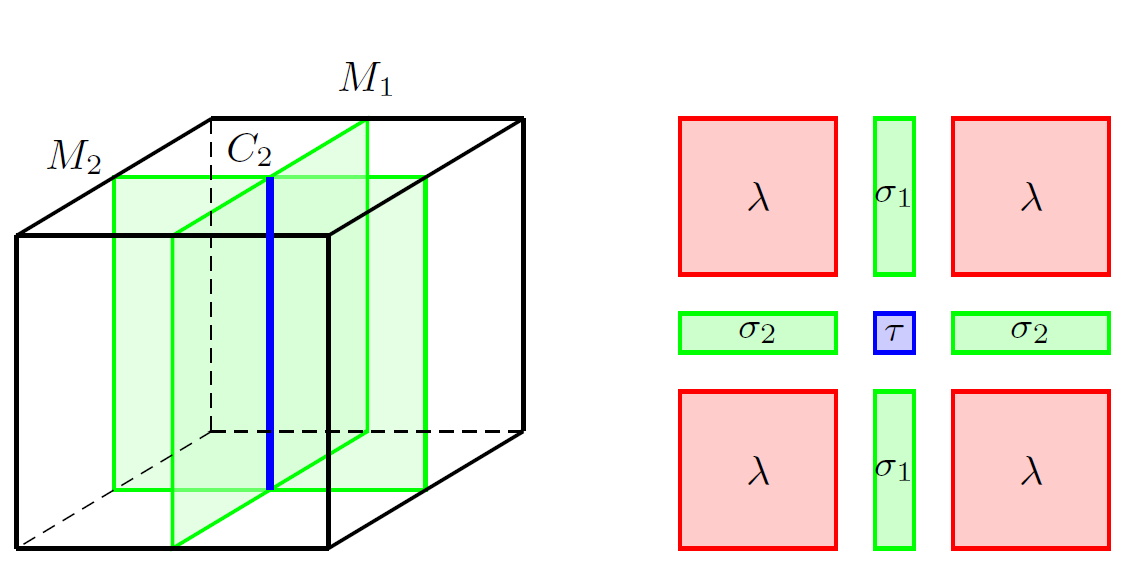}
    \caption{$C_{2v}$ lattice}
    \label{fig:eg_c2v}
\end{figure}

The analogy presented above also extends to the construction of the cell decomposition. First, the intersection line at the center of the unit cell contains a 1D block, $\tau$, invariant under the action of either reflection and thus having an onsite symmetry of $\bzt \times \bzt$. The two reflection planes host two 2D blocks each separated by the intersection line. The two $\sigma_1$ blocks on the plane $M_1$ are invariant under the action of $M_1$ but are connected to one another by the $M_2$ symmetry, and vice-versa. The two planes divide the rest of the unit cell into 4 3D $\lambda$ blocks which are connected to each other by the action of reflections. While there are no 0D blocks or rotation axes in this particular example, such elements can appear in other 3D point groups.

\subsubsection{Block state decorations}
We begin by considering the possible block states in 2D. The 2D blocks possess an onsite \(\mathbb{Z}_2\) symmetry by virtue of being reflection planes. In the spinless case, the total symmetry group for a 2D block is \(\mathbb{Z}_2 \times \mathbb{Z}_2^f\). Two types of SPTs can be decorated on these 2D blocks:
\begin{enumerate}
    \item The fermionic Levin-Gu (fLG) state (or \(p \pm ip\) superconductors)\cite{levingu2014fermionic}, which is classified by \(\mathbb{Z}_8\).
    \item The 2D chiral \(p+ip\) superconductor, which is classified by \(\mathbb{Z}\).
\end{enumerate}

The edge theory of the \(p+ip\) state is described by a chiral Majorana mode along the 1D boundary. The edge theory of the fLG state is described by one left-moving and one right-moving mode. While the edge is achiral, it cannot be gapped out as each chiral mode carries a different charge of the bosonic $\bzt$ symmetry. This indicates that the state carries a mixed anomaly of the \(\mathbb{Z}_2 \times \mathbb{Z}_2^f\) symmetry. This state is further analyzed using the K-matrix formalism in Sec. \ref{sec:element of construction}.

One can show that, in the clean case, a symmetric gap can be achieved by stacking 8 copies of this root phase, which leads to the \(\mathbb{Z}_8\) classification. However, when 4 layers of the fLG state are stacked, the resulting state is equivalent to the bosonic Levin-Gu state that carries only the bosonic \(\mathbb{Z}_2\) anomaly\cite{levin2012braiding}. Under averaged bosonic symmetry, this Levin-Gu state is trivialized, reducing the classification of the average block state decoration to \(\mathbb{Z}_4\).

In summary, there are two sets of 2D blocks, each set can be decorated with an arbitrary number of \(p+ip\) superconducting layers or up to 4 layers of fLG states. As we will elaborate later, most of these decorations are obstructed. 

For the 1D block \(\tau\), the onsite symmetry is \(\mathbb{Z}_2 \times \mathbb{Z}_2 \times \mathbb{Z}_2^f\). The possible block state decorations include the Majorana chain and the fermionic SPT (fSPT) decoration. The latter can be constructed by combining the fSPT phases corresponding to each \(\mathbb{Z}_2\) symmetry independently, yielding a classification by \(\mathbb{Z}_2^2\). Finally, in the clean system, there is a 1D bosonic $\bzt\times\bzt$ SPT, namely the Haldane phase\cite{pollmann2010entanglement}. However, in the decohered (and the disordered system), this phase is trivialized as it is protected solely by average symmetries. 

\subsubsection{Obstruction-free States}
Since there are no 0D blocks in our system, any decoration on the 1D block state is obstruction-free. It suffices to check what 2D decorations are obstruction-free on $\tau$. The obstruction condition is easy to understand from the folded SPT perspective: the $\sigma$ blocks sharing the same edge $\tau$ must fold into a trivial SPT in 2d. In the clean case, this means that the edge modes can be symmetrically gapped. Under decoherence, the edge theory can also be that of a bosonic Levin-Gu state since it is trivialized. This relaxed obstruction-free condition leads to the emergence of intrinsic ASPTs. 
 
First, a $p+ip$ SC decoration on $\sigma_1$ blocks introduces two Majorana modes on $\tau_1$ with the same chirality, which cannot be gapped, and hence is obstructed. However, this chiral anomaly may be canceled by stacking $p-ip$ SC on $\sigma_2$. We will show later decoration in this configuration is still obstructed. And using a bubble equivalence argument we can show that two copies of $p+ip$ decoration is equivalent to one copy of fLG decoration phase which is also obstructed. For more copies of $p+ip$ decoration, we can utilize this equivalence relation to map it to the fLG decoration case. 

Now let us consider decorating one set of blocks, say $\sigma_1$, with layers of fLG state. The mirror symmetries along the $\sigma_1$ and $\sigma_2$ blocks are labeled by $M_1$ and $M_2$. These two symmetries act on-site for the edge modes on 1D block $\tau$. We will introduce an anomaly indicator in Section \ref{sec:element of construction} through which we can show that the edge modes do not have a $M_2$-anomaly. Since the $M_1$ symmetry is just the onsite symmetry on $\sigma_1$, the extra symmetry does not pose an additional restriction in this case. The obstruction then simplifies to counting the number of fLG edge modes on $\tau$. Decorating one fLG state on the $\sigma_1$ blocks leads to two copies of fLG edge modes on $\tau$, and is hence obstructed since the classification is $\bbz_4$. On the other hand, the 2-fLG decoration on $\sigma_1$ leads to 4-fLG edge states on $\tau$ which is the edge state of bosonic Levin-Gu state, and is hence anomaly-free under decoherence (but still obstructed in the clean case). Therefore, the decoration of two fLG states on $\sigma_1$ (or $\sigma_2$ by the same argument) is obstruction-free and corresponds to an intrinsic average decoration. This decoration pattern is $\bzt$-classified as two copies of the decoration means 4 copies of fLG states on the 2D block and 4 fLG states with average $\bzt$ symmetry is trivial by itself. Counting decorations both on $\sigma_1$ and $\sigma_2$ gives us an obstruction-free classification of $\bzt^2$.

\subsubsection{Trivializations}

First we consider trivializations of the 1D obstruction-free states. The two root phases are the $\bzt$ fSPT corresponding to $M_1$, and the $\bzt$ fSPT corresponding to $M_2$. Now consider a 2D Majorana bubble on $\sigma_1$. This leaves two Majorana chains on $\tau$ which transform to each other by the action of $M_2$. This is precisely the $M_2$ $\bzt$ fSPT. Therefore, the fSPT decoration on $M_2$ is trivialized. Similarly, the Majorana bubble on $\sigma_2$ trivializes the $M_1$ $\bzt$ fSPT.  

Now we consider trivializations of the 2D states. The higher-dimensional block is $\lambda$, which has no onsite symmetry. Hence the only possible bubble is the 3D $p+ip$ superconductor bubble, which leaves a $p+ip$ SC on each 2D face of the block. As we demonstrate in Sec. \ref{sec:elems_spinless_bubble}, when this bubble is introduced on $\lambda$, it gives an equivalence between 2 layers of $p+ip$ (or $p-ip$) decoration and fLG state on $\sigma_1$ and $\sigma_2$. Hence the $\bbz$ classification of $p+ip$ block state decoration reduces to $\bzt$. This equivalence appears in all cases of 2D blocks with onsite $\bzt$ symmetry.

As mentioned in the previous subsection, by decorating $\sigma_1$ with $p+ip$-SC and $\sigma_2$ with $p-ip$ SC, we obtain an edge state on $\tau$ with no chiral anomaly. However, through the equivalence above, one can show that this decoration is obstructed by the on-site $\bzt$ symmetries from the mirror symmetries. We argue this by contradiction. Let us assume that this decoration is obstruction-free. Then, two copies of this decoration should also be obstruction-free. By bubble equivalence, the two-copy decoration is equivalent to fLG states decorated both on $\sigma_1$ and $\sigma_2$, which is known to be obstructed as in the previous section. Therefore, the decoration we started with is also obstructed.

In summary, this section has demonstrated examples of block state construction and classifications in the spinless case with decoherence. The general strategy works for other cases, however, there will be crucial differences in terms of their implementations. In the next section, we discuss systematically the obstruction-free conditions and the trivialization conditions, and their variation in the spinful case and disordered cases. 

\section{Elements of construction}
\label{sec:element of construction}

In this section, we first comprehensively describe the block states and obstruction-free conditions as well as the bubble equivalence relation employed in the classification of 2D and 3D crystalline average SPTs in the spinless case with decoherence. Then we will discuss the differences for decohered spin-1/2 cases and the disordered cases.

\subsection{Decorations and obstruction-free conditions}
Now we focus on the case of decohered spinless models, with an average bosonic symmetry $G_b$ which has a trivial extension with the exact fermion parity symmetry. We organize the discussion by dimensions.

\subsubsection{0D and 1D block states}\label{subsec:elem_0d1d_decorations}

\noindent\textbf{0D decoration} The 0D block states are given by the cohomology data:
\begin{align}
    n_0 &\in h^1(\bzt^f) = \bzt \label{eq:cohom_0d_n0}\\
    \nu_1 &\in \mathcal{H}^1(G_b,U(1)) \label{eq:cohom_0d_n1}
\end{align}
The $\nu_1$ layer is a bosonic 0D SPT classified by linear representations of $G_b$, and hence is trivialized under decoherence. Therefore, the only nontrivial class is given by the $n_0$ layer, which labels the fermion parity on the 0D block (with the nontrivial phase being odd fermion parity state). As the 0D block does not have a boundary, this decoration is automatically obstruction-free in the spinless case.

\vspace{0.05in}
\noindent\textbf{1D decoration} For the 1D blocks with an onsite bosonic symmetry $G_b$, we have two fermionic layers:
\begin{align}
    n_0 &\in h^2(\bzt^f) = \bzt \ \  \text{(Kitaev chain)} \\
    n_1 &\in \mathcal{H}^1(G_b,h^1(\bzt^f)) = \mathcal{H}^1(G_b,\bzt) \label{eq:n1layer_1d}
\end{align}
For all the 1D blocks we consider, the bosonic symmetry group \( G_b \) is either \( \mathbb{Z}_N \) or the dihedral group \( D_N \). As mentioned earlier, the $n_1$ layer can be constructed by decorating $G_b$ domain walls with complex fermions. 

We will first consider the rotation symmetry $\bbz_N$ case. The $n_1$ layer classification for a $\bbz_N$ rotation symmetry can be calculated from group cohomology\cite{QingRuiWangPRX2020} 
\begin{equation}
    \mathcal{H}^1(\bbz_N,\bzt) = \bbz_{gcd(N,2)} = \begin{cases}
        \bbz_1, &\text{odd } N\\ 
        \bzt, &\text{even } N.
    \end{cases}
\end{equation}

For even $N$, the $n_1$ layer SPT can be constructed by stacking \( N \) copies of Majorana chains, with the \( \mathbb{Z}_N \) symmetry acting as a permutation among the copies. While for odd \( N \), there is no $n_1$ SPT. A similar procedure yields only the \( n_0 \)-layer SPT for odd $N$. We now examine the details of this construction.

We start with the even $N$ case. Consider the Majorana zero modes $\gamma_0,\ldots,\gamma_{2n-1}$ from the blocks, where $G_b=\bbz_{2n}$ cyclically permutes the modes. We can form linear permutations of these modes:
\begin{equation}
    \gamma^R_{m} = \sum_{j=0}^{2n-1} \omega^{mj} \gamma_j, \omega = \exp\left(\frac{2\pi i}{2n}\right),\ m=-n-1,\ldots,n
\end{equation}
These transformed modes carry charges of the $G_b$ symmetry: $U_G\gamma^R_mU_G^\dagger = \omega^m \gamma^R_m$. As a consequence pairs of the form $\gamma^R_m\gamma^R_{-m}$ are symmetric and can be gapped. The only remaining modes are $\gamma^R_0$ and $\gamma^R_n$ with charges $1$ and $-1$. This effectively gives us the $\bzt$ fSPT model, which is the $n_1$-layer.

For odd $N$, the distinction emerges from the lack of a -1 charge for $G_b=\bbz_{2n+1}$. To demonstrate this, we consider the example of $G_b=\bbz_3$. To have nontrivial action under the symmetry, we need at least three Majorana chains which are rotated under the $\bbz_3$ symmetry. Just as in the even $N$ case, we perform a linear transformation to obtain modes which are symmetric under rotation with the charges:
\begin{align}
    \gamma_0&: U_G \gamma^R_0 U_G^\dagger = \gamma^R_0 \\
    \gamma_1&: U_G \gamma^R_1 U_G^\dagger = \omega\gamma^R_1 \\
    \gamma_2&: U_G \gamma^R_2 U_G^\dagger = \omega^2\gamma^R_2
\end{align}
where $\omega = \exp(i2\pi/3)$. Then it is easy to see that the term $i\gamma^R_1\gamma^R_2$ symmetrically gaps out two of the modes:
\begin{align}
    U_G {i\gamma^R_1\gamma^R_2} U_G^\dagger = i\omega^3 \gamma^R_1\gamma^R_2 = i\gamma^R_1\gamma^R_2
\end{align}
leaving only one Majorana chain with trivial symmetry action. Since this is precisely the Kitaev chain, this is classified by the $n_0$ layer, and hence $n_1$ is trivial. This procedure generalizes to all odd rotational bosonic symmetries ($\bbz_{2n+1}$).

The 1D block may also lie on the axis of a dihedral rotation. Since any dihedral group can be generated by a rotation and reflection group, the classification of the $n_1$ layer is given by $\mathcal{H}^1(D_N,\bzt) = \bbz_2\times\bbz_{gcd(N,2)}$\cite{QingRuiWangPRX2020}. The two root states can be understood as $n_1$-layer SPTs protected individually by reflection symmetry and rotation symmetry, respectively.

\vspace{0.05in}
\noindent\textbf{Obstruction-free conditions for 1D decorations} 
Equipped with a complete set of possible decorations on the 1D blocks, we now determine which of these possible configurations are obstruction-free. The obstruction-free condition of these 1D block states can be tested by 1) checking if there is an even number of Majorana zero modes on the 0D intersection, 2) checking if the fermion parity commutes with the bosonic symmetry at the 0D intersection of the 1D blocks. If both checks pass, then the edge mode is bosonic (hosting no mixed anomaly with fermion parity) and hence can be trivialized in the decohered setting.

To understand how this criterion can be applied in practice, we start by considering a set of 1D blocks $\tau$ with Majorana chain ($n_0$ layer) decoration. This leaves a set of Majorana zero modes (MZMs) at the neighboring 0D blocks. Consider one such block $\mu$ with onsite symmetry $G_\mu$, surrounded by $m$ $\tau$ blocks, consequently hosting $m$ MZMs. If $m$ is odd, then the state at $\mu$ hosts odd number of MZMs, which is anomalous. If $m$ is even, one can construct a representation of the fermion parity operator at $\mu$ by taking the product:
\begin{equation}\label{eq:elems_1dobs_Pf}
    P_f = i^{m/2} \gamma_1 \gamma_2\ldots \gamma_m
\end{equation}
The symmetries of $G_\mu$ permute these $m$ MZMs, as in the examples in Sec.~\ref{subsubsec:pmm_obs} and Sec.~\ref{subsubsec:p2_deco_and_obs}. If one of these permutations flips the sign of $P_f$, then $G_\mu$ does not commute with the fermion parity, and hence the state is obstructed. If all generators of $G_\mu$ preserve the sign of $P_f$, then there is no mixed anomaly, and hence the state is obstruction-free. 

It is straightforward to extend this test to decorations of the $n_1$ layer since we have demonstrated earlier that these states can be constructed by stacking Majorana chains with $G_\tau$ acting as a permutation symmetry between the chains, and hence the obstruction check can be performed on the Majorana modes from this construction. This is the general principle for determining obstruction-free conditions; however, in our setting, the analysis simplifies significantly since all $n_1$ decorations fall into $\bbz_2$ classifications.

When the $n_1$ data is $\bzt$-valued, the obstruction-free condition reduces to a simple rule: for nontrivial $n_1$ decoration on $m \ \tau$ blocks, with the edge modes meeting at the common boundary $\mu$, the decoration is obstruction-free if and only if $m$ is even. This can be explained by considering the folded state at $\mu$, formed by stacking $m$ copies of the $n_1$ decorations. Recall that for the decoration to be obstruction-free, the folded state must be trivial ASPT under the full folded symmetry labeled by $G_\mu$. The $G_\mu$ contains the original onsite symmetry $G_\tau$ on each block and the permutation symmetry between them, generated by $G_{\mu/\tau} \equiv G_\mu/G_\tau$. If $m$ is odd, and since $n_1$ is $\bzt$-valued, the folded state remains a $G_\tau$ ASPT, resulting in an obstructed decoration. If $m$ is even, then the folded state is not a $G_\tau$ SPT, but may still be protected by the additional symmetry at $\mu$, namely $G_{\mu/\tau}$. Hence, we need to check if this symmetry commutes with the fermion parity as well.

From our previous demonstration, we see that, for a single 1D block, the $\bzt$-valued $n_1$ decoration for spinless $G_\tau$ onsite symmetry is constructed by two Majorana chains at each block carrying charges $\pm 1$ of the $G_\tau$ symmetry. We collect the edge Majorana modes of all 1D blocks meeting at the 0D intersection into the sets $\{\gamma_+\}$ and $\{\gamma_-\}$ respectively. The $G_{\mu/\tau}$ symmetry acts as permutations in the elements in the set $\{\gamma_+\}$, and also has identical permutation action on $\{\gamma_-\}$. Hence, the total fermion parity (as the product of the operators in $\{\gamma_+\}$ and $\{\gamma_-\}$ together) at $\mu$ remains invariant under the action of $G_{\mu/\tau}$. Thus, the folded state is a trivial ASPT when $G_\mu$ is an average symmetry. Therefore, the decoration is obstruction-free. 

This can be generalized to $D_N$ symmetries where the $n_1$ data can be $\bzt^2$ valued for even $N$. In such a case, one can apply the odd/even criterion to each of the two $\bzt$-valued SPT indices separately. 

The strength of this argument lies in its applicability even to the spin-1/2 case, where the edge theory is not simply a collection of Majorana modes. This will be discussed in detail in the section on spin-1/2 systems.

\subsubsection{2D Block States and the K-matrix formalism}
\label{sec:elems_2d}
Now we consider the possible 2D states. For 2D wallpaper groups, there is no onsite bosonic symmetry acting onsite and for 3D point groups, the only possible onsite symmetry is $\bzt$ by virtue of the block being defined on a mirror plane. The layers are classified in cohomology data by
\begin{align}
    n_0 &\in h^3(\bzt^f) = \bbz \ \ \text{($p+ip$ superconductor)} \\
    n_1 &\in \mathcal{H}^1(G_b,h^2(\bzt^f)) = \mathcal{H}^1(\bzt,\bzt) = \bzt \\
    n_2 &\in \mathcal{H}^2(G_b,h^1(\bzt^f)) = \mathcal{H}^2(\bzt,\bzt) = \bzt \\
    \nu_3 &\in \mathcal{H}^3(G_b,U(1)) = \bzt.
\end{align}
The integer-classified $n_0$ data describes $p+ip$ SC with chiral edge modes. When 2D block states are decorated with $n_0$ data, we first check for the chiral anomaly at the 1D edge blocks (i.e., check that the state on the 1D block has equal number of left- and right-moving modes). If we find that it is not obstructed by the chiral anomaly, we must check that it is not anomalous with respect to the on-site symmetry of the 1D block (i.e. the folded SPT has trivial $n_1,n_2,\nu_3$ data), which we will discuss below. Besides the invertible topological phase $n_0$, the other decorations have a total classification of $\bbz_8$ given by nontrivial stacking between the three layers. The bosonic phase, described by the $\bzt$-valued $\nu_3$ layer is the bosonic Levin-Gu state. The $n_1$ layer, given by decorating 1D $G_b$ defects with Kitaev chains, is the fermionic Levin-Gu (fLG) state. A physical realization of the fermionic Levin-Gu state is the $p\pm ip$ superconductor with the edge state possessing a pair of counter-propagating Majorana modes. These can be gapped out in the absence of symmetry, but not when they carry different charges of the $G_b=\bzt$ symmetry. 

To systematically construct these phases, we can adopt the K-matrix formalism. The topological quantum field theory (TQFT) describing of a 2D SPT involves a Chern-Simons term whose edge theory can be described by the Luttinger liquid formalism:
\begin{equation}
    \mathcal{L} = \int \dd^2 x K_{IJ}\partial_t \phi_I \partial_x \phi_J - \int \dd^2 x V_{IJ}\partial_x \phi_I \partial_x \phi_J
\end{equation}
with $N$ fields $\{ \phi_I,I=1,\ldots,N\}$ that can be constructed by bosonizing the edge fermion modes. Furthermore, the constraint $\abs{\det K}=1$ to ensure that the state is non-degenerate (does not carry any anyonic excitations). The algebra of these fields is determined by the $N\times N$ K-matrix:
\begin{equation}\label{KmatAlg}
    \left[  \phi_I(x'), \partial_x \phi_J(x) \right] = 2\pi i K^{-1}_{IJ} \delta(x-x')
\end{equation}
The group action of a symmetry on $\Phi = \left( \phi_1, \ldots, \phi_N \right)$ can be implemented by:
\begin{equation}\label{eq:Kmat_phi_G}
    g\in G: \Phi \rightarrow W_g \Phi + \chi_g
\end{equation}
For $G$ to be a unitary symmetry of the edge theory, we require:
\begin{equation}
    W_g^T K W_g = K
\end{equation}
The 2D state is a nontrivial SPT if the edge theory is anomalous. Conversely, if the edge theory can be symmetrically gapped, then the bulk SPT has to be trivial. To check if a luttinger theory can be gapped or not, we can add backscattering (or Higgs) terms of the form:
\begin{equation}
    \mathcal{H'} = U \sum_{i=1}^{N/2} \cos({\Lambda^{i}}^T K\Phi + \alpha_i)
\end{equation}
If there are $N/2$ independent back-scattering terms satisfying the following conditions, then the theory can be symmetrically gapped. The condition for these $N/2$ terms is: 1)
\begin{equation}
    {\Lambda^{i}}^T K \Lambda^{j} = 0\ \forall\ i,j
\end{equation}
2) we must demand that the gapping terms do not break the symmetry explicitly:
\begin{equation}
    g[\mathcal{H}'] = \mathcal{H}'
\end{equation}
and 3) the symmetry is not spontaneously broken. The diagnostic for this is to construct the matrix $M =\left( \Lambda^{1} \ldots \Lambda^{N/2}\right)$ and check that the greatest common divisor of all the $N/2 \times N/2$ minors of $M$ is 1\cite{lu2012sptchernsimons,ning2021edge,Heinrich2018Z2Anomaly}. 
 
If we can find $N/2$ such terms, the edge can be symmetrically gapped hence the bulk is a trivial SPT. If the system turns out to be a nontrivial SPT, we can detect its classification using the luttinger liquid formalism as well. Suppose the SPT has $\bbz_M$ classification. This means for $M$ copies of the root phase we should be able to find $MN/2$ gap terms that satisfy the conditions. 
 
For the root state, namely the fermionic Levin-Gu state, we can use a 2-dimensional K-matrix (associated with two bosonic field $\phi_1$ and $\phi_2$), with a $\bzt$ symmetry $(W,\chi)$:
\begin{equation}\label{eq:fLG_Kmat}
    K = W = \begin{pmatrix}
        1 & 0 \\  0 & -1
    \end{pmatrix},\ \chi = \begin{pmatrix}
        0 \\ 0
    \end{pmatrix},
\end{equation}
and a cos term 
\begin{equation}
    H_{int} = g \int dx \cos(\phi_1+\phi_2) + \cos(\phi_1-\phi_2).
\end{equation}
This term ensures that we gap out half of the degrees of freedom and leave only a pair of counter-propagating Majorana fermion\cite{ning2021edge}. To classify average SPTs, we can extend this formalism to the doubled state. The doubled K-matrix is a direct sum of $K$ and $-K$. The strong symmetry becomes a doubled symmetry, acting independently but identically on each copy of the space, while the weak symmetry becomes a diagonal symmetry, acting in the same way across both copies simultaneously.

Most of the bosonic symmetries (which are average symmetries) encountered in our classification are $\bzt$ symmetries. In these cases, we can use the $\bzt$ anomaly indicator introduced in Ref.~\cite{Heinrich2018Z2Anomaly} as a way to detect anomaly and nontrivial SPTs. In this method, one needs to first perform a linear transformation on the K-matrix and the $\bzt$ symmetry operations to the block diagonal canonical form:
\begin{align}
    K &= \begin{pmatrix}
        A & 0 & B & -B \\
        0 & C & D & D \\
        B^T & D^T & E & F \\
        -B^T & D^T & F^T & E
    \end{pmatrix}
    \\ W &= \begin{pmatrix}
        -1_{n_--m} & 0 & 0 & 0 \\
        0 & 1_{n_+-m}& 0 & 0 \\
        0 & 0 & 0 & 1_m \\
        0 & 0 & 1_m & 0
    \end{pmatrix} \\  \chi &= \begin{pmatrix}
        0 \\ \chi_2 \\ 0 \\ 0
    \end{pmatrix}
\end{align}
With this form, we can evaluate the indicator: 
\begin{equation}\label{eq:anom_indicator}
    \nu_{\bzt} \equiv \frac{1}{2} \chi_+^T K^{-1} \chi_+ + \frac{1}{4} \text{sig}(K(1-W)) \pmod{2}
\end{equation}
where $\chi_+$ is the vector
\begin{equation}
    \chi_+ = \begin{pmatrix}
        0 \\ \chi_2 \\ \text{diag}(E+F)/2 \\ \text{diag}(E+F)/2
    \end{pmatrix}
\end{equation}
$\nu_{\bzt}$ takes values in multiples of $1/4 \pmod{2}$. The value $0 \pmod{2}$ corresponds to a trivial phase, giving the $\bbz_8$ classification in pure states. $\nu = 1 \pmod{2}$ corresponds to a bosonic Levin-Gu SPT, which becomes trivial if the $\bzt$ is averaged. Therefore, if a folded SPT has $\nu=1$, then the decoration will be obstruction-free and it can potentially lead to an intrinsic ASPT. Another advantage of the anomaly indicator approach is that it conveniently accommodates stacking: if some decoration gives an indicator value $\nu$, then $m$ layers of the decoration give an anomaly $m \nu$. Hence, calculating the indicator for one layer of fLG decoration is sufficient for finding the indicator values for multiple-fLG decoration.
  
Consider the example of the 3D point group $C_{2v}$, shown in Fig.~\ref{fig:eg_c2v}, with blocks $\sigma_1,\sigma_2$ intersecting at a 1D block $\tau$. First consider the decoration of $p+ip$-SC on one set of blocks, say $\sigma_1$. By the reflection symmetry that connects the two $\sigma_1$ blocks, this leaves two Majorana edge modes on $\tau$ that are directed in the same direction. Hence this decoration is obstructed by the chiral anomaly. By the same argument, $p+ip$-SC decoration on $\sigma_2$ is also obstructed. However, if we simultaneously decorate $\sigma_1$ and $\sigma_2$ by $p+ip$ and $p-ip$ respectively, then there is no chiral anomaly on $\tau$. This particular decoration turns out to be obstructed as well as we will show later.
 
Next, we consider the decoration of one fLG layer on $\sigma_1$ (with trivial decoration on $\sigma_2$). At $\tau$, where the edge modes meet, from Eq. \ref{eq:fLG_Kmat}, we have $K = \sigma_z\oplus\sigma_z$. The $\bzt$-reflection $M_1$, which is the onsite symmetry of the $\sigma_1$ blocks, has the implementation $W_1 = \sigma_z\oplus\sigma_z$. The other reflection symmetry $M_2$ exchanges the two layers and hence has the implementation $W_2 = \mathbb{I}_2\otimes \sigma_x$. The anomaly indicator for either symmetry can be obtained as $\nu_{M_1} = 1/2, \nu_{M_2} = 0$. Similarly, a decoration of fLG layer on $\sigma_2$ gives anomaly indicators $\nu_{M_1} = 0, \nu_{M_2} = 1/2$. For an obstruction-free decoration in the decohered setting, both anomaly indicators must be an integer modulo 2. The obstruction-free decorations then are 2 layers fLG on $\sigma_1$ or $\sigma_2$. Each decoration is $\bzt$ classified, since stacking them twice results in bosonic Levin-Gu decorations which are trivial under decoherence, resulting in total classification data $\bzt^2$. 
Furthermore, since these decorations give anomaly indicator $1 \pmod{2}$ for at least one of the symmetries, they are obstructed in the clean case and hence intrinsic.

This concludes all relevant decorations in 0D, 1D, and 2D blocks that we will encounter in our constructions of ACSPT of decohered spinless systems in 2D and 3D. Next, we will systematically discuss the bubble equivalence relation.

\subsection{Bubble equivalence}\label{sec:elems_spinless_bubble}

\noindent\textbf{0D bubble} Since 0D blocks do not have a boundary, there is no bubble equivalence relation coming from the 0D blocks. 

\noindent\textbf{1D bubble} For 1D blocks, as bosonic SPTs are trivialized when the symmetry is averaged, the only nontrivial bubble equivalence is a fermionic bubble. We can introduce two complex fermions in the bulk of the 1D trivial state: $c_L^\dagger c_R^\dagger \ket{0}$ which preserves fermion parity and has trivial action under all other symmetries. Then the $L$ and $R$ fermions can be moved adiabatically to the two ends of the 1D block. Hence, if a 0D block $\mu$ is surrounded by an odd number of such 1D blocks, then this bubble equivalence changes the fermion parity on $\mu$. Thus, odd fermion parity decoration on $\mu$ is trivialized.

\noindent\textbf{2D bubble} Then we can consider 2D blocks. In this case, we can have a Majorana chain bubble, a closed Majorana chain with anti-periodic boundary conditions, which can be created locally on the 2D blocks. The Majorana bubble on the 2D block may trivialize decoration on the 1D blocks on its boundary. In our initial discussion of bubble equivalence, this decoration was demonstrated to trivialize $\bzt$ fSPT on the 1D blocks in the $pmm$ case. The same logic can be generalized to axes of simple and dihedral rotation. Consider a 1D block $\tau$ with some onsite bosonic symmetry $G$. Consider popping Majorana bubbles on the 2D blocks surrounding $\tau$ and deforming them to the boundary of these 2D blocks. This extends Majorana chains to $\tau$ with the symmetry $G$ generally acting as a rotation between these Majorana chains. If the rotation is even-fold, this is precisely the 1D $G$-fSPT obtained from the $n_1$ layer decoration, which is hence trivialized by the bubble equivalence. If the rotation is odd-fold, this bubble equivalence trivializes the Majorana chain decoration ($n_0$ layer) on the axis. 
 
The 2D bubble equivalence can also affect the 0D block decoration. If deforming these Majorana bubble can leave a Majorana chain with periodic-boundary conditions surrounding the 0D block, this means the fermion parity on the 0D blocks can be flipped by the bubble process. However, such a decoration is incompatible with any reflection lines passing through $\mu$. But in the absence of reflections, this process trivializes odd complex fermion parity on $\mu$. We have seen such an example for the $p2$ wallpaper group in Sec.~\ref{subsec:p2_triv}, while it is not possible in the $pmm$ group because of the presence of reflection symmetries.
 
\noindent\textbf{3D bubble} Next, we consider bubble equivalence from the 3D blocks. Since 3D blocks have no onsite symmetry, the bubble that can be added is the $p+ip$ or $p-ip$ bubble. And we need to consider the effect of extending the $p+ip$ state onto the two-dimensional boundary of the 3D block. Since 2D blocks can only have $\bzt$ onsite symmetry, in most cases, such as the $C_{2v}$ example in Fig.~\ref{fig:eg_c2v}, a 2D block $\sigma$ will be the shared edge of two 3D blocks connected by the $\bzt$ reflection symmetry. This bubble gives an equivalence between the decoration of two copies of $p+ip$-states on $\sigma$ and the decoration of the fermionic Levin-Gu state. To see this, we start with two layers of $p+ip$ on a block $\sigma_M$ defined on a mirror plane $M$. The edge consists of two left-moving Majorana modes $\gamma^1_L,\gamma^2_L$ and the $\bzt$ mirror symmetry has trivial action on these modes. The two 3D blocks that share the boundary $\sigma_M$ can then push $p-ip$ bubble to the $\sigma_M$ block, which leaves two right-moving Majorana modes on the edge of $\sigma_M$: $\gamma^1_R,\gamma^2_R$. Note that these two modes  are exchanged by the reflection symmetry $M: \gamma^1_R\leftrightarrow \gamma^2_R$. We can then define $\gamma^{\pm}_R = (\gamma^1_R \pm \gamma^2_R)/\sqrt{2}$ with $M:\gamma^{\pm}_R\rightarrow \pm \gamma^{\pm}_R$. Since $\gamma^2_L$ and $\gamma^+_R$ have opposite chiralities and trivial action under $M$, they can be gapped out pairwise without breaking any symmetry. The two remaining modes $\gamma^1_L, \gamma^-_R$ have opposite chiralities and different $\bzt$ charges and hence form a fermionic Levin-Gu state. 
 
Now we can use this logic to understand the anomaly of the decoration of $n_0$ layers on $\sigma_{1,2}$ with opposite chiralities discussed earlier in the case of $C_{2v}$ group (Sec. \ref{subsec:3dC2v}). While this decoration does not possess a chiral anomaly in the bulk, it could possess a $\bzt$ anomaly of the mirror symmetries. This can be argued as the following. Let us consider stacking this decoration twice. By the 3D bubble equivalence discussed in the last paragraph, this is equivalent to the decoration of the fLG state on both $\sigma_1$ and $\sigma_2$, which gives anomaly indicators of $1 \pmod{2}$ as discussed earlier. The additive property of the indicator implies that the original single copy decoration must have an indicator value $1/2 \pmod{2}$, and is hence anomalous and obstructed. One can also directly compute the anomaly indicator by analyzing the edge states arising from the \( p \pm ip \) decoration; this approach leads to the same conclusion.

\begin{figure}[!htbp]
    \centering
    \includegraphics[width=0.9\linewidth]{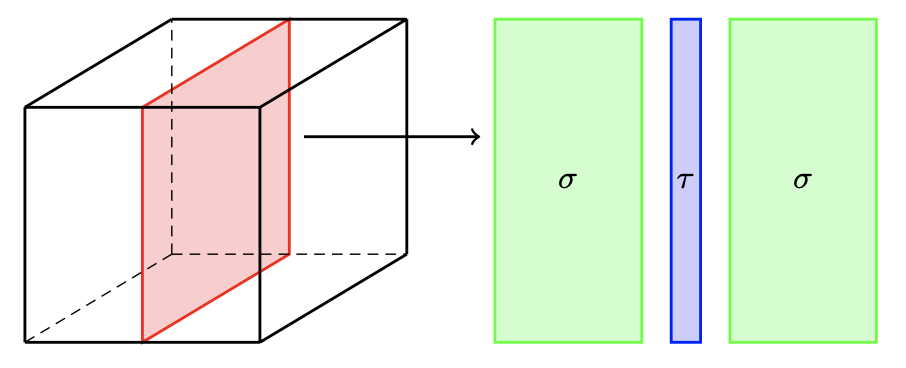}
    \caption{Cell decomposition for 3D $C_{2}$. The right hand side is the cell decomposition on the red plane.}\label{fig:latt_c2_main}
\end{figure}

The final possible trivialization appears on the open surface of a 3D crystal. Consider the case of the $C_2$ wallpaper group shown in Fig.~\ref{fig:latt_c2_main}. Here there are no 0D blocks, hence a Majorana chain decoration on the 1D block $\tau$ is obstruction-free, and cannot be trivialized by bubble equivalence. However, this leaves a Majorana mode on the open surface of the crystal, which can be trivialized by introducing a $p+ip$ SC on the surface with a trapped Majorana mode\cite{cheng2022rotation}. Any surface modification that respects the system's symmetry is considered an allowed trivialization. We end by noting that such a decoration is only possible as there are no reflection planes passing through the surface. 

We note that although the surface trivialization argument may appear to be independent of the bubble equivalence principle, it can in fact be viewed as a different manifestation of the same idea. One can imagine nucleating a symmetric bubble of \( p+ip \) superconductor at the center of the 3D bulk and expanding it outward toward the boundary in a symmetry-respecting manner. In the spinless case, consistency with \( C_2 \) symmetry requires that the two points where the \( C_2 \) axis pierces the bubble host Majorana zero modes. As the bubble reaches the boundary, the Majorana zero modes -- originating from the endpoints of Majorana chain decorations -- can be canceled (or trivialized) by the zero modes of the \( p+ip \) superconductor. 

One can likewise nucleate a bubble of Chern insulator that respects the point-group symmetry and then expand it until it meets the surface. In the spinless case this ``bubble equivalence’’ has an interesting effect: it cancels any complex-fermion decoration at the center. A Chern insulator may be viewed as a completely filled Landau level on a sphere. For spinless fermions each Landau level has an odd degeneracy \((2\ell+1)\); filling it therefore adds an odd number of electrons and flips the local fermion parity. Nucleating the Chern insulator bubble thus changes the fermion parity at the origin and annihilates the central complex-fermion decoration. By contrast, for spin-1/2 fermions, each Landau level is even-degenerate, forming half-integer spin representations of the SO(3) spatial rotation symmetry. As a result, filling a Landau level does not change the fermion parity. Consequently, the same bubble equivalence argument does \emph{not} trivialize the decoration in the spin-1/2 case. Furthermore, if reflection symmetry is present, there is no consistent way to generate a Chern bubble, since reflection reverses orientation and maps the Chern number to its negative. Consequently, this trivialization procedure is forbidden under reflection symmetry.

\subsection{Decohered Spin-1/2 ASPTs}\label{subsec:elems_dec_spinful_aspt}
We have described all relevant decorations in the decohered spinless case up to 3D. We will now move on to the decohered spin-1/2 case. 

\subsubsection{Decoration and obstruction-free conditions}

We start by identifying the nontrivial decorations on 0D blocks. The onsite symmetry group is $G_b \times_{\omega_2^f} \bzt^f$ with nontrivial $\omega_2^f$. In the clean case, the domain wall decorations are
\begin{align}
    n_0 &\in h^1(\bzt^f) = \bzt \\
    \nu_1 &\in \mathcal{H}^1(G_b,U_T(1)),
\end{align}
with the nontrivial extension imposing the obstruction in the bosonic layer
\begin{align}
    d\nu_1 = (-1)^{n_0\cup\omega_2}
\end{align} 
which forbids odd fermion decoration for spin-1/2. 
However, in the decoherence setting, the bosonic decorations and obstructions vanish (see discussion in Sec.~\ref{subsec:intrinsic_aspts}) and hence the odd fermion state is an allowed (and the only allowed) nontrivial state. As this state is only allowed in the decohered setting, it is by definition an intrinsic decoration.

For 1D blocks, we can generalize the discussion from Sec.~\ref{subsec:intrinsic_aspts}. The domain-wall decorations in the decohered setting are 
\begin{align}
    n_0 &\in h^2(\bzt^f) = \bzt \\
    n_1 &\in \mathcal{H}^1(G_b,\bzt)
\end{align}
with the obstruction condition
\begin{equation}
    dn_1 = \omega_2 \cup n_0
\end{equation}
As $\omega_2$ is nontrivial in the spin-1/2 case, this obstruction forbids nontrivial $n_0$ decoration, namely the Majorana chain. However, there is no obstruction corresponding to the $n_1$ decoration. Note, in the clean case, this would be generally obstructed in the bosonic layer (see Eq.~\ref{eq:obs_1d_nu2}). Hence, this decoration ends up being intrinsic in the spin-1/2 case.

Obstruction-free check for intrinsic ASPT decoration is subtle. In the spinless case, we could construct the $n_1$ phase (such as the 1D $\bzt$ fSPT) by stacking Majorana chains with additional symmetry constraints. This allowed us to write the 0D edge state as a collection of Majorana modes, from which the commutation relation between fermion parity and the $G_{0D}$ symmetry could be employed as an obstruction condition.
However, such a decoration does not work for intrinsic phases. For example, when \(G_b = \bzt\), the intrinsic-\(\bbz_4^f\) phase cannot be written as two Majorana chains. Indeed one can check that the mass term $i\gamma_1\gamma_2$ commutes with \(G_{\bbz_4^f}: \gamma_1\rightarrow\gamma_2,\ \gamma_2\rightarrow -\gamma_1\).  

However, we note that the 1D $n_1$ layer in the spin-1/2 case has the same obstruction-free condition as the spinless case with decoherence. This is due to the fact that the only difference between the two cases lies in the obstruction condition in the bosonic layer,
\begin{equation}
    d\nu_2 = (-1)^{\omega_2\cup n_1},
\end{equation}
which is trivialized by decoherence. This mathematical equivalence implies that any obstruction-free decoration must have the same $n_1$ data as in the spinless case -- although the resulting phases, while sharing the same decoration pattern, carry distinct physical interpretations. Specifically, we claim that to check if intrinsic $\bbz_4^f$ ASPT decoration on a set of blocks $\tau$ is obstructed at a 0D block $\mu$, it suffices to check if the $\bzt$ fSPT decoration was obstructed for the spinless system. Hence, we can once again use the odd/even criterion we had established in Sec. \ref{subsec:elem_0d1d_decorations}.

Now we consider 2D decorations. For 3D point groups, the only nontrivial spatial symmetry that can act onsite is $\bzt$.
In the spin-1/2 case, there are no nontrivial decorations on 2D blocks with onsite $\bzt$ symmetry. To see why this is the case, we consider the obstruction function (following Ref.~\cite{QingRuiWangPRX2020}) for each layer. 
\begin{enumerate}
    \item $n_1$ layer: $dn_1 = \omega_2 \cup n_0 \Rightarrow$ Obstruction-free: $n_0 = 0$
    \item $n_2$ layer: $dn_2 = \omega_2 \cup n_1 \Rightarrow$ Obstruction-free: $n_1 = 0$
    \item $\nu_3$ layer: $d\nu_3 = \mathcal{O}_4[n_2]$. However, since this is the obstruction condition on the bosonic layer, it vanishes under decoherence. Hence, $n_2$ is seemingly obstruction-free. However, it turns out that this can be trivialized as shown in Ref.\cite{QingRuiWangPRX2020}.
\end{enumerate}

2D blocks with no onsite spatial symmetry can be decorated with layers of $p+ip$ superconductor. An important distinction from the spinless case is that this decoration is now compatible with $n$-fold rotation. Hence, if we have a set of 2D blocks with no onsite symmetry that transform to each other under rotation, a $p+ip$ superconductor decoration would be allowed in the spin-1/2 case, but not in the spinless case. Hence, the only obstruction condition for such a decoration is a chiral anomaly on the surrounding 1D blocks. 

\subsubsection{Trivialization}
As with ASPT decorations on the blocks, the conditions for trivialization by bubble equivalence also differ for spin-1/2 systems. This is expected as the possible bubble states that can be inserted on a $d$-dimensional block with onsite symmetry $G_b^{ave}\times_{\omega_2^f} \bzt^f$ are determined by the $d-1$ ASPTs protected by the same symmetry.

This distinction does not emerge on one-dimensional blocks: even in the spin-1/2 case, the 1D fermion bubble is the only possible candidate since the odd fermion parity state is the nontrivial 0D ASPT for any average $G_b$. As in the spinless case, the odd fermion parity state on a 0D block surrounded by an odd number of 1D blocks can be trivialized by this bubble equivalence.

Next, we consider possible 2D bubbles. In the case where the 2D block does not have any spatial symmetry acting onsite, there is no nontrivial extension, and hence we can decorate the Majorana bubble. But while the bubble state remains the same, the set of decorations it can trivialize may differ from the spinless case.
For example, we find that this decoration does not trivialize the odd fermion parity state for point groups involving rotation symmetry, as the adiabatic transformation demonstrated in Sec.~\ref{subsec:p2_triv} is not compatible with spin-1/2 rotation. Furthermore, while this bubble could trivialize $\bzt$ fSPT decorations in the spinless case (as in the $pmm$ example in Sec.~\ref{subsec:pmm_triv}), it cannot trivialize any intrinsic $\bbz_4^f$ ASPT decorations as the latter cannot be constructed from stacking Majorana chains. 

From the above discussion it seems like the 2D Majorana bubble does not play any role in trivialization of decorations in spin-1/2 systems. However, this is not the case. If a 1D block $\tau$ is the common boundary of an odd number of $\sigma$ blocks (for example, when $\tau$ is the axis of 3-fold rotation), then the Majorana chain on $\tau$ will be trivialized by Majorana bubble on $\sigma$. 

Since spin-1/2 systems can be decorated with intrinsic onsite ASPTs, they can also have intrinsic bubble states. When the 2D block is a reflection plane, it has average onsite $\bbz_4^f$ symmetry and hence can host a 2D intrinsic $\bbz_4^f$-bubble. In analogy with the Majorana bubble, this can be thought of as the intrinsic $\bbz_4^f$ ASPT on the boundary of the 2D block with boundary conditions chosen such that it trivializes when shrunk to a point. If an odd number of these bubbles surround a 1D block $\tau$, it can trivialize the intrinsic ASPT decoration on $\tau$. Importantly, this is the only bubble that can be inserted on reflection planes: the ordinary Majorana bubble is forbidden since the Majorana chain is not compatible with onsite (weak or strong) $\bbz_4^f$ symmetry.

3D blocks, having no onsite spatial symmetry, can only host the 3D $p+ip$ bubble. Recall that in the spinless case, this was responsible for an equivalence between 2 copies of p+ip superconductor and the fermionic Levin-Gu state on blocks with onsite $\bzt\times\bzt^f$ symmetry. However, since there is no nontrivial 2D $\bbz_4^f$ block state in the decohered system, this bubble is inconsequential to the classification.

Moreover, the surface trivialization of Majorana chains decorated along the rotation axis does not apply in the spin-1/2 case. This is because a \( p+ip \) superconductor compatible with rotational symmetry in the spin-1/2 setting does not host a Majorana zero mode at the rotation center. As a result, it cannot trivialize the zero modes arising from Majorana chain decorations along the rotation axis. One might also consider a Chern insulator bubble. However, for spin-1/2 fermions, this does not trivialize configurations with odd fermion parity decoration, since Landau levels in this case have even degeneracy, as discussed in Sec.~\ref{sec:elems_spinless_bubble}. Therefore, these bubble equivalences do not impose additional trivialization conditions in the spin-1/2 setting.

\subsubsection{Non-trivial stacking relation}\label{subsec:elems_dec_spinful_stack}
We have discussed the subtle distinction for the obstruction-free decorations and trivializations for the decohered spin-1/2 case. These give us the root phases of the ACSPT classification. However, some of these root phases can be related each other by stacking, which gives rise to nontrivial extension of the classification. This occurs for the spin-1/2 cases, and we will consider the spin-1/2 $p2$ group as an example.

For the 0D block, the decoration is complex fermion decoration, just as in the spinless case. A key distinction is that this is de facto an intrinsic decoration, as it is forbidden in the clean case. Since 0D blocks have no boundary, the decoration is naturally obstruction free. Furthermore, the 2D Majorana bubble on $\sigma$ does not trivialize any of these decorations. Hence we obtain a $\bzt^4$ classification of 0D block states from complex fermion decorations on any of $\mu_{1,2,3,4}$. 

The 1D blocks $\tau_{1,2,3}$ have no onsite symmetry and hence the only possible decoration is Majorana chain. Let us consider decorating Majorana chain on $\tau_1$ which leaves two Majorana modes $\gamma^L, \gamma^R$ at $\mu_3$. The rotation symmetry compatible with spin-1/2 is $R: \gamma^L \rightarrow \gamma^R, \gamma^R\rightarrow -\gamma^L$. It is easy to verify that this transformation satisfies $R^2 = P_F: \gamma\rightarrow -\gamma$. The mass term $i\gamma^L\gamma^R$ then commutes with the parity. Therefore, the decoration is obstruction-free with decoherence. This logic generalizes to decoration on any of the $\tau$ blocks with the anomaly canceling on all the $\mu$ blocks. Furthermore, since each $\tau$ block is surrounded by two $\sigma$ blocks, this decoration cannot be trivialized by 2D Majorana bubble. The 1D classification is $\bzt^3$, corresponding to any configuration of Majorana chain decorations on $\tau_{1,2,3}$ blocks.

However, we will now show that these states are not independent. We find that the total classification is not simply a direct product of $E_{1d}^{1/2} = \bzt^3$ and $E_{0d}^{1/2}=\bzt^4$, but a nontrivial extension of these two groups. The nontrivial group extension emerges from a stacking rule: two copies of the Majorana chain decoration on a 1D block results in a complex fermion on the neighboring 0D blocks\cite{cheng2022rotation}. To see this, first consider Majorana chain decoration on one set of 1D blocks connected by a $\bzt$ rotation symmetry. In the context of $p2$, let us decorate $\tau_1$ with a Majorana chain, which leaves two Majorana modes at $\mu_1$ (and $\mu_2$), labeled by $\gamma^L$ and $\gamma^R$. These two majorana fermions can be viewed as coming from complex fermions $c^L,c^R$. For the sake of simplicity let us also label the adjacent Majoranas $\gamma^{LL},\gamma^{RR}$ (these fermions are gapped out with the bulk of the Majorana chain), defined by:
\begin{equation}
    c^L = \gamma^L+i\gamma^{LL}, c^R = \gamma^R+i\gamma^{RR} 
\end{equation}
The $\bzt$ symmetry of $\mu_1$ acts as $\gamma^L \rightarrow \gamma^R, \gamma^R\rightarrow-\gamma^L$, and similarly for the pairs $\gamma^{LL},\gamma^{RR}$. We can gap out the boundary Majorana fermions using $H=-i\gamma^L\gamma^R$. So this is an obstruction-free decoration. 

\begin{figure}[!htbp]
    \centering
    \includegraphics[width=0.7\linewidth]{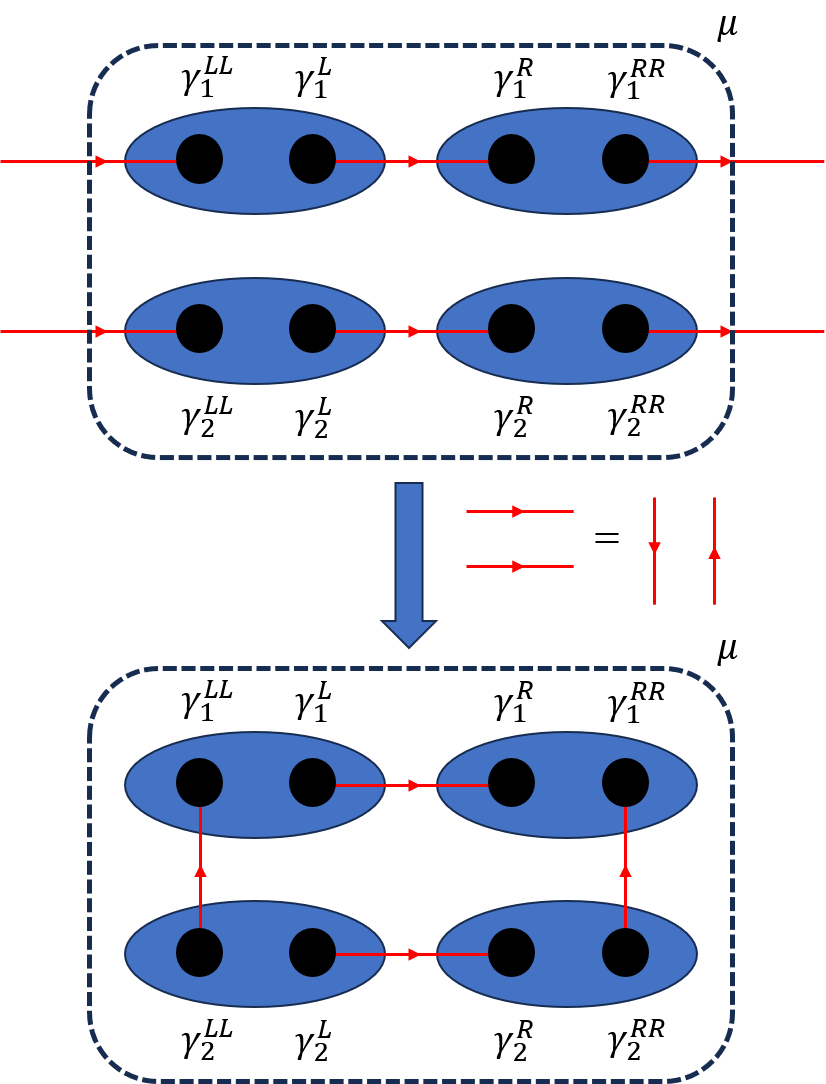}
    \caption{Two copies of Majorana chain transform to an odd fermion parity state on the rotation center.}
    \label{fig:stacking_twocopy_Majorana}
\end{figure}

Now let us consider two copies of this decoration, with two sets of the Majorana modes, indexed by subscript numbering: $\gamma\rightarrow\gamma_{1,2}$, as shown in Fig. \ref{fig:stacking_twocopy_Majorana}.
Understanding the state at $\mu$ becomes easier if we can restrict these 8 Majorana modes to pair among themselves. To achieve this, we use the adiabatic transformation that rotates Majorana pairings, introduced previously in Eq.~\ref{eq:Hadiabatic_theta},
to pair $i\gamma_1^{LL}\gamma_2^{LL}=1$, thus freeing them from their bindings to fermions in $\tau$. Note that this adiabatic transformation is valid since there is no symmetry between the two copies. The $\bzt$ symmetry of rotation then constrains $i\gamma_1^{RR}\gamma_2^{RR}=1$. We can now compute the total fermion parity of this state:
 \begin{align}
&i\gamma_1^{LL}\gamma_2^{LL}=i\gamma_1^{RR}\gamma_2^{RR}=i\gamma_1^L\gamma_1^R=i\gamma_2^L\gamma_2^R = 1 \\
\Rightarrow &P_f = (i\gamma_1^{LL}\gamma_1^{L})(i\gamma_1^{R}\gamma_1^{RR})(i\gamma_2^{RR}\gamma_2^{R})(i\gamma_2^{L}\gamma_2^{LL}) = -1
 \end{align}
To interpret this more physically, the new pairings (Fig.~\ref{fig:stacking_twocopy_Majorana} 
form a Majorana chain with \emph{periodic} boundary conditions after an even number of link flips and hence is an odd fermion parity state. Similar argument works for the other 0D block $\mu_2$. The rest of the Majorana chain can shrink to vacuum. Therefore, two copies of the Majorana chain can be deformed into a root state with $\mu_1$ and $\mu_2$ decorated with complex fermion. This concludes that the classification is $\bbz_4$ instead of $\bbz_2\times \bbz_2$.

In the case of $p2$, two copies of the Majorana chain on $\tau_1$ leaves odd fermion state on $\mu_2$ as well as $\mu_1$. Similarly, two copy decoration on $\tau_2$ leaves odd fermion state on $\mu_1$ and $\mu_3$, and decoration on $\tau_3$ leaves odd fermion state on $\mu_2$ and $\mu_4$. Therefore, we cannot reach a single 0D block decoration by the 1D decoration. As a result, the classification becomes $\bzt\times\bbz_4^3$, where the $\bzt$ corresponds to the single 0D block decoration.

We note that this interesting stacking phenomenon only occurs for the case of spin-1/2 fermions. 

\subsection{Disordered ASPTs}

\subsubsection{Decoration and obstruction-free conditions}

We now turn to the cases of disordered systems, treating the spinless and spin-1/2 scenarios in a unified framework. We will distinguish between the two cases as needed throughout the discussion to clarify their key differences.

\noindent\textbf{0D block decoration} -- The main distinction for the disordered systems from the decohered systems is that charge decorations on 0D defects are always trivialized under disorder due to localization physics. As discussed in Ref.~\cite{Ma2025ASPT2}, we can always shift the position of a charge by symmetrically changing the potential landscape for the charge without disrupting the localization of the system. Mathematically, this means that the layer corresponding to $\mathcal{H}^d(G_b,h^1(\bzt^f))$ vanishes. As a consequence, there are no nontrivial decorations on 0D block states, since the odd fermion parity is a 0D decoration of $\bzt^f$ charge.

The same localization physics implies that the obstruction condition for the \( n_d \) layer is also lifted, as this obstruction—arising from charge inconsistency during the deformation of symmetry defects—is effectively trivialized in the presence of disorder.

The above argument is valid for both spinless and spin-1/2 cases.

\noindent\textbf{1D block decoration} -- In the clean case, 1D blocks admit three layers of decoration: \( n_0 \), \( n_1 \), and \( \nu_2 \). Recall that \( \nu_2 \), which captures Berry phase contributions, becomes trivial in the mixed-state setting, as mixed states do not support nontrivial Berry phases. In the presence of disorder, the \( n_1 \) layer—corresponding to decorations of odd fermion parity on 0D \( G_b \)-domain walls—is also trivialized due to localization effects. As a result, the only remaining nontrivial decoration layer is \( n_0 \in \mathcal{H}^1(\mathbb{Z}_2^f) \). Both the spinless and spin-1/2 cases support nontrivial 1D states arising from this \( n_0 \) layer, though they differ in a subtle yet important way. We now elaborate on this difference.

In the spinless case, the \( n_0 \) layer corresponds directly to the Majorana chain. The on-site bosonic symmetry on the 1D block acts trivially on this decoration. This is the only nontrivial decoration in this case, implying that the only possible obstruction at a 0D intersection is the presence of an odd number of Majorana zero modes. Therefore, to determine whether the decoration is obstructed, we simply count the number of Majorana zero modes localized at the 0D intersections. If an odd (even) number of adjacent 1D blocks are decorated with Majorana chains, the configuration is obstructed (obstruction-free).

The spin-1/2 case requires a more subtle analysis. In both the clean and decohered settings, the ordinary Majorana chain -- arising from the \( n_0 \) layer -- is obstructed due to the \( dn_1 \) condition given in Eq.~(\ref{eq:obs_1d_nu2}). However, as noted earlier, this \( n_1 \) layer obstruction vanishes in the presence of disorder. As a result, a 1D intrinsic ASPT phase can emerge as a valid block-state decoration. In the specific case where \( G_b = \mathbb{Z}_2 \) (so that \( G_f = \mathbb{Z}_4^f \)), this construction corresponds to the localization-enabled, compressible ASPT phase described in Ref.~\cite{Ma2025ASPT2}, which is physically realized by a disordered Kitaev chain.

While this state is distinct from the Majorana chain, they share the same layer of decoration. This proves very useful for our purposes of classification and construction since they have analogous obstruction and trivialization conditions. First, as $n_0 = h^2(\bzt^f) = \bzt$ describes both the spinless and spin-1/2 cases, the intrinsic state is also $\bzt$-classified. Second, the obstruction-free condition in the crystalline system is that there must be an even number of $n_0$ decorations surrounding a 0D block. This is analogous to the spinless condition that there must be an even number of Majorana modes at the 0D center, as both conditions emerge from the same cohomological data. However, we crucially note that this still describes a different physical construction from the Kitaev chain as the state requires disorders to stabilize a short-range entangled (localized) bulk. 

\noindent\textbf{2D block decoration} -- Just as with the 1D case, the 2D decorations have similar layer-wise classification but are physically distinct states. 
If the 2D block has no onsite symmetry, the classification is the $p+ip$ SC with $\bbz$ classification (for both spinless and spin-1/2). This decoration may be obstructed by chiral anomaly at the 1D edge. Furthermore, just as in the clean and decohered systems, the $p+ip$ SC is incompatible (compatible) with rotational symmetry in the spinless (spin-1/2) system.

In the disordered spinless case, a 2D block with weak onsite $\bzt$ symmetry can also be decorated with the fermionic Levin-Gu state, just as in the decohered setting. We recall that this corresponds to the $n_1$ layer in the cohomological data, describing 1D domain walls decorated with Majorana chains. However, the distinction from the decohered systems is that the fermionic Levin-Gu state in the disordered setting is $\bzt$ classified instead of $\bbz_4$. The physical reasoning behind this claim is that the decoration pattern of 2 copies of fermionic Levin-Gu state is 0D complex fermion decoration\cite{QingRuiWangPRX2020}, which is trivialized by localization physics. 

In the spin-1/2 case, a nontrivial state emerges under disorder, which we call the disorder-intrinsic $\bbz_4^f$ fSPT state. While this state also appears in the $n_1$ layer of the classification, it is not the fLG state, as the $p\pm ip$-SC edge state can be gapped with $\bbz_4^f$-symmetric mass term in the K-matrix formalism. Furthermore, the same decoration pattern is obstructed in the decohered setting as well, by the condition
\begin{equation}
    dn_2 = \omega_2\cup n_1
\end{equation}
since $n_1=1$ (nontrivial decoration) and $\omega_2=1$ (spin-1/2) lead to a nontrivial value $dn_2 = 1 \pmod 2$. Nonetheless, this obstruction vanishes in the disordered system; thus, the $n_1$ decoration is a valid decoration.

As a consequence, the obstruction condition at the 1D edge is also simple. For a set of decorated 2D blocks $\{\sigma\}$ decorated with this state, the only condition at the shared 1D edge is if the number of $\sigma$ blocks is odd (obstructed) or even (obstruction-free). We note that the same applies for the spinless decoration, namely a single copy of the fLG state. If a 1D block is surrounded by an even number of fLG decorations, it is anomaly-free in the disordered system, as the anomaly indicator for each $\bzt$ symmetry gives $\nu=0\pmod 2$. The mathematical justification of the similarity for both cases emerges from the trivialization of the $dn_2$ condition.
This is analogous to the 1D case, where the disorder-intrinsic $\bbz_4^f$ state has identical obstruction conditions as the Majorana chain.

\subsubsection{Trivializations}
For 1D blocks, there are no possible bubbles as there are no nontrivial 0D states in the disordered setting.

As we have discussed above, the only nontrivial 1D state is Majorana chain, independent of the onsite symmetry and spin.
Hence the only possible bubble on 2D blocks is the anti-PBC Majorana chain on the 1D boundary. Just as in the clean and decoherence regimes, if a 1D block $\tau$ is surrounded by an odd number of 2D blocks $\{\sigma_i\}$, then the Majorana chain on $\tau$ is trivialized by the 2D Majorana bubble on $\sigma_i$. 

For 3D blocks, we have no onsite symmetry and hence the only possible decoration is the 3D $p \pm ip$ bubble discussed in Sec.~\ref{sec:elems_spinless_bubble}, just as in the decohered setting. Similarly, we can also have the open surface trivialization of edge Majorana modes in the spinless case, as discussed in the decohered setting. 

\subsection{Spectrum sequence method}

We also compute the results of ASPT classification, where the crystalline symmetries are replaced by onsite symmetries with the same group structure where the orientation reversing symmetries are treated as antiunitary onsite symmetries, and a modified $\omega_2$ (where spinless ACSPTs correspond to spin-1/2 onsite-symmetry ASPTs, and vise versa).
We conjecture that there is a crystalline equivalent principle similar to the case of clean SPTs~\cite{thorngren2018gauging}.
Therefore, the results of ASPT classification with onsite symmetries can be cross-checked with the results obtained in this paper.

As reviewed in Sec.~\ref{sec:review-aspt-onsite}, onsite-symmetry ASPTs can be constructed by decorating symmetry domains and domain-walls by invertible topological states.
In particular, when $\mathbb Z_2^f$ is the only exact symmetry, the decorated invertible topological states are the ones protected by $\mathbb Z_2^f$ only.
Mathematically, such decorations, and therefore the ASPTs, are classified by a generalized cohomology theory, which can be viewed as a modified version of the generalized cohomology theory for clean SPTs.
To reveal the domain-decoration picture, the generalized cohomology theory for clean SPTs, which we denote by $h^D(G_f)$, can be computed using an Atiyah-Hirzebruch spectral sequence (AHSS),
\begin{equation}
    \label{eq:ahss}
    E_2^{pq}\simeq \mathcal{H}^p(G_b, h^q(\mathbb Z_2^f))\Rightarrow h^{p+q}(G_f),
\end{equation}
where $h^q(\mathbb Z_2^f)$ denotes the invertible topological orders protected by $\mathbb Z_2^f$ only in $q$ space-time dimensions.
In particular, the cohomology group $\mathcal{H}^p(G_b, h^q(\mathbb Z_2^f))$ denotes the decoration of $q$-dimensional invertible states on symmetry domain (domain-walls) of codimension $p$.
Eq.~\eqref{eq:ahss} only shows the results on the second page of the spectral sequence: to compute the higher pages and eventually $h^D(G_f)$, one needs to use the higher derivatives of the spectral sequence, which have been computed in Ref.~\cite{QingRuiWangPRX2020}.

When $G_b$ becomes average, we no longer consider the bosonic-phase layer of the decoration for the decoherence case, and both the bosonic-phase and the complex-fermion layers for the disorder case.
This is achieved by setting $h^0(\mathbb Z_2^f)=0$ for the decoherence case and $h^0(\mathbb Z_2^f)=h^1(\mathbb Z_2^f)=0$ for the disorder case, respectively, in the AHSS.
This means that we not only discard the corresponding layers when counting possible decorations, but also ignore potential obstructions that fall into the corresponding layer.
The higher-page calculations are then carried out with the same higher derivatives as in the clean case.
We perform the calculation using a package~\cite{SptSet} for computing the spectral sequence in \eqref{eq:ahss} implementing the algorithm described in Ref.~\cite{Ouyang2021}, which is based on the GAP program~\cite{GAP4} and the HAP package~\cite{HAP}.

\section{Results}
\label{sec:results}
We list our classification results in this section. In the tables, if the state is an intrinsic ACSPT, we label them in blue. In this section, we only list results. For more details for each case, we have an extensive appendix treating each case.

The results for the 2D wallpaper group are presented in Tables \ref{tab:2d_decohered} (decohered) and \ref{tab:2d_disordered} (disordered). 

The results for the decohered and disordered classification in 3D are presented in Tables \ref{tab:3d_decohered} and \ref{tab:3d_disordered} respectively. 

\begin{table}[!htbp]   
\begin{center}
\begin{tabular}{|c|c||c|c|c||c|c|c|}
\hline
    \multicolumn{8}{|c|}{Decohered 2D crystal ASPTs}\\ \hline
    &  & \multicolumn{3}{|c||}{Spinless} & \multicolumn{3}{|c|}{Spin-1/2} \\ \hline
    & $G_b$ & $E_{0}^{1D}$ & $E_{0}^{0D}$ & $\mathcal{G}_{0,dec}$  & $E_{1/2}^{1D}$ & $E_{1/2}^{0D}$ & $\mathcal{G}_{1/2,dec}$\\
    \hline \hline
     1 & p1 & $\bzt^2$ & $\bzt$ & $\bzt^3$ & $\bzt^2$ & $\bzt$ & $\bzt^3$\\ \hline
     2 & p2 & $\bbz_1$ & $\bzt^3$ & $\bzt^3$ & $\bzt^3$ & $\anb{\bzt^4}$ & $\anb{\bzt\times\bbz_4^3}$\\ \hline
     3 & pm & $\bzt^3$ & $\bzt^2$ & $\bzt^5$ & $\bzt \times \anb{\bzt^2}$ & $\anb{\bzt^2}$ & $\anb{\bzt^3\times\bbz_4}$\\ \hline
     4 & pg & $\bzt^2$ & $\bzt$ & $\bzt^3$& $\bzt^2$ & $\bzt$ & $\bzt^3$ \\ \hline
     5 & cm & $\bzt^2$ & $\bzt$ & $\bzt^3$ & $\bzt \times\anb{\bzt}$ & $\anb{\bzt}$ & $\bzt\times\anb{\bzt^2}$\\ \hline
     6 & pmm & \anb{$\bzt^3$} & $\bzt^4$ & $\bzt^4\times\anb{\bzt^3}$ & $\anb{\bzt^4}$ & $\anb{\bzt^4}$ & $\anb{\bzt^8}$\\ \hline
     7 & pmg & $\bzt^2$ & $\bzt^2$ & $\bzt^4$ & $\bzt^2\times\anb{\bzt}$ & $\anb{\bzt^3}$ & $\anb{\bzt^2\times\bbz_4^2}$ \\ \hline
     8 & pgg & $\bzt$ & $\bzt$ & $\bzt^2$ & $\bzt^2$ & $\anb{\bzt^2}$ & $\bzt\times \anb{\bzt\times\bbz_4}$ \\ \hline
     9 & cmm & \anb{$\bzt^2$} & $\bzt^2$ & $\bzt^2\times\anb{\bzt^2}$ & $\bzt\times\anb{\bzt^2}$ & $\anb{\bzt^3}$ & $\anb{\bzt^4 \times \bbz_4}$ \\ \hline
     10 & p4 & $\bbz_1$ & $\bzt^2$ & $\bzt^2$& $\bzt^2$ & $\anb{\bzt^3}$ & $\bzt\times\anb{\bzt^2\times\bbz_4}$ \\ \hline
     11 & p4m & \anb{$\bzt^2$} & $\bzt^3$ & $\bzt^2\times\anb{\bzt^3}$ & $\anb{\bzt^3}$ & $\anb{\bzt^3}$ & $\anb{\bzt^6}$\\ \hline
     12 & p4g & $\bzt$ & $\bzt$ & $\bzt^2$ & $\bzt\times\anb{\bzt}$ & $\anb{\bzt^2}$ & $\bzt\times\anb{\bzt^3}$\\ \hline
     13 & p3 & $\bbz_1$ & $\bzt$ & $\bzt$& $\bbz_1$ & $\anb{\bzt}$ & $\anb{\bzt}$\\ \hline
     14 & p3m1 & $\bzt$ & $\bzt$ & $\bzt^2$ & $\anb{\bzt}$ & $\anb{\bzt}$ & $\anb{\bzt^2}$\\ \hline
     15 & p31m & $\bzt$ & $\bzt$ & $\bzt^2$& $\anb{\bzt}$ & $\anb{\bzt}$ & $\anb{\bzt^2}$  \\ \hline
     16 & p6 & $\bbz_1$ & $\bzt$  & $\bzt$ & $\bzt$ & $\anb{\bzt^2}$  & $\anb{\bzt\times\bbz_4}$ \\ \hline
     17 & p6m & \anb{$\bzt$} & $\bzt^2$ & $\bzt^2\times\anb{\bzt}$ & $\anb{\bzt^2}$ & $\anb{\bzt^2}$ & $\anb{\bzt^4}$ \\ \hline
    \end{tabular}
    \caption{Classification table for spinless and spin-1/2 fermions with decohered spatial symmetry. Cases with intrinsic decorations are labeled in blue.}
    \label{tab:2d_decohered}
\end{center}
\end{table}

\begin{table}[!htbp]
    \begin{center}
        \begin{tabular}{|c|c||c||c|}
            \hline
            \multicolumn{4}{|c|}{Disordered 2D crystal ASPTs}\\ \hline
            &  & Spinless & Spin-1/2 \\ \hline
            & $G_b$ & $\mathcal{G}_{0,dis}^{1D}$ & $\mathcal{G}_{1/2,dis}$ \\ \hline
            1 & p1 & $\bzt^2$ & $\bzt^2$\\ \hline
            2 & p2 & $\anb{\bzt^3}$ & $\bzt^3$\\ \hline
            3 & pm & $\bzt^2\times\anb{\bzt}$ & $\bzt\times\anb{\bzt^2}$ \\ \hline
            4 & pg & $\bzt^2$&$\bzt^2$ \\ \hline
            5 & cm & $\bzt^2$ & $\bzt\times\anb{\bzt}$\\ \hline
            6 & pmm & $\anb{\bzt^4}$ & $\anb{\bzt^4}$ \\ \hline
            7 & pmg & $\bzt\times\anb{\bzt^2}$ & $\bzt^2\times\anb{\bzt}$\\ \hline
            8 & pgg & $\bzt\times\anb{\bzt}$ & $\bzt^2$\\ \hline
            9 & cmm & $\anb{\bzt^3}$&$\bzt\times\anb{\bzt^2}$ \\ \hline
            10 & p4 & $\anb{\bzt^2}$ & $\bzt^2$ \\ \hline
            11 & p4m & $\anb{\bzt^3}$ & $\anb{\bzt^3}$ \\ \hline
            12 & p4g & $\anb{\bzt^2}$ & $\bzt\times\anb{\bzt}$ \\ \hline
            13 & p3 & $\bbz_1$ & $\bbz_1$ \\ \hline
            14 & p3m1 & $\bzt$ & $\anb{\bzt}$ \\ \hline
            15 & p31m & $\bzt$ & $\anb{\bzt}$ \\ \hline
            16 & p6 & $\anb{\bzt}$ & $\bzt$\\ \hline
            17 & p6m & $\anb{\bzt^2}$ & $\anb{\bzt^2}$\\ \hline
        \end{tabular}
        \caption{Classification of 2D wallpaper groups with disordered spatial symmetries. Spinless and spin-1/2 classifications are identical for 1D classification, and 0D classification is trivial.}
        \label{tab:2d_disordered}
    \end{center}
\end{table}

\begin{table*}[!htbp]
\begin{center}
    \begin{tabular}{|c|c||c|c|c|c||c|c|c|c|}
        \hline 
    \multicolumn{10}{|c|}{Decohered 3D crystal ASPTs}\\ \hline
    &  & \multicolumn{4}{||c||}{Spinless} & \multicolumn{4}{|c|}{Spin-1/2} \\ \hline
    & $G_b$ & $E_{0}^{2D}$ & $E_{0}^{1D}$ & $E_{0}^{0D}$ & $\mathcal{G}_{0}$ & $E_{1/2}^{2D}$ & $E_{1/2}^{1D}$ & $E_{1/2}^{0D}$ & $\mathcal{G}_{1/2}$\\
    \hline \hline 
    1 & $C_1$ & $\bbz_1$ & $\bbz_1$ & $\bbz_1$ & $\bbz_1$ & $\bbz_1$ & $\bbz_1$ & $\bbz_1$ & $\bbz_1$\\ \hline
    2 & $C_i$ & $\bbz_1$ & $\bbz_1$ & $\bbz_1$ & $\bbz_1$& $\bbz_1$ & $\bbz_1$ & $\bbz_1$ & $\bbz_1$\\ \hline
    3 & $C_2$ & $\bbz_1$ & $\bbz_1$ & $\bbz_1$ & $\bbz_1$ & $\bbz_1$ & $\anb{\bzt}$ & $\bbz_1$ & $\anb{\bzt}$\\ \hline 
    4 & $C_{1h}$ & $\bbz_8$ & $\bbz_1$ & $\bbz_1$&$\bbz_8$ & $\bbz_1$ & $\bbz_1$ & $\bbz_1$ & $\bbz_1$\\ \hline    
    5 & $C_{2h}$ & $\bbz_4$ &\anb{$\bzt$} &$\bbz_1$ & $\bbz_4\times\anb{\bzt}$ & $\bbz_1$ & $\anb{\bzt^2}$ & $\anb{\bzt}$ & $\anb{\bzt^3}$\\ \hline
    6 & $D_2=V$ & $\bbz_1$ & $\anb{\bzt}$ & $\bbz_1$ & $\anb{\bzt}$ & $\bbz_1$ & $\anb{\bzt^3}$ & $\anb{\bzt}$ & $\anb{\bzt^4}$\\ \hline
    7 & $C_{2v}$ & $\anb{\bzt^2}$ & $\bzt$ & $\bbz_1$ & $\bzt\times\anb{\bzt^2}$ & $\bbz_1$ & $\anb{\bzt^2}$ & $\bbz_1$ & $\anb{\bzt^2}$\\ \hline
    8 & $D_{2h}=V_h$ & \anb{$\bzt^3$} & $\anb{\bzt^3}$ & $\bzt$ & $\bzt\times\anb{\bzt^6}$ & $\bbz_1$ & $\anb{\bzt^6}$ & $\anb{\bzt}$ & $\anb{\bzt^7}$\\ \hline 
    9 & $C_4$ & $\bbz_1$ & $\bbz_1$ & $\bbz_1$ & $\bbz_1$ & $\bbz_1$ & $\anb{\bzt}$ & $\bbz_1$ & $\anb{\bzt}$\\ \hline
    10 & $S_4$ & $\bbz_1$ & $\bzt$ &$\bbz_1$ &$\bzt$ & $\bzt$ & $\anb{\bzt}$ & $\bbz_1$ & $\bzt\times\anb{\bzt}$ \\ \hline
    11 & $C_{4h}$ & $\bbz_4$ & $\anb{\bzt}$ & $\bbz_1$ & $\bbz_4\times\anb{\bzt}$ & $\bbz_1$ & $\anb{\bzt^2}$ & $\anb{\bzt}$ & $\anb{\bzt^3}$\\ \hline
    12 & $D_4$ & $\bbz_1$ & $\bzt$ & $\bbz_1$ & $\bzt$ & $\bbz_1$ & $\anb{\bzt^3}$ & $\anb{\bzt}$ & $\anb{\bzt^4}$\\ \hline
    13 & ${C_{4v}}$ & $\anb{\bzt^2}$ & $\bzt$ & $\bbz_1$ & $\bzt\times\anb{\bzt^2}$ & $\bbz_1$ & $\anb{\bzt^2}$ & $\bbz_1$ & $\anb{\bzt^2}$\\ \hline
    14 & ${D_{2d}=V_d}$ & $\anb{\bzt}$ & $\anb{\bzt}$ & $\bzt$ & $\anb{\bzt\times\bbz_4}$ & $\bbz_1$ & $\anb{\bzt^3}$ & $\bbz_1$ & $\anb{\bzt^3}$\\ \hline
    15 & ${D_{4h}}$ & $\anb{\bzt^3}$ & $\anb{\bzt^3}$ & $\bzt$ & $\bzt\times\anb{\bzt^6}$ & $\bbz_1$ & $\anb{\bzt^6}$ & $\anb{\bzt}$ & $\anb{\bzt^7}$\\ \hline
    16 & $C_3$ & $\bbz_1$ & $\bbz_1$ & $\bbz_1$ & $\bbz_1$ & $\bbz_1$ & $\bbz_1$ & $\bbz_1$ & $\bbz_1$\\ \hline
    17 & $S_6$ & $\bbz_1$ & $\bbz_1$ & $\bbz_1$ & $\bbz_1$ & $\bbz_1$ & $\bbz_1$ & $\bbz_1$ & $\bbz_1$\\ \hline
    18 & $D_3$ & $\bbz_1$ & $\bbz_1$ & $\bbz_1$ & $\bbz_1$ & $\bbz_1$ & $\anb{\bzt}$ & $\bbz_1$ & $\anb{\bzt}$\\ \hline
    19 & $C_{3v}$ & $\bbz_8$ & $\bbz_1$ & $\bbz_1$ & $\bbz_8$ & $\bbz_1$ & $\bbz_1$ & $\bbz_1$ & $\bbz_1$\\ \hline
    20 & $D_{3d}$ & $\anb{\bzt}$ & $\anb{\bzt}$ & $\bzt$ & $\anb{\bzt\times\bbz_4}$ & $\bbz_1$ & $\anb{\bzt^2}$ & $\bbz_1$ & $\anb{\bzt^2}$\\ \hline
    21 & $C_6$ & $\bbz_1$ & $\bbz_1$ & $\bbz_1$ & $\bbz_1$ & $\bbz_1$ & $\anb{\bzt}$ & $\bbz_1$ & $\anb{\bzt}$\\ \hline
    22 & $C_{3h}$ & $\bbz_4$ & $\bbz_1$ & $\bbz_1$ & $\bbz_4$ & $\bbz_1$ & $\bbz_1$ & $\bbz_1$ & $\bbz_1$\\ \hline
    23 & $C_{6h}$ & $\bbz_4$ & $\bbz_1$ & $\bzt$ & $\bzt\times\bbz_4$ & $\bbz_1$ & $\anb{\bzt^2}$ & $\bbz_1$ & $\anb{\bzt^2}$ \\ \hline
    24 & $D_6$ & $\bbz_1$ & $\anb{\bzt}$ & $\bbz_1$ & $\anb{\bzt}$ & $\bbz_1$ & $\anb{\bzt^3}$ & $\anb{\bzt}$ & $\anb{\bzt^4}$\\ \hline
    25 & $C_{6v}$ & $\anb{\bzt^2}$ & $\bzt$ & $\bbz_1$ & $\bzt\times\anb{\bzt^2}$ & $\bbz_1$ & $\anb{\bzt^2}$ & $\bbz_1$ & $\anb{\bzt^2}$\\ \hline
    26 & $D_{3h}$ & $\anb{\bzt^2}$ & $\bzt$ & $\bbz_1$ & $\bzt\times\anb{\bzt^2}$ & $\bbz_1$ & $\anb{\bzt^2}$ & $\bbz_1$ & $\anb{\bzt^2}$\\ \hline
    27 & $D_{6h}$ & $\anb{\bzt^3}$ & $\anb{\bzt^3}$ & $\bzt$ & $\bzt\times\anb{\bzt^6}$ & $\bbz_1$ & $\anb{\bzt^6}$ & $\anb{\bzt}$ & $\anb{\bzt^7}$\\ \hline
    28 & $T$ & $\bbz_1$ & $\anb{\bzt}$ & $\bbz_1$ & $\anb{\bzt}$& $\bbz_1$ & $\anb{\bzt}$ & $\anb{\bzt}$ & $\anb{\bzt^2}$ \\ \hline
    29 & $T_h$ & \anb{$\bzt$} & $\anb{\bzt}$ & $\bzt$ & $\bzt\times\anb{\bzt^2}$ & $\bbz_1$ & $\anb{\bzt^2}$ & $\anb{\bzt}$ & $\anb{\bzt^3}$ \\ \hline
    30 & $T_d$ & $\anb{\bzt}$ & $\anb{\bzt}$ & $\bzt$ & $\anb{\bzt\times\bbz_4}$ & $\bbz_1$ & $\anb{\bzt}$ & $\anb{\bzt}$ & $\anb{\bzt^2}$ \\ \hline
    31 & $O$ & $\bbz_1$ & $\anb{\bzt}$ & $\bbz_1$ & $\anb{\bzt}$& $\bbz_1$ & $\anb{\bzt^2}$ & $\anb{\bzt}$ & $\anb{\bzt^3}$ \\ \hline
    32 & $O_h$ & $\anb{\bzt^2}$ & $\bzt\times\anb{\bzt}$ & $\bzt$ & $\bzt^2\times\anb{\bzt^3}$& $\bbz_1$ & $\anb{\bzt^4}$ & $\anb{\bzt}$ & $\anb{\bzt^5}$ \\ \hline
    \end{tabular}
    \caption{Classification data for the 32 point groups under decoherence. Data in \anb{blue} indicates presence of intrinsic ASPTs (which may not necessarily change the classification from the clean case). Details for each case can be found in Sec. \ref{sec:3d_class_details}.}
    \label{tab:3d_decohered}
    \end{center}
\end{table*}

\begin{table}[!htbp]
    \begin{center}
        \begin{tabular}{|c|c||c|c|c||c|c|c|}
            \hline 
        \multicolumn{8}{|c|}{Disordered 3D crystal ASPTs}\\ \hline
        &  & \multicolumn{3}{|c||}{Spinless} & \multicolumn{3}{|c|}{Spin-1/2} \\ \hline
        & $G_b$ & $E_{0}^{2D}$ & $E_{0}^{1D}$ &$\mathcal{G}_0$ & $E_{1/2}^{2D}$ & $E_{1/2}^{1D}$ & $\mathcal{G}_{1/2}$\\ \hline
        1 & $C_1$ & $\bbz_1$ & $\bbz_1$ & $\bbz_1$ & $\bbz_1$ & $\bbz_1$ & $\bbz_1$ \\ \hline
        2 & $C_i$ & $\bbz_1$ & $\bbz_1$ & $\bbz_1$ & $\bbz_1$ & $\bbz_1$ & $\bbz_1$ \\ \hline
        3 & $C_2$ & $\bbz_1$ & $\bbz_1$ & $\bbz_1$ & $\bbz_1$ & $\anb{\bzt}$ & $\anb{\bzt}$  \\ \hline
        4 & $C_{1h}$ & $\bbz_4$ & $\bbz_1$ & $\bbz_4$ & $\anb{\bzt}$ & $\bbz_1$ & $\anb{\bzt}$  \\ \hline
        5 & $C_{2h}$ & $\bzt$ & $\anb{\bzt^2}$ & $\bzt\times\anb{\bzt^2}$ & $\anb{\bzt}$ & $\anb{\bzt^2}$ & $\anb{\bzt^3}$  \\ \hline
        6 & $D_{2}=V$ & $\bbz_1$ & $\anb{\bzt^2}$ & $\anb{\bzt^2}$ & $\bbz_1$ & $\anb{\bzt^3}$ & $\anb{\bzt^3}$  \\ \hline
        7 & $C_{2v}$ & $\anb{\bzt^2}$ & $\bzt$ & $\bzt\times\anb{\bzt^2}$ & $\anb{\bzt^2}$ & $\anb{\bzt}$ & $\anb{\bzt^3}$  \\ \hline
        8 & $D_{2h}=V_h$ & $\anb{\bzt^3}$ & $\anb{\bzt^3}$ & $\anb{\bzt^6}$ & $\anb{\bzt^3}$ & $\anb{\bzt^3}$ & $\anb{\bzt^6}$  \\ \hline
        9 & $C_4$ & $\bbz_1$ & $\bbz_1$ & $\bbz_1$ & $\bbz_1$ & $\anb{\bzt}$ & $\anb{\bzt}$  \\ \hline
        10 & $S_4$ & $\bbz_1$ & $\anb{\bzt}$ & $\anb{\bzt}$ & $\bzt$ & $\anb{\bzt}$ & $\bzt\times\anb{\bzt}$  \\ \hline
        11 & $C_{4h}$ & $\bzt$ & $\anb{\bzt^2}$ & $\bzt\times\anb{\bzt^2}$ & $\anb{\bzt}$ & $\anb{\bzt^2}$ & $\anb{\bzt^3}$  \\ \hline
        12 & $D_4$ & $\bbz_1$ & $\anb{\bzt^2}$ & $\anb{\bzt^2}$ & $\bbz_1$ & $\anb{\bzt^3}$ & $\anb{\bzt^3}$  \\ \hline
        13 & $C_{4v}$ & $\anb{\bzt^2}$ & $\bzt$ & $\bzt\times\anb{\bzt^2}$ & $\anb{\bzt^2}$ & $\anb{\bzt}$ & $\anb{\bzt^3}$  \\ \hline
        14 & $D_{2d}=V_d$ & $\anb{\bzt}$ & $\anb{\bzt^2}$ & $\anb{\bzt^3}$ & $\anb{\bzt}$ & $\anb{\bzt^2}$ & $\anb{\bzt^3}$  \\ \hline
        15 & $D_{4h}$ & $\anb{\bzt^3}$ & $\anb{\bzt^3}$ & $\anb{\bzt^6}$ & $\anb{\bzt^3}$ & $\anb{\bzt^3}$ & $\anb{\bzt^6}$  \\ \hline
        16 & $C_3$ & $\bbz_1$ & $\bbz_1$ & $\bbz_1$ & $\bbz_1$ & $\bbz_1$ & $\bbz_1$  \\ \hline
        17 & $S_6$ & $\bbz_1$ & $\anb{\bzt}$ & $\anb{\bzt}$ & $\bbz_1$ & $\anb{\bzt}$ & $\anb{\bzt}$  \\ \hline
        18 & $D_3$ & $\bbz_1$ & $\bbz_1$ & $\bbz_1$ & $\bbz_1$ & $\anb{\bzt}$ & $\anb{\bzt}$  \\ \hline
        19 & $C_{3v}$ & $\bbz_4$ & $\bbz_1$ & $\bbz_4$ & $\anb{\bzt}$ & $\bbz_1$ & $\anb{\bzt}$  \\ \hline
        20 & $D_{3d}$ & $\anb{\bzt}$ & $\anb{\bzt^2}$ & $\anb{\bzt^3}$ & $\anb{\bzt}$ & $\anb{\bzt^2}$ & $\anb{\bzt^3}$  \\ \hline
        21 & $C_6$ & $\bbz_1$ & $\bbz_1$ & $\bbz_1$ & $\bbz_1$ & $\anb{\bzt}$ & $\anb{\bzt}$  \\ \hline
        22 & $C_{3h}$ & $\bzt$ & $\bbz_1$ & $\bzt$ & $\anb{\bzt}$ & $\bbz_1$ & $\anb{\bzt}$  \\ \hline
        23 & $C_{6h}$ & $\bzt$ & $\anb{\bzt^2}$ & $\bzt\times\anb{\bzt^2}$ & $\anb{\bzt}$ & $\anb{\bzt^2}$ & $\anb{\bzt^3}$  \\ \hline
        24 & $D_6$ & $\bbz_1$ & $\anb{\bzt^2}$ & $\anb{\bzt^2}$ & $\bbz_1$ & $\anb{\bzt^3}$ & $\anb{\bzt^3}$  \\ \hline
        25 & $C_{6v}$ & $\anb{\bzt^2}$ & $\bzt$ & $\bzt\times\anb{\bzt^2}$ & $\anb{\bzt^2}$ & $\anb{\bzt}$ & $\anb{\bzt^3}$  \\ \hline
        26 & $D_{3h}$ & $\anb{\bzt^2}$ & $\bzt$ & $\bzt\times\anb{\bzt^2}$ & $\anb{\bzt^2}$ & $\anb{\bzt}$ & $\anb{\bzt^3}$  \\ \hline
        27 & $D_{6h}$ & $\anb{\bzt^3}$ & $\anb{\bzt^3}$ & $\anb{\bzt^6}$ & $\anb{\bzt^3}$ & $\anb{\bzt^3}$ & $\anb{\bzt^6}$  \\ \hline
        28 & $T$ & $\bbz_1$ & $\bbz_1$ & $\bbz_1$ & $\bbz_1$ & $\anb{\bzt}$ & $\anb{\bzt}$  \\ \hline
        29 & $T_h$ & $\anb{\bzt}$ & $\anb{\bzt}$ & $\anb{\bzt^2}$ & $\anb{\bzt}$ & $\anb{\bzt}$ & $\anb{\bzt^2}$  \\ \hline
        30 & $T_d$ & $\anb{\bzt}$ & $\anb{\bzt}$ & $\anb{\bzt^2}$ & $\anb{\bzt}$ & $\anb{\bzt}$ & $\anb{\bzt^2}$  \\ \hline
        31 & $O$ & $\bbz_1$ & $\anb{\bzt}$ &  $\anb{\bzt}$ & $\bbz_1$ & $\anb{\bzt^2}$  & $\anb{\bzt^2}$  \\ \hline
        32 & $O_h$ & $\anb{\bzt^2}$ & $\anb{\bzt^2}$ & $\anb{\bzt^4}$ & $\anb{\bzt^2}$ & $\anb{\bzt^2}$ & $\anb{\bzt^4}$  \\ \hline
        \end{tabular}
        \caption{Classification data for the 32 point groups under disorder.}
        \label{tab:3d_disordered}
    \end{center}
\end{table}

\section{Conclusion and Discussion}
\label{sec:conclusion}
In this work, we have extended the framework of average symmetry-protected topological phases to crystalline systems, developing the classification of average crystalline symmetry-protected topological phases. Our approach is based on a generalized block-state construction, wherein a crystalline system is decomposed into a hierarchy of lower-dimensional cells that inherit effective average onsite symmetry from the underlying spatial symmetries. By decorating these cells with appropriate ASPT states and enforcing consistency via generalized obstruction-free conditions and bubble equivalence relations, we have established a classification scheme for ACSPTs in fermionic systems subject to disorder or decoherence. Notably, the block state construction does not rely on a free-fermion limit but remains valid even in the presence of strong interactions. Additionally, we employ a generalized spectral sequence method, adapted to average symmetry cases, to mathematically derive the classification. The agreement between these two independent methods confirms the robustness of our results.

We note that the stacking relations among certain decoration patterns are subtle. Although we can construct continuous, gapped paths that establish stacking rules in most situations, three 3D point-group symmetries -- \(D_{3d}\), \(D_{2d}\), and \(T_d\) -- remain unresolved. In the \(D_{2d}\) and \(T_d\) cases, stacking necessarily involves intrinsic ASPTs, i.e., decorations that cannot appear in clean systems; any ``adiabatic'' interpolation must therefore involve paths in mixed-state or doubled Hilbert space, adding considerable complexity. While we can verify that each decoration is individually obstruction-free, their mutual stacking relations are still undetermined from the block state construction. For these symmetries, we currently rely on the generalized spectral-sequence results alone for their stacking relations. 

One of the highlight of our study is the identification of many intrinsic average crystalline symmetry-protected topological phases that arise only in the presence of disorder or decoherence. We find two distinct mechanisms for their emergence\cite{Zhang2023fractonic,Guo2025lpdo}: (1) the obstruction-free conditions that constrain block decorations in clean systems are relaxed under disorder or decoherence, enabling otherwise forbidden decoration patterns; (2) the decorated state itself is an intrinsic ASPT residing on lower-dimensional blocks and protected by an average onsite symmetry. Together, these mechanisms enable the stabilization of nontrivial topological phases that rely on imperfections—phases that would be strictly prohibited in clean or pure-state settings. As disorder or decoherence is gradually removed, such intrinsic ASPTs are expected to evolve into \emph{intrinsic gapless phases}~\cite{thorngren2021intrinsically,Ma2025ASPT2}, a remarkable class of quantum critical states whose classification, particularly in the presence of crystalline symmetries, remains largely unexplored. Our findings not only suggest the existence of these phases but also highlight their potential richness. A notable case study of an intrinsic ACSPT phase appears in Ref.~\cite{Chen2025intrinsic}, where a spinless fermion system with exact \( U(1) \) charge conservation and average anti-unitary \( C_4T \) crystalline symmetry exhibits similar behavior. However, this example lies outside the scope of our current framework, which does not yet incorporate anti-unitary or continuous symmetries such as \( U(1) \). Extending the classification to include time-reversal and charge conservation symmetries is therefore a natural and important direction for future work, bringing the theory closer to experimentally relevant systems. Another key direction is the construction of explicit lattice models that realize the new ASPT phases predicted by our classification. Through numerical simulations, one can explore the phase transitions between disorder-stabilized topological phases and trivial ones, offering deeper insight into how these novel phases emerge and persist under realistic imperfections.

A further implication of our classification concerns the generalization of Lieb--Schultz--Mattis (LSM) theorems\cite{LIEB1961407,Oshikawa2000,Hastings2004} to disordered and open quantum systems.  It is well established that the boundary of a $d$-dimensional SPT phase enforces an LSM‐type anomaly in $(d-1)$ dimensions\cite{Cho2017LSM,Jian2018LSM,Cheng2016LSM,Metlitski2018,Cheng2019LSM}.  By the same logic, the boundary of an average SPT should impose an \emph{average‐symmetry LSM constraint}, thereby extending the conventional theorem to settings with quenched disorder or decoherence.  Only a handful of such ``disordered LSM" examples involving average crystalline symmetry have been analyzed in the literature~\cite{Kimchi2018,Ma2022ASPT1,Ma2025ASPT2,Kawabata2024LSMopen,Zhou2024lsm}.  Our results reveal an entire zoo of average crystalline SPTs, suggesting a correspondingly large family of LSM constraints that remain robust under disorder.  An intriguing open question is whether intrinsic ACSPTs generate boundary anomalies that are qualitatively different from those produced by extrinsic ASPTs.  Clarifying the resulting LSM constraints -- and identifying experimental signatures -- will be an exciting direction for future work.

The block-state (or, more generally, defect-network) construction can also be applied to \emph{crystalline symmetry–enriched topological orders}\cite{else2019defectnetworks}. It is therefore natural to ask how our analysis extends to average crystalline symmetry–enriched topological order -- a question that may be particularly relevant for experiments, for example, in fractional quantum Hall systems where disorder is often substantial. Understanding how residual average crystalline symmetries shape the interplay between topological order and spatial symmetries in such disordered settings is an interesting avenue for future research.

\section*{Acknowledgment}
We thank Zhida Song for stimulating discussion and related collaboration. JHZ thanks Carolyn Zhang for helpful discussions. YQ acknowledges support from the National Key R\&D Program of China (Grant No. 2022YFA1403402), from the National Natural Science Foundation of China (Grant No. 12174068), from the Science and Technology Commission of Shanghai Municipality (Grant No. 24LZ1400100 and 23JC1400600), and from the Shuguang Program of Shanghai Education Development Foundation and Shanghai Municipal Education Commission. JHZ, SS, and ZB are supported by a startup fund from the Pennsylvania State University. SS and ZB are also supported by the NSF CAREER Grant DMR-2339319.

\bibliographystyle{apsrev4-2}
\bibliography{refs}

\appendix
\section{Classification details for 2D wallpaper groups}\label{app:2d_class_details}

In this appendix, we comprehensively detail the construction of the crystalline ASPTs for all 2D space-group symmetries (i.e., the 17 wallpaper groups) and show how it leads to the classification listed in Tables ~\ref{tab:2d_decohered} and \ref{tab:2d_disordered}. For each case (while distinguishing between decoherence/disorder and spinless/spin-1/2), we list the block state decorations, obstructions, the obstruction-free states, trivializations, and the final classification. For obstruction-free states and the final classification, we separate the extrinsic and intrinsic ASPTs with the labels $(E)$ and $(I)$ respectively. Extrinsic ASPTs are defined as ASPTs which are not intrinsic, or equivalently, ASPTs which are also SPTs in the clean case.

\subsection{p1}
\begin{figure}[!htbp]
    \centering
    \includegraphics[width=0.5\linewidth]{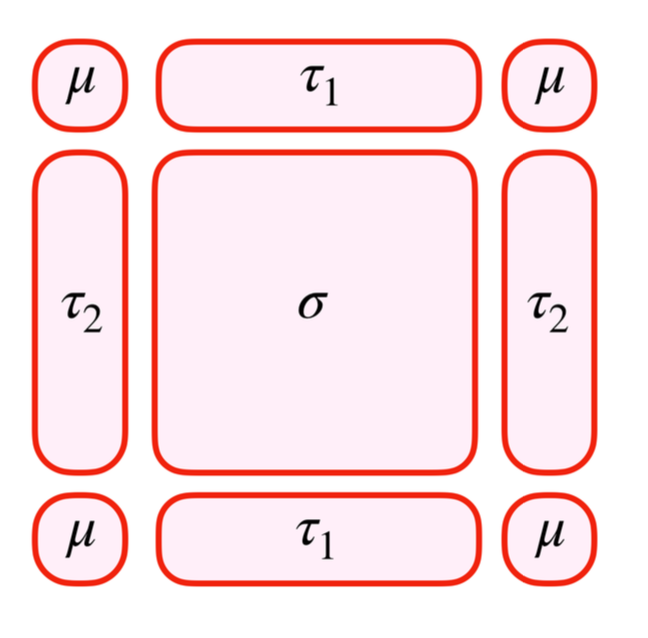}
    \caption{$p1$ lattice}
\end{figure}
Blocks and symmetry actions:
\begin{itemize}
    \item $\tau_1,\tau_2 = I$
    \item $\mu = I$
\end{itemize}
\subsubsection*{Decohered}
Since there are no onsite symmetries, the result is the same as the clean case, and there is no distinction between spinless and spin-1/2 cases. The obstruction- and trivialization-free block state decorations are Majorana chains on $\tau_1$ or $\tau_2$, and complex fermion on $\mu$ blocks.

\subsubsection*{Disordered}
Since $\mu$ is surrounded by an even number of $\tau_1$ and $\tau_2$ blocks, Majorana chains on $\tau_1$ or $\tau_2$ are obstruction-free. These states cannot be trivialized by Majorana bubble on $\sigma$.

\subsection{p2}
\subsubsection*{Cell Decomposition}
\begin{figure}[!htbp]
    \centering
    \includegraphics[width=0.9\linewidth]{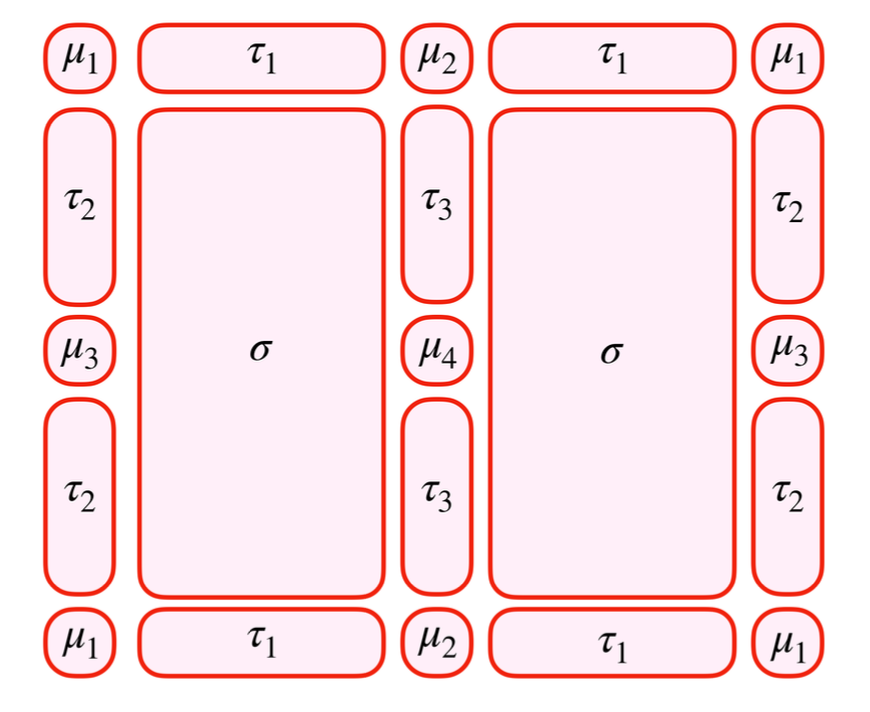}
    \caption{$p2$ lattice}
\end{figure}
\textbf{Blocks and onsite symmetries}:
\begin{itemize}
    \item[] 1D: $G_{\tau_1},G_{\tau_2},G_{\tau_3} = I$
    \item[] 0D: $G_{\mu_1},G_{\mu_2},G_{\mu_3},G_{\mu_4} = \bzt$
\end{itemize}
\subsubsection*{Decohered Spinless}
\textbf{Block state decorations}:
\begin{itemize}
    \item[] 0D
    \begin{itemize}
        \item $\mu_1,\ \mu_2,\ \mu_3,\ \mu_4$: Odd fermion
    \end{itemize}
    \item[] 1D
    \begin{itemize}
        \item $\tau_1,\ \tau_2,\ \tau_3$ : Majorana chain
    \end{itemize}
\end{itemize}

\textbf{Obstructions}
\begin{itemize}
    \item Majorana chain on $\tau_1$ is obstructed at $\mu_1$ and $\mu_2$ by 2-fold rotational symmetry.
    \item Majorana chain on $\tau_2$ is obstructed at $\mu_3$
    \item Majorana chain on $\tau_3$ is obstructed at $\mu_4$
    \item Simultaneous decorations are also obstructed.
\end{itemize}

\textbf{Obstruction-free states}:
\begin{itemize}
\item[] 0D states ($\bzt^4$) are obstruction-free (E). 

\item[] 1D: No obstruction-free states ($\bbz_1$)
\end{itemize}

\textbf{Trivializations}:
\begin{itemize}
    \item Majorana bubble on $\sigma$ $\Rightarrow$ Simultaneous decoration of odd fermions on $\mu_{1,2,3,4}$. Therefore, 0D classification reduces to $\bzt^3$
\end{itemize}

\textbf{Final classification:}
\begin{itemize}
    \item[] $E_{0,dec}^{0D} = (E) \bzt^3$
    \item[] $E_{0,dec}^{1D} = \bbz_1$
    \item[] $\mathcal{G}_{0,dec} = E_{0,dec}^{0D} \times E_{0,dec}^{1D}  = (E) \bzt^3$
\end{itemize}

\subsubsection*{Decohered Spin-1/2}
\textbf{Block state decorations}:
\begin{itemize}
    \item[] 0D
    \begin{itemize}
        \item $\mu_1,\ \mu_2,\ \mu_3,\ \mu_4$: Odd fermion (intrinsic)
    \end{itemize}
    \item[] 1D
    \begin{itemize}
        \item $\tau_1,\ \tau_2,\ \tau_3$: Majorana chain
    \end{itemize}
    \item[] These states are all \textbf{obstruction-free}.
\end{itemize}

\textbf{Trivializations and Stacking}:
\begin{itemize}
\item[] Bubble equivalence does not reduce classification. However, there is nontrivial stacking between 1D and 0D states:
\begin{itemize}
    \item Decoration of two copies of Majorana chain on $\tau_1$ is equivalent to odd fermions on $\mu_1$ and $\mu_2$.
    \item Decoration of two copies of Majorana chain on $\tau_2$ is equivalent to odd fermions on $\mu_1$ and $\mu_3$.
    \item Decoration of two copies of Majorana chain on $\tau_3$ is equivalent to odd fermions on $\mu_2$ and $\mu_4$.
\end{itemize}
\end{itemize}

\textbf{Final classification:}
\begin{itemize}
    \item[] $E_{1/2,dec}^{0D} = \bzt^4 (I)$
    \item[] $E_{1/2,dec}^{1D} = \bzt^3 (E)$
    \item[] Non-trivial stacking $\Rightarrow$ $\mathcal{G}_{1/2,dec} = E_{1/2,dec}^{0D} \rtimes E_{1/2,dec}^{1D}  = \bzt (I) \times \bbz_4^3 (I)$
\end{itemize}

\subsubsection*{Disordered Spinless}
\textbf{Block state decorations}:
\begin{itemize}
    \item[] 1D
    \begin{itemize}
        \item $\tau_1, \tau_2, \tau_3$: Majorana chain 
    \end{itemize}
    \item[] These states are all \textbf{obstruction-free} and \textbf{trivialization-free}.
    
\end{itemize}

\textbf{Final classification:}
\begin{itemize}
    \item[] $\mathcal{G}_{0,dis} =  E_{0,dis}^{1D}  = \bzt^3 (I) $
    \item[] $\mathcal{G}_{1/2,dis} = E_{1,dis}^{1D} = \bzt^3 (E) $
\end{itemize}

\subsection{pm}
\subsubsection*{Cell Decomposition}
\begin{figure}[!htbp]
    \centering
    \includegraphics[width=0.9\linewidth]{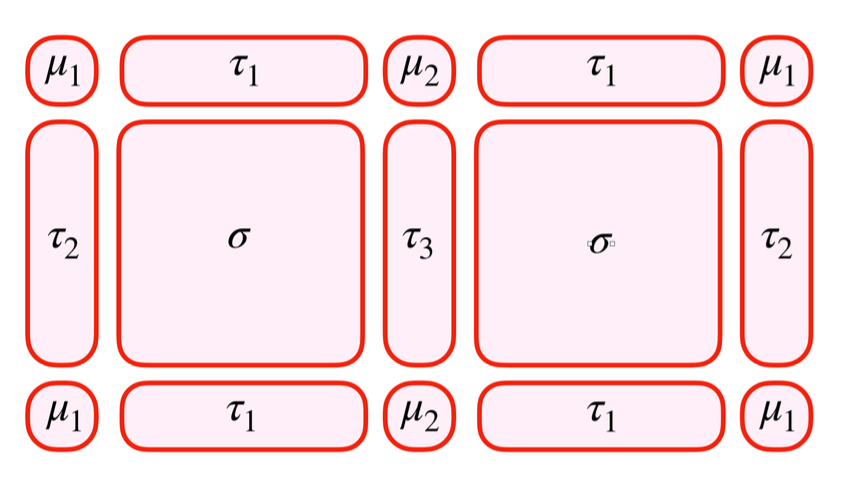}
    \caption{$pm$ lattice}
\end{figure}
\textbf{Blocks and onsite symmetries}:
\begin{itemize}
    \item[] 1D: $G_{\tau_1} = I,\ G_{\tau_2} = \bzt,\ G_{\tau_3} = \bzt$
    \item[] 0D: $G_{\mu_1},\ G_{\mu_2} = \bzt$
\end{itemize}
\subsubsection*{Decohered Spinless}

\textbf{Block state decorations}:
\begin{itemize}
    \item[] 0D
    \begin{itemize}
        \item $\mu_1$,$\mu_2$: Odd fermion
    \end{itemize}
    \item[] 1D
    \begin{itemize}
        \item $\tau_1$: Majorana chain
        \item $\tau_2$, $\tau_3$: Majorana chain, $\bzt$ fSPT
    \end{itemize}
\end{itemize}

\textbf{Obstructions}
\begin{itemize}
    \item Majorana chain on $\tau_1$ is obstructed at $\mu_1$ and $\mu_2$ by 2-fold rotational symmetry.
\end{itemize}

\textbf{Obstruction-free states}:
\begin{itemize}
\item[] 0D states ($\bzt^2$) are obstruction-free (E). 

\item[] 1D ($\bzt^4$)
\begin{enumerate}
    \item Majorana chain on $\tau_2$ (E)
    \item Majorana chain on $\tau_3$ (E)
    \item $\bzt$ fSPT on $\tau_2$ (E)
    \item $\bzt$ fSPT on $\tau_3$ (E)
\end{enumerate}
    
\end{itemize}

\textbf{Trivializations}:
\begin{itemize}
    \item Majorana bubble on $\sigma$ $\Rightarrow$ Simultaneous decoration of $\bzt$ fSPT on $\tau_2$ and $\bzt$ fSPT on $\tau_3$. Therefore, 1D classification reduces to $\bzt^3$.
\end{itemize}

\textbf{Final classification:}
\begin{itemize}
    \item[] $E_{0,dec}^{0D} = \bzt^2$
    \item[] $E_{0,dec}^{1D} = \bzt^3$
    \item[]  $\mathcal{G}_{0,dec} = E_{0,dec}^{0D} \times E_{0,dec}^{1D}  =\bzt^5 (E)$
\end{itemize}

\subsubsection*{Decohered Spin-1/2}
\textbf{Block state decorations}:
\begin{itemize}
    \item[] 0D
    \begin{itemize}
        \item $\mu_1$,$\mu_2$: Odd fermion (I)
    \end{itemize}
    \item[] 1D
    \begin{itemize}
        \item $\tau_1$: Majorana chain (E)
        \item $\tau_2$, $\tau_3$: $\bbz_4^f$ ASPT (I)
    \end{itemize}
    \item[] These states are all \textbf{obstruction-free}.
\end{itemize}

\textbf{Trivializations and Stacking}:
\begin{itemize}
\item[] Bubble equivalence does not reduce classification. However, there is nontrivial stacking between 1D and 0D states:
\begin{itemize}
    \item Decoration of two copies of Majorana chain on $\tau_1$ is equivalent to odd fermions on $\mu_1$ and $\mu_2$.
\end{itemize}
\end{itemize}

\textbf{Final classification:}
\begin{itemize}
    \item[] $E_{1/2,dec}^{0D} = \bzt^2$
    \item[] $E_{1/2,dec}^{1D} = \bzt^3$
    \item[] Non-trivial stacking $\Rightarrow$ $\mathcal{G}_{1/2,dec} = E_{1/2,dec}^{0D} \rtimes E_{1/2,dec}^{1D}  = \bzt^3 \times \bbz_4$
\end{itemize}

\subsubsection*{Disordered Spinless}
\textbf{Block state decorations}:
\begin{itemize}
    \item[] 1D
    \begin{itemize}
        \item $\tau_1, \tau_2, \tau_3$: Majorana chain (E) 
    \end{itemize}
    \item[] These states are all \textbf{obstruction-free} and \textbf{trivialization-free}. Majorana chain on $\tau_1$ is only obstruction-free in the disordered case and is hence intrinsic phase.
    
\end{itemize}

\textbf{Final classification:}
\begin{itemize}
    \item[] $\mathcal{G}_{0,dis} =  E_{0,dis}^{1D}  = \bzt^2 (E)\times\bzt (I)$
\end{itemize}

\subsubsection*{Disordered Spin-1/2}
\textbf{Block state decorations}:
\begin{itemize}
    \item[] 1D
    \begin{itemize}
        \item $\tau_1$: Majorana chain (E)
        \item $\tau_2, \tau_3$: \placeholder (I)
    \end{itemize}
    \item[] These states are all \textbf{obstruction-free} and \textbf{trivialization-free}.
    
\end{itemize}

\textbf{Final classification:}
\begin{itemize}
    \item[] $\mathcal{G}_{1/2,dis} =  E_{1/2,dis}^{1D}  = \bzt (E)\times\bzt^2 (I)$
\end{itemize}

\subsection{pg}
\subsubsection*{Cell Decomposition}
\begin{figure}[!htbp]
    \centering
    \includegraphics[width=0.9\linewidth]{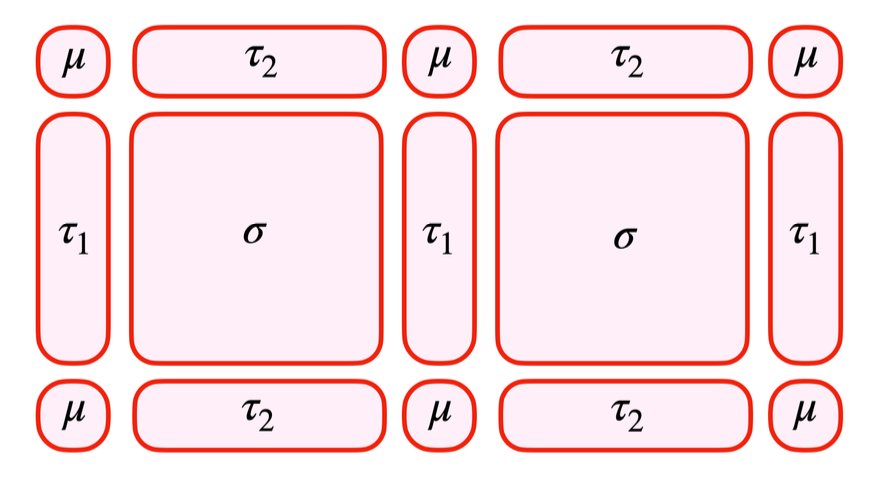}
    \caption{$pg$ lattice}
\end{figure}
\textbf{Blocks and onsite symmetries}:
\begin{itemize}
    \item[] 1D: $G_{\tau_1},\ G_{\tau_2} = I$
    \item[] 0D: $G_{\mu} = I$
    \item[] Since glide symmetries do not act onsite, all blocks have trivial onsite symmetry. Hence there is no distinction between spinless and spin-1/2 cases.
\end{itemize}
\subsubsection*{Decohered}

\textbf{Block state decorations}:
\begin{itemize}
    \item[] 0D
    \begin{itemize}
        \item $\mu$: Odd fermion
    \end{itemize}
    \item[] 1D
    \begin{itemize}
        \item $\tau_1,\ \tau_2$: Majorana chain
    \end{itemize}
    \item[] These states are all \textbf{obstruction-free} and \textbf{trivialization-free}.
\end{itemize}

\textbf{Final classification:}
\begin{itemize}
    \item[] $E_{dec}^{0D} = \bzt$
    \item[] $E_{dec}^{1D} = \bzt^2$
    \item[]  $\mathcal{G}_{dec} = E_{dec}^{0D} \times E_{dec}^{1D}  =\bzt^3 (E)$
\end{itemize}

\subsubsection*{Disordered}
\textbf{Block state decorations}:
\begin{itemize}
    \item[] 1D
    \begin{itemize}
        \item $\tau_1, \tau_2$: Majorana chain 
    \end{itemize}
    \item[] These states are all \textbf{obstruction-free} and \textbf{trivialization-free}.
    
\end{itemize}

\textbf{Final classification:}
\begin{itemize}
    \item[] $\mathcal{G}_{dis} =  E_{dis}^{1D}  = \bzt^2 (E)$
\end{itemize}

\subsection{cm}
\subsubsection*{Cell Decomposition}
\begin{figure}[!htbp]
    \centering
    \includegraphics[width=0.5\linewidth]{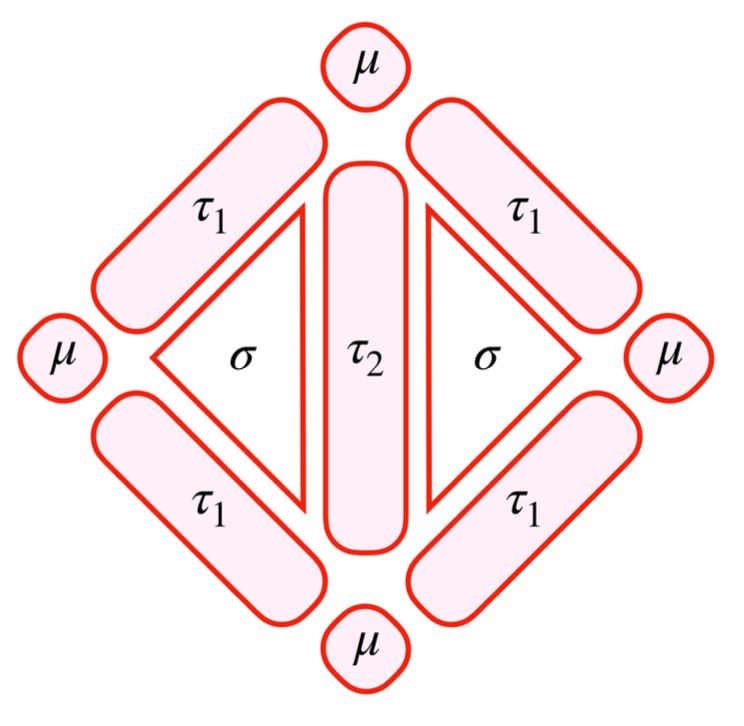}
    \caption{$cm$ lattice}
\end{figure}
\textbf{Blocks and onsite symmetries}:
\begin{itemize}
    \item[] 1D: $G_{\tau_1} = I,\ G_{\tau_2} = \bzt$
    \item[] 0D: $G_{\mu} = \bzt$
\end{itemize}
\subsubsection*{Decohered Spinless}

\textbf{Block state decorations}:
\begin{itemize}
    \item[] 0D
    \begin{itemize}
        \item $\mu$: Odd fermion
    \end{itemize}
    \item[] 1D
    \begin{itemize}
        \item $\tau_1$: Majorana chain (E)
        \item $\tau_2$: Majorana chain, $\bzt$ fSPT (E)
    \end{itemize}
    \item[] These states are all \textbf{obstruction-free}.
\end{itemize}

\textbf{Trivializations}:
\begin{itemize}
    \item Majorana bubble on $\sigma$ $\Rightarrow$ Decoration of $\bzt$ fSPT on $\tau_2$. Therefore, 1D classification reduces to $\bzt^2$.
\end{itemize}

\textbf{Final classification:}
\begin{itemize}
    \item[] $E_{0,dec}^{0D} = \bzt$
    \item[] $E_{0,dec}^{1D} = \bzt^2$
    \item[]  $\mathcal{G}_{0,dec} = E_{0,dec}^{0D} \times E_{0,dec}^{1D}  =\bzt^3 (E)$
\end{itemize}

\subsubsection*{Decohered Spin-1/2}

\textbf{Block state decorations}:
\begin{itemize}
    \item[] 0D
    \begin{itemize}
        \item $\mu$: Odd fermion (I)
    \end{itemize}
    \item[] 1D
    \begin{itemize}
        \item $\tau_1$: Majorana chain (E)
        \item $\tau_2$: $\bbz_4^f$ ASPT (I)
    \end{itemize}
    \item[] These states are all \textbf{obstruction-free} and \textbf{trivialization-free}.
\end{itemize}

\textbf{Final classification:}
\begin{itemize}
    \item[] $E_{1/2,dec}^{0D} = \bzt (I)$
    \item[] $E_{1/2,dec}^{1D} = \bzt(E)\times\bzt(I)$
    \item[]  $\mathcal{G}_{1/2,dec} = E_{1/2,dec}^{0D} \times E_{1/2,dec}^{1D}  = \bzt (E)\times\bzt^2(I)$
\end{itemize}

\subsubsection*{Disordered Spinless}
\textbf{Block state decorations}:
\begin{itemize}
    \item[] 1D
    \begin{itemize}
        \item $\tau_1, \tau_2$: Majorana chain (E)
    \end{itemize}
    \item[] These states are all \textbf{obstruction-free} and \textbf{trivialization-free}.
    
\end{itemize}

\textbf{Final classification:}
\begin{itemize}
    \item[] $\mathcal{G}_{0,dis} =  E_{0,dis}^{1D}  = \bzt^2 (E)$
\end{itemize}

\subsubsection*{Disordered Spin-1/2}
\textbf{Block state decorations}:
\begin{itemize}
    \item[] 1D
    \begin{itemize}
        \item $\tau_1$: Majorana chain (E)
        \item $\tau_2$: \placeholder (I)
    \end{itemize}
    \item[] These states are all \textbf{obstruction-free} and \textbf{trivialization-free}.
    
\end{itemize}

\textbf{Final classification:}
\begin{itemize}
    \item[] $\mathcal{G}_{1/2,dis} =  E_{1/2,dis}^{1D}  = \bzt(E)\times\bzt(I)$
\end{itemize}

\subsection{pmm}
\subsubsection*{Cell Decomposition}
\begin{figure}[!htbp]
    \centering
    \includegraphics[width=0.6\linewidth]{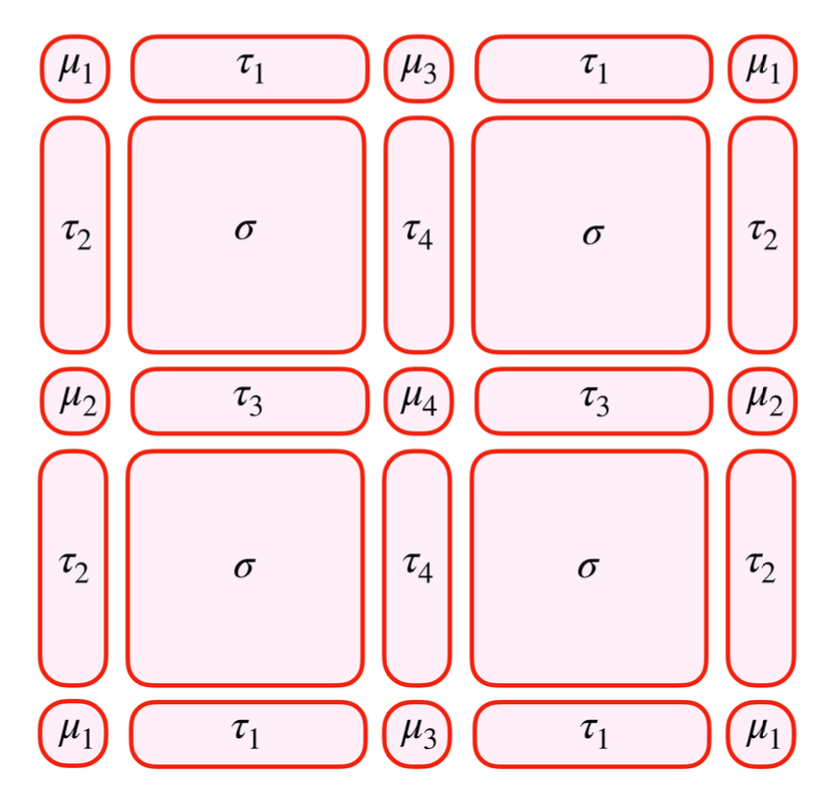}
    \caption{$pmm$ lattice}
\end{figure}
\textbf{Blocks and onsite symmetries}:
\begin{itemize}
    \item $G_{\tau_1},\ G_{\tau_2},\ G_{\tau_3},\ G_{\tau_4} = \bzt$
    \item $G_{\mu_1}, G_{\mu_2}, G_{\mu_3}, G_{\mu_4}  = D_2 = \bzt\rtimes \bzt$
\end{itemize}
\subsubsection*{Decohered Spinless}

\textbf{Block state decorations}:
\begin{itemize}
    \item[] 0D
    \begin{itemize}
        \item $\mu_1,\ \mu_2,\ \mu_3,\ \mu_4$: Odd fermion
    \end{itemize}
    \item[] 1D
    \begin{itemize}
        \item $\tau_1,\ \tau_2,\ \tau_3,\ \tau_4$: Majorana chain, $\bzt$ fSPT
    \end{itemize}
\end{itemize}

\textbf{Obstructions}
\begin{itemize}
    \item Majorana chain on $\tau_1$ is obstructed at $\mu_1$ and $\mu_3$.
    \item Majorana chain on $\tau_2$ is obstructed at $\mu_1$ and $\mu_2$.
    \item Majorana chain on $\tau_3$ is obstructed at $\mu_2$ and $\mu_4$.
    \item Majorana chain on $\tau_4$ is obstructed at $\mu_3$ and $\mu_4$.
    \item[] Simultaneous decorations of the above states are also obstructed.
\end{itemize}

\textbf{Obstruction-free states}:
\begin{itemize}
\item[] 0D states ($\bzt^4$) are obstruction-free (E). 

\item[] 1D ($\bzt^4$)
\begin{enumerate}
    \item $\bzt$ fSPT on any of $\tau_1$, $\tau_2$, $\tau_3$ or $\tau_4$ (I)
\end{enumerate}
    
\end{itemize}

\textbf{Trivializations}:
\begin{itemize}
    \item Majorana bubble on $\sigma$ $\Rightarrow$ Simultaneous decoration of $\bzt$ fSPT on $\tau_1,\ \tau_2,\ \tau_3,$ and $\tau_4$. Therefore, the 1D classification reduces to $\bzt^3$.
\end{itemize}

\textbf{Final classification:}
\begin{itemize}
    \item[] $E_{0,dec}^{0D} = \bzt^4 (E)$
    \item[] $E_{0,dec}^{1D} = \bzt^3 (I)$
    \item[]  $\mathcal{G}_{0,dec} = E_{0,dec}^{0D} \times E_{0,dec}^{1D}  =\bzt^4 (E) \times \bzt^3(I)$
\end{itemize}

\subsubsection*{Decohered Spin-1/2}
\textbf{Block state decorations}:
\begin{itemize}
    \item[] 0D
    \begin{itemize}
        \item $\mu_1,\ \mu_2,\ \mu_3,\ \mu_4$: Odd fermion (I)
    \end{itemize}
    \item[] 1D
    \begin{itemize}
        \item $\tau_1,\ \tau_2,\ \tau_3,\ \tau_4$: $\bbz_4^f$ ASPT (I)
    \end{itemize}
    \item[] These states are all \textbf{obstruction-free} and \textbf{trivialization-free}.
\end{itemize}

\textbf{Final classification:}
\begin{itemize}
    \item[] $E_{1/2,dec}^{0D} = \bzt^4 (I)$
    \item[] $E_{1/2,dec}^{1D} = \bzt^4 (I)$
    \item[]  $\mathcal{G}_{1/2,dec} = E_{1/2,dec}^{0D} \times E_{1/2,dec}^{1D}  = \bzt^8 (I)$
\end{itemize}

\subsubsection*{Disordered Spinless}
\textbf{Block state decorations}:
\begin{itemize}
    \item[] 1D
    \begin{itemize}
        \item $\tau_1, \tau_2, \tau_3,\ \tau_4$: Majorana chain 
    \end{itemize}
    \item[] These states are all \textbf{obstruction-free} and \textbf{trivialization-free}. As they are obstructed in the clean case, these are intrinsic phases.
\end{itemize}

\textbf{Final classification:}
\begin{itemize}
    \item[] $\mathcal{G}_{0,dis} =  E_{0,dis}^{1D}  = \bzt^4 (I)$
\end{itemize}

\subsubsection*{Disordered Spin-1/2}
\textbf{Block state decorations}:
\begin{itemize}
    \item[] 1D
    \begin{itemize}
        \item $\tau_1, \tau_2, \tau_3,\ \tau_4$: \placeholder 
    \end{itemize}
    \item[] These states are all \textbf{obstruction-free} and \textbf{trivialization-free}.
\end{itemize}

\textbf{Final classification:}
\begin{itemize}
    \item[] $\mathcal{G}_{1/2,dis} =  E_{1/2,dis}^{1D}  = \bzt^4 (I)$
\end{itemize}

\subsection{pmg}
\subsubsection*{Cell decomposition}
\begin{figure}[!htbp]
    \centering
    \includegraphics[width=0.9\linewidth]{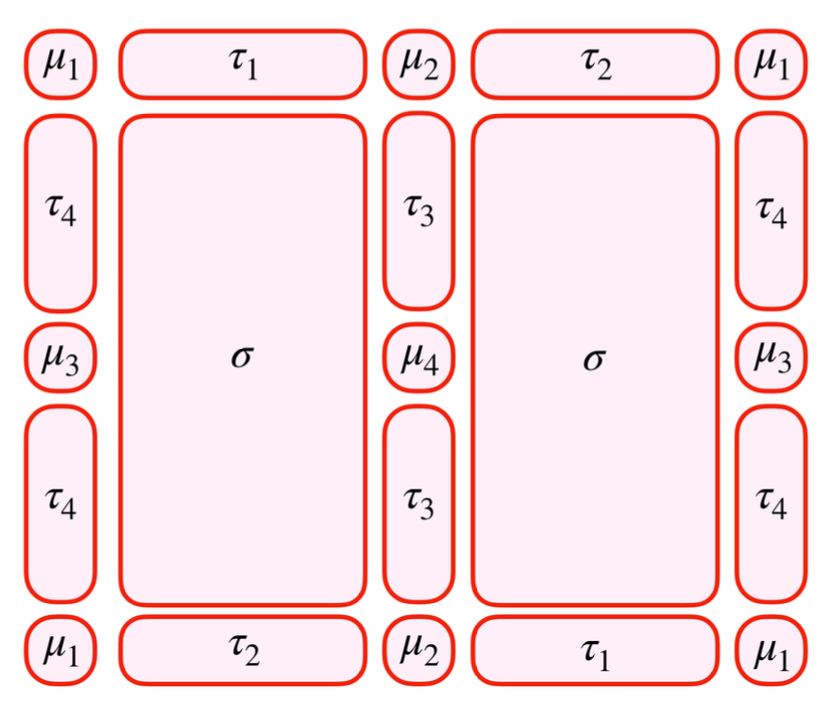}
    \caption{$pmg$ lattice}
\end{figure}
\textbf{Blocks and onsite symmetries}:
\begin{itemize}
    \item[] 1D: $G_{\tau_1},\ G_{\tau_2} = \bzt,\ G_{\tau_3} = I,\ G_{\tau_4} = I$
    \item[] 0D: $G_{\mu_1},\ G_{\mu_2} = \bzt$ (reflection), $G_{\mu_3},\ G_{\mu_4} = \bzt$ (rotation)
\end{itemize}
\subsubsection*{Decohered Spinless}

\textbf{Block state decorations}:
\begin{itemize}
    \item[] 0D
    \begin{itemize}
        \item $\mu_1,\ \mu_2,\ \mu_3,\ \mu_4$: Odd fermion
    \end{itemize}
    \item[] 1D
    \begin{itemize}
        \item $\tau_1,\ \tau_2$: Majorana chain, $\bzt$ fSPT
        \item $\tau_3,\ \tau_4$: Majorana chain
    \end{itemize}
\end{itemize}

\textbf{Obstructions}
\begin{itemize}
    \item Majorana chain on $\tau_1$ or $\tau_2$ are obstructed at $\mu_1$ and $\mu_2$. However, simultaneous decoration of both these phases is obstruction-free.
    \item Similarly, $\bzt$ phase on $\tau_1$ or $\tau_2$ is obstructed, but simultaneous decoration on both is obstruction-free.
    \item Majorana chain on $\tau_3$ is obstructed at $\mu_4$.
    \item Majorana chain on $\tau_4$ is obstructed at $\mu_3$.
\end{itemize}

\textbf{Obstruction-free states}:
\begin{itemize}
\item[] 0D states ($\bzt^4$) are obstruction-free (E). 

\item[] 1D ($\bzt^2$)
\begin{enumerate}
    \item Simultaneous decoration of Majorana chain on $\tau_1$ and $\tau_2$ (E)
    \item Simultaneous decoration of $\bzt$ fSPT on $\tau_1$ and $\tau_2$ (E)
\end{enumerate}
    
\end{itemize}

\textbf{Trivializations}:
\begin{itemize}
    \item Majorana bubble on $\sigma$ $\Rightarrow$ Simultaneous decoration of $\bzt$ fSPT on $\tau_1$ and $\tau_2$, and odd fermions on $\mu_3$ and $\mu_4$. This establishes an equivalence between a 1D and a 0D block state. We choose the convention of treating this as a reduction of the 0D classification, from $\bzt^4$ to $\bzt^3$, while the 1D classification remains the same.
    \item 1D fermionic insulator bubble on one of $\tau_1$ or $\tau_2$ $\Rightarrow$ simultaneous decoration of odd fermions on $\mu_1$ and $\mu_2$. Therefore, the 0D classification further reduces to $\bzt^2$
\end{itemize}

\textbf{Final classification:}
\begin{itemize}
    \item[] $E_{0,dec}^{0D} = \bzt^2 (E)$
    \item[] $E_{0,dec}^{1D} = \bzt^2 (E)$
    \item[]  $\mathcal{G}_{0,dec} = E_{0,dec}^{0D} \times E_{0,dec}^{1D}  =\bzt^4 (E)$
\end{itemize}

\subsubsection*{Decohered Spin-1/2}
\textbf{Block state decorations}:
\begin{itemize}
    \item[] 0D
    \begin{itemize}
        \item $\mu_1,\ \mu_2,\ \mu_3,\ \mu_4$: Odd fermion (I)
    \end{itemize}
    \item[] 1D
    \begin{itemize}
        \item $\tau_1,\ \tau_2$: $\bbz_4^f$ ASPT (I)
        \item $\tau_3,\ \tau_4$: Majorana chain (E)
    \end{itemize}
\end{itemize}

\textbf{Obstructions}:
\begin{itemize}
    \item Decoration of intrinsic $\bbz_4^f$ ASPT on only one of $\tau_1$ or $\tau_2$ is obstructed at $\mu_1$ and $\mu_2$. However, simultaneous decoration of both cancels the anomaly and is obstruction-free. 
\end{itemize}

\textbf{Obstruction-free states}:
\begin{itemize}
\item[] 0D states ($\bzt^4$) are obstruction-free (I). 

\item[] 1D ($\bzt^3$)
\begin{enumerate}
    \item Simultaneous decoration of $\bbz_4^f$ ASPT on $\tau_1$ and $\tau_2$ (I)
    \item Majorana chain on $\tau_3$ (E)
    \item Majorana chain on $\tau_4$ (E)
\end{enumerate}
    
\end{itemize}

\textbf{Trivializations and Stacking}:
\begin{itemize}
    \item 1D fermionic insulator bubble on one of $\tau_1$ or $\tau_2$ $\Rightarrow$ simultaneous decoration of odd fermions on $\mu_1$ and $\mu_2$. Therefore, the 0D classification further reduces to $\bzt^3$.
    \item[] We have two nontrivial extensions between the 1D and 0D decorations. 
    \item Stacking two Majorana chains on $\tau_3$ $\Rightarrow$ odd fermion on $\mu_4$.
    \item Stacking two Majorana chains on $\tau_4$ $\Rightarrow$ odd fermion on $\mu_3$. 
\end{itemize}

\textbf{Final classification:}
\begin{itemize}
    \item[] $E_{1/2,dec}^{0D} = \bzt^3 (I)$
    \item[] $E_{1/2,dec}^{1D} = \bzt^2(E)\times\bzt(I)$
    \item[] Non-trivial stacking $\Rightarrow$ $\mathcal{G}_{1/2,dec} = E_{1/2,dec}^{0D} \rtimes E_{1/2,dec}^{1D}  = \bzt^2(I)\times\bbz_4^2(I)$
\end{itemize}

\subsubsection*{Disordered Spinless}
\textbf{Block state decorations}:
\begin{itemize}
    \item[] 1D
    \begin{itemize}
        \item $\tau_1, \tau_2, \tau_3,\ \tau_4$: Majorana chain 
    \end{itemize}
\end{itemize}

\textbf{Obstructions}:
\begin{itemize}
    \item Decoration of Majorana chain on only one of $\tau_1$ or $\tau_2$ is obstructed at $\mu_1$ and $\mu_2$ (odd number of Majorana modes). However, simultaneous decoration of both cancels the anomaly and is obstruction-free. 
\end{itemize}

\textbf{Obstruction-free states}:
\begin{itemize}
\item[] 1D ($\bzt^3$)
\begin{enumerate}
    \item Simultaneous decoration of Majorana chain on $\tau_1$ and $\tau_2$ (E)
    \item Majorana chain on $\tau_3$ or $\tau_4$ (I)
\end{enumerate}
\item[] These states are also \textbf{trivialization-free}.
\end{itemize}

\textbf{Final classification:}
\begin{itemize}
    \item[] $\mathcal{G}_{0,dis} =  E_{0,dis}^{1D}  = \bzt(E)\times\bzt^2(I)$
\end{itemize}

\subsubsection*{Disordered Spin-1/2}
\textbf{Block state decorations}:
\begin{itemize}
    \item[] 1D
    \begin{itemize}
        \item $\tau_1, \tau_2$: \placeholder 
        \item $\tau_3,\ \tau_4$: Majorana chain 
    \end{itemize}
\end{itemize}

\textbf{Obstructions}:
\begin{itemize}
    \item Decoration of \placeholder on only one of $\tau_1$ or $\tau_2$ is obstructed at $\mu_1$ and $\mu_2$ (odd number of \placeholdermodes). However, simultaneous decoration of both cancels the anomaly and is obstruction-free. 
\end{itemize}

\textbf{Obstruction-free states}:
\begin{itemize}
\item[] 1D ($\bzt^3$)
\begin{enumerate}
    \item \placeholders on $\tau_1$ and $\tau_2$ (I)
    \item Majorana chain on $\tau_3$ or $\tau_4$ (E)
\end{enumerate}
\item[] These states are also \textbf{trivialization-free}.
\end{itemize}

\textbf{Final classification:}
\begin{itemize}
    \item[] $\mathcal{G}_{1/2,dis} =  E_{1/2,dis}^{1D}  = \bzt^2(E)\times\bzt(I)$
\end{itemize}

\subsection{pgg}
\subsubsection*{Cell Decomposition}
\begin{figure}[!htbp]
    \centering
    \includegraphics[width=0.9\linewidth]{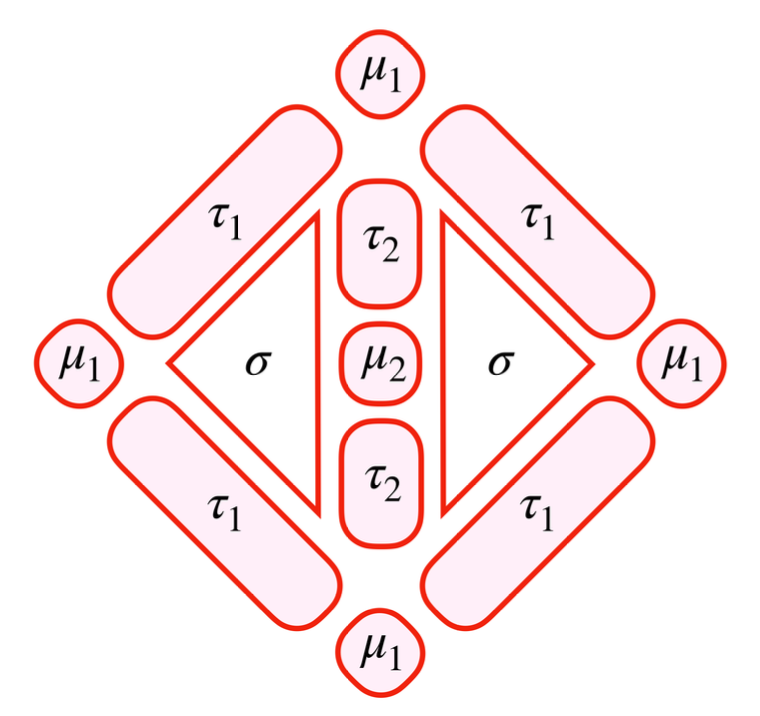}
    \caption{$pgg$ lattice}
\end{figure}

\textbf{Blocks and onsite symmetries}:
\begin{itemize}
    \item[] 1D: $G_{\tau_1},\ G_{\tau_2} = I$
    \item[] 0D: $G_{\mu_1},\ G_{\mu_2} = \bzt$ 
\end{itemize}
\subsubsection*{Decohered Spinless}

\textbf{Block state decorations}:
\begin{itemize}
    \item[] 0D
    \begin{itemize}
        \item $\mu_1,\ \mu_2$: Odd fermion
    \end{itemize}
    \item[] 1D
    \begin{itemize}
        \item $\tau_1,\ \tau_2$: Majorana chain
    \end{itemize}
\end{itemize}

\textbf{Obstructions}
\begin{itemize}
    \item Majorana chain on $\tau_2$ is obstructed at $\mu_2$
\end{itemize}

\textbf{Obstruction-free states}:
\begin{itemize}
\item[] 0D states ($\bzt^2$) are obstruction-free (E). 

\item[] 1D ($\bzt$)
\begin{enumerate}
    \item Majorana chain on $\tau_1$ (E)
\end{enumerate}
    
\end{itemize}

\textbf{Trivializations}:
\begin{itemize}
    \item Majorana bubble on $\sigma$ $\Rightarrow$ Simultaneous decoration of odd fermions on $\mu_1$ and $\mu_2$. This reduces the 0D classification to $\bzt$
\end{itemize}

\textbf{Final classification:}
\begin{itemize}
    \item[] $E_{0,dec}^{0D} = \bzt (E)$
    \item[] $E_{0,dec}^{1D} = \bzt (E)$
    \item[]  $\mathcal{G}_{0,dec} = E_{0,dec}^{0D} \times E_{0,dec}^{1D}  =\bzt^2 (E)$
\end{itemize}

\subsubsection*{Decohered Spin-1/2}
\textbf{Block state decorations}:
\begin{itemize}
    \item[] 0D
    \begin{itemize}
        \item $\mu_1,\ \mu_2$: Odd fermion (I)
    \end{itemize}
    \item[] 1D
    \begin{itemize}
        \item $\tau_1,\ \tau_2$: Majorana chain (E) 
    \end{itemize}
    \item[] These states are all \textbf{obstruction-free}.
\end{itemize}

\textbf{Trivialization and stacking};
\begin{itemize}
    \item[] While bubble equivalence does not reduce the classification, we have nontrivial stacking:
    \item Stacking two Majorana chains on $\tau_2$ $\Rightarrow$ Simultaneous decoration of odd fermions on $\mu_1$ and $\mu_2$
\end{itemize}

\textbf{Final classification:}
\begin{itemize}
    \item[] $E_{1/2,dec}^{0D} = \bzt^2 (I)$
    \item[] $E_{1/2,dec}^{1D} = \bzt^2 (E)$
    \item[] Non-trivial stacking $\Rightarrow$ $\mathcal{G}_{1/2,dec} = E_{1/2,dec}^{0D} \rtimes E_{1/2,dec}^{1D}  = \bzt(E)\times\bzt(I)\times\bbz_4(I)$
\end{itemize}

\subsubsection*{Disordered}
\textbf{Block state decorations}:
\begin{itemize}
    \item[] 1D
    \begin{itemize}
        \item $\tau_1, \tau_2$: Majorana chain 
    \end{itemize}
    \item[] These states are \textbf{obstruction-free} and \textbf{trivialization-free}. Majorana chain on $\tau_2$ is obstructed in the spinless clean case, and hence is an intrinsic phase.
\end{itemize}

\textbf{Final classification:}
\begin{itemize}
    \item[] $\mathcal{G}_{0,dis} =  E_{dis}^{0,1D}  = \bzt(E)\times\bzt(I)$
    \item[] $\mathcal{G}_{1/2,dis} =  E_{dis}^{1/2,1D}  = \bzt^2(E)$
\end{itemize}

\subsection{cmm}
\subsubsection*{Cell Decomposition}
\begin{figure}[!htbp]
    \centering
    \includegraphics[width=0.9\linewidth]{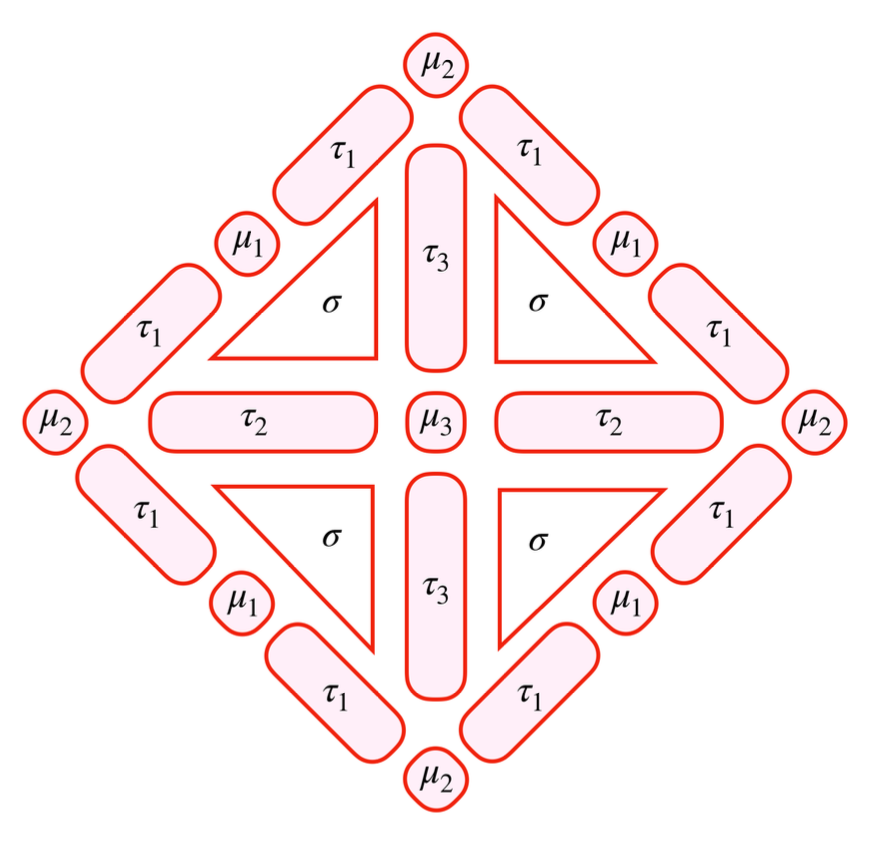}
    \caption{$cmm$ lattice}
\end{figure}
\textbf{Blocks and onsite symmetries}:
\begin{itemize}
    \item[] 1D: $G_{\tau_1} = I,\ G_{\tau_2} = \bzt,\ G_{\tau_3} = \bzt$
    \item[] 0D: $G_{\mu_1} = \bzt,\ G_{\mu_2} = \bzt\rtimes \bzt,\ G_{\mu_3} = \bzt\rtimes \bzt$ 
\end{itemize}
\subsubsection*{Decohered Spinless}

\textbf{Block state decorations}:
\begin{itemize}
    \item[] 0D
    \begin{itemize}
        \item $\mu_1,\ \mu_2,\ \mu_3$: Odd fermion
    \end{itemize}
    \item[] 1D
    \begin{itemize}
        \item $\tau_1$: Majorana chain
        \item $\tau_2,\ \tau_3$: Majorana chain and $\bzt$ fSPT
    \end{itemize}
\end{itemize}

\textbf{Obstructions}
\begin{itemize}
    \item Majorana chain on $\tau_1$ is obstructed at $\mu_1$
    \item Majorana chain on $\tau_2$ or $\tau_3$ is obstructed at $\mu_2$. Simultaneous decoration of these phases are also obstructed.
\end{itemize}

\textbf{Obstruction-free states}:
\begin{itemize}
\item[] 0D states ($\bzt^3$) are obstruction-free (E). 

\item[] 1D ($\bzt^2$)
\begin{enumerate}
    \item $\bzt$ fSPT on $\tau_2$ or $\tau_3$ (I)
\end{enumerate}
    
\end{itemize}

\textbf{Trivializations}:
\begin{itemize}
    \item Majorana bubble on $\sigma$ $\Rightarrow$ Simultaneous decoration of $\bzt$ fSPT on $\tau_2$ and $\tau_3$ and odd fermion on $\mu_1$. This reduces the 0D classification to $\bzt^2$.
\end{itemize}

\textbf{Final classification:}
\begin{itemize}
    \item[] $E_{0,dec}^{0D} = \bzt^2 (E)$
    \item[] $E_{0,dec}^{1D} = \bzt^2 (I)$
    \item[]  $\mathcal{G}_{0,dec} = E_{0,dec}^{0D} \times E_{0,dec}^{1D}  =\bzt^2(E)\times\bzt^2(I)$
\end{itemize}

\subsubsection*{Decohered Spin-1/2}
\textbf{Block state decorations}:
\begin{itemize}
    \item[] 0D
    \begin{itemize}
        \item $\mu_1,\ \mu_2,\ \mu_3$: Odd fermion (I)
    \end{itemize}
    \item[] 1D
    \begin{itemize}
        \item $\tau_1$: Majorana chain (E)
        \item $\tau_2,\ \tau_3$: $\bbz_4^f$ ASPT (I) 
    \end{itemize}
    \item[] These states are all \textbf{obstruction-free} and \textbf{trivialization-free}.
\end{itemize}

\textbf{Trivialization and stacking};
\begin{itemize}
    \item[] While bubble equivalence does not reduce the classification, we have nontrivial stacking:
    \item Stacking two Majorana chains on $\tau_1$ $\Rightarrow$ Odd fermions on $\mu_1$ 
\end{itemize}

\textbf{Final classification:}
\begin{itemize}
    \item[] $E_{1/2,dec}^{0D} = \bzt^3 (I)$
    \item[] $E_{1/2,dec}^{1D} = \bzt(E)\times\bzt^2(I)$
    \item[] Non-trivial stacking $\Rightarrow$ $\mathcal{G}_{1/2,dec} = E_{1/2,dec}^{0D} \rtimes E_{1/2,dec}^{1D}  = \bzt^4(I)\times\bbz_4(I)$
\end{itemize}

\subsubsection*{Disordered Spinless}
\textbf{Block state decorations}:
\begin{itemize}
    \item[] 1D
    \begin{itemize}
        \item $\tau_1,\ \tau_2,\ \tau_3$: Majorana chain 
    \end{itemize}
    \item[] These states are \textbf{obstruction-free} and \textbf{trivialization-free}. All these decorations are obstructed in the clean case, and hence are intrinsic phases.
\end{itemize}

\textbf{Final classification:}
\begin{itemize}
    \item[] $\mathcal{G}_{0,dis} =  E_{0,dis}^{1D}  = \bzt^3 (I)$
\end{itemize}

\subsubsection*{Disordered Spin-1/2}
\textbf{Block state decorations}:
\begin{itemize}
    \item[] 1D
    \begin{itemize}
        \item $\tau_1$: Majorana chain (E)
        \item $\tau_2,\ \tau_3$: \placeholder (I) 
    \end{itemize}
    \item[] These states are \textbf{obstruction-free} and \textbf{trivialization-free}.
\end{itemize}

\textbf{Final classification:}
\begin{itemize}
    \item[] $\mathcal{G}_{1/2,dis} =  E_{1/2,dis}^{1D}  = \bzt(E)\times\bzt^2(I)$
\end{itemize}

\subsection{p4}
\subsubsection*{Cell Decomposition}
\begin{figure}[!htbp]
    \centering
    \includegraphics[width=0.7\linewidth]{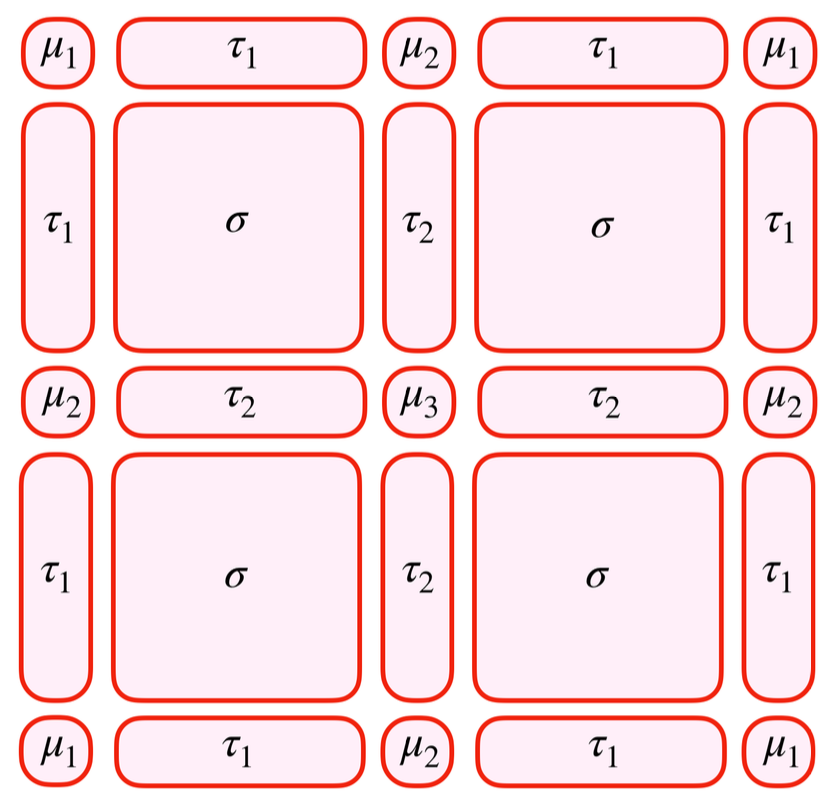}
    \caption{$p4$ lattice}
\end{figure}
\textbf{Blocks and onsite symmetries}:
\begin{itemize}
    \item[] 1D: $G_{\tau_1},\ G_{\tau_2} = I$
    \item[] 0D: $G_{\mu_1} = \bbz_4,\ G_{\mu_2} = \bzt,\ G_{\mu_3} = \bbz_4$ 
\end{itemize}
\subsubsection*{Decohered Spinless}

\textbf{Block state decorations}:
\begin{itemize}
    \item[] 0D
    \begin{itemize}
        \item $\mu_1,\ \mu_2,\ \mu_3$: Odd fermion
    \end{itemize}
    \item[] 1D
    \begin{itemize}
        \item $\tau_1,\ \tau_2$: Majorana chain
    \end{itemize}
\end{itemize}

\textbf{Obstructions}
\begin{itemize}
    \item Majorana chain on $\tau_1$ or $\tau_2$ as well as simultaneous decorations are obstructed at all the rotation centers $\mu_1,\ \mu_2$, and $\mu_3$.
\end{itemize}

\textbf{Obstruction-free states}:
\begin{itemize}
\item[] 0D states ($\bzt^3$) are obstruction-free (E). 

\item[] 1D: No obstruction-free block states ($\bbz_1$)
\end{itemize}

\textbf{Trivializations}:
\begin{itemize}
    \item Majorana bubble on $\sigma$ $\Rightarrow$ Simultaneous decoration of odd fermions on $\mu_1,\ \mu_2$, and $\mu_3$. This reduces the 0D classification to $\bzt^2$.
\end{itemize}

\textbf{Final classification:}
\begin{itemize}
    \item[] $E_{0,dec}^{0D} = \bzt^2 (E)$
    \item[] $E_{0,dec}^{1D} = \bbz_1$
    \item[]  $\mathcal{G}_{0,dec} = E_{0,dec}^{0D} \times E_{0,dec}^{1D}  =\bzt^2 (E)$
\end{itemize}

\subsubsection*{Decohered Spin-1/2}
\textbf{Block state decorations}:
\begin{itemize}
    \item[] 0D
    \begin{itemize}
        \item $\mu_1,\ \mu_2,\ \mu_3$: Odd fermion (I)
    \end{itemize}
    \item[] 1D
    \begin{itemize}
        \item $\tau_1,\ \tau_2$: Majorana chain (E)
    \end{itemize}
    \item[] These states are all \textbf{obstruction-free} and \textbf{trivialization-free}.
\end{itemize}

\textbf{Trivialization and stacking};
\begin{itemize}
    \item[] While bubble equivalence does not reduce the classification, we have nontrivial stacking:
    \item Stacking two Majorana chains on $\tau_1$ $\Rightarrow$ Odd fermions on $\mu_2$ 
\end{itemize}

\textbf{Final classification:}
\begin{itemize}
    \item[] $E_{1/2,dec}^{0D} = \bzt^3 (I)$
    \item[] $E_{1/2,dec}^{1D} = \bzt^2 (E)$
    \item[] Non-trivial stacking $\Rightarrow$ $\mathcal{G}_{1/2,dec} = E_{1/2,dec}^{0D} \rtimes E_{1/2,dec}^{1D}  = \bzt(E)\times\bzt^2(I)\times\bbz_4(I)$
\end{itemize}

\subsubsection*{Disordered}
\textbf{Block state decorations}:
\begin{itemize}
    \item[] 1D
    \begin{itemize}
        \item $\tau_1,\ \tau_2$: Majorana chain 
    \end{itemize}
    \item[] These states are \textbf{obstruction-free} and \textbf{trivialization-free}. All of these decorations are obstructed in the spinless clean case, and hence are intrinsic phases.
\end{itemize}

\textbf{Final classification:}
\begin{itemize}
    \item[] $\mathcal{G}_{0,dis} =  E_{0,dis}^{1D}  = \bzt^2 (I)$
    \item[] \item[] $\mathcal{G}_{1/2,dis} =  E_{1/2,dis}^{1D}  = \bzt^2 (E)$
\end{itemize}

\subsection{p4m}
\subsubsection*{Cell Decomposition}
\begin{figure}[!htbp]
    \centering
    \includegraphics[width=0.7\linewidth]{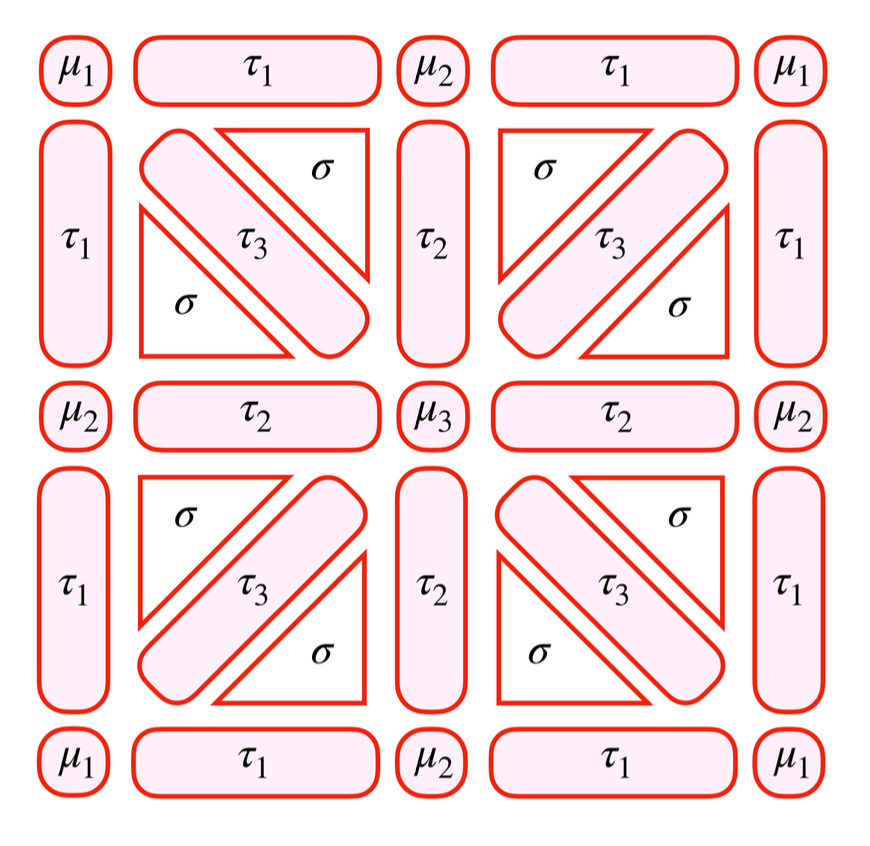}
    \caption{$p4m$ lattice}
\end{figure}
\textbf{Blocks and onsite symmetries}:
\begin{itemize}
    \item[] 1D: $G_{\tau_1},\ G_{\tau_2},\ G_{\tau_3} = \bzt$
    \item[] 0D: $G_{\mu_1} = \bbz_4\rtimes\bzt,\ G_{\mu_2} = \bzt\rtimes\bzt,\ G_{\mu_3} = \bbz_4\rtimes\bzt$ 
\end{itemize}
\subsubsection*{Decohered Spinless}

\textbf{Block state decorations}:
\begin{itemize}
    \item[] 0D
    \begin{itemize}
        \item $\mu_1,\ \mu_2,\ \mu_3$: Odd fermion
    \end{itemize}
    \item[] 1D
    \begin{itemize}
        \item $\tau_1,\ \tau_2,\ \tau_3$: Majorana chain, $\bzt$ fSPT
    \end{itemize}
\end{itemize}

\textbf{Obstructions}
\begin{itemize}
    \item Majorana chain on $\tau_1$, $\tau_2$, or $\tau_3$ as well as any simultaneous decorations of the above are obstructed at all the dihedral centers $\mu_1,\ \mu_2$, and $\mu_3$.
\end{itemize}

\textbf{Obstruction-free states}:
\begin{itemize}
\item[] 0D states ($\bzt^3$) are obstruction-free (E). 

\item[] 1D ($\bzt^3$) \begin{enumerate}
    \item $\bzt$ fSPT on $\tau_1$, $\tau_2$, or $\tau_3$ (I)
\end{enumerate}
\end{itemize}

\textbf{Trivializations}:
\begin{itemize}
    \item Majorana bubble on $\sigma$ $\Rightarrow$ Simultaneous decoration of $\bzt$ fSPT on $\tau_1,\ \tau_2$, and $\tau_3$. This reduces the 1D classification to $\bzt^2$.
\end{itemize}

\textbf{Final classification:}
\begin{itemize}
    \item[] $E_{0,dec}^{0D} = \bzt^2 (E)$
    \item[] $E_{0,dec}^{1D} = \bzt^3 (I)$
    \item[]  $\mathcal{G}_{0,dec} = E_{0,dec}^{0D} \times E_{0,dec}^{1D}  =\bzt^2(E)\times\bzt^3(I)$
\end{itemize}

\subsubsection*{Decohered Spin-1/2}
\textbf{Block state decorations}:
\begin{itemize}
    \item[] 0D
    \begin{itemize}
        \item $\mu_1,\ \mu_2,\ \mu_3$: Odd fermion (I)
    \end{itemize}
    \item[] 1D
    \begin{itemize}
        \item $\tau_1,\ \tau_2,\ \tau_3$: $\bbz_4^f$ ASPT (I)
    \end{itemize}
    \item[] These states are all \textbf{obstruction-free} and \textbf{trivialization-free}.
\end{itemize}

\textbf{Final classification:}
\begin{itemize}
    \item[] $E_{1/2,dec}^{0D} = \bzt^3(I)$
    \item[] $E_{1/2,dec}^{1D} = \bzt^3(I)$
    \item[]  $\mathcal{G}_{1/2,dec} = E_{1/2,dec}^{0D} \times E_{1/2,dec}^{1D}  = \bzt^6(I)$
\end{itemize}

\subsubsection*{Disordered Spinless}
\textbf{Block state decorations}:
\begin{itemize}
    \item[] 1D
    \begin{itemize}
        \item $\tau_1,\ \tau_2, \tau_3$: Majorana chain 
    \end{itemize}
    \item[] These states are \textbf{obstruction-free} and \textbf{trivialization-free}. These decorations are all obstructed in the clean case, and are hence intrinsic phases.
\end{itemize}

\textbf{Final classification:}
\begin{itemize}
    \item[] $\mathcal{G}_{0,dis} =  E_{0,dis}^{1D}  = \bzt^3 (I)$
\end{itemize}

\subsubsection*{Disordered Spin-1/2}
\textbf{Block state decorations}:
\begin{itemize}
    \item[] 1D
    \begin{itemize}
        \item $\tau_1,\ \tau_2, \tau_3$: \placeholder (I)
    \end{itemize}
    \item[] These states are \textbf{obstruction-free} and \textbf{trivialization-free}.
\end{itemize}

\textbf{Final classification:}
\begin{itemize}
    \item[] $\mathcal{G}_{1/2,dis} =  E_{1/2,dis}^{1D}  = \bzt^3 (I)$
\end{itemize}

\subsection{p4g}
\subsubsection*{Cell Decomposition}
\begin{figure}[!htbp]
    \centering
    \includegraphics[width=0.9\linewidth]{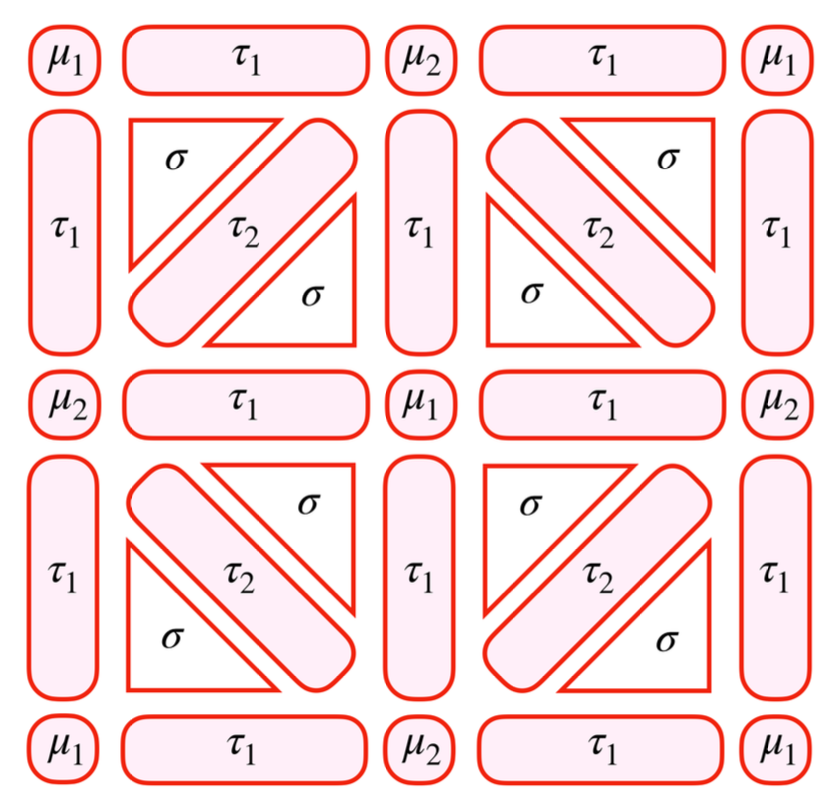}
    \caption{$p4g$ lattice}
\end{figure}
\textbf{Blocks and onsite symmetries}:
\begin{itemize}
    \item[] 1D: $G_{\tau_1} = I,\ G_{\tau_2} = \bzt$
    \item[] 0D: $G_{\mu_1} = \bbz_4,\ G_{\mu_2} = \bzt\rtimes\bzt$ 
\end{itemize}
\subsubsection*{Decohered Spinless}

\textbf{Block state decorations}:
\begin{itemize}
    \item[] 0D
    \begin{itemize}
        \item $\mu_1,\ \mu_2$: Odd fermion
    \end{itemize}
    \item[] 1D
    \begin{itemize}
        \item $\tau_1$: Majorana chain
        \item $\tau_2$: Majorana chain, $\bzt$ fSPT
    \end{itemize}
\end{itemize}

\textbf{Obstructions}
\begin{itemize}
    \item Majorana chain on $\tau_1$ is obstructed at rotation center $\mu_1$
    \item Majorana chain on $\tau_2$ is obstructed at dihedral center $\mu_2$
\end{itemize}

\textbf{Obstruction-free states}:
\begin{itemize}
\item[] 0D states ($\bzt^2$) are obstruction-free (E). 

\item[] 1D ($\bzt$) \begin{enumerate}
    \item $\bzt$ fSPT on $\tau_2$ (E)
\end{enumerate}
\end{itemize}

\textbf{Trivializations}:
\begin{itemize}
    \item Majorana bubble on $\sigma$ $\Rightarrow$ Simultaneous decoration of $\bzt$ fSPT on $\tau_2$ and odd fermion on $\mu_1$. This reduces the 0D classification to $\bzt$.
\end{itemize}

\textbf{Final classification:}
\begin{itemize}
    \item[] $E_{0,dec}^{0D} = \bzt (E)$
    \item[] $E_{0,dec}^{1D} = \bzt (E)$
    \item[]  $\mathcal{G}_{0,dec} = E_{0,dec}^{0D} \times E_{0,dec}^{1D}  =\bzt^2 (E)$
\end{itemize}

\subsubsection*{Decohered Spin-1/2}
\textbf{Block state decorations}:
\begin{itemize}
    \item[] 0D
    \begin{itemize}
        \item $\mu_1,\ \mu_2$: Odd fermion (I)
    \end{itemize}
    \item[] 1D
    \begin{itemize}
        \item $\tau_1$: Majorana chain (E)
        \item $\tau_2$: $\bbz_4^f$ ASPT (I)
    \end{itemize}
    \item[] These states are all \textbf{obstruction-free} and \textbf{trivialization-free}.
\end{itemize}

\textbf{Final classification:}
\begin{itemize}
    \item[] $E_{1/2,dec}^{0D} = \bzt^2(I)$
    \item[] $E_{1/2,dec}^{1D} = \bzt(E)\times\bzt(I)$
    \item[]  $\mathcal{G}_{1/2,dec} = E_{1/2,dec}^{0D} \times E_{1/2,dec}^{1D}  = \bzt(E)\times\bzt^3(I)$
\end{itemize}

\subsubsection*{Disordered Spinless}
\textbf{Block state decorations}:
\begin{itemize}
    \item[] 1D
    \begin{itemize}
        \item $\tau_1,\ \tau_2$: Majorana chain 
    \end{itemize}
    \item[] These states are \textbf{obstruction-free} and \textbf{trivialization-free}. All these decorations are obstructed in the clean case, and are hence intrinsic phases.
\end{itemize}

\textbf{Final classification:}
\begin{itemize}
    \item[] $\mathcal{G}_{0,dis} =  E_{0,dis}^{1D}  = \bzt^2(I)$ 
\end{itemize}

\subsubsection*{Disordered Spin-1/2}
\textbf{Block state decorations}:
\begin{itemize}
    \item[] 1D
    \begin{itemize}
        \item $\tau_1$: Majorana chain (E)
        \item $\tau_2$: \placeholder (I)
    \end{itemize}
    \item[] These states are \textbf{obstruction-free} and \textbf{trivialization-free}.
\end{itemize}

\textbf{Final classification:}
\begin{itemize}
    \item[] $\mathcal{G}_{1/2,dis} =  E_{1/2,dis}^{1D}  = \bzt(E)\times\bzt(I)$
\end{itemize}

\subsection{p3}
\subsubsection*{Cell Decomposition}
\begin{figure}[!htbp]
    \centering
    \includegraphics[width=0.6\linewidth]{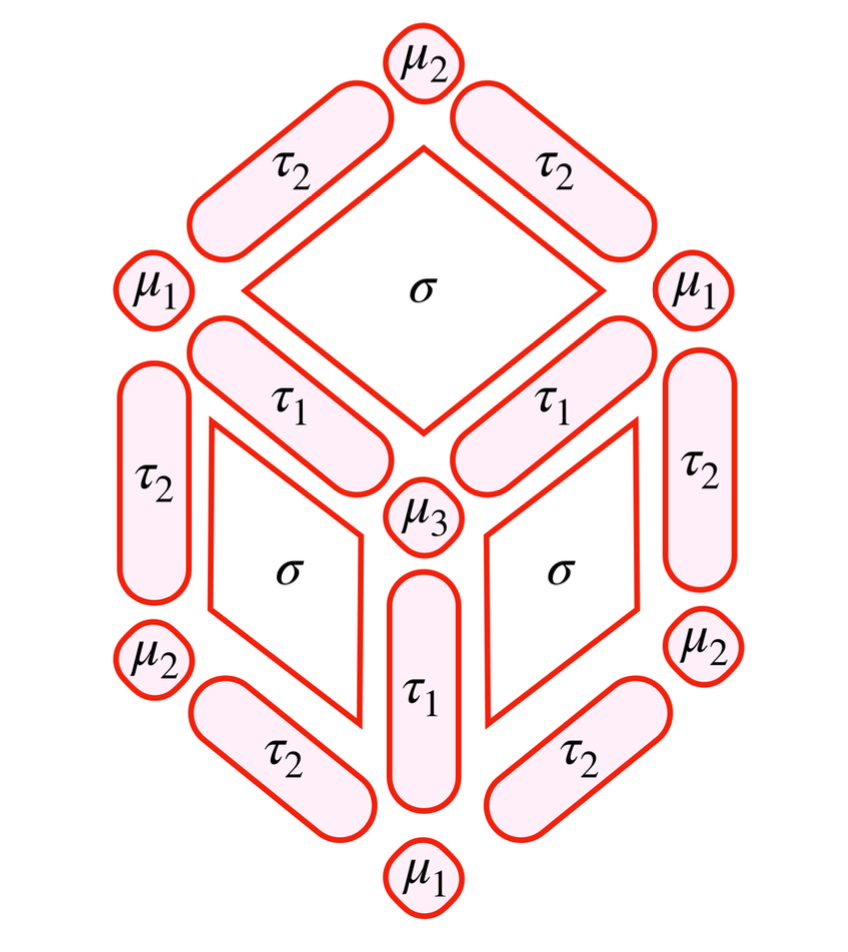}
    \caption{$p3$ lattice}
\end{figure}
B\textbf{Blocks and onsite symmetries}:
\begin{itemize}
    \item[] 1D: $G_{\tau_1},\ G_{\tau_2} = I$
    \item[] 0D: $G_{\mu_1},\ G_{\mu_2},\ G_{\mu_3} = \bbz_3$ 
\end{itemize}

For this point group, the obstruction-free and trivialization-states are identical for both spinless and spin-1/2 cases, and are hence discussed together. 
\subsubsection*{Decohered}

\textbf{Block state decorations}:
\begin{itemize}
    \item[] 0D
    \begin{itemize}
        \item $\mu_1,\ \mu_2,\ \mu_3$: Odd fermion
    \end{itemize}
    \item[] 1D
    \begin{itemize}
        \item $\tau_1,\ \tau_2$: Majorana chain 
    \end{itemize}
\end{itemize}

\textbf{Obstructions}
\begin{itemize}
    \item Majorana chain on $\tau_1$ is obstructed as it leaves odd number of Majorana modes on $\mu_3$.
    \item Majorana chain on $\tau_2$ is obstructed as it leaves odd number of Majorana modes on $\mu_2$.
\end{itemize}

\textbf{Obstruction-free states}:
\begin{itemize}
\item[] 0D states ($\bzt^3$) are obstruction-free. 

\item[] 1D: No obstruction-free states ($\bbz_1$)

\end{itemize}

\textbf{Trivializations}:
\begin{itemize}
    \item Fermionic insulator bubble on $\tau_1$ $\Rightarrow$ Odd fermions on $\mu_1$ and $\mu_3$. This reduces 0D classification to $\bzt^2$
    \item Fermionic insulator bubble on $\tau_2$ $\Rightarrow$ Odd fermions on $\mu_1$ and $\mu_2$. This further reduces 0D classification to $\bzt$.
\end{itemize}

\textbf{Final classification:}
\begin{itemize}
    \item[] $E_{0,dec}^{0D} = \bzt(E)$
    \item[] $E_{0,dec}^{1D} = \bbz_1$
    \item[]  $\mathcal{G}_{0,dec} = E_{0,dec}^{0D} \times E_{0,dec}^{1D}  =\bzt(E)$
    \item The 0D decoration is obstructed in the spin-1/2 clean case, and is hence an intrinsic phase.
    \item[] $E_{1/2,dec}^{0D} = \bzt(I)$
    \item[] $E_{1/2,dec}^{1D} = \bbz_1$
    \item[]  $\mathcal{G}_{1/2,dec} = E_{1/2,dec}^{0D} \times E_{1/2,dec}^{1D}  =\bzt(I)$
\end{itemize}

\subsubsection*{Disordered}
\textbf{Block state decorations}:
\begin{itemize}
    \item[] 1D
    \begin{itemize}
        \item $\tau_1,\ \tau_2$: Majorana chain 
    \end{itemize}
\end{itemize}

\textbf{Obstructions}
\begin{itemize}
    \item[] Same as for decohered system.
    \item Majorana chain on $\tau_1$ is obstructed as it leaves odd number of Majorana modes on $\mu_3$.
    \item Majorana chain on $\tau_2$ is obstructed as it leaves odd number of Majorana modes on $\mu_2$.
    \item Hence there are \textbf{no obstruction-free states}.
\end{itemize}

\textbf{Final classification:}
\begin{itemize}
    \item[] $\mathcal{G}_{dis} =  E_{dis}^{1D}  = \bbz_1$
\end{itemize}

\subsection{p3m1}
\subsubsection*{Cell Decomposition}
\begin{figure}[!htbp]
    \centering
    \includegraphics[width=0.9\linewidth]{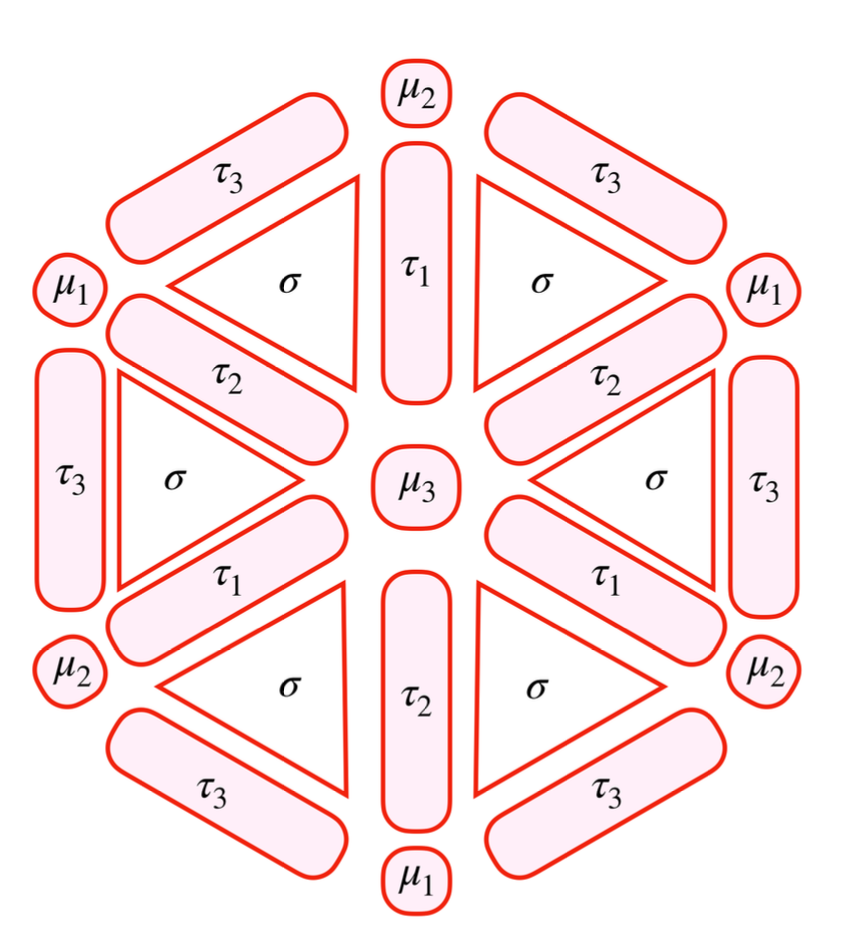}
    \caption{$p3m1$ lattice}
\end{figure}
\textbf{Blocks and onsite symmetries}:
\begin{itemize}
    \item 1D: $G_{\tau_1}.\ G_{\tau_2},\ G_{\tau_3} = \bzt$
    \item 0D: $G_{\mu_1},\ G_{\mu_2},\ G_{\mu_3} = \bbz_3\rtimes\bzt$ ($D_3$)
\end{itemize}

\subsubsection*{Decoherence Spinless}
\textbf{Block state decorations}:
\begin{itemize}
    \item[] 0D
    \begin{itemize}
        \item $\mu_1,\ \mu_2,\ \mu_3$: Odd fermion
    \end{itemize}
    \item[] 1D
    \begin{itemize}
        \item $\tau_1,\ \tau_2, \tau_3$: Majorana chain, $\bzt$ fSPT
    \end{itemize}
\end{itemize}

\textbf{Obstructions}
\begin{itemize}
    \item Majorana chain on $\tau_1$ leaves odd number of Majorana modes on $\mu_2$ and $\mu_3$
    \item Majorana chain on $\tau_2$ leaves odd number of Majorana modes on $\mu_1$ and $\mu_3$
    \item Majorana chain on $\tau_3$ leaves odd number of Majorana modes on $\mu_3$ and $\mu_2$
    \item To obtain an obstruction-free state, there must be even number of Majorana modes on all 0D blocks $\mu$, hence the only obstruction-free state is the simultaneous decoration of all of the above phases.
    \item Similarly, any decoration of $\bzt$ fSPT on $\tau_1,\ \tau_2,\ \tau_3$ is obstructed except for the simultaneous decoration of all three states. 
\end{itemize}

\textbf{Obstruction-free states}:
\begin{itemize}
\item[] 0D states ($\bzt^3$) are obstruction-free (E). 

\item[] 1D ($\bzt^2$) \begin{enumerate}
    \item Simultaneous decoration of Majorana chain on $\tau_1,\ \tau_2,\ \tau_3$ (E).
    \item Simultaneous decoration of $\bzt$ fSPT on $\tau_1,\ \tau_2,\ \tau_3$. (E)
\end{enumerate}
\end{itemize}

\textbf{Trivializations}:
\begin{itemize}
    \item Majorana bubble on $\sigma$ $\Rightarrow$ $\bzt$ fSPT on $\tau_1,\ \tau_2$, and $\tau_3$. This reduces 1D classification to $\bzt$
    \item Fermionic insulator bubble on $\tau_1$ $\Rightarrow$ Odd fermions on $\mu_2$ and $\mu_3$. This reduces 0D classification to $\bzt^2$
    \item Fermionic insulator bubble on $\tau_2$ $\Rightarrow$ Odd fermions on $\mu_1$ and $\mu_3$. This further reduces 0D classification to $\bzt$.
\end{itemize}

\textbf{Final classification:}
\begin{itemize}
    \item[] $E_{0,dec}^{0D} = \bzt(E)$
    \item[] $E_{0,dec}^{1D} = \bzt(E)$
    \item[]  $\mathcal{G}_{0,dec} = E_{0,dec}^{0D} \times E_{0,dec}^{1D}  =\bzt^2(E)$
\end{itemize}

\subsubsection*{Decohered Spin-1/2}

\textbf{Block state decorations}:
\begin{itemize}
    \item[] 0D
    \begin{itemize}
        \item $\mu_1,\ \mu_2,\ \mu_3$: Odd fermion (I)
    \end{itemize}
    \item[] 1D
    \begin{itemize}
        \item $\tau_1,\ \tau_2$: $\bbz_4^f$ ASPT (I)
    \end{itemize}
\end{itemize}

\textbf{Obstructions}
\begin{itemize}
    \item $\bbz_4^f$ ASPT on $\tau_1$ is obstructed at $\mu_2$ and $\mu_3$
    \item $\bbz_4^f$ ASPT on $\tau_2$ is obstructed at $\mu_1$ and $\mu_3$
    \item $\bbz_4^f$ ASPT on $\tau_3$ is obstructed at $\mu_1$ and $\mu_2$
    \item Only decoration that cancels anomaly on all 0D blocks is the simultaneous decoration of the above 3 phases. 
\end{itemize}

\textbf{Obstruction-free states}:
\begin{itemize}
\item[] 0D states ($\bzt^3$) are obstruction-free (I). 

\item[] 1D ($\bzt$) \begin{enumerate}
    \item Simultaneous decoration of $\bbz_4^f$ ASPT on $\tau_1,\ \tau_2,\ \tau_3$ (I).
\end{enumerate}
\end{itemize}

\textbf{Trivializations}:
\begin{itemize}
    \item Fermionic insulator bubble on $\tau_1$ $\Rightarrow$ Odd fermions on $\mu_2$ and $\mu_3$. This reduces 0D classification to $\bzt^2$
    \item Fermionic insulator bubble on $\tau_2$ $\Rightarrow$ Odd fermions on $\mu_1$ and $\mu_3$. This further reduces 0D classification to $\bzt$.
\end{itemize}

\textbf{Final classification:}
\begin{itemize}
    \item[] $E_{1/2,dec}^{0D} = \bzt (I)$
    \item[] $E_{1/2,dec}^{1D} = \bzt (I)$
    \item[]  $\mathcal{G}_{1/2,dec} = E_{1/2,dec}^{0D} \times E_{1/2,dec}^{1D}  =\bzt^2(I)$
\end{itemize}

\subsubsection*{Disordered Spinless}
\textbf{Block state decorations}:
\begin{itemize}
    \item[] 1D
    \begin{itemize}
        \item $\tau_1,\ \tau_2,\ \tau_3$: Majorana chain
    \end{itemize}
\end{itemize}

\textbf{Obstructions}
\begin{itemize}
    \item Majorana chain on $\tau_1$ leaves odd number of Majorana modes on $\mu_2$ and $\mu_3$
    \item Majorana chain on $\tau_2$ leaves odd number of Majorana modes on $\mu_1$ and $\mu_3$
    \item Majorana chain on $\tau_3$ leaves odd number of Majorana modes on $\mu_3$ and $\mu_2$
\end{itemize}

\textbf{Obstruction-free states}:
\begin{itemize}

\item[] 1D ($\bzt$)
\begin{enumerate}
    \item Simultaneous decoration of Majorana chain on $\tau_1,\ \tau_2,\ \tau_3$ (E)
\end{enumerate}
\item[] This state is \textbf{trivialization-free}. 
\end{itemize}

\textbf{Final classification:}
\begin{itemize}
    \item[] $\mathcal{G}_{0,dis} =  E_{0,dis}^{1D}  = \bzt(E)$
\end{itemize}

\subsubsection*{Disordered Spin-1/2}
\textbf{Block state decorations}:
\begin{itemize}
    \item[] 1D
    \begin{itemize}
        \item $\tau_1,\ \tau_2,\ \tau_3$: \placeholder (I)
    \end{itemize}
\end{itemize}

\textbf{Obstructions}
\begin{itemize}
    \item \placeholder on $\tau_1$ leaves odd number of \placeholdermodes on $\mu_2$ and $\mu_3$
    \item \placeholder on $\tau_2$ leaves odd number of \placeholdermodes on $\mu_1$ and $\mu_3$
    \item \placeholder on $\tau_3$ leaves odd number of \placeholdermodes on $\mu_3$ and $\mu_2$
\end{itemize}

\textbf{Obstruction-free states}:
\begin{itemize}

\item[] 1D ($\bzt$)
\begin{enumerate}
    \item \placeholders on $\tau_1,\ \tau_2$ and $\tau_3$
\end{enumerate}
\item[] This state is \textbf{trivialization-free}.
\end{itemize}

\textbf{Final classification:}
\begin{itemize}
    \item[] $\mathcal{G}_{1/2,dis} =  E_{1/2,dis}^{1D}  = \bzt(I)$
\end{itemize}

\subsection{p31m}
\subsubsection*{Cell Decomposition}
\begin{figure}[!htbp]
    \centering
    \includegraphics[width=0.9\linewidth]{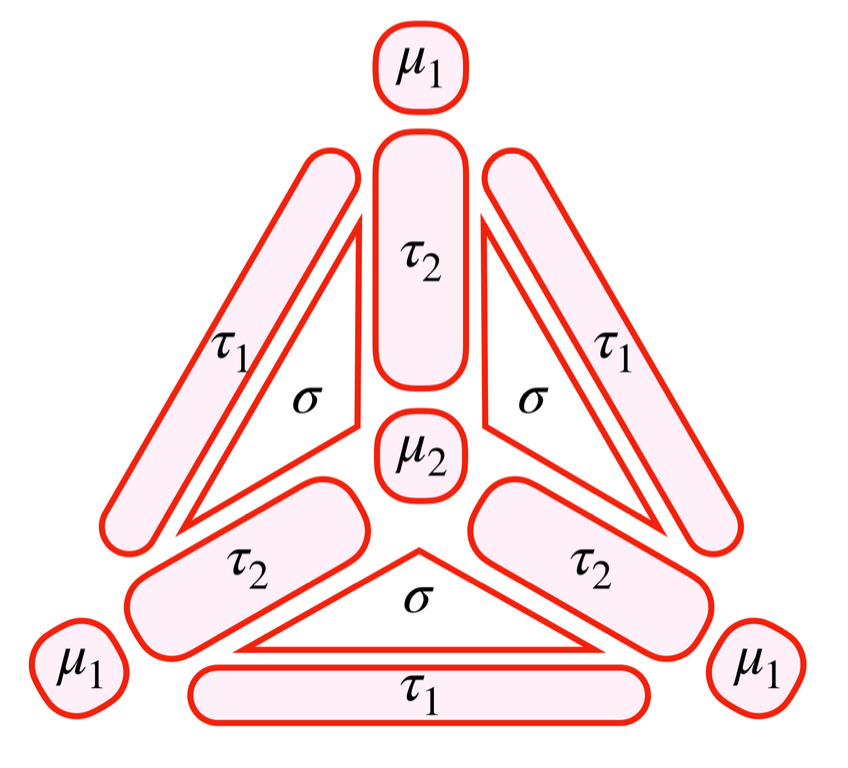}
    \caption{$p31m$ lattice}
\end{figure}
\textbf{Blocks and onsite symmetries}:
\begin{itemize}
    \item 1D: $G_{\tau_1} = \bzt,\ G_{\tau_2} = I$
    \item 0D: $G_{\mu_1} = \bbz_3\rtimes \bzt,\ G_{\mu_2} = \bbz_3$
\end{itemize}
\subsubsection*{Decohered Spinless}
\textbf{Block state decorations}:
\begin{itemize}
    \item[] 0D
    \begin{itemize}
        \item $\mu_1,\ \mu_2$: Odd fermion
    \end{itemize}
    \item[] 1D
    \begin{itemize}
        \item $\tau_1$: Majorana chain, $\bzt$ fSPT
        \item $\tau_2$: Majorana chain
    \end{itemize}
\end{itemize}

\textbf{Obstructions}
\begin{itemize}
    \item Majorana chain on $\tau_2$ is obstructed as it leaves odd number of Majorana modes on $\mu_2$
\end{itemize}

\textbf{Obstruction-free states}:
\begin{itemize}
\item[] 0D states ($\bzt^2$) are obstruction-free (E). 

\item[] 1D ($\bzt^2$) \begin{enumerate}
    \item Majorana chain on $\tau_1$ (E)
    \item $\bzt$ fSPT on $\tau_1$ (E)
\end{enumerate}
\end{itemize}

\textbf{Trivializations}:
\begin{itemize}
    \item Majorana bubble on $\sigma$ $\Rightarrow$ $\bzt$ fSPT on $\tau_1$. This reduces 1D classification to $\bzt$
    \item Fermionic insulator bubble on $\tau_2$ $\Rightarrow$ Odd fermions on $\mu_2$. This reduces 0D classification to $\bzt$.
\end{itemize}

\textbf{Final classification:}
\begin{itemize}
    \item[] $E_{0,dec}^{0D} = \bzt(E)$
    \item[] $E_{0,dec}^{1D} = \bzt(E)$
    \item[]  $\mathcal{G}_{0,dec} = E_{0,dec}^{0D} \times E_{0,dec}^{1D}  =\bzt^2(E)$
\end{itemize}

\subsubsection*{Decohered Spin-1/2}

\textbf{Block state decorations}:
\begin{itemize}
    \item[] 0D
    \begin{itemize}
        \item $\mu_1,\ \mu_2$: Odd fermion
    \end{itemize}
    \item[] 1D
    \begin{itemize}
        \item $\tau_1$: $\bbz_4^f$ ASPT
        \item $\tau_2$: Majorana chain
    \end{itemize}
\end{itemize}

\textbf{Obstructions}:
\begin{itemize}
    \item Majorana chain on $\tau_2$ is obstructed as it leaves odd number of Majorana modes on $\mu_2$
\end{itemize}

\textbf{Obstruction-free states}:
\begin{itemize}
\item[] 0D states ($\bzt^2$) are obstruction-free (I). 

\item[] 1D ($\bzt$) \begin{enumerate}
    \item $\bbz_4^f$ ASPT on $\tau_1$ (I)
\end{enumerate}
\end{itemize}

\textbf{Trivializations}:
\begin{itemize}
    \item Fermionic insulator bubble on $\tau_2$ $\Rightarrow$ Odd fermions on $\mu_2$. This reduces 0D classification to $\bzt$.
\end{itemize}

\textbf{Final classification:}
\begin{itemize}
    \item[] $E_{1/2,dec}^{0D} = \bzt(I)$
    \item[] $E_{1/2,dec}^{1D} = \bzt(I)$
    \item[]  $\mathcal{G}_{1/2,dec} = E_{1/2,dec}^{0D} \times E_{1/2,dec}^{1D}  =\bzt^2(I)$
\end{itemize}

\subsubsection*{Disordered Spinless}
\textbf{Block state decorations}:
\begin{itemize}
    \item[] 1D
    \begin{itemize}
        \item $\tau_1,\ \tau_2$: Majorana chain
    \end{itemize}
\end{itemize}

\textbf{Obstructions}
\begin{itemize}
    \item Majorana chain on $\tau_2$ is obstructed as it leaves odd number of Majorana modes on $\mu_2$
\end{itemize}

\textbf{Obstruction-free states}:
\begin{itemize}

\item[] 1D ($\bzt$)
\begin{enumerate}
    \item Majorana chain on $\tau_1$ (E)
\end{enumerate}
\item[] This state is \textbf{trivialization-free}.
\end{itemize}

\textbf{Final classification:}
\begin{itemize}
    \item[] $\mathcal{G}_{0,dis} =  E_{0,dis}^{1D}  = \bzt(E)$
\end{itemize}

\subsubsection*{Disordered Spin-1/2}
\textbf{Block state decorations}:
\begin{itemize}
    \item[] 1D
    \begin{itemize}
        \item $\tau_1$: \placeholder
        \item $\tau_2$: Majorana chain
    \end{itemize}
\end{itemize}

\textbf{Obstructions}
\begin{itemize}
    \item Majorana chain on $\tau_2$ is obstructed as it leaves odd number of Majorana modes on $\mu_2$
\end{itemize}

\textbf{Obstruction-free states}:
\begin{itemize}

\item[] 1D ($\bzt$)
\begin{enumerate}
    \item \placeholder on $\tau_1$ (I)
\end{enumerate}
\item[] This state is \textbf{trivialization-free}.
\end{itemize}

\textbf{Final classification:}
\begin{itemize}
    \item[] $\mathcal{G}_{1/2,dis} =  E_{1/2,dis}^{1D}  = \bzt(I)$
\end{itemize}

\subsection{p6}
\subsubsection*{Cell Decomposition}
\begin{figure}[!htbp]
    \centering
    \includegraphics[width=0.9\linewidth]{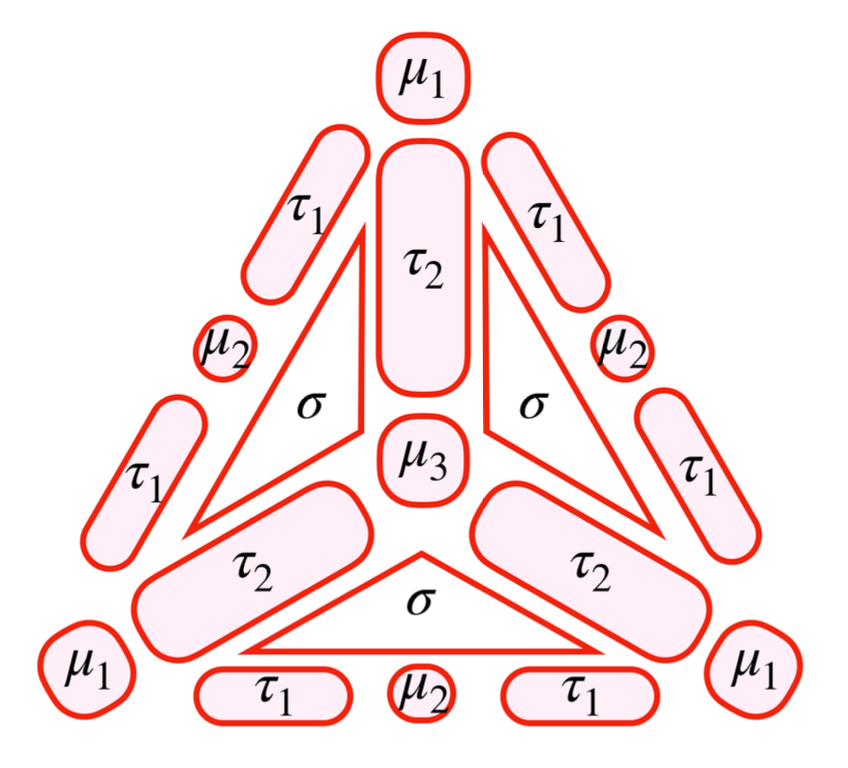}
    \caption{$p6$ lattice}
\end{figure}
\textbf{Blocks and onsite symmetries}:
\begin{itemize}
    \item 1D: $G_{\tau_1},\ G_{\tau_2} = I$
    \item 0D: $G_{\mu_1} = \bbz_6,\ G_{\mu_2} = \bzt,\ G_{\mu_3} = \bbz_3$
\end{itemize}
\subsubsection*{Decohered Spinless}
\textbf{Block state decorations}:
\begin{itemize}
    \item[] 0D
    \begin{itemize}
        \item $\mu_1,\ \mu_2, \mu_3$: Odd fermion
    \end{itemize}
    \item[] 1D
    \begin{itemize}
        \item $\tau_1,\ \tau_2$: Majorana chain
    \end{itemize}
\end{itemize}

\textbf{Obstructions}
\begin{itemize}
    \item Majorana chain on $\tau_1$ is obstructed at $\mu_2$
    \item Majorana chain on $\tau_2$ is obstructed at $\mu_3$
\end{itemize}

\textbf{Obstruction-free states}:
\begin{itemize}
\item[] 0D states ($\bzt^3$) are obstruction-free (E). 

\item[] 1D: No obstruction-free states ($\bbz_1$)
\end{itemize}

\textbf{Trivializations}:
\begin{itemize}
    \item Majorana bubble on $\sigma$ $\Rightarrow$ Simultaneous decoration of odd fermions on $\mu_1$ and $\mu_2$. This reduces 0D classification to $\bzt^2$
    \item Fermionic insulator bubble on $\tau_2$ $\Rightarrow$ Odd fermions on $\mu_3$. This further reduces 0D classification to $\bzt$.
\end{itemize}

\textbf{Final classification:}
\begin{itemize}
    \item[] $E_{0,dec}^{0D} = \bzt(E)$
    \item[] $E_{0,dec}^{1D} = \bbz_1$
    \item[]  $\mathcal{G}_{0,dec} = E_{0,dec}^{0D} \times E_{0,dec}^{1D}  =\bzt(E)$
\end{itemize}

\subsubsection*{Decohered Spin-1/2}

\textbf{Block state decorations}:
\begin{itemize}
    \item[] 0D
    \begin{itemize}
        \item $\mu_1,\ \mu_2, \mu_3$: Odd fermion
    \end{itemize}
    \item[] 1D
    \begin{itemize}
        \item $\tau_1,\ \tau_2$: Majorana chain
    \end{itemize}
\end{itemize}

\textbf{Obstructions}:
\begin{itemize}
    \item Majorana chain on $\tau_2$ is obstructed at $\mu_3$
\end{itemize}

\textbf{Obstruction-free states}:
\begin{itemize}
\item[] 0D states ($\bzt^3$) are obstruction-free (I). 

\item[] 1D ($\bzt$) \begin{enumerate}
    \item Majorana chain on $\tau_1$ (E)
\end{enumerate}
\end{itemize}

\textbf{Trivializations and Stacking}:
\begin{itemize}
    \item Fermionic insulator bubble on $\tau_2$ $\Rightarrow$ Odd fermions on $\mu_3$. This reduces 0D classification to $\bzt^2$.
    \item Nontrivial stacking: Two copies of Majorana chain on $\tau_1$ $\Rightarrow$ Odd fermion on $\mu_2$ 
\end{itemize}

\textbf{Final classification:}
\begin{itemize}
    \item[] $E_{1/2,dec}^{0D} = \bzt^2(I)$
    \item[] $E_{1/2,dec}^{1D} = \bzt(E)$
    \item[] Non-trivial stacking $\Rightarrow$ $\mathcal{G}_{1/2,dec} = E_{1/2,dec}^{0D} \rtimes E_{1/2,dec}^{1D}  =\bzt(I)\times \bbz_4(I)$
\end{itemize}

\subsubsection*{Disordered}
\textbf{Block state decorations}:
\begin{itemize}
    \item[] 1D
    \begin{itemize}
        \item $\tau_1,\ \tau_2$: Majorana chain
    \end{itemize}
\end{itemize}

\textbf{Obstructions}
\begin{itemize}
    \item Majorana chain on $\tau_2$ is obstructed as it leaves odd number of Majorana modes on $\mu_3$
\end{itemize}

\textbf{Obstruction-free states}:
\begin{itemize}

\item[] 1D ($\bzt$)
\begin{enumerate}
    \item Majorana chain on $\tau_1$
\end{enumerate}
\item[] This state is \textbf{trivialization-free}. This decoration is obstructed in the spinless clean case, and is hence an intrinsic phase.
\end{itemize}

\textbf{Final classification:}
\begin{itemize}
    \item[] $\mathcal{G}_{0,dis} =  E_{0,dis}^{1D}  = \bzt(I)$
    \item[] $\mathcal{G}_{1/2,dis} = E_{1/2,dis}^{1D} = \bzt(E)$
\end{itemize}

\subsection{p6m}
\subsubsection*{Cell Decomposition}
\begin{figure}[!htbp]
    \centering
    \includegraphics[width=0.9\linewidth]{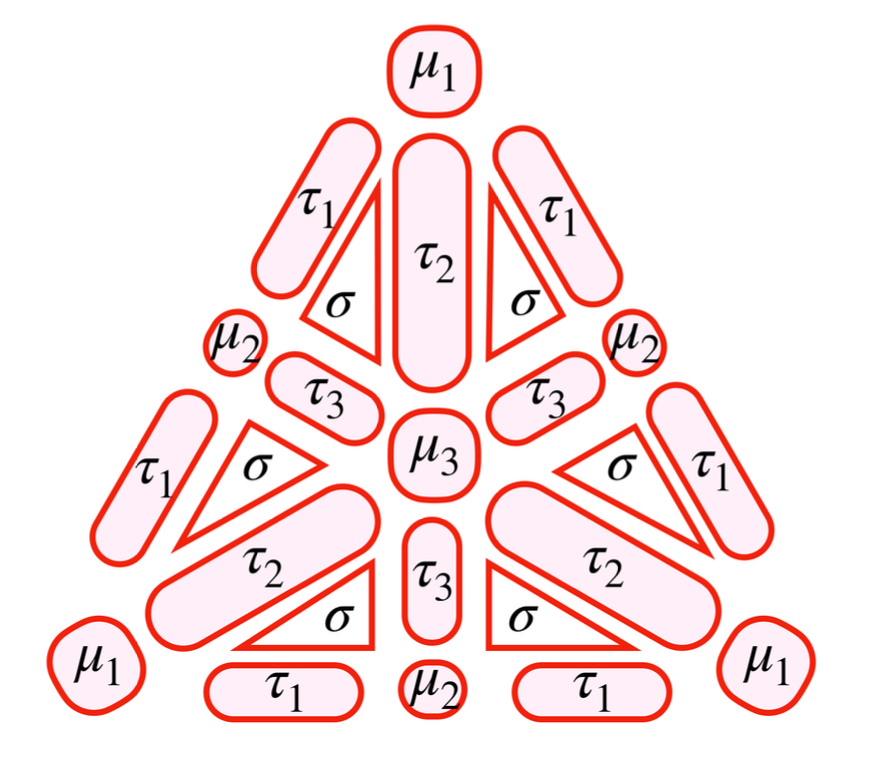}
    \caption{$p6m$ lattice}
\end{figure}
\textbf{Blocks and onsite symmetries}:
\begin{itemize}
    \item $G_{\tau_1},\ G_{\tau_2}\ G_{\tau_3} = \bzt$
    \item $G_{\mu_1} = \bbz_6\rtimes\bzt,\ G_{\mu_2} = \bzt \rtimes \bzt,\ G_{\mu_3} = \bbz_3\rtimes \bzt$
\end{itemize}
\subsubsection*{Decohered Spinless}
\textbf{Block state decorations}:
\begin{itemize}
    \item[] 0D
    \begin{itemize}
        \item $\mu_1,\ \mu_2, \mu_3$: Odd fermion
    \end{itemize}
    \item[] 1D
    \begin{itemize}
        \item $\tau_1,\ \tau_2$: Majorana chain, $\bzt$ fSPT
    \end{itemize}
\end{itemize}

\textbf{Obstructions}
\begin{itemize}
    \item Majorana chain on $\tau_1$ is obstructed at $\mu_1$ and $\mu_2$
    \item Majorana chain on $\tau_2$ is obstructed at $\mu_1$ and $\mu_3$
    \item Majorana chain on $\tau_3$ is obstructed at $\mu_2$ and $\mu_3$
    \item Any simultaneous decorations of the above phases is also obstructed
    \item $\bzt$ fSPT on one of $\tau_2$ or $\tau_3$ is obstructed at $\mu_3$ (odd dihedral center), while the simultaneous decoration of both is obstruction-free.
\end{itemize}

\textbf{Obstruction-free states}:
\begin{itemize}
\item[] 0D states ($\bzt^3$) are obstruction-free (E). 

\item[] 1D: ($\bzt^2$) \begin{enumerate}
    \item $\bzt$ fSPT on $\tau_1$ (I)  
    \item Simultaneous decoration of $\bzt$ fSPT on $\tau_2$ and $\tau_3$ (E)
\end{enumerate}
\end{itemize}

\textbf{Trivializations}:
\begin{itemize}
    \item Majorana bubble on $\sigma$ $\Rightarrow$ Simultaneous decoration of $\bzt$ fSPT on $\tau_1,\ \tau_2$, and $\tau_3$. This reduces 1D classification to $\bzt$.
    \item Fermionic insulator bubble on $\tau_2$ (or $\tau_3$) $\Rightarrow$ Odd fermion on $\mu_3$. This reduces 0D classification to $\bzt^2$.
\end{itemize}

\textbf{Final classification:}
\begin{itemize}
    \item[] $E_{0,dec}^{0D} = \bzt^2 (E)$
    \item[] $E_{0,dec}^{1D} = \bzt(I)$
    \item[]  $\mathcal{G}_{0,dec} = E_{0,dec}^{0D} \times E_{0,dec}^{1D}  =\bzt^2(E)\times\bzt(I)$
\end{itemize}

\subsubsection*{Decohered Spin-1/2}

\textbf{Block state decorations}:
\begin{itemize}
    \item[] 0D
    \begin{itemize}
        \item $\mu_1,\ \mu_2, \mu_3$: Odd fermion
    \end{itemize}
    \item[] 1D
    \begin{itemize}
        \item $\tau_1,\ \tau_2, \tau_3$: $\bbz_4^f$ ASPT 
    \end{itemize}
\end{itemize}

\textbf{Obstructions}:
\begin{itemize}
    \item $\bbz_4^f$ ASPT on one of $\tau_2$ or $\tau_3$ is obstructed at $\mu_3$ (odd dihedral center), while the simultaneous decoration of both is obstruction-free.
\end{itemize}

\textbf{Obstruction-free states}:
\begin{itemize}
\item[] 0D states ($\bzt^3$) are obstruction-free (I). 

\item[] 1D ($\bzt^2$) \begin{enumerate}
    \item $\bbz_4^f$ ASPT on $\tau_1$  (I)
    \item Simultaneous decoration of $\bbz_4^f$ ASPT on $\tau_2$ and $\tau_3$ (I)
\end{enumerate}
\end{itemize}

\textbf{Trivializations}:
\begin{itemize}
    \item Fermionic insulator bubble on $\tau_2$ (or $\tau_3$) $\Rightarrow$ Odd fermion on $\mu_3$. This reduces 0D classification to $\bzt^2$.
\end{itemize}

\textbf{Final classification:}
\begin{itemize}
    \item[] $E_{1/2,dec}^{0D} = \bzt^2 (I)$
    \item[] $E_{1/2,dec}^{1D} = \bzt^2 (I)$
    \item[]  $\mathcal{G}_{1/2,dec} = E_{1/2,dec}^{0D} \rtimes E_{1/2,dec}^{1D}  =\bzt^4(I)$
\end{itemize}

\subsubsection*{Disordered Spinless}
\textbf{Block state decorations}:
\begin{itemize}
    \item[] 1D
    \begin{itemize}
        \item $\tau_1,\ \tau_2, \tau_3$: Majorana chain
    \end{itemize}
\end{itemize}

\textbf{Obstructions}
\begin{itemize}
    \item Majorana chain on one of $\tau_2$ or $\tau_3$ is obstructed as it leaves odd number of Majorana modes at $\mu_3$, while the simultaneous decoration of both is obstruction-free.
\end{itemize}

\textbf{Obstruction-free states}:
\begin{itemize}

\item[] 1D ($\bzt^2$)
\begin{enumerate}
    \item Majorana chain on $\tau_1$ (I)
    \item Simultaneous decoration of Majorana chain on $\tau_2$ and $\tau_3$ (I)
\end{enumerate}
\item[] These states are all \textbf{trivialization-free}.
\end{itemize}

\textbf{Final classification:}
\begin{itemize}
    \item[] $\mathcal{G}_{0,dis} =  E_{0,dis}^{1D}  = \bzt^2(I)$
\end{itemize}

\subsubsection*{Disordered Spin-1/2}
\textbf{Block state decorations}:
\begin{itemize}
    \item[] 1D
    \begin{itemize}
        \item $\tau_1,\ \tau_2, \tau_3$: \placeholder
    \end{itemize}
\end{itemize}

\textbf{Obstructions}
\begin{itemize}
    \item \placeholder on one of $\tau_2$ or $\tau_3$ is obstructed as it leaves odd number of \placeholdermodes at $\mu_3$, while the simultaneous decoration of both is obstruction-free.
\end{itemize}

\textbf{Obstruction-free states}:
\begin{itemize}

\item[] 1D ($\bzt^2$)
\begin{enumerate}
    \item \placeholder on $\tau_1$ (I)
    \item \placeholders on $\tau_2$ and $\tau_3$ (I)
\end{enumerate}
\item[] These states are all \textbf{trivialization-free}.
\end{itemize}

\textbf{Final classification:}
\begin{itemize}
    \item[] $\mathcal{G}_{1/2,dis} =  E_{1/2,dis}^{1D}  = \bzt^2 (I)$
\end{itemize}

\section{Classification details for 3D point groups}\label{sec:3d_class_details}

In this appendix, we describe the detailed construction and classification of ASPTs for all 32 3D point groups. Just as for the 2D space-groups above, we list all elements of construction and separately label the extrinsic and intrinsic classification data as $(E)$ and $(I)$ respectively.

\subsection{$C_1$ - Trivial lattice}
No symmetries inside the unit cell, hence no SPT.
\subsection{$C_i$}
The 1D and 2D blocks do not have any internal symmetry, and hence nothing changes in the average case.

\subsection{$C_2$}
\subsubsection*{Cell Decomposition}
\begin{figure}[!htbp]
    \centering
    \includegraphics[width=0.9\linewidth]{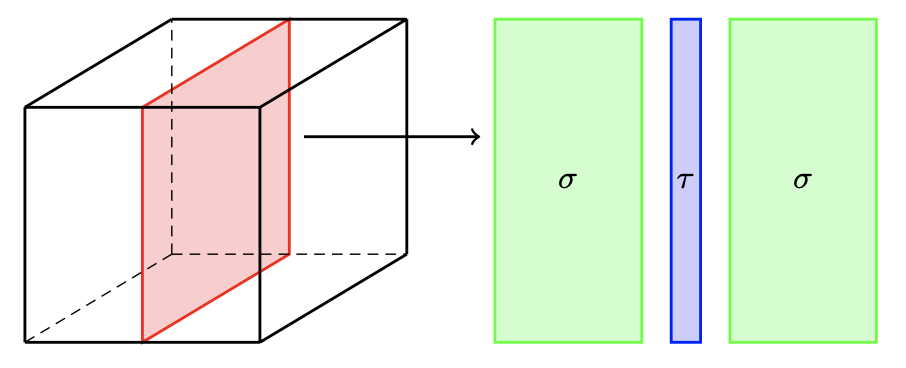}
    \caption{$C_{2}$ lattice}
\end{figure}

\textbf{Blocks and onsite symmetries}:
\begin{itemize}
    \item 2D: $G_{\sigma} = I$    
    \item 1D: $G_{\tau} = \bzt$
\end{itemize}

\subsubsection*{Decohered Spinless}
\textbf{Block state decorations}:
\begin{itemize}
    \item[] 1D
    \begin{itemize}
        \item $\tau$: Majorana chain, $\bzt$ fSPT
    \end{itemize}
    \item[] 2D
    \begin{itemize}
        \item $\sigma$:\( p+ip \) SC
    \end{itemize}
\end{itemize}

\textbf{Obstructions}
\begin{itemize}
    \item[] 2D
    \begin{itemize}
        \item\( p+ip \) SC on $\sigma$ is obstructed by chiral anomaly at $\tau$
    \end{itemize}
\end{itemize}

\textbf{Obstruction-free states}:
\begin{itemize}
\item[] 1D: ($\bzt^2$) \begin{enumerate}
    \item Majorana chain on $\tau$ (E)
    \item $\bzt$ fSPT on $\tau$ (E)
\end{enumerate}

\item[] 2D: No obstruction-free states ($\bbz_1$). 
\end{itemize}

\textbf{Trivializations}:
\begin{itemize}
    \item Open surface decoration trivializes Majorana chain on $\tau$. Therefore, the 1D classification reduces to $\bzt$.
    \item Majorana bubble on $\sigma$ $\Rightarrow$ $\bzt$ fSPT on $\tau$. Therefore, the 1D classification further reduces to $\bbz_1$ (trivial)
\end{itemize}

\textbf{Final classification:}
\begin{itemize}
    \item[] $E_{0,dec}^{1D} = \bbz_1$
    \item[] $E_{0,dec}^{2D} = \bbz_1$
    \item[]  $\mathcal{G}_{0,dec} =  E_{0,dec}^{1D} \times E_{0,dec}^{2D} = \bbz_1$
\end{itemize}

\subsubsection*{Decohered Spin-1/2}
\textbf{Block state decorations}:
\begin{itemize}
    \item[] 1D
    \begin{itemize}
        \item $\tau$: $\bbz_4^f$ ASPT
    \end{itemize}
    \item[] 2D
    \begin{itemize}
        \item $\sigma$:\( p+ip \) SC
    \end{itemize}
    
\end{itemize}

\textbf{Obstructions}
\begin{itemize}
    \item[] 2D
    \begin{itemize}
        \item\( p+ip \) SC on $\sigma$ is obstructed by chiral anomaly at $\tau$
    \end{itemize}
\end{itemize}

\textbf{Obstruction-free states}:
\begin{itemize}
\item[] 1D: ($\bzt$) \begin{enumerate}
    \item $\bbz_4^f$ ASPT on $\tau$ (I)
\end{enumerate} 
\item[] This state is also \textbf{trivialization-free}.

\item[] 2D: No obstruction-free states ($\bbz_1$). 
\end{itemize}

\textbf{Final classification:}
\begin{itemize}
    \item[] $E_{1/2,dec}^{1D} = \bzt(I)$
    \item[] $E_{1/2,dec}^{2D} = \bbz_1$
    \item[]  $\mathcal{G}_{1/2,dec} =E_{1/2,dec}^{1D} \times E_{1/2,dec}^{2D} = \bzt(I)$
\end{itemize}

\subsubsection*{Disordered Spinless}
\textbf{Block state decorations}:
\begin{itemize}
    \item[] 1D
    \begin{itemize}
        \item $\tau_1$: Majorana chain
    \end{itemize}
    \item[] 2D
    \begin{itemize}
        \item $\sigma$:\( p+ip \) SC
    \end{itemize}
    
\end{itemize}

\textbf{Obstructions}
\begin{itemize}
    \item[] 2D
    \begin{itemize}
        \item\( p+ip \) SC on $\sigma$ is obstructed by chiral anomaly at $\tau$
    \end{itemize}
\end{itemize}

\textbf{Obstruction-free states}:
\begin{itemize}
\item[] 1D: ($\bzt$) \begin{enumerate}
    \item Majorana chain on $\tau$ (E)
\end{enumerate} 
\item[] This state is also \textbf{trivialization-free}.

\item[] 2D: No obstruction-free states ($\bbz_1$). 
\end{itemize}

\textbf{Trivializations}:
\begin{itemize}
    \item Open surface decoration trivializes Majorana chain on $\tau$. Therefore, the 1D classification reduces to $\bbz_1$ (trivial).
\end{itemize}

\textbf{Final classification:}
\begin{itemize}
    \item[] $E_{0,dis}^{1D} = \bbz_1$
    \item[] $E_{0,dis}^{2D} = \bbz_1$
    \item[]  $\mathcal{G}_{0,dis} = E_{0,dis}^{0D} \times E_{0,dis}^{1D} \times E_{0,dis}^{2D} = \bbz_1$
\end{itemize}

\subsubsection*{Disordered Spin-1/2}
\textbf{Block state decorations}:
\begin{itemize}
    \item[] 1D
    \begin{itemize}
        \item $\tau_1$: \placeholder
    \end{itemize}
    \item[] 2D
    \begin{itemize}
        \item $\sigma$:\( p+ip \) SC
    \end{itemize}
    
\end{itemize}

\textbf{Obstructions}
\begin{itemize}
    \item[] 2D
    \begin{itemize}
        \item\( p+ip \) SC on $\sigma$ is obstructed by chiral anomaly at $\tau$
    \end{itemize}
\end{itemize}

\textbf{Obstruction-free states}:
\begin{itemize}
\item[] 1D: ($\bzt$) \begin{enumerate}
    \item \placeholder on $\tau$ (I)
\end{enumerate} 
\item[] This state is also \textbf{trivialization-free}.

\item[] 2D: No obstruction-free states ($\bbz_1$). 
\end{itemize}

\textbf{Final classification:}
\begin{itemize}
    \item[] $E_{1/2,dis}^{1D} = \bzt(I)$
    \item[] $E_{1/2,dis}^{2D} = \bbz_1$
    \item[]  $\mathcal{G}_{1/2,dis} = E_{1/2,dis}^{1D} \times E_{1/2,dis}^{2D} = \bzt(I)$
\end{itemize}

\subsection{$C_{1h}$}
\subsubsection*{Cell Decomposition}
\begin{figure}[!htbp]
    \centering
    \includegraphics[width=0.9\linewidth]{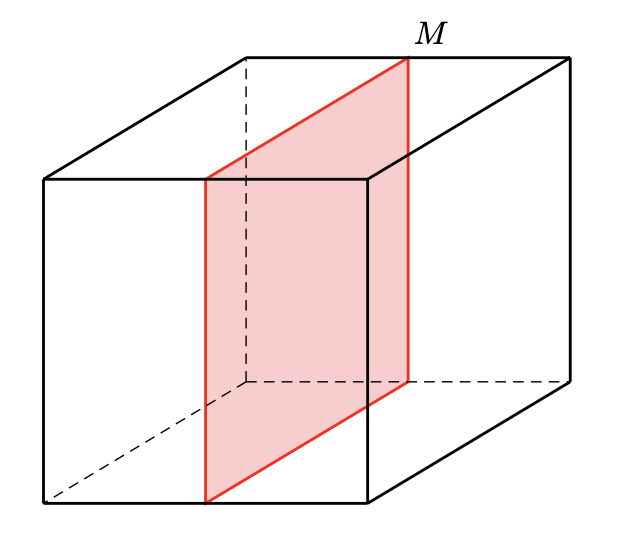}
    \caption{$C_{1h}$ lattice}
\end{figure}
\textbf{Blocks and onsite symmetries}:
\begin{itemize}
    \item $G_{\sigma} = \bzt$
\end{itemize}

\subsubsection*{Decohered Spinless}
\textbf{Block state decorations}:
\begin{itemize}
    \item[] 2D ($\bbz\times \bzt^4$)
    \begin{itemize}
        \item $\sigma$:\( p+ip \) SC, fLG (E)
    \end{itemize}
    \item[] These states are all \textbf{obstruction-free}.
\end{itemize}

\textbf{Trivializations}:
\begin{itemize}
\item 3D p+ip-SC on $\lambda$ $\Rightarrow$ Equivalence between two layers of\( p+ip \) SC and fLG on $\sigma$. This reduces the 2D classification to $\bbz_8$. 
\end{itemize}

\textbf{Final classification:}
\begin{itemize}
    \item[] $\mathcal{G}_{0,dec} =  E_{0,dec}^{2D} = \bbz_8(E)$
\end{itemize}

\subsubsection*{Decohered Spin-1/2}
\textbf{Block state decorations}:
\begin{itemize}
    \item[] 2D
    \begin{itemize}
        \item $\sigma$: No nontrivial block state
    \end{itemize}
    
\end{itemize}

\textbf{Final classification}:
\begin{itemize}
    \item[] $\mathcal{G}_{1/2,dec} = E_{1/2,dec}^{2D} = \bbz_1$
\end{itemize}

\subsubsection*{Disordered Spinless}
\textbf{Block state decorations}:
\begin{itemize}
    \item[] 2D ($\bbz \times \bzt^2$)
    \begin{itemize}
        \item $\sigma$:\( p+ip \) SC, fLG (E)
    \end{itemize}
    These states are all \textbf{obstruction-free}.
\end{itemize}

\textbf{Trivializations}:
\begin{itemize}
\item 3D p+ip-SC on $\lambda$ $\Rightarrow$ Equivalence between two layers of\( p+ip \) SC and fLG on $\sigma$. This reduces the 2D classification to $\bbz_4$. 
\end{itemize}

\textbf{Final classification:}
\begin{itemize}
    \item[] $\mathcal{G}_{0,dis} = E_{0,dis}^{2D} = \bbz_4(E)$
\end{itemize}

\subsubsection*{Disordered Spin-1/2}
\textbf{Block state decorations}:
\begin{itemize}
    \item[] 2D
    \begin{itemize}
        \item $\sigma$: 2D $\bbz_4^f$ ASPT (I)
    \end{itemize}
    This state is \textbf{obstruction-free} and \textbf{trivialization-free}.
\end{itemize}

\textbf{Final classification:}
\begin{itemize}
    \item[] $\mathcal{G}_{1/2,dis} = E_{1/2,dis}^{2D} = \bzt(I)$
\end{itemize}

\subsection{$C_{2h}$}
\subsubsection*{Cell Decomposition}
\begin{figure}[!htbp]
    \centering
    \includegraphics[width=0.9\linewidth]{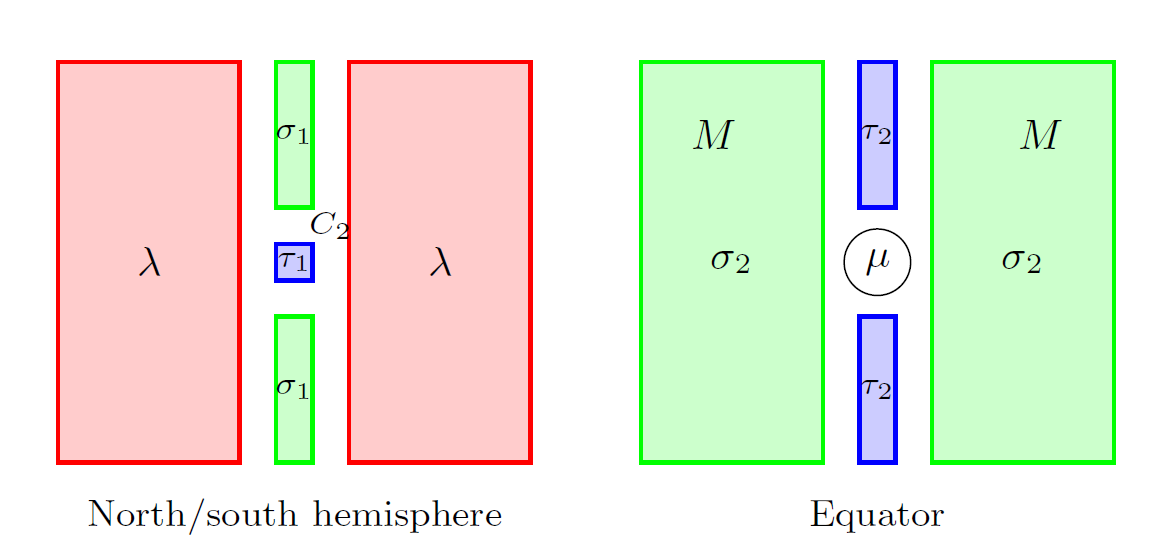}
    \caption{$C_{2h}$ lattice}
\end{figure}
\textbf{Blocks and onsite symmetries}:
\begin{itemize}
    \item 2D: $G_{\sigma_1} = I,\ G_{\sigma_2} = \bzt$
    \item 1D: $G_{\tau_1},\ G_{\tau_2} = \bzt$
    \item 0D: $G_{\mu} = C_{2h}$
\end{itemize}
\subsubsection*{Decohered Spinless}
\textbf{Block state decorations}:
\begin{itemize}
    \item[] 0D
    \begin{itemize}
        \item $\mu$: Odd fermion
    \end{itemize}
    \item[] 1D
    \begin{itemize}
        \item $\tau_1,\ \tau_2$: Majorana chain, $\bzt$ fSPT
    \end{itemize}
    \item[] 2D
    \begin{itemize}
        \item $\sigma_1$:\( p+ip \) SC
        \item $\sigma_2$:\( p+ip \) SC, fLG
    \end{itemize}
\end{itemize}

\textbf{Obstructions}
\begin{itemize}
    \item[] 1D
    \begin{itemize}
        \item Majorana chain on $\tau_1$ is obstructed at $\mu$
        \item Majorana chain on $\tau_2$ is obstructed at $\mu$
        \item[] Simultaneous decoration of the above phases is also obstructed.
    \end{itemize}
    \item[] 2D
    \begin{itemize}
        \item\( p+ip \) SC on $\sigma_1$ is obstructed by chiral anomaly at $\tau_1$
        \item\( p+ip \) SC on $\sigma_2$ is obstructed by chiral anomaly at $\tau_2$
    \end{itemize}
\end{itemize}

\textbf{Obstruction-free states}:
\begin{itemize}
\item[] 0D state ($\bzt$) is obstruction-free (E)

\item[] 1D: ($\bzt^2$) \begin{enumerate}
    \item $\bzt$ fSPT on $\tau_1$ or $\tau_2$ (I)
\end{enumerate}

\item[] 2D: ($\bbz_4$) \begin{enumerate}
    \item fLG on $\sigma_2$ (E)
\end{enumerate} 
\end{itemize}

\textbf{Trivializations}:
\begin{itemize}
    \item Majorana bubble on $\sigma_1$ $\Rightarrow$ Simultaneous decoration of $\bzt$ fSPT on $\tau_1$ and $\tau_2$. Therefore, the 1D classification reduces to $\bzt$.
    \item Majorana bubble on $\sigma_2$ $\Rightarrow$ Odd fermion on $\mu$. Therefore, the 0D classification reduces to $\bbz_1$ (trivial). 
\end{itemize}

\textbf{Final classification:}
\begin{itemize}
    \item[] $E_{0,dec}^{0D} = \bbz_1$
    \item[] $E_{0,dec}^{1D} = \bzt(I)$
    \item[] $E_{0,dec}^{2D} = \bbz_4(E)$
    \item[]  $\mathcal{G}_{0,dec} =  E_{0,dec}^{1D} \times E_{0,dec}^{2D} = \bbz_4(E) \times \bzt(I)$
\end{itemize}

\subsubsection*{Decohered Spin-1/2}
\textbf{Block state decorations}:
\begin{itemize}
    \item[] 0D
    \begin{itemize}
        \item $\mu$: Odd fermion
    \end{itemize}
    \item[] 1D
    \begin{itemize}
        \item $\tau_1,\ \tau_2$: $\bbz_4^f$ ASPT
    \end{itemize}
    \item[] 2D
    \begin{itemize}
        \item $\sigma_1$:\( p+ip \) SC
        \item $\sigma_2$: No nontrivial block state
    \end{itemize}
\end{itemize}

\textbf{Obstructions}
\begin{itemize}
    \item[] 2D
    \begin{itemize}
        \item\( p+ip \) SC on $\sigma_1$ is obstructed by chiral anomaly at $\tau_1$
    \end{itemize}
\end{itemize}

\textbf{Obstruction-free states}:
\begin{itemize}
\item[] 0D state ($\bzt$) is obstruction-free (I)
\item[] 1D: ($\bzt^2$) \begin{enumerate}
    \item $\bbz_4^f$ ASPT on $\tau_1$ or $\tau_2$ (I)
\end{enumerate} 
\item[] These states are also \textbf{trivialization-free}.

\item[] 2D: No obstruction-free states ($\bbz_1$). 
\end{itemize}

\textbf{Final classification:}
\begin{itemize}
    \item[] $E_{1/2,dec}^{0D} = \bzt (I)$
    \item[] $E_{1/2,dec}^{1D} = \bzt^2(I)$
    \item[] $E_{1/2,dec}^{2D} = \bbz_1$
    \item[]  $\mathcal{G}_{1/2,dec} =E_{1/2,dec}^{1D} \times E_{1/2,dec}^{2D} = \bzt^3(I)$
\end{itemize}

\subsubsection*{Disordered Spinless}
\textbf{Block state decorations}:
\begin{itemize}
    \item[] 1D
    \begin{itemize}
        \item $\tau_1, \tau_2$: Majorana chain
    \end{itemize}
    \item[] 2D
    \begin{itemize}
        \item $\sigma_1$:\( p+ip \) SC
        \item $\sigma_2$:\( p+ip \) SC, fLG
    \end{itemize}
\end{itemize}

\textbf{Obstructions}
\begin{itemize}
    \item[] 2D
    \begin{itemize}
        \item\( p+ip \) SC on $\sigma_1$ is obstructed by chiral anomaly at $\tau_1$
        \item\( p+ip \) SC on $\sigma_2$ is obstructed by chiral anomaly at $\tau_2$
    \end{itemize}
\end{itemize}

\textbf{Obstruction-free states}:
\begin{itemize}
\item[] 1D ($\bzt^2$) \begin{enumerate}
    \item Majorana chain on $\tau_1$ or $\tau_2$ (I)
\end{enumerate} 

\item[] 2D ($\bzt$) \begin{enumerate}
    \item fLG on $\sigma_2$ (E)
\end{enumerate} 
\item[] These states are all \textbf{trivialization-free}.
\end{itemize}

\textbf{Final classification:}
\begin{itemize}
    \item[] $E_{0,dis}^{1D} = \bzt^2 (I)$
    \item[] $E_{0,dis}^{2D} = \bzt(E)$
    \item[]  $\mathcal{G}_{0,dis} = E_{0,dis}^{1D} \times E_{0,dis}^{2D} = \bzt(E)\times\bzt^2(I)$
\end{itemize}

\subsubsection*{Disordered Spin-1/2}
\textbf{Block state decorations}:
\begin{itemize}
    \item[] 1D
    \begin{itemize}
        \item $\tau_1, \tau_2$: \placeholder 
    \end{itemize}
    \item[] 2D
    \begin{itemize}
        \item $\sigma_1$:\( p+ip \) SC
        \item $\sigma_2$: 2D $\bbz_4^f$ ASPT
    \end{itemize}
\end{itemize}

\textbf{Obstructions}
\begin{itemize}
    \item[] 2D
    \begin{itemize}
        \item\( p+ip \) SC on $\sigma_1$ is obstructed by chiral anomaly at $\tau_1$
    \end{itemize}
\end{itemize}

\textbf{Obstruction-free states}:
\begin{itemize}
\item[] 1D ($\bzt^2$) \begin{enumerate}
    \item \placeholder on $\tau_1$ or $\tau_2$ (I)
\end{enumerate} 

\item[] 2D ($\bzt$) \begin{enumerate}
    \item 2D $\bbz_4^f$ ASPT on $\sigma_2$ (I)
\end{enumerate} 
\item[] These states are all \textbf{trivialization-free}.
\end{itemize}

\textbf{Final classification:}
\begin{itemize}
    \item[] $E_{1/2,dis}^{1D} = \bzt^2(I)$
    \item[] $E_{1/2,dis}^{2D} = \bzt(I)$
    \item[]  $\mathcal{G}_{1/2,dis} = E_{1/2,dis}^{1D} \times E_{1/2,dis}^{2D} = \bzt^3(I)$
\end{itemize}

\subsection{$D_2=V$}
\subsubsection*{Cell Decomposition}
\begin{figure}[!htbp]
    \centering
    \includegraphics[width=0.9\linewidth]{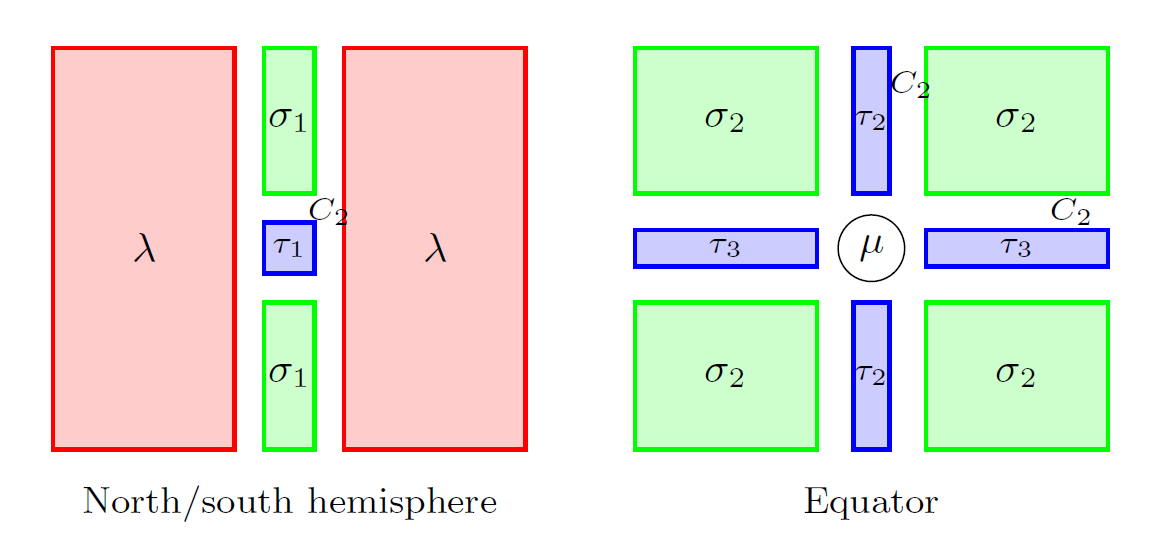}
    \caption{$D_2$ lattice}
\end{figure}
\textbf{Blocks and onsite symmetries}:
\begin{itemize}
    \item 2D: $G_{\sigma_1},\ G_{\sigma_2} =I$
    \item 1D: $G_{\tau_1},\ G_{\tau_2},\ G_{\tau_3}=\bzt$ (rotation)
    \item 0D: $G_{\mu} = D_2$
\end{itemize}
\subsubsection*{Decohered Spinless}
\textbf{Block state decorations}:
\begin{itemize}
    \item[] 0D
    \begin{itemize}
        \item $\mu$: Odd fermion
    \end{itemize}
    \item[] 1D
    \begin{itemize}
        \item $\tau_1,\ \tau_2,\ \tau_3$: Majorana chain, $\bzt$ fSPT
    \end{itemize}
    \item[] 2D
    \begin{itemize}
        \item $\sigma_1,\ \sigma_2$:\( p+ip \) SC
    \end{itemize}
\end{itemize}

\textbf{Obstructions}
\begin{itemize}
    \item[] 1D
    \begin{itemize}
        \item Majorana chain on $\tau_1$, $\tau_2$, or $\tau_3$ are obstructed at $\mu$. 
    \end{itemize}
    \item[] 2D
    \begin{itemize}
        \item\( p+ip \) SC on $\sigma_1$ is obstructed by chiral anomaly at $\tau_1$
        \item\( p+ip \) SC on $\sigma_2$ is obstructed by chiral anomaly at $\tau_2$
    \end{itemize}
\end{itemize}

\textbf{Obstruction-free states}:
\begin{itemize}
\item[] 0D state ($\bzt$) is obstruction-free (E)

\item[] 1D: ($\bzt^3$) \begin{enumerate}
    \item $\bzt$ fSPT on $\tau_1$, $\tau_2$, or $\tau_3$ (I)
\end{enumerate}

\item[] 2D: No obstruction-free states ($\bbz_1$)
\end{itemize}

\textbf{Trivializations}:
\begin{itemize}
    \item Majorana bubble on $\sigma_1$ $\Rightarrow$ Simultaneous decoration of $\bzt$ fSPT on $\tau_1$ and $\tau_2$. Therefore, the 1D classification reduces to $\bzt^2$.
    \item Majorana bubble on $\sigma_2$ $\Rightarrow$ Simultaneous decoration of $\bzt$ fSPT on $\tau_2$ and $\tau_3$. Therefore, the 1D classification further reduces to $\bzt$. 
    \item Chern insulator bubble trivializes odd fermion parity on $\mu$. The 0D classification reduces to $\bbz_1$.
\end{itemize}

\textbf{Final classification:}
\begin{itemize}
    \item[] $E_{0,dec}^{0D} = \bbz_1$
    \item[] $E_{0,dec}^{1D} = \bzt(I)$
    \item[] $E_{0,dec}^{2D} = \bbz_1$
    \item[]  $\mathcal{G}_{0,dec} =  E_{0,dec}^{0D}\times E_{0,dec}^{1D} \times E_{0,dec}^{2D} = \bzt(I)$
\end{itemize}

\subsubsection*{Decohered Spin-1/2}
\textbf{Block state decorations}:
\begin{itemize}
    \item[] 0D
    \begin{itemize}
        \item $\mu$: Odd fermion (I)
    \end{itemize}
    \item[] 1D
    \begin{itemize}
        \item $\tau_1,\ \tau_2,\ \tau_3$: $\bbz_4^f$ ASPT
    \end{itemize}
    \item[] 2D
    \begin{itemize}
        \item $\sigma_1,\sigma_2$:\( p+ip \) SC
    \end{itemize}
\end{itemize}

\textbf{Obstructions}
\begin{itemize}
    \item[] 2D
    \begin{itemize}
        \item\( p+ip \) SC on $\sigma_1$ is obstructed by chiral anomaly at $\tau_1$
        \item\( p+ip \) SC on $\sigma_2$ is obstructed by chiral anomaly at $\tau_2$
    \end{itemize}
\end{itemize}

\textbf{Obstruction-free states}:
\begin{itemize}
\item[] 0D state ($\bzt$) is obstruction-free (I)
\item[] 1D: ($\bzt^3$) \begin{enumerate}
    \item $\bbz_4^f$ ASPT on $\tau_1$,$\tau_2$, or $\tau_3$ (I)
\end{enumerate} 
\item[] These states are also \textbf{trivialization-free}.

\item[] 2D: No obstruction-free states ($\bbz_1$). 
\end{itemize}

\textbf{Final classification:}
\begin{itemize}
    \item[] $E_{1/2,dec}^{0D} = \bzt(I)$
    \item[] $E_{1/2,dec}^{1D} = \bzt^3(I)$
    \item[] $E_{1/2,dec}^{2D} = \bbz_1$
    \item[]  $\mathcal{G}_{1/2,dec} = E_{1/2,dec}^{0D}\times E_{1/2,dec}^{1D} \times E_{1/2,dec}^{2D} = \bzt^4(I)$
\end{itemize}

\subsubsection*{Disordered Spinless}
\textbf{Block state decorations}:
\begin{itemize}
    \item[] 1D
    \begin{itemize}
        \item $\tau_1,\ \tau_2,\ \tau_3$: Majorana chain
    \end{itemize}
    \item[] 2D
    \begin{itemize}
        \item $\sigma_1,\ \sigma_2$:\( p+ip \) SC
    \end{itemize}
\end{itemize}

\textbf{Obstructions}
\begin{itemize}
    \item[] 2D
    \begin{itemize}
        \item\( p+ip \) SC on $\sigma_1$ is obstructed by chiral anomaly at $\tau_1$
        \item\( p+ip \) SC on $\sigma_2$ is obstructed by chiral anomaly at $\tau_2$
    \end{itemize}
\end{itemize}

\textbf{Obstruction-free states}:
\begin{itemize}
\item[] 1D ($\bzt^3$) \begin{enumerate}
    \item Majorana chain on $\tau_1$, $\tau_2$ or $\tau_3$ (I)
\end{enumerate} 

\item[] 2D: No obstruction-free states ($\bbz_1$)
\end{itemize}

\textbf{Trivializations}:
\begin{itemize}
    \item Open surface decoration $\Rightarrow$ Simultaneous decoration of Majorana chains on $\tau_1,\ \tau_2$, and $\tau_3$. This reduces the 1D classification to $\bzt^2$.
\end{itemize}

\textbf{Final classification:}
\begin{itemize}
    \item[] $E_{0,dis}^{1D} = \bzt^2(I)$
    \item[] $E_{0,dis}^{2D} = \bbz_1$
    \item[]  $\mathcal{G}_{0,dis} = E_{0,dis}^{1D} \times E_{0,dis}^{2D} = \bzt^2(I)$
\end{itemize}

\subsubsection*{Disordered Spin-1/2}
\textbf{Block state decorations}:
\begin{itemize}
    \item[] 1D
    \begin{itemize}
        \item $\tau_1,\ \tau_2,\ \tau_3$: \placeholder
    \end{itemize}
    \item[] 2D
    \begin{itemize}
        \item $\sigma_1,\ \sigma_2$:\( p+ip \) SC
    \end{itemize}
\end{itemize}

\textbf{Obstructions}
\begin{itemize}
    \item[] 2D
    \begin{itemize}
        \item\( p+ip \) SC on $\sigma_1$ is obstructed by chiral anomaly at $\tau_1$
        \item\( p+ip \) SC on $\sigma_2$ is obstructed by chiral anomaly at $\tau_2$
    \end{itemize}
\end{itemize}

\textbf{Obstruction-free states}:
\begin{itemize}
\item[] 1D ($\bzt^3$) \begin{enumerate}
    \item \placeholder on $\tau_1$, $\tau_2$ or $\tau_3$ (I)
\end{enumerate} 
\item[] These states are also \textbf{trivialization-free}.
\item[] 2D: No obstruction-free states ($\bbz_1$)
\end{itemize}

\textbf{Final classification:}
\begin{itemize}
    \item[] $E_{1/2,dis}^{1D} = \bzt^3(I)$
    \item[] $E_{1/2,dis}^{2D} = \bbz_1$
    \item[]  $\mathcal{G}_{1/2,dis} = E_{1/2,dis}^{1D} \times E_{1/2,dis}^{2D} = \bzt^3(I)$
\end{itemize}

\subsection{$C_{2v}$}

\subsubsection*{Cell Decomposition}
\begin{figure}[!htbp]
    \centering
    \includegraphics[width=0.9\linewidth]{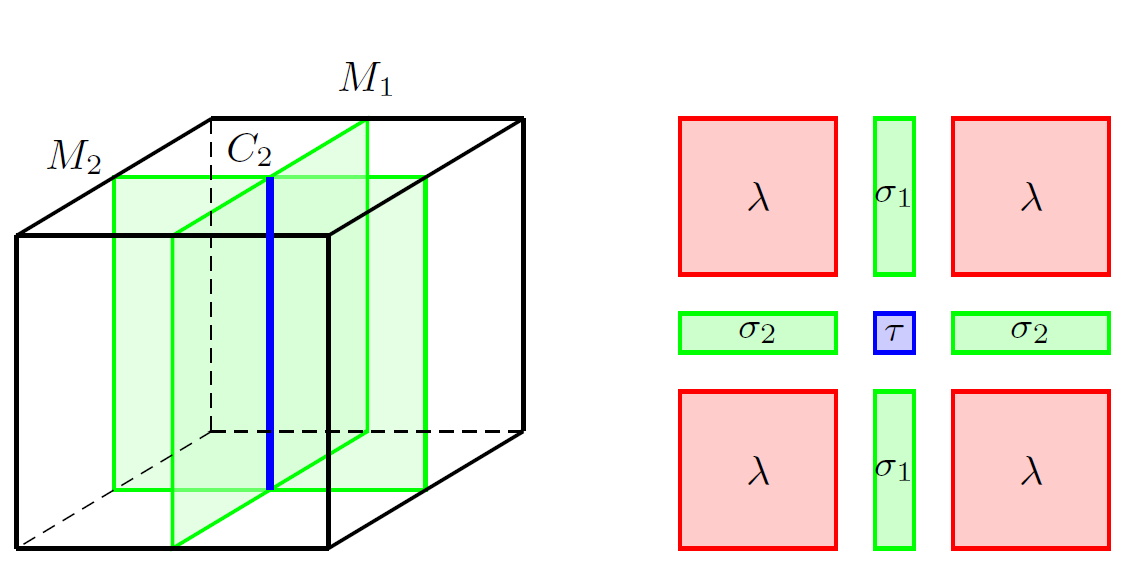}
    \caption{$C_{2v}$ lattice}
\end{figure}
\textbf{Blocks and onsite symmetries}:
\begin{itemize}
    \item $G_{\sigma_1},\ G_{\sigma_2}=\bzt$
    \item $G_{\tau}=\bzt\times\bzt$
\end{itemize}
\subsubsection*{Decohered Spinless}
\textbf{Block state decorations}:
\begin{itemize}
    \item[] 1D
    \begin{itemize}
        \item $\tau$: Majorana chain $\bzt^{M_1}$ fSPT, $\bzt^{M_2}$ fSPT
    \end{itemize}
    \item[] 2D
    \begin{itemize}
        \item $\sigma_1,\ \sigma_2$:\( p+ip \) SC, fLG
    \end{itemize}
\end{itemize}

\textbf{Obstructions}
\begin{itemize}
    \item[] 2D
    \begin{itemize}
        \item\( p+ip \) SC on $\sigma_1$ is obstructed by chiral anomaly at $\tau$
        \item\( p+ip \) SC on $\sigma_2$ is obstructed by chiral anomaly at $\tau$
        \item Decoration of $p\pm ip$-SC with opposite chiralities on $\sigma_1$ and $\sigma_2$ is obstructed at $\tau$ (anomaly indicator $\nu_{M_1}=1/4$) 
        \item fLG on $\sigma_1$ is obstructed at $\tau$ ($\nu_{M_1}=1/2$)
        \item fLG on $\sigma_2$ is obstructed at $\tau$ ($\nu_{M_2}=1/2$)
    \end{itemize}
\end{itemize}

\textbf{Obstruction-free states}:
\begin{itemize}

\item[] 1D: ($\bzt^3$) \begin{enumerate}
    \item Majorana chain on $\tau$ (E)
    \item $\bzt \times \bzt$ fSPT on $\tau$ (E)
\end{enumerate}

\item[] 2D: ($\bzt^2$) \begin{enumerate}
    \item $n=2$ fLG on $\sigma_1$ or $\sigma_2$ (I)
\end{enumerate}
\end{itemize}

\textbf{Trivializations}:
\begin{itemize}
    \item Majorana bubble on $\sigma_1$ $\Rightarrow$ $\bzt^{M_1}$ fSPT on $\tau$. Therefore, the 1D classification reduces to $\bzt^2$.
    \item Majorana bubble on $\sigma_2$ $\Rightarrow$ $\bzt^{M_2}$ fSPT on $\tau$. Therefore, the 1D classification further reduces to $\bzt$. 
\end{itemize}

\textbf{Final classification:}
\begin{itemize}
    \item[] $E_{0,dec}^{1D} = \bzt(E)$
    \item[] $E_{0,dec}^{2D} = \bzt^2(I)$
    \item[]  $\mathcal{G}_{0,dec} =  E_{0,dec}^{1D} \times E_{0,dec}^{2D} = \bzt(E)\times\bzt^2(I)$
\end{itemize}

\subsubsection*{Decohered Spin-1/2}
\textbf{Block state decorations}:
\begin{itemize}
    \item[] 1D
    \begin{itemize}
        \item $\tau$: $\bbz_4^{f,M_1}$ ASPT, $\bbz_4^{f,M_2}$ ASPT (I)
    \end{itemize}
    \item[] These states are all \textbf{obstruction-free} and \textbf{trivialization-free}
    \item[] 2D
    \begin{itemize}
        \item $\sigma_1,\ \sigma_2$: No nontrivial block state
    \end{itemize}
\end{itemize}

\textbf{Final classification:}
\begin{itemize}
    \item[] $E_{1/2,dec}^{1D} = \bzt^2(I)$
    \item[] $E_{1/2,dec}^{2D} = \bbz_1$
    \item[]  $\mathcal{G}_{1/2,dec} = E_{1/2,dec}^{1D} \times E_{1/2,dec}^{2D} = \bzt^2(I)$
\end{itemize}

\subsubsection*{Disordered Spinless}
\textbf{Block state decorations}:
\begin{itemize}
    \item[] 1D
    \begin{itemize}
        \item $\tau$: Majorana chain
    \end{itemize}
    \item[] 2D
    \begin{itemize}
        \item $\sigma_1,\ \sigma_2$:\( p+ip \) SC, fLG
    \end{itemize}
\end{itemize}

\textbf{Obstructions}
\begin{itemize}
    \item[] 2D
    \begin{itemize}
        \item\( p+ip \) SC on $\sigma_1$ is obstructed by chiral anomaly at $\tau$
        \item\( p+ip \) SC on $\sigma_2$ is obstructed by chiral anomaly at $\tau$
        \item Decoration of $p\pm ip$-SC with opposite chiralities on $\sigma_1$ and $\sigma_2$ is obstructed at $\tau$ (anomaly indicator $\nu_{M_1}=1/4$) 
    \end{itemize}
\end{itemize}

\textbf{Obstruction-free states}:
\begin{itemize}
\item[] 1D ($\bzt$) \begin{enumerate}
    \item Majorana chain on $\tau$ (E)
\end{enumerate} 

\item[] 2D ($\bzt^2$) \begin{enumerate}
    \item fLG on $\sigma_1$ or $\sigma_2$ (I)
\end{enumerate}
\item[] These states are all \textbf{trivialization-free}
\end{itemize}

\textbf{Final classification:}
\begin{itemize}
    \item[] $E_{0,dis}^{1D} = \bzt(E)$
    \item[] $E_{0,dis}^{2D} = \bzt^2(I)$
    \item[]  $\mathcal{G}_{0,dis} = E_{0,dis}^{1D} \times E_{0,dis}^{2D} = \bzt(E)\times\bzt^2(I)$
\end{itemize}

\subsubsection*{Disordered Spin-1/2}
\textbf{Block state decorations}:
\begin{itemize}
    \item[] 1D
    \begin{itemize}
        \item $\tau$: \placeholder (I)
    \end{itemize}
    \item[] 2D
    \begin{itemize}
        \item $\sigma_1,\ \sigma_2$: 2D $\bbz_4^f$ ASPT (I)
    \end{itemize}
    \item[] These states are all \textbf{obstruction-free} and \textbf{trivialization-free}.
\end{itemize}

\textbf{Final classification:}
\begin{itemize}
    \item[] $E_{1/2,dis}^{1D} = \bzt(I)$
    \item[] $E_{1/2,dis}^{2D} = \bzt^2(I)$
    \item[]  $\mathcal{G}_{1/2,dis} = E_{1/2,dis}^{1D} \times E_{1/2,dis}^{2D} = \bzt^3(I)$
\end{itemize}

\subsection{$D_{2h}/V_h$}
\subsubsection*{Cell Decomposition}
\begin{figure}[!htbp]
    \centering
    \includegraphics[width=0.9\linewidth]{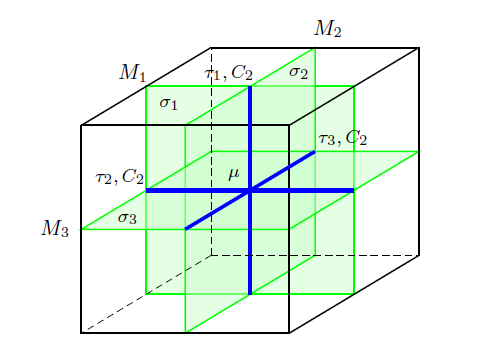}
    \caption{$D_{2h}/V_h$ lattice}
\end{figure}
\textbf{Blocks and onsite symmetries}:
\begin{itemize}
    \item 2D: $G_{\sigma_1},\ G_{\sigma_2},\ G_{\sigma_3} = \bzt$
    \item 1D: $G_{\tau_1},\ G_{\tau_2},\ G_{\tau_3} = \bzt\times\bzt$
    \item 0D: $G_{\mu} = \bzt\times\bzt\times\bzt$ 
\end{itemize}

\subsubsection*{Decohered Spinless}
\textbf{Block state decorations}:
\begin{itemize}
    \item[] 0D
    \begin{itemize}
        \item $\mu$: Odd fermion
    \end{itemize}
    \item[] 1D
    \begin{itemize}
        \item $\tau_1$: Majorana chain, $\bzt^{M_1}$ fSPT, $\bzt^{M_2}$ fSPT
        \item $\tau_2$: Majorana chain, $\bzt^{M_1}$ fSPT, $\bzt^{M_3}$ fSPT
        \item $\tau_3$: Majorana chain, $\bzt^{M_2}$ fSPT, $\bzt^{M_3}$ fSPT
    \end{itemize}
    \item[] 2D
    \begin{itemize}
        \item $\sigma_1,\ \sigma_2,\ \sigma_3$:\( p+ip \) SC, fLG
    \end{itemize}
\end{itemize}

\textbf{Obstructions}
\begin{itemize}
    \item[] 1D
    \begin{itemize}
        \item Majorana chain on $\tau_1$, $\tau_2$, or $\tau_3$ is obstructed at $\mu$
    \end{itemize}
    \item[] 2D
    \begin{itemize}
        \item\( p+ip \) SC on $\sigma_1$, $\sigma_2$, $\sigma_3$ is obstructed by chiral anomaly at $\tau_1$, $\tau_2$, $\tau_3$ respectively
        \item Simultaneous decorations are obstructed as well
        \item fLG on $\sigma_1$ is obstructed at $\tau_1$ ($\nu_{M_1}=1/2$)
        \item fLG on $\sigma_2$ is obstructed at $\tau_2$ ($\nu_{M_2}=1/2$)
        \item fLG on $\sigma_3$ is obstructed at $\tau_3$ ($\nu_{M_3}=1/2$)   
    \end{itemize}
\end{itemize}

\textbf{Obstruction-free states}:
\begin{itemize}
\item[] 0D states are obstruction-free ($\bzt$) (E)
 
\item[] 1D: ($\bzt^6$) \begin{enumerate}
    \item Any $\bzt$ fSPT on $\tau_1$, $\tau_2$ or $\tau_3$. 
    \item[] Individual decorations are intrinsic whereas decoration on any pair of $\tau$ blocks is extrinsic.
\end{enumerate}

\item[] 2D: ($\bzt^3$) \begin{enumerate}
    \item $n=2$ fLG on $\sigma_1$, $\sigma_2$, or $\sigma_3$ (I)
\end{enumerate}
\end{itemize}

\textbf{Trivializations}:
\begin{itemize}
    \item Majorana bubble on $\sigma_1$ $\Rightarrow$ Simultaneous decoration of $\bzt^{M_1}$ fSPT on $\tau_1$ and $\tau_2$. Therefore, the 1D classification reduces to $\bzt^5$ .
    \item Majorana bubble on $\sigma_2$ $\Rightarrow$ Simultaneous decoration of $\bzt^{M_2}$ fSPT on $\tau_1$ and $\tau_3$. Therefore, the 1D classification further reduces to $\bzt^4$.
    \item Majorana bubble on $\sigma_3$ $\Rightarrow$ Simultaneous decoration of $\bzt^{M_3}$ fSPT on $\tau_2$ and $\tau_3$. Therefore, the 1D classification further reduces to $\bzt^3$. 
\end{itemize}

\textbf{Final classification:}
\begin{itemize}
    \item[] $E_{0,dec}^{0D} = \bzt(E)$
    \item[] $E_{0,dec}^{1D} = \bzt^3(I)$
    \item[] $E_{0,dec}^{2D} = \bzt^3(I)$
    \item[]  $\mathcal{G}_{0,dec} = E_{0,dec}^{0D} \times E_{0,dec}^{1D} \times E_{0,dec}^{2D} = \bzt(E)\times\bzt^6(I)$
\end{itemize}

\subsubsection*{Decohered Spin-1/2}
\textbf{Block state decorations}:
\begin{itemize}
    \item[] 0D
    \begin{itemize}
        \item $\mu$: Odd fermion
    \end{itemize}
    \item[] 1D
    \begin{itemize}
        \item $\tau_1$: $\bbz_4^{f,M_1}$ ASPT, $\bbz_4^{f,M_2}$ ASPT
        \item $\tau_2$: $\bbz_4^{f,M_1}$ ASPT, $\bbz_4^{f,M_3}$ ASPT
        \item $\tau_3$: $\bbz_4^{f,M_2}$ ASPT, $\bbz_4^{f,M_3}$ ASPT
    \end{itemize}
    \item[] These states are all intrinsic, \textbf{obstruction-free}, and \textbf{trivialization-free}
    \item[] 2D
    \begin{itemize}
        \item $\sigma_1,\ \sigma_2,\ \sigma_3$: No nontrivial block state
    \end{itemize}
\end{itemize}

\textbf{Final classification:}
\begin{itemize}
    \item[] $E_{1/2,dec}^{0D} = \bzt(I)$
    \item[] $E_{1/2,dec}^{1D} = \bzt^6(I)$
    \item[] $E_{1/2,dec}^{2D} = \bbz_1$
    \item[]  $\mathcal{G}_{1/2,dec} = E_{1/2,dec}^{0D} \times E_{1/2,dec}^{1D} \times E_{1/2,dec}^{2D} = \bzt^7(I)$
\end{itemize}

\subsubsection*{Disordered Spinless}
\textbf{Block state decorations}:
\begin{itemize}
    \item[] 1D
    \begin{itemize}
        \item $\tau_1,\ \tau_2,\ \tau_3$: Majorana chain
    \end{itemize}
    \item[] 2D
    \begin{itemize}
        \item $\sigma_1,\ \sigma_2,\ \sigma_3$:\( p+ip \) SC, fLG
    \end{itemize}
\end{itemize}

\textbf{Obstructions}
\begin{itemize}
    \item[] 2D
    \begin{itemize}
        \item\( p+ip \) SC on $\sigma_1$, $\sigma_2$, $\sigma_3$ is obstructed by chiral anomaly at $\tau_1$, $\tau_2$, $\tau_3$ respectively.
    \end{itemize}
\end{itemize}

\textbf{Obstruction-free states}:
\begin{itemize}
\item[] 1D ($\bzt^3$) \begin{enumerate}
    \item Majorana chain on $\tau_1$,$\tau_2$, or $\tau_3$ (I) 
\end{enumerate} 

\item[] 2D ($\bzt^3$) \begin{enumerate}
    \item fLG on $\sigma_1$, $\sigma_2$ or $\sigma_3$ (I)
\end{enumerate}
\item[] These states are all \textbf{trivialization-free}
\end{itemize}

\textbf{Final classification:}
\begin{itemize}
    \item[] $E_{0,dis}^{1D} = \bzt^3(I)$
    \item[] $E_{0,dis}^{2D} = \bzt^3(I)$
    \item[]  $\mathcal{G}_{0,dis} = E_{0,dis}^{1D} \times E_{0,dis}^{2D} = \bzt^6(I)$
\end{itemize}

\subsubsection*{Disordered Spin-1/2}
\textbf{Block state decorations}:
\begin{itemize}
    \item[] 1D
    \begin{itemize}
        \item $\tau_1,\ \tau_2,\ \tau_3$: \placeholder (I)
    \end{itemize}
    \item[] 2D
    \begin{itemize}
        \item $\sigma_1,\ \sigma_2,\ \sigma_3$: 2D $\bbz_4^f$ ASPT (I)
    \end{itemize}
    \item[] These states are all \textbf{obstruction-free} and \textbf{trivialization-free}.
\end{itemize}

\textbf{Final classification:}
\begin{itemize}
    \item[] $E_{1/2,dis}^{1D} = \bzt^3(I)$
    \item[] $E_{1/2,dis}^{2D} = \bzt^3(I)$
    \item[]  $\mathcal{G}_{1/2,dis} = E_{1/2,dis}^{1D} \times E_{1/2,dis}^{2D} = \bzt^6(I)$
\end{itemize}

\subsection{$C_4$}
\subsubsection*{Cell Decomposition}
\begin{figure}[!htbp]
    \centering
    \includegraphics[width=0.9\linewidth]{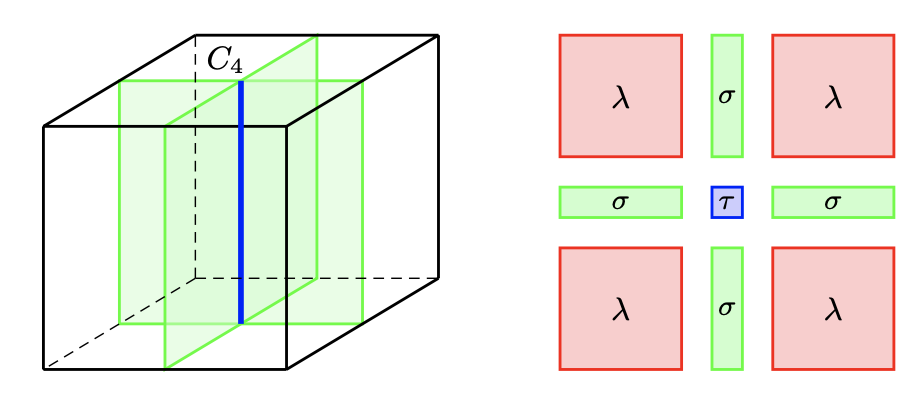}
    \caption{$C_4$ lattice}
\end{figure}
\textbf{Blocks and onsite symmetries}:
\begin{itemize}
    \item $G_{\sigma} = I$
    \item $G_{\tau} = \bbz_4$
\end{itemize}
\subsubsection*{Decohered Spinless}
\textbf{Block state decorations}:
\begin{itemize}
    \item[] 1D
    \begin{itemize}
        \item $\tau$: Majorana chain, $\bbz_4$ fSPT
    \end{itemize}
    \item[] 2D
    \begin{itemize}
        \item $\sigma_1$:\( p+ip \) SC
    \end{itemize}
\end{itemize}

\textbf{Obstructions}
\begin{itemize}
    \item[] 2D
    \begin{itemize}
        \item\( p+ip \) SC on $\sigma$ is obstructed by chiral anomaly at $\tau$
    \end{itemize}
\end{itemize}

\textbf{Obstruction-free states}:
\begin{itemize}

\item[] 1D: ($\bzt^2$) \begin{enumerate}
    \item Majorana chain on $\tau$ (E)
    \item $\bbz_4$ fSPT on $\tau$ (E)
\end{enumerate}

\item[] 2D: No obstruction-free states ($\bbz_1$)
\end{itemize}

\textbf{Trivializations}:
\begin{itemize}
    \item Majorana bubble on $\sigma$ $\Rightarrow$ $\bbz_4$ ASPT on $\tau$. Therefore, the 1D classification reduces to $\bzt^2$
    \item Open surface decorations trivializes Majorana chain on $\tau$. Therefore, the 1D classification further reduces to $\bbz_1$ (trivial).
\end{itemize}

\textbf{Final classification:}
\begin{itemize}
    \item[] $E_{0,dec}^{1D} = \bbz_1$
    \item[] $E_{0,dec}^{2D} = \bbz_1$
    \item[]  $\mathcal{G}_{0,dec} =  E_{0,dec}^{1D} \times E_{0,dec}^{2D} = \bbz_1$
\end{itemize}

\subsubsection*{Decohered Spin-1/2}
\textbf{Block state decorations}:
\begin{itemize}
    \item[] 1D
    \begin{itemize}
        \item $\tau$: $\bbz_4\times_{\omega_2^f} \bzt^f$ ASPT
    \end{itemize}
    \item[] 2D
    \begin{itemize}
        \item $\sigma$:\( p+ip \) SC
    \end{itemize}
\end{itemize}

\textbf{Obstructions}
\begin{itemize}
    \item[] 2D
    \begin{itemize}
        \item\( p+ip \) SC on $\sigma$ is obstructed by chiral anomaly at $\tau$
    \end{itemize}
\end{itemize}

\textbf{Obstruction-free states}:
\begin{itemize}

\item[] 1D: ($\bzt$) \begin{enumerate}
   \item $\bbz_4\times_{\omega_2^f} \bzt^f$ ASPT on $\tau$ (I)
\end{enumerate} 

This state is also \textbf{trivialization-free}.

\item[] 2D: No obstruction-free states ($\bbz_1$)
\end{itemize}

\textbf{Final classification:}
\begin{itemize}
    \item[] $E_{1/2,dec}^{1D} = \bzt(I)$
    \item[] $E_{1/2,dec}^{2D} = \bbz_1$
    \item[]  $\mathcal{G}_{1/2,dec} = E_{1/2,dec}^{1D} \times E_{1/2,dec}^{2D} = \bzt(I)$
\end{itemize}

\subsubsection*{Disordered Spinless}
\textbf{Block state decorations}:
\begin{itemize}
    \item[] 1D
    \begin{itemize}
        \item $\tau$: Majorana chain
    \end{itemize}
    \item[] 2D
    \begin{itemize}
        \item $\sigma$:\( p+ip \) SC
    \end{itemize}
\end{itemize}

\textbf{Obstructions}
\begin{itemize}
    \item[] 2D
    \begin{itemize}
        \item\( p+ip \) SC on $\sigma$ is obstructed by chiral anomaly at $\tau$
    \end{itemize}
\end{itemize}

\textbf{Obstruction-free states}:
\begin{itemize}

\item[] 1D: ($\bzt$) \begin{enumerate}
    \item Majorana chain on $\tau$ (E)
\end{enumerate}

\item[] 2D: No obstruction-free states ($\bbz_1$)
\end{itemize}

\textbf{Trivializations}:
\begin{itemize}
    \item Open surface decorations trivializes Majorana chain on $\tau$. Therefore, the 1D classification reduces to $\bbz_1$ (trivial).
\end{itemize}

\textbf{Final classification:}
\begin{itemize}
    \item[] $E_{0,dis}^{1D} = \bbz_1$
    \item[] $E_{0,dis}^{2D} = \bbz_1$
    \item[]  $\mathcal{G}_{0,dis} = E_{0,dis}^{1D} \times E_{0,dis}^{2D} = \bbz_1$
\end{itemize}

\subsubsection*{Disordered Spin-1/2}
\textbf{Block state decorations}:
\begin{itemize}
    \item[] 1D
    \begin{itemize}
        \item $\tau$: \placeholder
    \end{itemize}
    \item[] 2D
    \begin{itemize}
        \item $\sigma$:\( p+ip \) SC
    \end{itemize}
\end{itemize}

\textbf{Obstructions}
\begin{itemize}
    \item[] 2D
    \begin{itemize}
        \item\( p+ip \) SC on $\sigma$ is obstructed by chiral anomaly at $\tau$
    \end{itemize}
\end{itemize}

\textbf{Obstruction-free states}:
\begin{itemize}

\item[] 1D: ($\bzt$) \begin{enumerate}
    \item \placeholder on $\tau$ (I)
\end{enumerate} 
\item[] This state is also \textbf{trivialization-free}.

\item[] 2D: No obstruction-free states ($\bbz_1$)
\end{itemize}

\textbf{Final classification:}
\begin{itemize}
    \item[] $E_{1/2,dis}^{1D} = \bzt(I)$
    \item[] $E_{1/2,dis}^{2D} = \bbz_1$
    \item[]  $\mathcal{G}_{1/2,dis} = E_{1/2,dis}^{1D} \times E_{1/2,dis}^{2D} = \bzt(I)$
\end{itemize}

\subsection{$S_4$}
\subsubsection*{Cell Decomposition}
\begin{figure}[!htbp]
    \centering
    \includegraphics[width=0.9\linewidth]{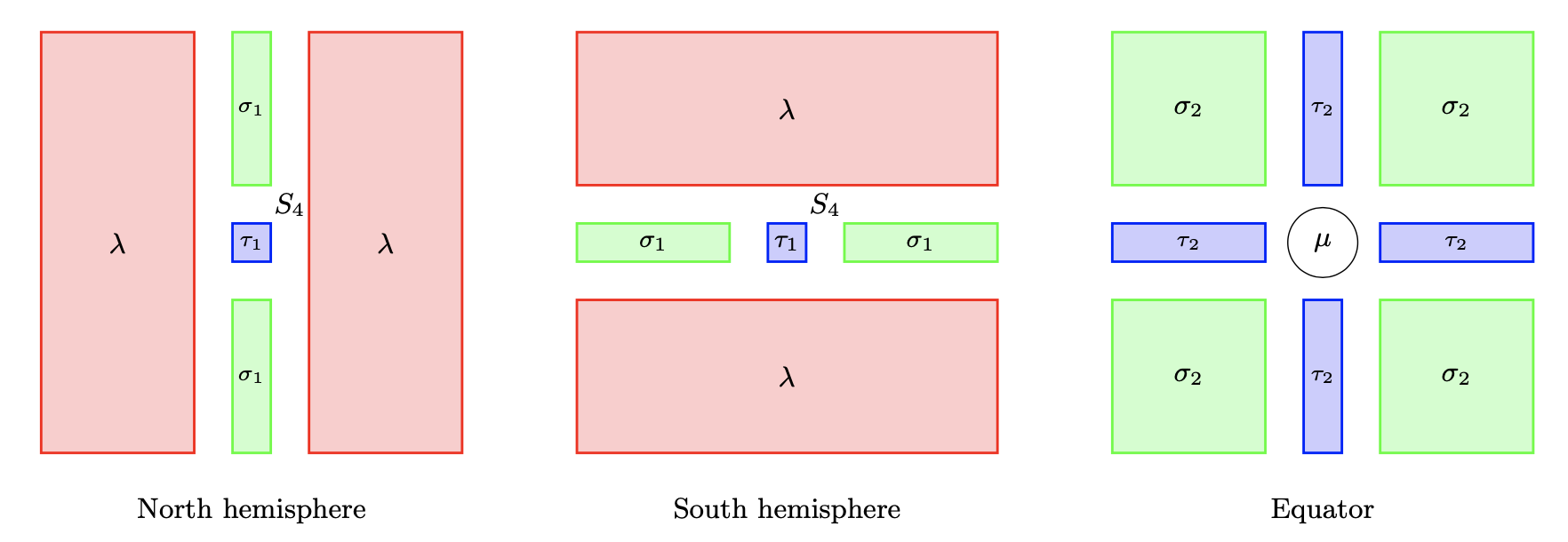}
    \caption{$S_4$ lattice}
\end{figure}
\textbf{Blocks and onsite symmetries}:
\begin{itemize}
    \item 2D: $G_{\sigma_1},\ G_{\sigma_2} = I$
    \item 1D: $G_{\tau_1} = \bzt,\ G_{\tau_2} = I$
    \item $G_{\mu} = S_4$
\end{itemize}
\subsubsection*{Decohered Spinless}
\textbf{Block state decorations}:
\begin{itemize}
    \item[] 0D
    \begin{itemize}
        \item $\mu$: Odd fermion
    \end{itemize}
    \item[] 1D
    \begin{itemize}
        \item $\tau_1$: Majorana chain, $\bzt$ fSPT
        \item $\tau_2$: Majorana chain
    \end{itemize}
    \item[] 2D
    \begin{itemize}
        \item $\sigma_1,\ \sigma_2$:\( p+ip \) SC
    \end{itemize}
\end{itemize}

\textbf{Obstructions}
\begin{itemize}
    \item[] 1D
    \begin{itemize}
        \item Majorana chain on $\tau_1$ or $\tau_2$ is obstructed at $\mu$, but the simultaneous decoration is obstruction-free.
        \item $\bzt$ fSPT on $\tau_1$ is obstructed, but simultaneous decoration with Majorana chain on $\tau_2$ is obstruction-free.
    \end{itemize}
    \item[] 2D
    \begin{itemize}
        \item\( p+ip \) SC on $\sigma_1$ is obstructed by chiral anomaly at $\tau_1$
        \item\( p+ip \) SC on $\sigma_2$ is incompatible with $C_4$ rotation on the equator
    \end{itemize}
\end{itemize}

\textbf{Obstruction-free states}:
\begin{itemize}
\item[] 0D state ($\bzt$) is obstruction-free.

\item[] 1D: ($\bzt^2$) \begin{enumerate}
    \item Majorana chains on $\tau_1$ and $\tau_2$ (E)
    \item $\bzt$ fSPT on $\tau_1$ and Majorana chain on $\tau_2$ (E)
\end{enumerate}

\item[] 2D: No obstruction-free states ($\bbz_1$)
\end{itemize}

\textbf{Trivializations}:
\begin{itemize}
    \item Majorana bubble on $\sigma_1$ $\Rightarrow$ $\bzt$ fSPT on $\tau_1$ and Majorana chain on $\tau_2$. Therefore, the 1D classification reduces to $\bzt$.
    \item Majorana bubble on $\sigma_2$ $\Rightarrow$ Odd fermion on $\mu$. Therefore, the 0D classification reduces to $\bbz_1$ (trivial).
\end{itemize}

\textbf{Final classification:}
\begin{itemize}
    \item[] $E_{0,dec}^{0D} = \bbz_1$
    \item[] $E_{0,dec}^{1D} = \bzt(E)$
    \item[] $E_{0,dec}^{2D} = \bbz_1$
    \item[]  $\mathcal{G}_{0,dec} =  E_{0,dec}^{0D} \times E_{0,dec}^{1D} \times E_{0,dec}^{2D} = \bzt(E)$
\end{itemize}

\subsubsection*{Decohered Spin-1/2}
\textbf{Block state decorations}:
\begin{itemize}
    \item[] 0D
    \begin{itemize}
        \item $\mu$: Odd fermion
    \end{itemize}
    \item[] 1D
    \begin{itemize}
        \item $\tau_1$: $\bbz_4^f$ ASPT
        \item $\tau_2$: Majorana chain
    \end{itemize}
    \item[] 2D
    \begin{itemize}
        \item $\sigma_1,\ \sigma_2$:\( p+ip \) SC
    \end{itemize}
\end{itemize}

\textbf{Obstructions}
\begin{itemize}
    \item[] 2D
    \begin{itemize}
        \item\( p+ip \) SC on $\sigma_1$ is obstructed by chiral anomaly at $\tau_1$
    \end{itemize}
\end{itemize}

\textbf{Obstruction-free states}:
\begin{itemize}
\item[] 0D state ($\bzt$) is obstruction-free (I).

\item[] 1D: ($\bzt^2$) \begin{enumerate}
    \item $\bbz_4^f$ ASPT on $\tau_1$ (I)
    \item Majorana chain on $\tau_2$ (E)
\end{enumerate}

\item[] 2D: ($\bzt$) \begin{enumerate}
    \item\( p+ip \) SC on $\sigma_2$ (E)
\end{enumerate}
\end{itemize}

\textbf{Trivializations}:
\begin{itemize}
    \item Majorana bubble on $\sigma_1$ $\Rightarrow$ Majorana chain on $\tau_2$. Therefore, the 1D classification reduces to $\bzt$.
    \item Fermion bubble on $\tau_1$ $\Rightarrow$ Odd fermion on $\mu$. Therefore, the 0D classification reduces to $\bbz_1$ (trivial).
\end{itemize}

\textbf{Final classification:}
\begin{itemize}
    \item[] $E_{1/2,dec}^{0D} = \bbz_1$
    \item[] $E_{1/2,dec}^{1D} = \bzt(I)$
    \item[] $E_{1/2,dec}^{2D} = \bzt(E)$
    \item[] Non-trivial stacking $\Rightarrow$ $\mathcal{G}_{1/2,dec} =  E_{1/2,dec}^{0D} \rtimes E_{1/2,dec}^{1D} \rtimes E_{1/2,dec}^{2D} = \bzt(E)\times\bzt(I)$
\end{itemize}

\subsubsection*{Disordered Spinless}
\textbf{Block state decorations}:
\begin{itemize}
    \item[] 1D
    \begin{itemize}
        \item $\tau_1,\ \tau_2$: Majorana chain
    \end{itemize}
    \item[] 2D
    \begin{itemize}
        \item $\sigma_1,\ \sigma_2$:\( p+ip \) SC
    \end{itemize}
\end{itemize}

\textbf{Obstructions}
\begin{itemize}
    \item[] 2D
    \begin{itemize}
        \item\( p+ip \) SC on $\sigma_1$ is obstructed by chiral anomaly at $\tau_1$
        \item\( p+ip \) SC on $\sigma_2$ is incompatible with $C_4$ rotation on the equator
    \end{itemize}
\end{itemize}

\textbf{Obstruction-free states}:
\begin{itemize}

\item[] 1D: ($\bzt^2$) \begin{enumerate}
    \item Majorana chain on $\tau_1$ (I)
    \item Majorana chain on $\tau_2$ (I)
\end{enumerate}

\item[] 2D: No obstruction-free states ($\bbz_1$)
\end{itemize}

\textbf{Trivializations}:
\begin{itemize}
    \item Majorana bubble on $\sigma_1$ $\Rightarrow$ Majorana chain on $\tau_2$. Therefore, the 1D classification reduces to $\bzt$.
\end{itemize}

\textbf{Final classification:}
\begin{itemize}
    \item[] $E_{0,dis}^{1D} = \bzt(I)$
    \item[] $E_{0,dis}^{2D} = \bbz_1$
    \item[]  $\mathcal{G}_{0,dis} = E_{0,dis}^{1D} \times E_{0,dis}^{2D} = \bzt(I)$
\end{itemize}

\subsubsection*{Disordered Spin-1/2}
\textbf{Block state decorations}:
\begin{itemize}
    \item[] 1D
    \begin{itemize}
        \item $\tau_1$: \placeholder
        \item $\tau_2$: Majorana chain
    \end{itemize}
    \item[] 2D
    \begin{itemize}
        \item $\sigma_1,\ \sigma_2$:\( p+ip \) SC
    \end{itemize}
\end{itemize}

\textbf{Obstructions}
\begin{itemize}
    \item[] 2D
    \begin{itemize}
        \item\( p+ip \) SC on $\sigma_1$ is obstructed by chiral anomaly at $\tau_1$
    \end{itemize}
\end{itemize}

\textbf{Obstruction-free states}:
\begin{itemize}

\item[] 1D: ($\bzt^2$) \begin{enumerate}
    \item \placeholder on $\tau_1$ (I)
    \item Majorana chain on $\tau_2$ (I)
\end{enumerate}

\item[] 2D: 2D: ($\bzt$) \begin{enumerate}
    \item\( p+ip \) SC on $\sigma_2$ (E)
\end{enumerate}
\end{itemize}

\textbf{Trivializations}:
\begin{itemize}
    \item Majorana bubble on $\sigma_1$ $\Rightarrow$ Majorana chain on $\tau_2$. Therefore, the 1D classification reduces to $\bzt$.
\end{itemize}

\textbf{Final classification:}
\begin{itemize}
    \item[] $E_{1/2,dis}^{1D} = \bzt(I)$
    \item[] $E_{1/2,dis}^{2D} = \bzt(E)$
    \item[]  $\mathcal{G}_{1/2,dis} = E_{1/2,dis}^{1D} \times E_{1/2,dis}^{2D} = \bzt(E)\times\bzt(I)$
\end{itemize}

\subsection{$C_{4h}$}
\subsubsection*{Cell Decomposition}
\begin{figure}[!htbp]
    \centering
    \includegraphics[width=0.9\linewidth]{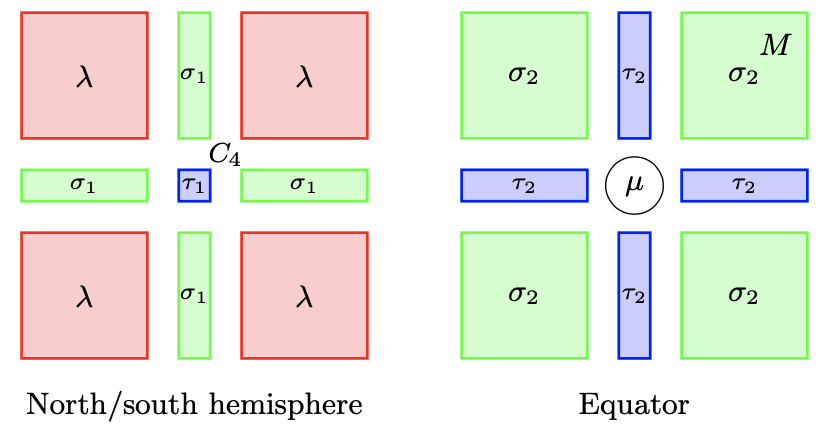}
    \caption{$C_{4h}$ lattice}
\end{figure}
\textbf{Blocks and onsite symmetries}:
\begin{itemize}
    \item $G_{\sigma_1} = I,\ G_{\sigma_2} = \bzt$
    \item $G_{\tau_1} = \bbz_4,\ G_{\tau_2} = \bzt$
    \item $G_{\mu} = C_4 \times \bzt$
\end{itemize}
\subsubsection*{Decohered Spinless}
\textbf{Block state decorations}:
\begin{itemize}
    \item[] 0D
    \begin{itemize}
        \item $\mu$: Odd fermion
    \end{itemize}
    \item[] 1D
    \begin{itemize}
        \item $\tau_1$: Majorana chain, $\bbz_4$ fSPT
        \item $\tau_2$: Majorana chain, $\bzt$ fSPT
    \end{itemize}
    \item[] 2D
    \begin{itemize}
        \item $\sigma_1$:\( p+ip \) SC
        \item $\sigma_2$:\( p+ip \) SC, fLG
    \end{itemize}
\end{itemize}

\textbf{Obstructions}
\begin{itemize}
    \item[] 1D
    \begin{itemize}
        \item Majorana chain on $\tau_1$ or $\tau_2$ is obstructed at $\mu$
        \item[] Simultaneous decoration of the above phases is also obstructed.
    \end{itemize}
    \item[] 2D
    \begin{itemize}
        \item\( p+ip \) SC on $\sigma_1$ is obstructed by chiral anomaly at $\tau_1$
        \item\( p+ip \) SC on $\sigma_2$ is incompatible with spinless rotational symmetry on equator.
    \end{itemize}
\end{itemize}

\textbf{Obstruction-free states}:
\begin{itemize}
\item[] 0D state ($\bzt$) is obstruction-free (E)

\item[] 1D: ($\bzt^2$) \begin{enumerate}
    \item $\bbz_4$ fSPT on $\tau_1$ (I) 
    \item $\bzt$ fSPT on $\tau_2$ (I)
\end{enumerate}

\item[] 2D: ($\bbz_4$) \begin{enumerate}
    \item fLG on $\sigma_2$ (E)
\end{enumerate} 
\end{itemize}

\textbf{Trivializations}:
\begin{itemize}
    \item Majorana bubble on $\sigma_1$ $\Rightarrow$ $\bbz_4$ fSPT on $\tau_1$. Therefore, the 1D classification reduces to $\bzt$.
    \item Majorana bubble on $\sigma_2$ $\Rightarrow$ Odd fermion on $\mu$. Therefore, the 0D classification reduces to $\bbz_1$ (trivial). 
\end{itemize}

\textbf{Final classification:}
\begin{itemize}
    \item[] $E_{0,dec}^{0D} = \bbz_1$
    \item[] $E_{0,dec}^{1D} = \bzt(I)$
    \item[] $E_{0,dec}^{2D} = \bbz_4(E)$
    \item[]  $\mathcal{G}_{0,dec} =  E_{0,dec}^{1D} \times E_{0,dec}^{2D} = \bbz_4(E)\times\bzt(I)$
\end{itemize}

\subsubsection*{Decohered Spin-1/2}
\textbf{Block state decorations}:
\begin{itemize}
    \item[] 0D
    \begin{itemize}
        \item $\mu$: Odd fermion
    \end{itemize}
    \item[] 1D
    \begin{itemize}
        \item $\tau_1$: $\bbz_8^f$ ASPT
        \item $\tau_2$: $\bbz_4^f$ ASPT
    \end{itemize}
    \item[] 2D
    \begin{itemize}
        \item $\sigma_1$:\( p+ip \) SC
        \item $\sigma_2$: No nontrivial block state
    \end{itemize}
\end{itemize}

\textbf{Obstructions}
\begin{itemize}
    \item[] 2D
    \begin{itemize}
        \item\( p+ip \) SC on $\sigma_1$ is obstructed by chiral anomaly at $\tau_1$
    \end{itemize}
\end{itemize}

\textbf{Obstruction-free states}:
\begin{itemize}
\item[] 0D state ($\bzt$) is obstruction-free (I)
\item[] 1D: ($\bzt^2$) \begin{enumerate}
    \item $\bbz_8^f$ ASPT on $\tau_1$ (I)
    \item $\bbz_4^f$ ASPT on $\tau_2$ (I)
\end{enumerate} 
\item[] These states are also \textbf{trivialization-free}.

\item[] 2D: No obstruction-free states ($\bbz_1$). 
\end{itemize}

\textbf{Final classification:}
\begin{itemize}
    \item[] $E_{1/2,dec}^{0D} = \bzt(I)$
    \item[] $E_{1/2,dec}^{1D} = \bzt^2(I)$
    \item[] $E_{1/2,dec}^{2D} = \bbz_1$
    \item[]  $\mathcal{G}_{1/2,dec} = E_{1/2,dec}^{0D} \times E_{1/2,dec}^{1D} \times E_{1/2,dec}^{2D} = \bzt^3(I)$
\end{itemize}

\subsubsection*{Disordered Spinless}
\textbf{Block state decorations}:
\begin{itemize}
    \item[] 1D
    \begin{itemize}
        \item $\tau_1, \tau_2$: Majorana chain
    \end{itemize}
    \item[] 2D
    \begin{itemize}
        \item $\sigma_1$:\( p+ip \) SC
        \item $\sigma_2$:\( p+ip \) SC, fLG
    \end{itemize}
\end{itemize}

\textbf{Obstructions}
\begin{itemize}
    \item[] 2D
    \begin{itemize}
        \item\( p+ip \) SC on $\sigma_1$ is obstructed by chiral anomaly at $\tau_1$
        \item\( p+ip \) SC on $\sigma_2$ is incompatible with spinless rotational symmetry on equator.
    \end{itemize}
\end{itemize}

\textbf{Obstruction-free states}:
\begin{itemize}
\item[] 1D ($\bzt^2$) \begin{enumerate}
    \item Majorana chain on $\tau_1$ or $\tau_2$ (I)
\end{enumerate} 

\item[] 2D ($\bzt$) \begin{enumerate}
    \item fLG on $\sigma_2$ (E)
\end{enumerate} 
\item[] These states are all \textbf{trivialization-free}.
\end{itemize}

\textbf{Final classification:}
\begin{itemize}
    \item[] $E_{0,dis}^{1D} = \bzt^2(I)$
    \item[] $E_{0,dis}^{2D} = \bzt(E)$
    \item[]  $\mathcal{G}_{0,dis} = E_{0,dis}^{1D} \times E_{0,dis}^{2D} = \bzt(E)\times\bzt^2(I)$
\end{itemize}

\subsubsection*{Disordered Spin-1/2}
\textbf{Block state decorations}:
\begin{itemize}
    \item[] 1D
    \begin{itemize}
        \item $\tau_1, \tau_2$: \placeholder
    \end{itemize}
    \item[] 2D
    \begin{itemize}
        \item $\sigma_1$:\( p+ip \) SC
        \item $\sigma_2$: 2D $\bbz_4^f$ ASPT
    \end{itemize}
\end{itemize}

\textbf{Obstructions}
\begin{itemize}
    \item[] 2D
    \begin{itemize}
        \item\( p+ip \) SC on $\sigma_1$ is obstructed by chiral anomaly at $\tau_1$
    \end{itemize}
\end{itemize}

\textbf{Obstruction-free states}:
\begin{itemize}
\item[] 1D ($\bzt^2$) \begin{enumerate}
    \item \placeholder on $\tau_1$ or $\tau_2$ (I)
\end{enumerate} 

\item[] 2D ($\bzt$) \begin{enumerate}
    \item 2D $\bbz_4^f$ ASPT on $\sigma_2$ (I)
\end{enumerate} 
\item[] These states are all \textbf{trivialization-free}.
\end{itemize}

\textbf{Final classification:}
\begin{itemize}
    \item[] $E_{1/2,dis}^{1D} = \bzt^2(I)$
    \item[] $E_{1/2,dis}^{2D} = \bzt(I)$
    \item[]  $\mathcal{G}_{1/2,dis} = E_{1/2,dis}^{1D} \times E_{1/2,dis}^{2D} = \bzt^3(I)$
\end{itemize}

\subsection{$D_4$}
\subsubsection*{Cell Decomposition}
\begin{figure}[!htbp]
    \centering
    \includegraphics[width=0.9\linewidth]{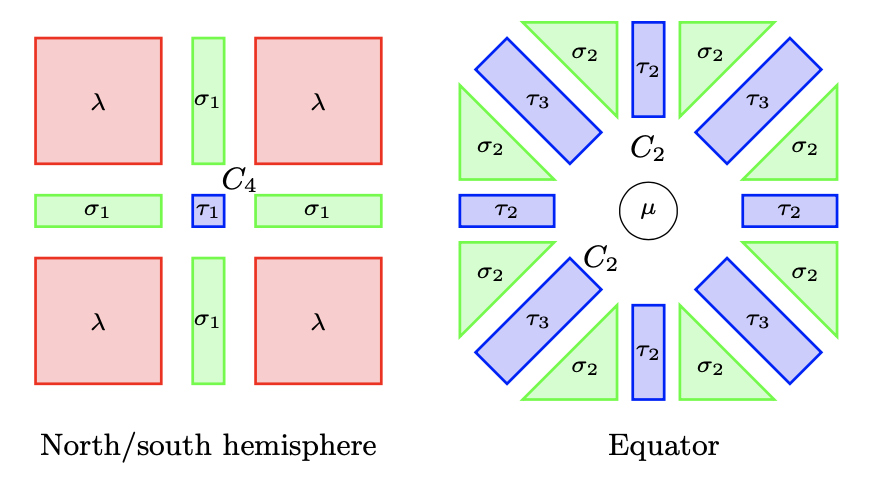}
    \caption{$D_{4}$ lattice}
\end{figure}
\textbf{Blocks and onsite symmetries}:
\begin{itemize}
    \item 2D: $G_{\sigma_1},\ G_{\sigma_2} = I$
    \item 1D: $G_{\tau_1} = \bbz_4,\ G_{\tau_2},\ G_{\tau_3} = \bzt$ 
    \item 0D: $G_{\mu} = D_4$
\end{itemize}

\subsubsection*{Decohered Spinless}
\textbf{Block state decorations}:
\begin{itemize}
    \item[] 0D
    \begin{itemize}
        \item $\mu$: Odd fermion
    \end{itemize}
    \item[] 1D
    \begin{itemize}
        \item $\tau_1$: Majorana chain, $\bbz_4$ fSPT 
        \item $\tau_2,\ \tau_3$: Majorana chain, $\bzt$ fSPT
    \end{itemize}
    \item[] 2D
    \begin{itemize}
        \item $\sigma_1,\ \sigma_2$:\( p+ip \) SC
    \end{itemize}
\end{itemize}

\textbf{Obstructions}
\begin{itemize}
    \item[] 1D
    \begin{itemize}
        \item Majorana chain on $\tau_1,\ \tau_2$, or $\tau_3$ is obstructed at $\mu$
    \end{itemize}
    \item[] 2D
    \begin{itemize}
        \item\( p+ip \) SC on $\sigma_1$ is obstructed by chiral anomaly at $\tau_1$
        \item\( p+ip \) SC on $\sigma_2$ is obstructed by chiral anomaly at $\tau_2$
    \end{itemize}
\end{itemize}

\textbf{Obstruction-free states}:
\begin{itemize}
\item[] 0D state ($\bzt$) is obstruction-free (E)

\item[] 1D: ($\bzt^3$) \begin{enumerate}
    \item $\bbz_4$ fSPT on $\tau_1$ (I)
    \item $\bzt$ fSPT on $\tau_2$ (E)
    \item $\bzt$ fSPT on $\tau_3$ (E)
\end{enumerate}

\item[] 2D: No obstruction-free states ($\bbz_1$)
\end{itemize}

\textbf{Trivializations}:
\begin{itemize}
    \item Majorana bubble on $\sigma_1$ $\Rightarrow$ Simultaneous decoration of $\bbz_4$ fSPT on $\tau_1$ and $\bzt$ fSPT on $\tau_2$. Therefore, the 1D classification reduces to $\bzt^2$.
    \item Majorana bubble on $\sigma_2$ $\Rightarrow$ Simultaneous decoration of $\bzt$ fSPT on $\tau_2$ and $\tau_3$. Therefore, the 1D classification further reduces to $\bzt$. 
    \item Chern insulator bubble trivializes odd fermion parity on $\mu$. The 0D classification reduces to $\bbz_1$.
\end{itemize}

\textbf{Final classification:}
\begin{itemize}
    \item[] $E_{0,dec}^{0D} = \bbz_1$
    \item[] $E_{0,dec}^{1D} = \bzt(E)$
    \item[] $E_{0,dec}^{2D} = \bbz_1$
    \item[]  $\mathcal{G}_{0,dec} =  E_{0,dec}^{0D} \times E_{0,dec}^{1D} \times E_{0,dec}^{2D} = \bzt(E)$
\end{itemize}

\subsubsection*{Decohered Spin-1/2}
\textbf{Block state decorations}:
\begin{itemize}
    \item[] 0D
    \begin{itemize}
        \item $\mu$: Odd fermion
    \end{itemize}
    \item[] 1D
    \begin{itemize}
        \item $\tau_1$: $\bbz_8^f$ ASPT
        \item $\tau_2,\ \tau_3$: $\bbz_4^f$ ASPT
    \end{itemize}
    \item[] 2D
    \begin{itemize}
        \item $\sigma_1,\sigma_2$:\( p+ip \) SC
    \end{itemize}
\end{itemize}

\textbf{Obstructions}
\begin{itemize}
    \item[] 2D
    \begin{itemize}
        \item\( p+ip \) SC on $\sigma_1$ is obstructed by chiral anomaly at $\tau_1$
        \item\( p+ip \) SC on $\sigma_2$ is obstructed by chiral anomaly at $\tau_2$
    \end{itemize}
\end{itemize}

\textbf{Obstruction-free states}:
\begin{itemize}
\item[] 0D state ($\bzt$) is obstruction-free (I)
\item[] 1D: ($\bzt^3$) \begin{enumerate}
    \item $\bbz_8^f$ ASPT on $\tau_1$ (I)
    \item $\bbz_4^f$ ASPT on $\tau_2$ or $\tau_3$ (I)
\end{enumerate} 
\item[] These states are also \textbf{trivialization-free}.

\item[] 2D: No obstruction-free states ($\bbz_1$). 
\end{itemize}

\textbf{Final classification:}
\begin{itemize}
    \item[] $E_{1/2,dec}^{0D} = \bzt(I)$
    \item[] $E_{1/2,dec}^{1D} = \bzt^3(I)$
    \item[] $E_{1/2,dec}^{2D} = \bbz_1$
    \item[]  $\mathcal{G}_{1/2,dec} = E_{1/2,dec}^{0D}\times E_{1/2,dec}^{1D} \times E_{1/2,dec}^{2D} = \bzt^4(I)$
\end{itemize}

\subsubsection*{Disordered Spinless}
\textbf{Block state decorations}:
\begin{itemize}
    \item[] 1D
    \begin{itemize}
        \item $\tau_1,\ \tau_2,\ \tau_3$: Majorana chain
    \end{itemize}
    \item[] 2D
    \begin{itemize}
        \item $\sigma_1,\ \sigma_2$:\( p+ip \) SC
    \end{itemize}
\end{itemize}

\textbf{Obstructions}
\begin{itemize}
    \item[] 2D
    \begin{itemize}
        \item\( p+ip \) SC on $\sigma_1$ is obstructed by chiral anomaly at $\tau_1$
        \item\( p+ip \) SC on $\sigma_2$ is obstructed by chiral anomaly at $\tau_2$
    \end{itemize}
\end{itemize}

\textbf{Obstruction-free states}:
\begin{itemize}
\item[] 1D ($\bzt^3$) \begin{enumerate}
    \item Majorana chain on $\tau_1$, $\tau_2$ or $\tau_3$ (I)
\end{enumerate} 

\item[] 2D: No obstruction-free states ($\bbz_1$)
\end{itemize}

\textbf{Trivializations}:
\begin{itemize}
    \item Open surface decoration $\Rightarrow$ Simultaneous decoration of Majorana chains on $\tau_1,\ \tau_2$, and $\tau_3$. This reduces the 1D classification to $\bzt^2$.
\end{itemize}

\textbf{Final classification:}
\begin{itemize}
    \item[] $E_{0,dis}^{1D} = \bzt^2(I)$
    \item[] $E_{0,dis}^{2D} = \bbz_1$
    \item[]  $\mathcal{G}_{0,dis} = E_{0,dis}^{1D} \times E_{0,dis}^{2D} = \bzt^2(I)$
\end{itemize}

\subsubsection*{Disordered Spin-1/2}
\textbf{Block state decorations}:
\begin{itemize}
    \item[] 1D
    \begin{itemize}
        \item $\tau_1,\ \tau_2,\ \tau_3$: \placeholder
    \end{itemize}
    \item[] 2D
    \begin{itemize}
        \item $\sigma_1,\ \sigma_2$:\( p+ip \) SC
    \end{itemize}
\end{itemize}

\textbf{Obstructions}
\begin{itemize}
    \item[] 2D
    \begin{itemize}
        \item\( p+ip \) SC on $\sigma_1$ is obstructed by chiral anomaly at $\tau_1$
        \item\( p+ip \) SC on $\sigma_2$ is obstructed by chiral anomaly at $\tau_2$
    \end{itemize}
\end{itemize}

\textbf{Obstruction-free states}:
\begin{itemize}
\item[] 1D ($\bzt^3$) \begin{enumerate}
    \item \placeholder on $\tau_1$,$\tau_2$, or $\tau_3$ (I)
\end{enumerate} 
\item[] These states are also \textbf{trivialization-free}.
\item[] 2D: No obstruction-free states ($\bbz_1$)
\end{itemize}

\textbf{Final classification:}
\begin{itemize}
    \item[] $E_{1/2,dis}^{1D} = \bzt^3(I)$
    \item[] $E_{1/2,dis}^{2D} = \bbz_1$
    \item[]  $\mathcal{G}_{1/2,dis} = E_{1/2,dis}^{1D} \times E_{1/2,dis}^{2D} = \bzt^3(I)$
\end{itemize}

\subsection{$C_{4v}$}
\subsubsection*{Cell Decomposition}
\begin{figure}[!htbp]
    \centering
    \includegraphics[width=0.9\linewidth]{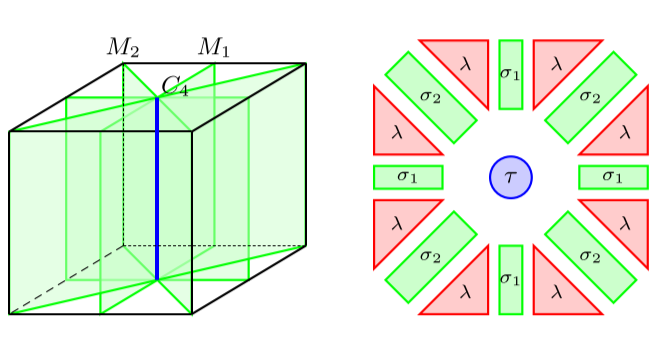}
    \caption{$C_{4v}$ lattice}
\end{figure}
\textbf{Blocks and onsite symmetries}:
\begin{itemize}
    \item 2D: $G_{\sigma_1},\ G_{\sigma_2} = \bzt$
    \item 1D: $G_{\tau} = \bbz_4\rtimes\bzt$
\end{itemize}
\subsubsection*{Decohered Spinless}
\textbf{Block state decorations}:
\begin{itemize}
    \item[] 1D
    \begin{itemize}
        \item $\tau$: Majorana chain, $\bzt^{M_1}$ fSPT, $\bzt^{M_2}$ fSPT
    \end{itemize}
    \item[] 2D
    \begin{itemize}
        \item $\sigma_1,\ \sigma_2$:\( p+ip \) SC, fLG
    \end{itemize}
\end{itemize}

\textbf{Obstructions}
\begin{itemize}
    \item[] 2D
    \begin{itemize}
        \item\( p+ip \) SC on $\sigma_1$ or $\sigma_2$ is obstructed by chiral anomaly at $\tau$
        \item Decoration of $p\pm ip$-SC with opposite chiralities on $\sigma_1$ and $\sigma_2$ is obstructed at $\tau$ (anomaly indicator $\nu_{M_1}=1/4$) 
        \item fLG on $\sigma_1$ is obstructed at $\tau$ ($\nu_{M_1}=1/2$)
        \item fLG on $\sigma_2$ is obstructed at $\tau$ ($\nu_{M_2}=1/2$)
    \end{itemize}
\end{itemize}

\textbf{Obstruction-free states}:
\begin{itemize}

\item[] 1D: ($\bzt^3$) \begin{enumerate}
    \item Majorana chain on $\tau$ (E)
    \item $\bbz_4 \rtimes \bzt$ fSPT on $\tau$ (E)
\end{enumerate}

\item[] 2D: ($\bzt^2$) \begin{enumerate}
    \item $n=2$ fLG on $\sigma_1$ or $\sigma_2$ (I)
\end{enumerate}
\end{itemize}

\textbf{Trivializations}:
\begin{itemize}
    \item Majorana bubble on $\sigma_1$ $\Rightarrow$ $\bzt^{M_1}$ fSPT on $\tau$. Therefore, the 1D classification reduces to $\bzt^2$.
    \item Majorana bubble on $\sigma_2$ $\Rightarrow$ $\bzt^{M_2}$ fSPT on $\tau$. Therefore, the 1D classification further reduces to $\bzt$. 
\end{itemize}

\textbf{Final classification:}
\begin{itemize}
    \item[] $E_{0,dec}^{1D} = \bzt(E)$
    \item[] $E_{0,dec}^{2D} = \bzt^2(I)$
    \item[]  $\mathcal{G}_{0,dec} =  E_{0,dec}^{1D} \times E_{0,dec}^{2D} = \bzt(E)\times\bzt^2(I)$
\end{itemize}

\subsubsection*{Decohered Spin-1/2}
\textbf{Block state decorations}:
\begin{itemize}
    \item[] 1D
    \begin{itemize}
        \item $\tau$: $\bbz_4^{f,M_1}$ ASPT, $\bbz_4^{f,M_2}$ ASPT (I)
    \end{itemize}
    \item[] These states are all \textbf{obstruction-free} and \textbf{trivialization-free}
    \item[] 2D
    \begin{itemize}
        \item $\sigma_1,\ \sigma_2$: No nontrivial block state
    \end{itemize}
\end{itemize}

\textbf{Final classification:}
\begin{itemize}
    \item[] $E_{1/2,dec}^{1D} = \bzt^2(I)$
    \item[] $E_{1/2,dec}^{2D} = \bbz_1$
    \item[]  $\mathcal{G}_{1/2,dec} = E_{1/2,dec}^{1D} \times E_{1/2,dec}^{2D} = \bzt^2(I)$
\end{itemize}

\subsubsection*{Disordered Spinless}
\textbf{Block state decorations}:
\begin{itemize}
    \item[] 1D
    \begin{itemize}
        \item $\tau$: Majorana chain
    \end{itemize}
    \item[] 2D
    \begin{itemize}
        \item $\sigma_1,\ \sigma_2$:\( p+ip \) SC, fLG
    \end{itemize}
\end{itemize}

\textbf{Obstructions}
\begin{itemize}
    \item[] 2D
    \begin{itemize}
        \item\( p+ip \) SC on $\sigma_1$ or $\sigma_2$ is obstructed by chiral anomaly at $\tau$
        \item\( p+ip \) SC on $\sigma_2$ is obstructed by chiral anomaly at $\tau$
        \item Decoration of $p\pm ip$-SC with opposite chiralities on $\sigma_1$ and $\sigma_2$ is obstructed at $\tau$ (anomaly indicator $\nu_{M_1}=1/4$) 
    \end{itemize}
\end{itemize}

\textbf{Obstruction-free states}:
\begin{itemize}
\item[] 1D ($\bzt$) \begin{enumerate}
    \item Majorana chain on $\tau$ (E)
\end{enumerate} 

\item[] 2D ($\bzt^2$) \begin{enumerate}
    \item fLG on $\sigma_1$ or $\sigma_2$ (I)
\end{enumerate}
\item[] These states are all \textbf{trivialization-free}
\end{itemize}

\textbf{Final classification:}
\begin{itemize}
    \item[] $E_{0,dis}^{1D} = \bzt(E)$
    \item[] $E_{0,dis}^{2D} = \bzt^2(I)$
    \item[]  $\mathcal{G}_{0,dis} = E_{0,dis}^{1D} \times E_{0,dis}^{2D} = \bzt(E)\times\bzt^2(I)$
\end{itemize}

\subsubsection*{Disordered Spin-1/2}
\textbf{Block state decorations}:
\begin{itemize}
    \item[] 1D
    \begin{itemize}
        \item $\tau$: \placeholder (I)
    \end{itemize}
    \item[] 2D
    \begin{itemize}
        \item $\sigma_1,\ \sigma_2$: 2D $\bbz_4^f$ ASPT (I)
    \end{itemize}
    \item[] These states are all \textbf{obstruction-free} and \textbf{trivialization-free}.
\end{itemize}

\textbf{Final classification:}
\begin{itemize}
    \item[] $E_{1/2,dis}^{1D} = \bzt(I)$
    \item[] $E_{1/2,dis}^{2D} = \bzt^2(I)$
    \item[]  $\mathcal{G}_{1/2,dis} = E_{1/2,dis}^{1D} \times E_{1/2,dis}^{2D} = \bzt^3(I)$
\end{itemize}

\subsection{$D_{2d}=V_d$}
\begin{figure}[!htbp]
    \centering
    \includegraphics[width=0.9\linewidth]{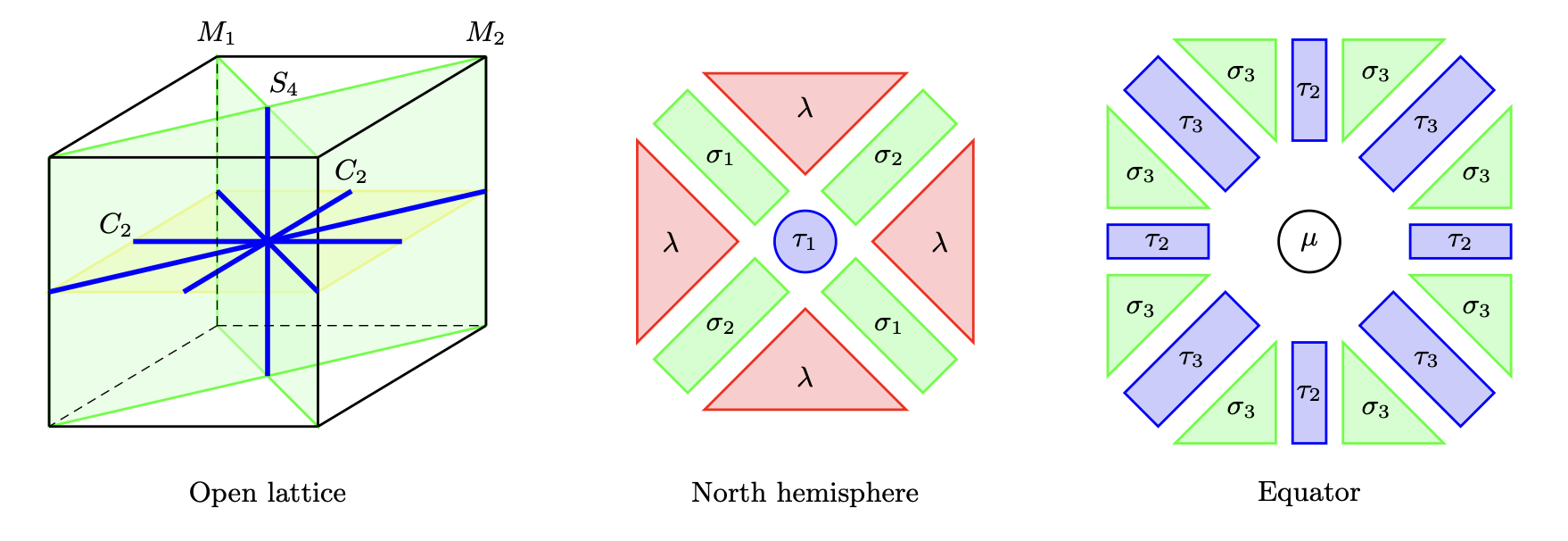}
    \caption{$D_{2d}=V_d$ lattice. On the southern hemisphere, the positions of $\sigma_1$ and $\sigma_2$ are exchanged by the $S_4$ rotoreflection symmetry. The $\tau_2$ blocks have $\bzt$ onsite symmetry as they are axes of $C_2$ rotation, while the $\tau_3$ blocks have $\bzt$ onsite symmetry since they lie on mirror planes $M_1/M_2$.}
\end{figure}
\textbf{Blocks and onsite symmetries}:
\begin{itemize}
    \item $G_{\sigma_1},\ G_{\sigma_2} = \bzt, G_{\sigma_3} = I$
    \item $G_{\tau_1} = \bzt\times\bzt,\ G_{\tau_2},\ G_{\tau_3} = \bzt$
    \item $G_{\mu} = \bbz_4 \rtimes \bzt$
\end{itemize}

\subsubsection*{Decohered Spinless}
\textbf{Block state decorations}:
\begin{itemize}
    \item[] 0D
    \begin{itemize}
        \item $\mu$: Odd fermion
    \end{itemize}
    \item[] 1D
    \begin{itemize}
        \item $\tau_1$: Majorana chain, $\bzt^{M_1}$ fSPT, $\bzt^{M_2}$ fSPT 
        \item $\tau_2,\ \tau_3$: Majorana chain, $\bzt$ fSPT
    \end{itemize}
    \item[] 2D
    \begin{itemize}
        \item $\sigma_1,\ \sigma_2$:\( p+ip \) SC, fLG
        \item $\sigma_3$:\( p+ip \) SC
    \end{itemize}
\end{itemize}

\textbf{Obstructions}
\begin{itemize}
    \item[] 1D
    \begin{itemize}
        \item Majorana chain on $\tau_1,\ \tau_2$, or $\tau_3$ is obstructed at $\mu$. Among these phases, only simultaneous decoration of Majorana chains on $\tau_1$ and $\tau_2$ is obstruction-free.
        \item $\bzt^{M_1}$ and $\bzt^{M_2}$ fSPT phases on $\tau_1$ are obstructed, but simultaneous decoration with Majorana chain on $\tau_2$ is obstruction-free. 
    \end{itemize}
    \item[] 2D
    \begin{itemize}
        \item\( p+ip \) SC on $\sigma_1$ is obstructed by chiral anomaly at $\tau_1$ and $\tau_3$.
        \item\( p+ip \) SC on $\sigma_2$ is obstructed by chiral anomaly at $\tau_1$ and $\tau_3$
        \item Decoration of $p\pm ip$-SC with opposite chiralities on $\sigma_1$ and $\sigma_2$ is obstructed at $\tau_2$ (anomaly indicator $\nu_{M_1}=1/4$)
        \item p+ip-SC on $\sigma_3$ is obstructed by chiral anomaly at $\tau_2$
        \item fLG on $\sigma_1,\ \sigma_2$, or $\sigma_3$ is obstructed.
        \item $n=2$ fLG on $\sigma_1$ or $\sigma_2$ is obstructed, unless both are simultaneously decorated.
    \end{itemize}
\end{itemize}

\textbf{Obstruction-free states}:
\begin{itemize}
\item[] 0D state ($\bzt$) is obstruction-free (E)

\item[] 1D: ($\bzt^5$) \begin{enumerate}
    \item Simultaneous decoration of Majorana chains on $\tau_1$ and $\tau_3$ (I)
    \item Simultaneous decoration of $\bzt^{M_1}$ fSPT on $\tau_1$ and Majorana chain on $\tau_3$ (I)
    \item Simultaneous decoration of $\bzt^{M_2}$ fSPT on $\tau_1$ and Majorana chain on $\tau_3$ (I)
    \item $\bzt$ fSPT on $\tau_2$ (E)
    \item $\bzt$ fSPT on $\tau_3$ (E)
\end{enumerate}

\item[] 2D: ($\bzt$) \begin{enumerate}
    \item $n=2$ fLG on $\sigma_1$ and $\sigma_2$ (I)
\end{enumerate}
\end{itemize}

\textbf{Trivializations}:
\begin{itemize}
    \item Majorana bubble on $\sigma_1$ $\Rightarrow$ Simultaneous decoration of $\bzt^{M_1}$ fSPT on $\tau_1$ and Majorana chain on $\tau_3$. Therefore, the 1D classification reduces to $\bzt^4$.
    \item Majorana bubble on $\sigma_2$ $\Rightarrow$ Simultaneous decoration of $\bzt^{M_2}$ fSPT on $\tau_1$ and Majorana chain on $\tau_3$. Therefore, the 1D classification reduces to $\bzt^3$.
    \item Majorana bubble on $\sigma_3$ $\Rightarrow$ Simultaneous decoration of $\bzt$ fSPT on $\tau_2$ and $\tau_3$. Therefore, the 1D classification reduces to $\bzt^2$.
    \item $\bzt$ fSPT bubble on $\sigma_1$ $\Rightarrow$ $\bzt$ fSPT on $\tau_3$. Therefore, the 1D classification reduces to $\bzt$.
\end{itemize}

\textbf{Final classification:}
\begin{itemize}
    \item[] $E_{0,dec}^{0D} = \bzt (E)$
    \item[] $E_{0,dec}^{1D} = \bzt (I)$
    \item[] $E_{0,dec}^{2D} = \bzt (I)$
    \item[] Non-trivial stacking (derived from spectral sequence) $\Rightarrow$ $\mathcal{G}_{0,dec} =  E_{0,dec}^{0D} \rtimes E_{0,dec}^{1D} \rtimes E_{0,dec}^{2D} = \bzt (I)\times\bbz_4 (I)$
\end{itemize}

\subsubsection*{Decohered Spin-1/2}
\textbf{Block state decorations}:
\begin{itemize}
    \item[] 0D
    \begin{itemize}
        \item $\mu$: Odd fermion
    \end{itemize}
    \item[] 1D
    \begin{itemize}
        \item $\tau_1$: $\bbz_4^{f,M_1}$ ASPT, $\bbz_4^{f,M_2}$ ASPT
        \item $\tau_2,\ \tau_3$: $\bbz_4^f$ ASPT
    \end{itemize}
    \item[] 2D
    \begin{itemize}
        \item $\sigma_1,\sigma_2$: No nontrivial block state
        \item $\sigma_3$:\( p+ip \) SC
    \end{itemize}
\end{itemize}

\textbf{Obstructions}
\begin{itemize}
    \item[] 2D
    \begin{itemize}
        \item\( p+ip \) SC on $\sigma_3$ is obstructed by chiral anomaly at $\tau_2$
    \end{itemize}
\end{itemize}

\textbf{Obstruction-free states}:
\begin{itemize}
\item[] 0D state ($\bzt$) is obstruction-free (I)
\item[] 1D: ($\bzt^4$) \begin{enumerate}
    \item $\bbz_4^{f,M_1}$ ASPT on $\tau_1$ (I)
    \item $\bbz_4^{f,M_2}$ ASPT on $\tau_1$ (I)
    \item $\bbz_4^f$ ASPT on $\tau_2$ (I)
    \item $\bbz_4^f$ ASPT on $\tau_3$ (I)
\end{enumerate} 

\item[] 2D: No obstruction-free states ($\bbz_1$). 
\end{itemize}

\textbf{Trivializations:}
\begin{itemize}
    \item $\bbz_4^f$ ASPT on $\sigma_1$ $\Rightarrow$ $\bbz_4^f$ ASPT on $\tau_3$. Therefore, the 1D classification reduces to $\bzt^3$.
    \item Fermion bubble on $\tau_1$ $\Rightarrow$ Odd fermion on $\mu$. Therefore, the 0D classification reduces to $\bbz_1$ (trivial).
\end{itemize}

\textbf{Final classification:}
\begin{itemize}
    \item[] $E_{1/2,dec}^{0D} = \bbz_1$
    \item[] $E_{1/2,dec}^{1D} = \bzt^3 (I)$
    \item[] $E_{1/2,dec}^{2D} = \bbz_1$
    \item[]  $\mathcal{G}_{1/2,dec} = E_{1/2,dec}^{0D}\times E_{1/2,dec}^{1D} \times E_{1/2,dec}^{2D} = \bzt^3 (I)$
\end{itemize}

\subsubsection*{Disordered Spinless}
\textbf{Block state decorations}:
\begin{itemize}
    \item[] 1D
    \begin{itemize}
        \item $\tau_1,\ \tau_2,\ \tau_3$: Majorana chain
    \end{itemize}
    \item[] 2D
    \begin{itemize}
        \item $\sigma_1,\ \sigma_2$:\( p+ip \) SC, fLG
        \item $\sigma_3$:\( p+ip \) SC
    \end{itemize}
\end{itemize}

\textbf{Obstructions}
\begin{itemize}
    \item[] 2D
    \begin{itemize}     
        \item\( p+ip \) SC on $\sigma_1$ is obstructed by chiral anomaly at $\tau_1$ and $\tau_3$.
        \item\( p+ip \) SC on $\sigma_2$ is obstructed by chiral anomaly at $\tau_1$ and $\tau_3$
        \item Decoration of $p\pm ip$-SC with opposite chiralities on $\sigma_1$ and $\sigma_2$ is obstructed at $\tau_2$ (anomaly indicator $\nu_{M_1}=1/4$)
        \item p+ip-SC on $\sigma_3$ is obstructed by chiral anomaly at $\tau_2$
        \item fLG on $\sigma_1$ or $\sigma_2$ is obstructed, unless both are simultaneously decorated.
    \end{itemize}
\end{itemize}

\textbf{Obstruction-free states}:
\begin{itemize}
\item[] 1D ($\bzt^3$) \begin{enumerate}
    \item Majorana chain on $\tau_1$, $\tau_2$, or $\tau_3$ (I)
\end{enumerate} 

\item[] 2D ($\bzt$) \begin{enumerate}
    \item fLG on $\sigma_1$ and $\sigma_2$ (I)
\end{enumerate}
\end{itemize}

\textbf{Trivializations}:
\begin{itemize}
    \item Majorana bubble on $\sigma_1$ $\Rightarrow$ Majorana chain on $\tau_2$. Therefore, the 1D classification reduces to $\bzt^2$.
\end{itemize}

\textbf{Final classification:}
\begin{itemize}
    \item[] $E_{0,dis}^{1D} = \bzt^2(I)$
    \item[] $E_{0,dis}^{2D} = \bzt(I)$
    \item[]  $\mathcal{G}_{0,dis} = E_{0,dis}^{1D} \times E_{0,dis}^{2D} = \bzt^3(I)$
\end{itemize}

\subsubsection*{Disordered Spin-1/2}
\textbf{Block state decorations}:
\begin{itemize}
    \item[] 1D
    \begin{itemize}
        \item $\tau_1,\ \tau_2,\ \tau_3$: \placeholder
    \end{itemize}
    \item[] 2D
    \begin{itemize}
        \item $\sigma_1,\ \sigma_2$: 2D $\bbz_4^f$ ASPT
        \item $\sigma_3$:\( p+ip \) SC
    \end{itemize}
\end{itemize}

\textbf{Obstructions}
\begin{itemize}
    \item[] 2D
    \begin{itemize}     
        \item p+ip-SC on $\sigma_3$ is obstructed by chiral anomaly at $\tau_2$
        \item 2D $\bbz_4^f$ ASPT on $\sigma_1$ or $\sigma_2$ is obstructed, unless both are simultaneously decorated.
    \end{itemize}
\end{itemize}

\textbf{Obstruction-free states}:
\begin{itemize}
\item[] 1D ($\bzt^3$) \begin{enumerate}
    \item \placeholder on $\tau_1$, $\tau_2$, or $\tau_3$ (I)
\end{enumerate} 

\item[] 2D ($\bzt$) \begin{enumerate}
    \item 2D $\bbz_4^f$ ASPT on $\sigma_1$ and $\sigma_2$ (I)
\end{enumerate}
\end{itemize}

\textbf{Trivializations}:
\begin{itemize}
    \item \bubbleplaceholder on $\sigma_1$ $\Rightarrow$ Majorana chain on $\tau_2$. Therefore, the 1D classification reduces to $\bzt^2$.
\end{itemize}

\textbf{Final classification:}
\begin{itemize}
    \item[] $E_{1/2,dis}^{1D} = \bzt^2(I)$
    \item[] $E_{1/2,dis}^{2D} = \bzt(I)$
    \item[]  $\mathcal{G}_{1/2,dis} = E_{1/2,dis}^{1D} \times E_{1/2,dis}^{2D} = \bzt^3(I)$
\end{itemize}

\subsection{$D_{4h}$}
\subsubsection*{Cell Decomposition}
\begin{figure}[!htbp]
    \centering
    \includegraphics[width=0.9\linewidth]{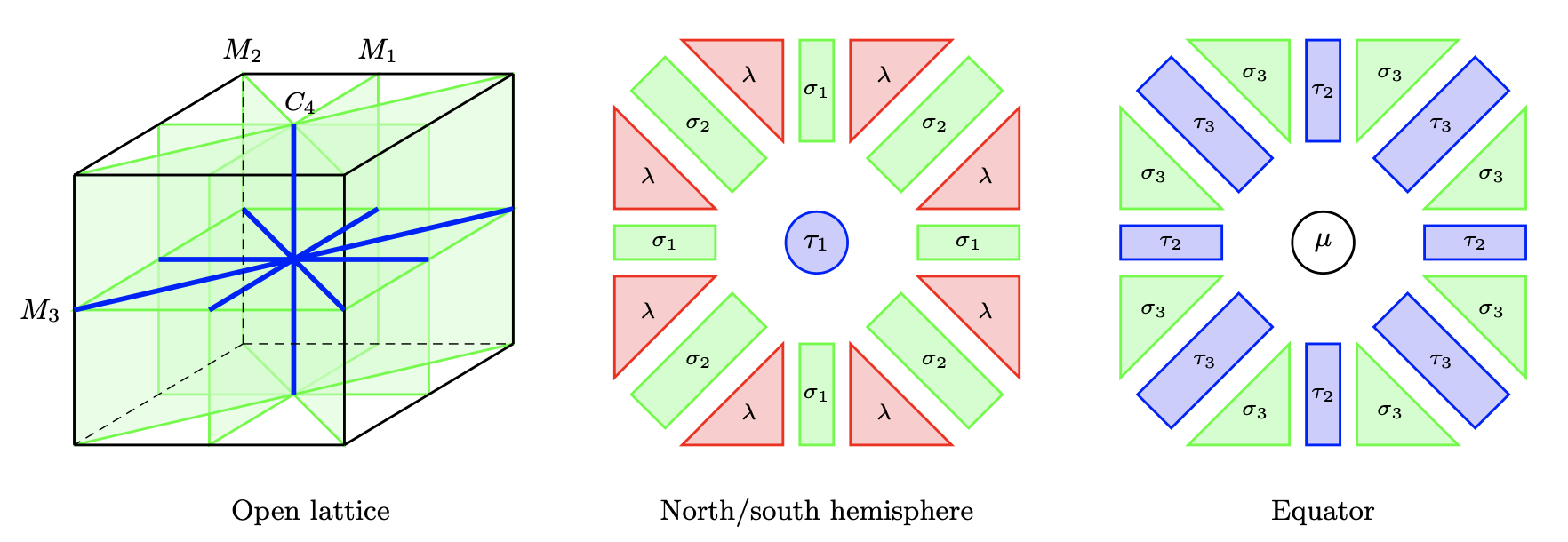}
    \caption{$D_{4h}$ lattice}
\end{figure}
\textbf{Blocks and onsite symmetries}:
\begin{itemize}
    \item 2D: $G_{\sigma_1},\ G_{\sigma_2},\ G_{\sigma_3} = \bzt$
    \item 1D: $G_{\tau_1} = \bbz_4\rtimes\bzt,\ G_{\tau_2},\ G_{\tau_3} = \bzt\times\bzt$
    \item 0D: $G_{\mu} = \bzt\times(\bbz_4\rtimes\bzt)$
\end{itemize}

\subsubsection*{Decohered Spinless}
\textbf{Block state decorations}:
\begin{itemize}
    \item[] 0D
    \begin{itemize}
        \item $\mu$: Odd fermion
    \end{itemize}
    \item[] 1D
    \begin{itemize}
        \item $\tau_1$: Majorana chain, $\bzt^{M_1}$ fSPT, $\bzt^{M_2}$ fSPT
        \item $\tau_2$: Majorana chain, $\bzt^{M_1}$ fSPT, $\bzt^{M_3}$ fSPT
        \item $\tau_3$: Majorana chain, $\bzt^{M_2}$ fSPT, $\bzt^{M_3}$ fSPT
    \end{itemize}
    \item[] 2D
    \begin{itemize}
        \item $\sigma_1,\ \sigma_2,\ \sigma_3$:\( p+ip \) SC, fLG
    \end{itemize}
\end{itemize}

\textbf{Obstructions}
\begin{itemize}
    \item[] 1D
    \begin{itemize}
        \item Majorana chain on $\tau_1$, $\tau_2$, or $\tau_3$ is obstructed at $\mu$
    \end{itemize}
    \item[] 2D
    \begin{itemize}
        \item\( p+ip \) SC on $\sigma_1$ is obstructed by chiral anomaly at $\tau_1$
        \item\( p+ip \) SC on $\sigma_2$ is obstructed by chiral anomaly at $\tau_2$
        \item\( p+ip \) SC on $\sigma_3$ is obstructed by chiral anomaly at $\tau_3$
        \item Simultaneous decorations are obstructed as well
        \item fLG on $\sigma_1$ is obstructed at $\tau_1$ ($\nu_{M_1}=1/2$)
        \item fLG on $\sigma_2$ is obstructed at $\tau_2$ ($\nu_{M_2}=1/2$)
        \item fLG on $\sigma_3$ is obstructed at $\tau_3$ ($\nu_{M_3}=1/2$)   
    \end{itemize}
\end{itemize}

\textbf{Obstruction-free states}:
\begin{itemize}
\item[] 0D states are obstruction-free ($\bzt$) (E)
 
\item[] 1D: ($\bzt^6$) \begin{enumerate}
    \item Any combination of $\bzt$ fSPTs on $\tau_1$, $\tau_2$, $\tau_3$ (I).
\end{enumerate}

\item[] 2D: ($\bzt^3$) \begin{enumerate}
    \item $n=2$ fLG on $\sigma_1$, $\sigma_2$ or $\sigma_3$ (I) 
\end{enumerate}
\end{itemize}

\textbf{Trivializations}:
\begin{itemize}
    \item Majorana bubble on $\sigma_1$ $\Rightarrow$ Simultaneous decoration of $\bzt^{M_1}$ fSPT on $\tau_1$ and $\tau_2$. Therefore, the 1D classification reduces to $\bzt^5$.
    \item Majorana bubble on $\sigma_2$ $\Rightarrow$ Simultaneous decoration of $\bzt^{M_2}$ fSPT on $\tau_1$ and $\tau_3$. Therefore, the 1D classification further reduces to $\bzt^4$.
    \item Majorana bubble on $\sigma_3$ $\Rightarrow$ Simultaneous decoration of $\bzt^{M_3}$ fSPT on $\tau_2$ and $\tau_3$. Therefore, the 1D classification further reduces to $\bzt^3$. 
\end{itemize}

\textbf{Final classification:}
\begin{itemize}
    \item[] $E_{0,dec}^{0D} = \bzt(E)$
    \item[] $E_{0,dec}^{1D} = \bzt^3(I)$
    \item[] $E_{0,dec}^{2D} = \bzt^3(I)$
    \item[]  $\mathcal{G}_{0,dec} = E_{0,dec}^{0D} \times E_{0,dec}^{1D} \times E_{0,dec}^{2D} = \bzt\times\bzt^6(I)$
\end{itemize}

\subsubsection*{Decohered Spin-1/2}
\textbf{Block state decorations}:
\begin{itemize}
    \item[] 0D
    \begin{itemize}
        \item $\mu$: Odd fermion (I)
    \end{itemize}
    \item[] 1D
    \begin{itemize}
        \item $\tau_1$: $\bbz_4^{f,M_1}$ ASPT, $\bbz_4^{f,M_2}$ ASPT (I)
        \item $\tau_2$: $\bbz_4^{f,M_1}$ ASPT, $\bbz_4^{f,M_3}$ ASPT (I)
        \item $\tau_3$: $\bbz_4^{f,M_2}$ ASPT, $\bbz_4^{f,M_3}$ ASPT (I)
    \end{itemize}
    \item[] These states are all \textbf{obstruction-free} and \textbf{trivialization-free}
    \item[] 2D
    \begin{itemize}
        \item $\sigma_1,\ \sigma_2,\ \sigma_3$: No nontrivial block state
    \end{itemize}
\end{itemize}

\textbf{Final classification:}
\begin{itemize}
    \item[] $E_{1/2,dec}^{0D} = \bzt(I)$
    \item[] $E_{1/2,dec}^{1D} = \bzt^6(I)$
    \item[] $E_{1/2,dec}^{2D} = \bbz_1$
    \item[]  $\mathcal{G}_{1/2,dec} = E_{1/2,dec}^{0D} \times E_{1/2,dec}^{1D} \times E_{1/2,dec}^{2D} = \bzt^7(I)$
\end{itemize}

\subsubsection*{Disordered Spinless}
\textbf{Block state decorations}:
\begin{itemize}
    \item[] 1D
    \begin{itemize}
        \item $\tau_1,\ \tau_2,\ \tau_3$: Majorana chain
    \end{itemize}
    \item[] 2D
    \begin{itemize}
        \item $\sigma_1,\ \sigma_2,\ \sigma_3$:\( p+ip \) SC, fLG
    \end{itemize}
\end{itemize}

\textbf{Obstructions}
\begin{itemize}
    \item[] 2D
    \begin{itemize}
        \item\( p+ip \) SC on $\sigma_1$, $\sigma_2$, $\sigma_3$ is obstructed by chiral anomaly at $\tau_1$, $\tau_2$, $\tau_3$ respectively.
    \end{itemize}
\end{itemize}

\textbf{Obstruction-free states}:
\begin{itemize}
\item[] 1D ($\bzt^3$) \begin{enumerate}
    \item Majorana chain on $\tau_1$, $\tau_2$, or $\tau_3$ (I) 
\end{enumerate} 

\item[] 2D ($\bzt^3$) \begin{enumerate}
    \item fLG on $\sigma_1$, $\sigma_2$, or $\sigma_3$ (I)
\end{enumerate}
\item[] These states are all \textbf{trivialization-free}
\end{itemize}

\textbf{Final classification:}
\begin{itemize}
    \item[] $E_{0,dis}^{1D} = \bzt^3(I)$
    \item[] $E_{0,dis}^{2D} = \bzt^3(I)$
    \item[]  $\mathcal{G}_{0,dis} = E_{0,dis}^{1D} \times E_{0,dis}^{2D} = \bzt^6(I)$
\end{itemize}

\subsubsection*{Disordered Spin-1/2}
\textbf{Block state decorations}:
\begin{itemize}
    \item[] 1D
    \begin{itemize}
        \item $\tau_1,\ \tau_2,\ \tau_3$: \placeholder (I)
    \end{itemize}
    \item[] 2D
    \begin{itemize}
        \item $\sigma_1,\ \sigma_2,\ \sigma_3$: 2D $\bbz_4^f$ ASPT (I)
    \end{itemize}
    \item[] These states are all \textbf{obstruction-free} and \textbf{trivialization-free}.
\end{itemize}

\textbf{Final classification:}
\begin{itemize}
    \item[] $E_{1/2,dis}^{1D} = \bzt^3(I)$
    \item[] $E_{1/2,dis}^{2D} = \bzt^3(I)$
    \item[]  $\mathcal{G}_{1/2,dis} = E_{1/2,dis}^{1D} \times E_{1/2,dis}^{2D} = \bzt^6(I)$
\end{itemize}

\subsection{$C_3$}
\subsubsection*{Cell Decomposition}
\begin{figure}[!htbp]
    \centering
    \includegraphics[width=0.9\linewidth]{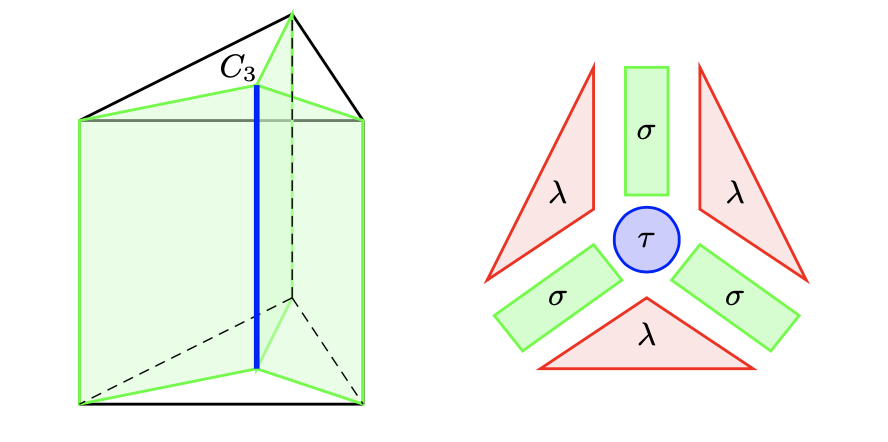}
    \caption{$C_3$ lattice}
\end{figure}
\textbf{Blocks and onsite symmetries}:
\begin{itemize}
    \item $G_{\sigma}=I$
    \item $G_{\tau}=\bbz_3$
\end{itemize}

For all four cases, the decorations, obstructions, and trivializations are the same for the $C_3$ point group.

\subsubsection*{Decohered/Disordered}
\textbf{Block state decorations}:
\begin{itemize}
    \item[] 1D
    \begin{itemize}
        \item $\tau$: Majorana chain
    \end{itemize}
    \item[] 2D
    \begin{itemize}
        \item $\sigma$:\( p+ip \) SC
    \end{itemize}
\end{itemize}

\textbf{Obstructions}
\begin{itemize}
    \item[] 2D
    \begin{itemize}
        \item\( p+ip \) SC on $\sigma$ is obstructed by chiral anomaly at $\tau$.
    \end{itemize}
\end{itemize}

\textbf{Obstruction-free states}:
\begin{itemize}
\item[] 1D: ($\bzt$) \begin{enumerate}
    \item Majorana chain on $\tau$
\end{enumerate}

\item[] 2D: No obstruction-free states ($\bbz_1$). 
\end{itemize}

\textbf{Trivializations}:
\begin{itemize}
    \item Majorana bubble on $\sigma$ $\Rightarrow$ Majorana on $\tau$. Therefore, the 1D classification reduces to $\bbz_1$ (trivial)
\end{itemize}

\textbf{Final classification:}
\begin{itemize}
    \item[] $E^{1D} = \bbz_1$
    \item[] $E^{2D} = \bbz_1$
    \item[]  $\mathcal{G} =  E^{1D} \times E^{2D} = \bbz_1$
\end{itemize}

\subsection{$S_6$}
\subsubsection*{Cell Decomposition} 
\begin{figure}[!htbp]
    \centering
    \includegraphics[width=0.9\linewidth]{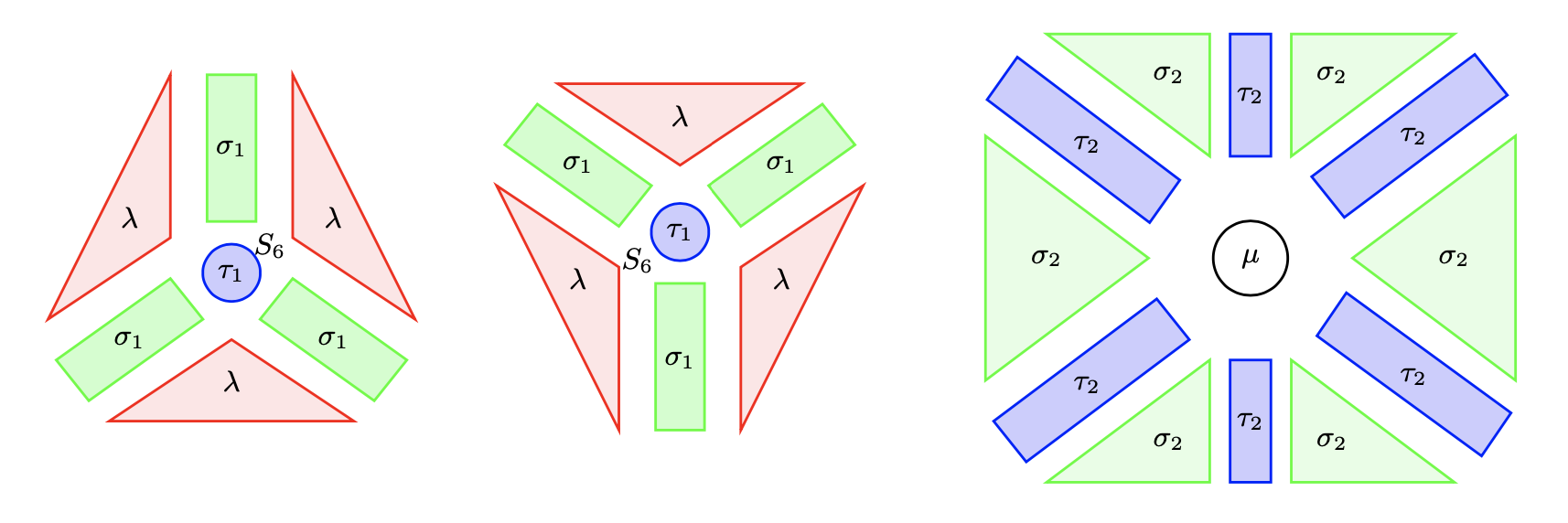}
    \caption{$S_6$ lattice}
\end{figure}
\textbf{Blocks and onsite symmetries}:
\begin{itemize}
    \item $\sigma_1,\sigma_2=I$
    \item $\tau_1 = \bbz_3,\tau_2 = I$
    \item $\mu = S_6$ center
\end{itemize}
An important distinction of this point group is that it is \textbf{inversion-symmetric}. If we denote the generating element of $S_6$ roto-reflection as $g_{S_6}$, it is easy to see that $g_{S_6}^3$ inverts any point in the unit cell. This is important, as inversion symmetry has no nontrivial extension with fermion parity, and hence the point group has no spin-1/2 case. To intuit this, we note that the inversion symmetry $g_{inv}$ can be written as the composition of a $\pi$-rotation about the z-axis, $R_{\pi,z}$ and reflection about the $xy$ plane, $M_{xy}$. In the spin-1/2 case, this would have implementation
\begin{equation}
    M_{xy}^2 = R_z^2= P_f
\end{equation}
However, this implies that $g_{inv}^2=I$. But since the inversion symmetry also has order 2, we require that $g_{S_6}^6 = g_{inv}^2=P_f$ in the spin-1/2 case. This inconsistency forbids the possibility of the spin-1/2 case.

Hence we only discuss the spinless classification. For the sake of consistency, we keep the spin-1/2 section in the classification table, but the entries are duplicated from the spinless case.

\subsubsection*{Decohered}
\textbf{Block state decorations}:
\begin{itemize}
    \item[] 0D \begin{itemize}
        \item $\mu$: Odd fermion
    \end{itemize}
    \item[] 1D \begin{itemize}
        \item $\tau_1,\ \tau_2$: Majorana chain 
    \end{itemize}
    \item[] 2D \begin{itemize}
        \item $\sigma_1,\ \sigma_2$:\( p+ip \) SC
    \end{itemize}
\end{itemize}
\textbf{Obstructions}
\begin{itemize}
    \item[] 1D
    \begin{itemize}
        \item Majorana chain on $\tau_1$ or $\tau_2$ is obstructed at $\mu$ by $S_6$ roto-reflection symmetry, unless simultaneously decorated.
    \end{itemize}
    \item[] 2D
    \begin{itemize}
        \item\( p+ip \) SC on $\sigma_1$ is obstructed by chiral anomaly at $\tau_1$
        \item\( p+ip \) SC on $\sigma_2$ is incompatible with rotational symmetry at equator
    \end{itemize}
\end{itemize}
\textbf{Obstruction-free states}:
\begin{itemize}
    \item[] 0D state ($\bzt$) is obstruction-free
    \item[] 1D ($\bzt$) \begin{enumerate}
        \item Majorana chains on $\tau_1$ and $\tau_2$ (E)
    \end{enumerate}
    \item[] 2D: No obstruction-free states ($\bbz_1$)
\end{itemize}
\textbf{Trivializations}:
\begin{itemize}
    \item Majorana bubble on $\sigma_1$ $\Rightarrow$ Majorana chains on $\tau_1$ and $\tau_2$. Therefore, the 1D classification reduces to $\bbz_1$ (trivial).
\end{itemize}

\textbf{Final classification:}
\begin{itemize}
    \item[] $E_{0,dec}^{0D} = \bbz_1$
    \item[] $E_{0,dec}^{1D} = \bbz_1$
    \item[] $E_{0,dec}^{2D} = \bbz_1$
    \item[]  $\mathcal{G}_{0,dec} =  E_{0,dec}^{1D} \times E_{0,dec}^{2D} = \bbz_1$
\end{itemize}

\subsubsection*{Disordered}
\textbf{Block state decorations}:
\begin{itemize}
    \item[] 1D
    \begin{itemize}
        \item $\tau_1,\ \tau_2$: Majorana chain
    \end{itemize}
    \item[] 2D
    \begin{itemize}
        \item $\sigma_1,\ \sigma_2$:\( p+ip \) SC
    \end{itemize}
\end{itemize}

\textbf{Obstructions}
\begin{itemize}
    \item[] 2D
    \begin{itemize}
        \item\( p+ip \) SC on $\sigma_1$ is obstructed by chiral anomaly at $\tau_1$
        \item\( p+ip \) SC on $\sigma_2$ is incompatible with rotational symmetry at equator
    \end{itemize}
\end{itemize}

\textbf{Obstruction-free states}:
\begin{itemize}
\item[] 1D ($\bzt^2$) \begin{enumerate}
    \item Majorana chain on $\tau_1$ or $\tau_2$ (I)
\end{enumerate} 

\item[] 2D: No obstruction-free states ($\bbz_1$)
\end{itemize}

\textbf{Trivializations}:
\begin{itemize}
    \item Majorana bubble on $\sigma_1$ $\Rightarrow$ Majorana chains on $\tau_1$ and $\tau_2$. Therefore, the 1D classification reduces to $\bzt$
\end{itemize}

\textbf{Final classification:}
\begin{itemize}
    \item[] $E_{0,dis}^{1D} = \bzt(I)$
    \item[] $E_{0,dis}^{2D} = \bbz_1$
    \item[]  $\mathcal{G}_{0,dis} = E_{0,dis}^{1D} \times E_{0,dis}^{2D} = \bzt(I)$
\end{itemize}

\subsection{$D_3$}
\subsubsection*{Cell Decomposition}
\begin{figure}[!htbp]
    \centering
    \includegraphics[width=0.9\linewidth]{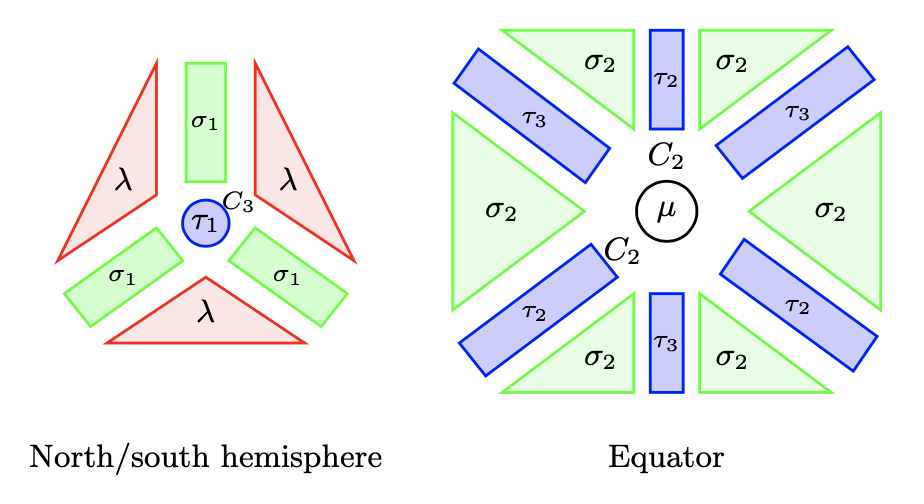}
    \caption{$D_3$ lattice}
\end{figure}
\textbf{Blocks and onsite symmetries}:
\begin{itemize}
    \item 2D: $G_{\sigma_1},\ G_{\sigma_2}=I$
    \item 1D: $G_{\tau_1}=I,\ G_{\tau_2},\ G_{\tau_3} = \bzt$
    \item 0D: $G_{\mu}=D_3$
\end{itemize}

\subsubsection*{Decohered Spinless}
\textbf{Block state decorations}:
\begin{itemize}
    \item[] 0D
    \begin{itemize}
        \item $\mu$: Odd fermion
    \end{itemize}
    \item[] 1D
    \begin{itemize}
        \item $\tau_1$: Majorana chain
        \item $\tau_2,\ \tau_3$: Majorana chain, $\bzt$ fSPT
    \end{itemize}
    \item[] 2D
    \begin{itemize}
        \item $\sigma_1,\ \sigma_2$:\( p+ip \) SC
    \end{itemize}
\end{itemize}
\textbf{Obstructions}
\begin{itemize}
    \item[] 1D
    \begin{itemize}
        \item Majorana chain on $\tau_1$ is obstructed at $\mu$ by two-fold rotational symmetry.
        \item Majorana chain or $\bzt$ fSPT on one of $\tau_2$ or $\tau_3$ is obstructed at $\mu$ (odd number of gapless modes)
    \end{itemize}
    \item[] 2D
    \begin{itemize}
        \item\( p+ip \) SC on $\sigma_1$, $\sigma_2$ is obstructed by chiral anomaly at $\tau_1$, $\tau_2$ respectively
    \end{itemize}
\end{itemize}

\textbf{Obstruction-free states}:
\begin{itemize}
\item[] 0D state ($\bzt$) is obstruction-free (E)

\item[] 1D: ($\bzt^2$) \begin{enumerate}
    \item Majorana chains on $\tau_2$ and $\tau_3$ (E)
    \item $\bzt$ fSPTs on $\tau_2$ and $\tau_3$ (E)
\end{enumerate}

\item[] 2D: No obstruction-free states ($\bbz_1$)
\end{itemize}

\textbf{Trivializations}:
\begin{itemize}
    \item Majorana bubble on $\sigma_1$ and open surface decoration $\Rightarrow$ Majorana chains on $\tau_2$ and $\tau_3$. Therefore, the 1D classification reduces to $\bzt$.
    \item Majorana bubble on $\sigma_2$ $\Rightarrow$ Simultaneous decoration of $\bzt$ fSPT on $\tau_2$ and $\tau_3$. Therefore, the 1D classification further reduces to $\bbz_1$ (trivial). 
    \item Fermion bubble on $\tau_2$ $\Rightarrow$ Odd fermion on $\mu$. Therefore, the 0D classification reduces to $\bbz_1$ (trivial).
\end{itemize}

\textbf{Final classification:}
\begin{itemize}
    \item[] $E_{0,dec}^{0D} = \bbz_1$
    \item[] $E_{0,dec}^{1D} = \bbz_1$
    \item[] $E_{0,dec}^{2D} = \bbz_1$
    \item[]  $\mathcal{G}_{0,dec} =  E_{0,dec}^{0D}\times E_{0,dec}^{1D} \times E_{0,dec}^{2D} = \bbz_1$
\end{itemize}

\subsubsection*{Decohered Spin-1/2}
\textbf{Block state decorations}:
\begin{itemize}
    \item[] 0D
    \begin{itemize}
        \item $\mu$: Odd fermion
    \end{itemize}
    \item[] 1D
    \begin{itemize}
        \item $\tau_1$: Majorana chain 
        \item $\tau_2,\ \tau_3$: $\bbz_4^f$ ASPT
    \end{itemize}
    \item[] 2D
    \begin{itemize}
        \item $\sigma_1,\sigma_2$:\( p+ip \) SC
    \end{itemize}
\end{itemize}

\textbf{Obstructions}
\begin{itemize}
    \item[] 2D
    \begin{itemize}
        \item\( p+ip \) SC on $\sigma_1$, $\sigma_2$ is obstructed by chiral anomaly at $\tau_1$, $\tau_2$ respectively.
    \end{itemize}
    \item[] 1D
    \begin{itemize}
        \item $\bbz_4^f$ ASPT on one of $\tau_2$ or $\tau_3$ leaves odd number of edge modes at $\mu$
    \end{itemize}
\end{itemize}

\textbf{Obstruction-free states}:
\begin{itemize}
\item[] 0D state ($\bzt$) is obstruction-free (I)
\item[] 1D: ($\bzt^3$) \begin{enumerate}
    \item Majorana chain on $\tau_1$ (E)
    \item $\bbz_4^f$ ASPT on $\tau_2$ and $\tau_3$(I)
\end{enumerate} 

\item[] 2D: No obstruction-free states ($\bbz_1$). 
\end{itemize}

\textbf{Trivializations}:
\begin{itemize}
\item Majorana bubble on $\sigma_1$ $\Rightarrow$ Majorana chain on $\tau_1$. Therefore, the 1D classification reduces to $\bzt$.
\item Fermion bubble on $\tau_2$ $\Rightarrow$ Odd fermion on $\mu$. Therefore, the 0D classification reduces to $\bbz_1$ (trivial).
\end{itemize}
\textbf{Final classification:}
\begin{itemize}
    \item[] $E_{1/2,dec}^{0D} = \bbz_1$
    \item[] $E_{1/2,dec}^{1D} = \bzt(I)$
    \item[] $E_{1/2,dec}^{2D} = \bbz_1$
    \item[]  $\mathcal{G}_{1/2,dec} = E_{1/2,dec}^{0D}\times E_{1/2,dec}^{1D} \times E_{1/2,dec}^{2D} = \bzt(I)$
\end{itemize}

\subsubsection*{Disordered Spinless}
\textbf{Block state decorations}:
\begin{itemize}
    \item[] 1D
    \begin{itemize}
        \item $\tau_1,\ \tau_2,\ \tau_3$: Majorana chain
    \end{itemize}
    \item[] 2D
    \begin{itemize}
        \item $\sigma_1,\ \sigma_2$:\( p+ip \) SC
    \end{itemize}
\end{itemize}

\textbf{Obstructions}
\begin{itemize}
    \item[] 2D
    \begin{itemize}
        \item\( p+ip \) SC on $\sigma_1$, $\sigma_2$ is obstructed by chiral anomaly at $\tau_1$, $\tau_2$ respectively.
    \end{itemize}
    \item[] 1D
    \begin{itemize}
        \item Majorana chain on one of $\tau_2$ or $\tau_3$ leaves odd number of Majorana modes at $\mu$
    \end{itemize}
\end{itemize}

\textbf{Obstruction-free states}:
\begin{itemize}
\item[] 1D ($\bzt^2$) \begin{enumerate}
    \item Majorana chain on $\tau_1$(I)
    \item Majorana chain on $\tau_2$ and $\tau_3$(I)
\end{enumerate} 

\item[] 2D: No obstruction-free states ($\bbz_1$)
\end{itemize}

\textbf{Trivializations}:
\begin{itemize}
    \item Open surface decoration $\Rightarrow$ Simultaneous decoration of Majorana chains on $\tau_1,\ \tau_2$, and $\tau_3$. This reduces the 1D classification to $\bzt$.
    \item Majorana bubble on $\sigma_1$ $\Rightarrow$ Majorana chain on $\tau_1$. This further reduces the 1D classification to $\bbz_1$ (trivial). 
\end{itemize}

\textbf{Final classification:}
\begin{itemize}
    \item[] $E_{0,dis}^{1D} = \bbz_1$
    \item[] $E_{0,dis}^{2D} = \bbz_1$
    \item[]  $\mathcal{G}_{0,dis} = E_{0,dis}^{1D} \times E_{0,dis}^{2D} = \bbz_1$
\end{itemize}

\subsubsection*{Disordered Spin-1/2}
\textbf{Block state decorations}:
\begin{itemize}
    \item[] 1D
    \begin{itemize}
        \item $\tau_1$: Majorana chain
        \item $\tau_2,\ \tau_3$: \placeholder
    \end{itemize}
    \item[] 2D
    \begin{itemize}
        \item $\sigma_1,\ \sigma_2$:\( p+ip \) SC
    \end{itemize}
\end{itemize}

\textbf{Obstructions}
\begin{itemize}
    \item[] 2D
    \begin{itemize}
        \item\( p+ip \) SC on $\sigma_1$, $\sigma_2$ is obstructed by chiral anomaly at $\tau_1$, $\tau_2$ respectively
    \end{itemize}
    \item[] 1D
    \begin{itemize}
        \item \placeholder on one of $\tau_2$ or $\tau_3$ leaves odd number of \placeholdermodes at $\mu$
    \end{itemize}
\end{itemize}

\textbf{Obstruction-free states}:
\begin{itemize}
\item[] 1D ($\bzt^3$) \begin{enumerate}
    \item Majorana chain on $\tau_1$ (E)
    \item \placeholder on $\tau_2$ and $\tau_3$ (I)
\end{enumerate} 
\item[] 2D: No obstruction-free states ($\bbz_1$)
\end{itemize}

\textbf{Trivializations}:
\begin{itemize}
    \item Majorana bubble on $\sigma_1$ $\Rightarrow$ Majorana chain on $\tau_1$. This reduces the 1D classification to $\bzt$
\end{itemize}

\textbf{Final classification:}
\begin{itemize}
    \item[] $E_{1/2,dis}^{1D} = \bzt(I)$
    \item[] $E_{1/2,dis}^{2D} = \bbz_1$
    \item[]  $\mathcal{G}_{1/2,dis} = E_{1/2,dis}^{1D} \times E_{1/2,dis}^{2D} = \bzt(I)$
\end{itemize}

\subsection{$C_{3v}$}
\subsubsection*{Cell Decomposition}
\begin{figure}[!htbp]
    \centering
    \includegraphics[width=0.9\linewidth]{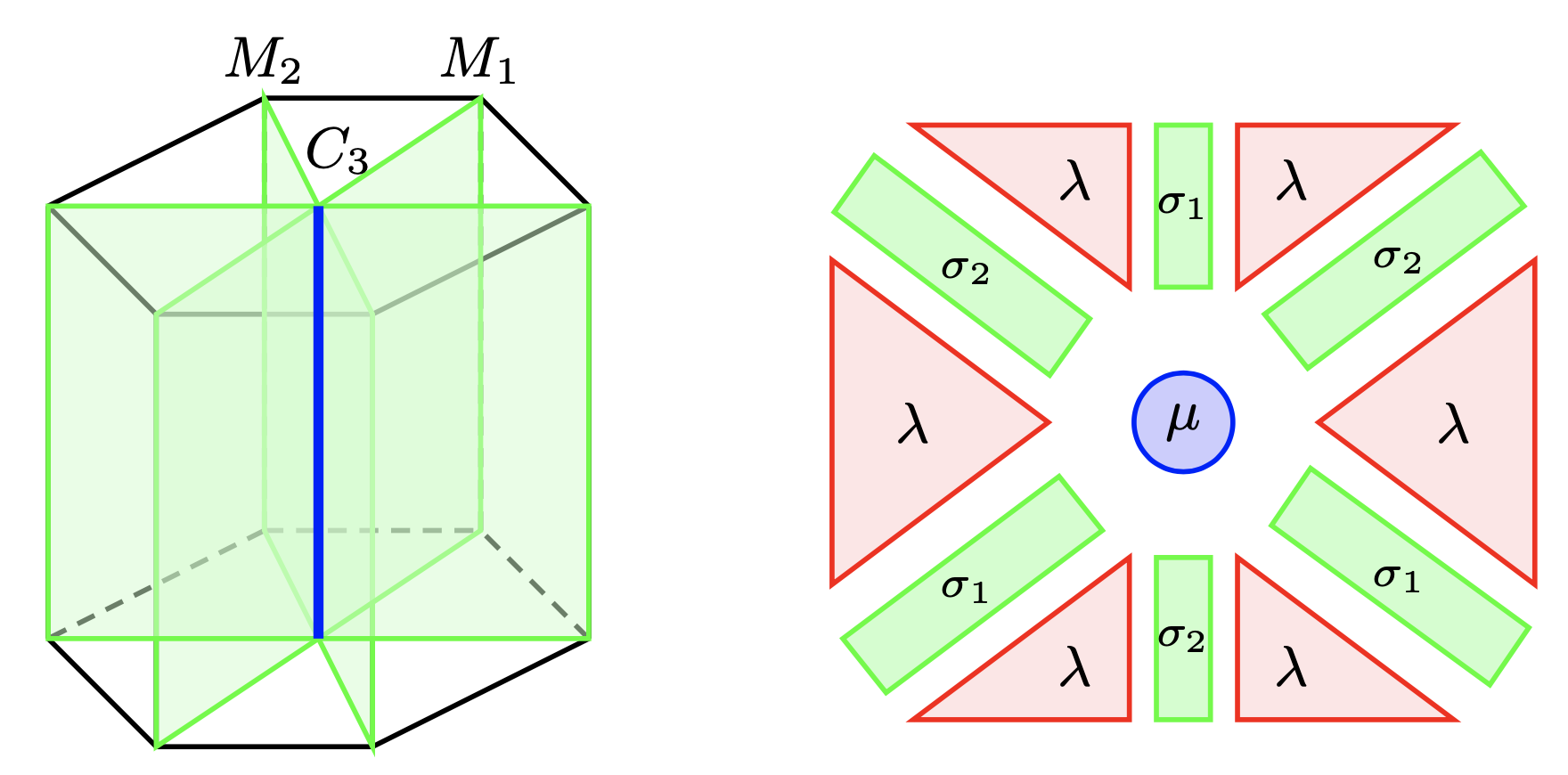}
    \caption{$C_{3v}$ lattice}
\end{figure}
\textbf{Blocks and onsite symmetries}:
\begin{itemize}
    \item 2D: $G_{\sigma_1},\ G_{\sigma_2}=\bzt$
    \item 1D: $G_{\tau} = \bbz_3 \rtimes \bzt$ 
\end{itemize}
\subsubsection*{Decohered Spinless}
\textbf{Block state decorations}:
\begin{itemize}
    \item[] 1D
    \begin{itemize}
        \item $\tau$: Majorana chain, $\bzt$ fSPT
    \end{itemize}
    \item[] 2D
    \begin{itemize}
        \item $\sigma_1,\ \sigma_2$:\( p+ip \) SC, fLG
    \end{itemize}
\end{itemize}

\textbf{Obstructions}
\begin{itemize}
    \item[] 2D
    \begin{itemize}
        \item\( p+ip \) SC on $\sigma_1$ or $\sigma_2$ is obstructed by chiral anomaly at $\tau$
        \item fLG on one of $\sigma_1$ or $\sigma_2$ results in fLG state at $\tau$ and is hence obstructed. Similarly, $n=2$ fLG and $n=3$ fLG on one of $\sigma_1$ or $\sigma_2$ is obstructed. 
    \end{itemize}
\end{itemize}

\textbf{Obstruction-free states}:
\begin{itemize}

\item[] 1D: ($\bzt^2$) \begin{enumerate}
    \item Majorana chain on $\tau$
    \item $\bzt$ fSPT on $\tau$
\end{enumerate}

\item[] 2D: ($\bbz \times \bbz_4$) \begin{enumerate}
    \item Simultaneous decoration of $p\pm ip$ SC on $\sigma_1$ and $\sigma_2$ with opposite chiralities. (E)
    \item Simultaneous decoration of (up to $n=4$) fLG on $\sigma_1$ and $\sigma_2$ (E)
\end{enumerate}
\end{itemize}

\textbf{Trivializations}:
\begin{itemize}
    \item Majorana chain on one of $\sigma_1$ or $\sigma_2$ $\Rightarrow$ Majorana chain on $\tau$. Therefore, the 1D classification reduces to $\bzt$.
    \item Majorana chain on $\sigma_1$ and $\sigma_2$ $\Rightarrow$ $\bzt$ fSPT on $\tau$. Therefore the 1D classification further reduces to $\bbz_1$ (trivial).
    \item\( p+ip \) SC bubble on $\lambda$ $\Rightarrow$ Two layers of $p\pm ip$ SC with opposite chiralities on $\sigma_1$ and $\sigma_2$ is equivalent to simultaneous decoration of fLG on both blocks. Therefore, the 2D classification reduces to $\bzt\rtimes \bbz_4 = \bbz_8$. 
\end{itemize}

\textbf{Final classification:}
\begin{itemize}
    \item[] $E_{0,dec}^{1D} = \bbz_1$
    \item[] $E_{0,dec}^{2D} = \bbz_8(E)$
    \item[]  $\mathcal{G}_{0,dec} =  E_{0,dec}^{1D} \times E_{0,dec}^{2D} = \bbz_8(E)$
\end{itemize}

\subsubsection*{Decohered Spin-1/2}
\textbf{Block state decorations}:
\begin{itemize}
    \item[] 1D
    \begin{itemize}
        \item $\tau$: $\bbz_4^f$ ASPT 
    \end{itemize}
    \item[] This state is \textbf{obstruction-free}
    \item[] 2D
    \begin{itemize}
        \item $\sigma_1,\ \sigma_2$: No nontrivial block state
    \end{itemize}
\end{itemize}

\textbf{Trivializations}
\begin{itemize}
    \item $\bbz_4^f$ ASPT on $\sigma_1$ or $\sigma_2$ $\Rightarrow$ $\bbz^4$ ASPT on $\tau$. Therefore the 1D classification reduces to $\bbz_1$ (trivial).
\end{itemize}

\textbf{Final classification:}
\begin{itemize}
    \item[] $E_{1/2,dec}^{1D} = \bbz_1$
    \item[] $E_{1/2,dec}^{2D} = \bbz_1$
    \item[]  $\mathcal{G}_{1/2,dec} = E_{1/2,dec}^{1D} \times E_{1/2,dec}^{2D} = \bbz_1$
\end{itemize}

\subsubsection*{Disordered Spinless}
\textbf{Block state decorations}:
\begin{itemize}
    \item[] 1D
    \begin{itemize}
        \item $\tau$: Majorana chain
    \end{itemize}
    \item[] 2D
    \begin{itemize}
        \item $\sigma_1,\ \sigma_2$:\( p+ip \) SC, fLG
    \end{itemize}
\end{itemize}

\textbf{Obstructions}
\begin{itemize}
    \item[] 2D
    \begin{itemize}
        \item\( p+ip \) SC on $\sigma_1$ or $\sigma_2$ is obstructed by chiral anomaly at $\tau$
        \item fLG on one of $\sigma_1$ or $\sigma_2$ results in fLG state at $\tau$ and is hence obstructed.  
    \end{itemize}
\end{itemize}

\textbf{Obstruction-free states}:
\begin{itemize}

\item[] 1D: ($\bzt$) \begin{enumerate}
    \item Majorana chain on $\tau$
\end{enumerate}

\item[] 2D: ($\bbz \times \bzt$) \begin{enumerate}
    \item Simultaneous decoration of $p\pm ip$ SC on $\sigma_1$ and $\sigma_2$ with opposite chiralities (E).
    \item Simultaneous decoration of fLG on $\sigma_1$ and $\sigma_2$  (E)
\end{enumerate}
\end{itemize}

\textbf{Trivializations}:
\begin{itemize}
    \item Majorana chain on one of $\sigma_1$ or $\sigma_2$ $\Rightarrow$ Majorana chain on $\tau$. Therefore, the 1D classification reduces to $\bbz_1$ (trivial).
    \item\( p+ip \) SC bubble on $\lambda$ $\Rightarrow$ Two layers of $p\pm ip$ SC with opposite chiralities on $\sigma_1$ and $\sigma_2$ is equivalent to simultaneous decoration of fLG on both blocks. Therefore, the 2D classification reduces to $\bzt\rtimes \bzt = \bbz_4$. 
\end{itemize}

\textbf{Final classification:}
\begin{itemize}
    \item[] $E_{0,dis}^{1D} = \bbz_1$
    \item[] $E_{0,dis}^{2D} = \bbz_4(E)$
    \item[]  $\mathcal{G}_{0,dis} = E_{0,dis}^{1D} \times E_{0,dis}^{2D} = \bbz_4(E)$
\end{itemize}

\subsubsection*{Disordered Spin-1/2}
\textbf{Block state decorations}:
\begin{itemize}
    \item[] 1D
    \begin{itemize}
        \item $\tau$: Majorana chain
    \end{itemize}
    \item[] 2D
    \begin{itemize}
        \item $\sigma_1,\ \sigma_2$: 2D $\bbz_4^f$ ASPT
    \end{itemize}
\end{itemize}

\textbf{Obstructions}
\begin{itemize}
    \item[] 2D
    \begin{itemize}
    \item 2D $\bbz_4^f$ ASPT on one of $\sigma_1$ or $\sigma_2$ results in 2D $\bbz_4^f$ ASPT edge state at $\tau$ and is hence obstructed. 
    \end{itemize}
\end{itemize}

\textbf{Obstruction-free states}:
\begin{itemize}

\item[] 1D: ($\bzt$) \begin{enumerate}
    \item \placeholder on $\tau$ (I)
\end{enumerate}

\item[] 2D: ($\bzt$) \begin{enumerate}
    \item Simultaneous decoration of 2D $\bbz_4^f$ ASPT on $\sigma_1$ and $\sigma_2$  (I)
\end{enumerate}
\end{itemize}

\textbf{Trivializations}:
\begin{itemize}
    \item \bubbleplaceholder on one of $\sigma_1$ or $\sigma_2$ $\Rightarrow$ \placeholder on $\tau$. Therefore, the 1D classification reduces to $\bbz_1$ (trivial). 
\end{itemize}

\textbf{Final classification:}
\begin{itemize}
    \item[] $E_{1/2,dis}^{1D} = \bbz_1$
    \item[] $E_{1/2,dis}^{2D} = \bzt(I)$
    \item[]  $\mathcal{G}_{1/2,dis} = E_{1/2,dis}^{1D} \times E_{1/2,dis}^{2D} = \bzt(I)$
\end{itemize}

\subsection{$D_{3d}$}
\subsubsection*{Cell Decomposition}
\begin{figure}[!htbp]
    \centering
    \includegraphics[width=0.9\linewidth]{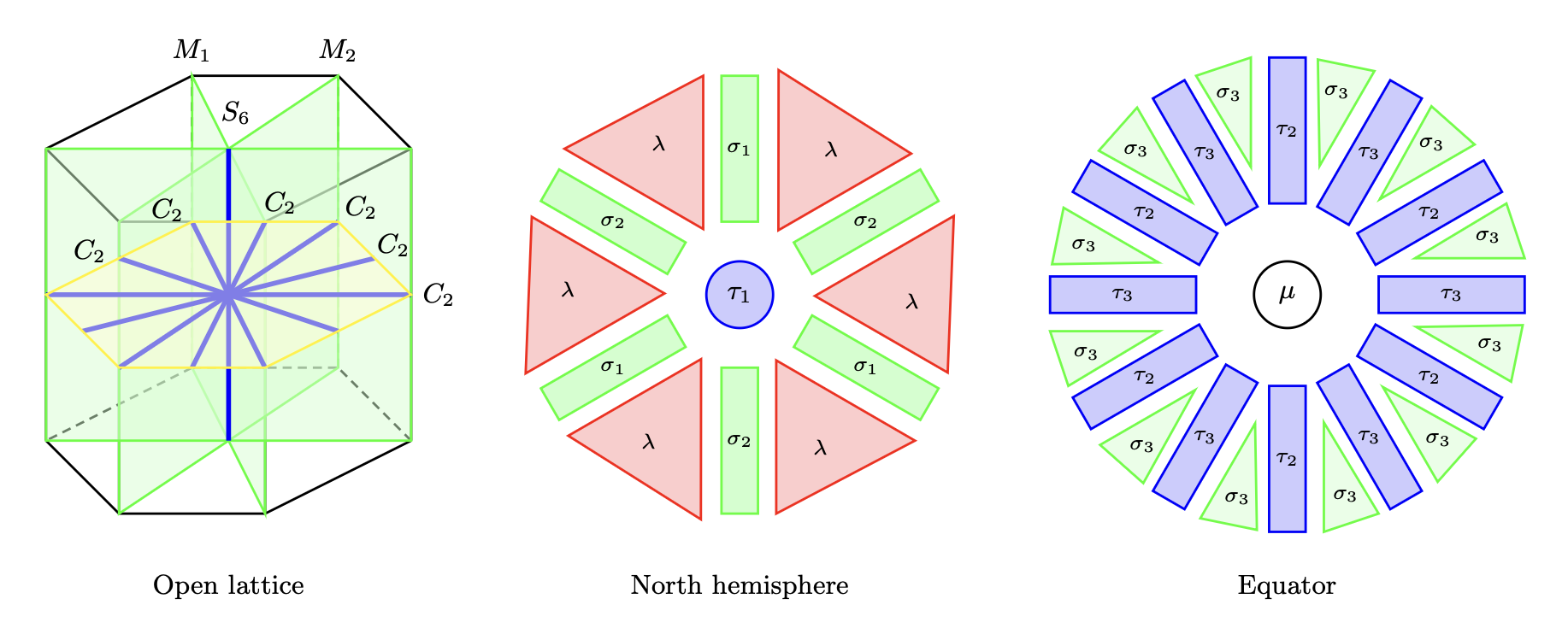}
    \caption{$D_{3d}$ lattice. On the southern hemisphere, the positions of $\sigma_1$ and $\sigma_2$ are exchanged by the $S_6$ rotoreflection symmetry. The $\tau_2$ blocks have $\bzt$ onsite symmetry as they are axes of $C_2$ rotation, while the $\tau_3$ blocks have $\bzt$ onsite symmetry since they lie on mirror planes $M_1/M_2$.}
\end{figure}
\textbf{Blocks and onsite symmetries}:
\begin{itemize}
    \item $G_{\sigma_1},\ G_{\sigma_2} = \bzt,\ G_{\sigma_3} = I$
    \item $G_{\tau_1} = \bbz_3 \rtimes \bzt,\ G_{\tau_2},\ G_{\tau_3} = \bzt$
    \item $G_{\mu} = \bbz_6\rtimes \bzt$
\end{itemize}

\subsubsection*{Decohered Spinless}
\textbf{Block state decorations}:
\begin{itemize}
    \item[] 0D
    \begin{itemize}
        \item $\mu$: Odd fermion
    \end{itemize}
    \item[] 1D
    \begin{itemize}
        \item $\tau_1$: Majorana chain, $\bbz_3\rtimes\bzt$ fSPT $\cong$ $\bzt$ fSPT
        \item $\tau_2,\ \tau_3$: Majorana chain, $\bzt$ fSPT
    \end{itemize}
    \item[] 2D
    \begin{itemize}
        \item $\sigma_1,\ \sigma_2$:\( p+ip \) SC, fLG
        \item $\sigma_3$:\( p+ip \) SC
    \end{itemize}
\end{itemize}

\textbf{Obstructions}
\begin{itemize}
    \item[] 1D
    \begin{itemize}
        \item Majorana chain on $\tau_1,\ \tau_2$, or $\tau_3$ is obstructed at $\mu$. Among these phases, only simultaneous decoration of Majorana chains on $\tau_1$ and $\tau_2$ is obstruction-free. 
    \end{itemize}
    \item[] 2D
    \begin{itemize}
        \item\( p+ip \) SC on $\sigma_1$ or $\sigma_2$ is obstructed by chiral anomaly at $\tau_1$
        \item Decoration of $p\pm ip$-SC with opposite chiralities on $\sigma_1$ and $\sigma_2$ is obstructed at $\tau_1$ (anomaly indicator $\nu_{M_1}=1/4$)
        \item\( p+ip \) SC on $\sigma_3$ is obstructed by chiral anomaly at $\tau_3$
        \item If $\sigma_1$ and $\sigma_2$ have non-identical fLG decorations, they are obstructed at $\tau_2$.
        \item Simultaneous decoration of fLG on $\sigma_1$ and $\sigma_2$ is obstructed at $\tau_1$ ($\nu=1/2$).
    \end{itemize}
\end{itemize}

\textbf{Obstruction-free states}:
\begin{itemize}
\item[] 0D state ($\bzt$) is obstruction-free (E)

\item[] 1D: ($\bzt^4$) \begin{enumerate}
    \item Simultaneous decoration of Majorana chains on $\tau_1$ and $\tau_3$ (E)
    \item $\bzt$ fSPT on $\tau_1$, $\tau_2$, $\tau_3$ (I)
\end{enumerate}

\item[] 2D: ($\bzt$) \begin{enumerate}
    \item $n=2$ fLG on $\sigma_1$ and $\sigma_2$ (I)
\end{enumerate}
\end{itemize}

\textbf{Trivializations}:
\begin{itemize}
    \item Majorana bubble on $\sigma_1$ $\Rightarrow$ Simultaneous decoration of Majorana chain on $\tau_1$ and $\tau_3$. Therefore, the 1D classification reduces to $\bzt^3$.
    \item Majorana bubbles on $\sigma_1$ and $\sigma_2$ $\Rightarrow$ Simultaneous decoration of $\bzt$ fSPT on $\tau_1$ and $\tau_2$. Therefore, the 1D classification further reduces to $\bzt^2$.
    \item Majorana bubble on $\sigma_3$ $\Rightarrow$ Simultaneous decoration of $\bzt$ fSPT on $\tau_2$ and $\tau_3$. Therefore, the 1D classification finally reduces to $\bzt$.
\end{itemize}

\textbf{Final classification:}
\begin{itemize}
    \item[] $E_{0,dec}^{0D} = \bzt(E)$
    \item[] $E_{0,dec}^{1D} = \bzt(I)$
    \item[] $E_{0,dec}^{2D} = \bzt(I)$
    \item[] Non-trivial stacking (derived from spectral sequence) $\Rightarrow$ $\mathcal{G}_{0,dec} =  E_{0,dec}^{0D} \rtimes E_{0,dec}^{1D} \rtimes E_{0,dec}^{2D} = \bzt(I)\times\bbz_4(I)$
\end{itemize}

\subsubsection*{Decohered Spin-1/2}
\textbf{Block state decorations}:
\begin{itemize}
    \item[] 0D
    \begin{itemize}
        \item $\mu$: Odd fermion
    \end{itemize}
    \item[] 1D
    \begin{itemize}
        \item $\tau_1$: $\bbz_3\rtimes\bzt\rtimes\bzt^f$ ASPT $\cong$ $\bbz_4^f$ ASPT
        \item $\tau_2,\ \tau_3$: $\bbz_4^f$ ASPT
    \end{itemize}
    \item[] 2D
    \begin{itemize}
        \item $\sigma_1,\sigma_2$: No nontrivial block state
        \item $\sigma_3$:\( p+ip \) SC
    \end{itemize}
\end{itemize}

\textbf{Obstructions}
\begin{itemize}
    \item[] 2D
    \begin{itemize}
        \item\( p+ip \) SC on $\sigma_3$ is obstructed by chiral anomaly at $\tau_3$
    \end{itemize}
\end{itemize}

\textbf{Obstruction-free states}:
\begin{itemize}
\item[] 0D state ($\bzt$) is obstruction-free (I)
\item[] 1D: ($\bzt^3$) \begin{enumerate}
    \item $\bbz_4^f$ ASPT on $\tau_1$, $\tau_2$, or $\tau_3$ (I)
\end{enumerate} 

\item[] 2D: No obstruction-free states ($\bbz_1$). 
\end{itemize}

\textbf{Trivializations:}
\begin{itemize}
    \item $\bbz_4^f$ ASPT on $\sigma_1$ $\Rightarrow$ Simultaneous decoration of $\bbz_4^f$ ASPT on $\tau_1$ and $\tau_2$. Therefore, the 1D classification reduces to $\bzt^2$.
    \item Fermion bubble on $\tau_1$ $\Rightarrow$ Odd fermion on $\mu$. Therefore, the 0D classification reduces to $\bbz_1$ (trivial).
\end{itemize}

\textbf{Final classification:}
\begin{itemize}
    \item[] $E_{1/2,dec}^{0D} = \bbz_1$
    \item[] $E_{1/2,dec}^{1D} = \bzt^2(I)$
    \item[] $E_{1/2,dec}^{2D} = \bbz_1$
    \item[]  $\mathcal{G}_{1/2,dec} = E_{1/2,dec}^{0D}\times E_{1/2,dec}^{1D} \times E_{1/2,dec}^{2D} = \bzt^2(I)$
\end{itemize}

\subsubsection*{Disordered Spinless}
\textbf{Block state decorations}:
\begin{itemize}
    \item[] 1D
    \begin{itemize}
        \item $\tau_1,\ \tau_2,\ \tau_3$: Majorana chain
    \end{itemize}
    \item[] 2D
    \begin{itemize}
        \item $\sigma_1,\ \sigma_2$:\( p+ip \) SC, fLG
        \item $\sigma_3$:\( p+ip \) SC
    \end{itemize}
\end{itemize}

\textbf{Obstructions}
\begin{itemize}
    \item[] 2D
    \begin{itemize}     
        \item\( p+ip \) SC on $\sigma_1$ or $\sigma_2$ is obstructed by chiral anomaly at $\tau_1$
        \item Decoration of $p\pm ip$-SC with opposite chiralities on $\sigma_1$ and $\sigma_2$ is obstructed at $\tau_1$ (anomaly indicator $\nu_{M_1}=1/4$)
        \item\( p+ip \) SC on $\sigma_3$ is obstructed by chiral anomaly at $\tau_3$
        \item If $\sigma_1$ and $\sigma_2$ have non-identical fLG decorations, they are obstructed at $\tau_2$.
    \end{itemize}
\end{itemize}

\textbf{Obstruction-free states}:
\begin{itemize}
\item[] 1D ($\bzt^3$) \begin{enumerate}
    \item Majorana chain on $\tau_1$, $\tau_2$, or $\tau_3$ (I)
\end{enumerate} 

\item[] 2D ($\bzt$) \begin{enumerate}
    \item fLG on $\sigma_1$ and $\sigma_2$ (I)
\end{enumerate}
\end{itemize}

\textbf{Trivializations}:
\begin{itemize}
    \item Majorana bubble on $\sigma_1$ $\Rightarrow$ Majorana chains on $\tau_1$ and $\tau_2$. Therefore, the 1D classification reduces to $\bzt^2$.
\end{itemize}

\textbf{Final classification:}
\begin{itemize}
    \item[] $E_{0,dis}^{1D} = \bzt^2(I)$
    \item[] $E_{0,dis}^{2D} = \bzt(I)$
    \item[]  $\mathcal{G}_{0,dis} = E_{0,dis}^{1D} \times E_{0,dis}^{2D} = \bzt^3(I)$
\end{itemize}

\subsubsection*{Disordered Spin-1/2}
\textbf{Block state decorations}:
\begin{itemize}
    \item[] 1D
    \begin{itemize}
        \item $\tau_1,\ \tau_2,\ \tau_3$: \placeholder
    \end{itemize}
    \item[] 2D
    \begin{itemize}
        \item $\sigma_1,\ \sigma_2$: 2D $\bbz_4^f$ ASPT
        \item $\sigma_3$:\( p+ip \) SC
    \end{itemize}
\end{itemize}

\textbf{Obstructions}
\begin{itemize}
    \item[] 2D
    \begin{itemize}     
        \item p+ip-SC on $\sigma_3$ is obstructed by chiral anomaly at $\tau_3$
        \item  If $\sigma_1$ and $\sigma_2$ have non-identical ASPT decorations, they are obstructed at $\tau_2$.
    \end{itemize}
\end{itemize}

\textbf{Obstruction-free states}:
\begin{itemize}
\item[] 1D ($\bzt^3$) \begin{enumerate}
    \item \placeholder on $\tau_1$, $\tau_2$, or $\tau_3$ (I)
\end{enumerate} 

\item[] 2D ($\bzt$) \begin{enumerate}
    \item 2D $\bbz_4^f$ ASPT on $\sigma_1$ and $\sigma_2$ (I)
\end{enumerate}
\end{itemize}

\textbf{Trivializations}:
\begin{itemize}
    \item \bubbleplaceholder on $\sigma_1$ $\Rightarrow$ \placeholder on $\tau_1$ and $\tau_2$. Therefore, the 1D classification reduces to $\bzt^2$.
\end{itemize}

\textbf{Final classification:}
\begin{itemize}
    \item[] $E_{1/2,dis}^{1D} = \bzt^2(I)$
    \item[] $E_{1/2,dis}^{2D} = \bzt(I)$
    \item[]  $\mathcal{G}_{1/2,dis} = E_{1/2,dis}^{1D} \times E_{1/2,dis}^{2D} = \bzt^3(I)$
\end{itemize}

\subsection{$C_6$}
\subsubsection*{Cell Decomposition}
\begin{figure}[!htbp]
    \centering
    \includegraphics[width=0.9\linewidth]{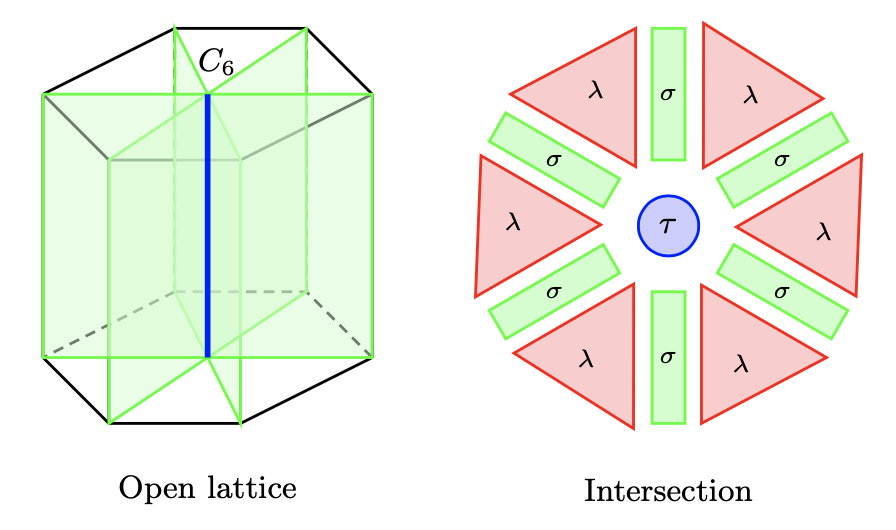}
    \caption{$C_6$ lattice}
\end{figure}
\textbf{Blocks and onsite symmetries}:
\begin{itemize}
    \item $G_{\sigma} = I$
    \item $G_{\tau} = \bbz_6$
\end{itemize}
\subsubsection*{Decohered Spinless}
\textbf{Block state decorations}:
\begin{itemize}
    \item[] 1D
    \begin{itemize}
        \item $\tau$: Majorana chain, $\bbz_6$ fSPT
    \end{itemize}
    \item[] 2D
    \begin{itemize}
        \item $\sigma$:\( p+ip \) SC
    \end{itemize}
\end{itemize}

\textbf{Obstructions}
\begin{itemize}
    \item[] 2D
    \begin{itemize}
        \item\( p+ip \) SC on $\sigma$ is obstructed by chiral anomaly at $\tau$
    \end{itemize}
\end{itemize}

\textbf{Obstruction-free states}:
\begin{itemize}

\item[] 1D: ($\bzt^2$) \begin{enumerate}
    \item Majorana chain on $\tau$
    \item $\bbz_6$ fSPT on $\tau$
\end{enumerate}

\item[] 2D: No obstruction-free states ($\bbz_1$)
\end{itemize}

\textbf{Trivializations}:
\begin{itemize}
    \item Majorana bubble on $\sigma$ $\Rightarrow$ $\bbz_6$ ASPT on $\tau$. Therefore, the 1D classification reduces to $\bzt$
    \item Open surface decorations trivializes Majorana chain on $\tau$. Therefore, the 1D classification further reduces to $\bbz_1$ (trivial).
\end{itemize}

\textbf{Final classification:}
\begin{itemize}
    \item[] $E_{0,dec}^{1D} = \bbz_1$
    \item[] $E_{0,dec}^{2D} = \bbz_1$
    \item[]  $\mathcal{G}_{0,dec} =  E_{0,dec}^{1D} \times E_{0,dec}^{2D} = \bbz_1$
\end{itemize}

\subsubsection*{Decohered Spin-1/2}
\textbf{Block state decorations}:
\begin{itemize}
    \item[] 1D
    \begin{itemize}
        \item $\tau$: $\bbz_6\times_{\omega_2^f} \bzt^f$ ASPT
    \end{itemize}
    \item[] 2D
    \begin{itemize}
        \item $\sigma$:\( p+ip \) SC
    \end{itemize}
\end{itemize}

\textbf{Obstructions}
\begin{itemize}
    \item[] 2D
    \begin{itemize}
        \item\( p+ip \) SC on $\sigma$ is obstructed by chiral anomaly at $\tau$
    \end{itemize}
\end{itemize}

\textbf{Obstruction-free states}:
\begin{itemize}

\item[] 1D: ($\bzt$) \begin{enumerate}
   \item $\bbz_6\times_{\omega_2^f} \bzt^f$ ASPT on $\tau$ (I)
\end{enumerate} 

This state is also \textbf{trivialization-free}.

\item[] 2D: No obstruction-free states ($\bbz_1$)
\end{itemize}

\textbf{Final classification:}
\begin{itemize}
    \item[] $E_{1/2,dec}^{1D} = \bzt(I)$
    \item[] $E_{1/2,dec}^{2D} = \bbz_1$
    \item[]  $\mathcal{G}_{1/2,dec} = E_{1/2,dec}^{1D} \times E_{1/2,dec}^{2D} = \bzt(I)$
\end{itemize}

\subsubsection*{Disordered Spinless}
\textbf{Block state decorations}:
\begin{itemize}
    \item[] 1D
    \begin{itemize}
        \item $\tau$: Majorana chain
    \end{itemize}
    \item[] 2D
    \begin{itemize}
        \item $\sigma$:\( p+ip \) SC
    \end{itemize}
\end{itemize}

\textbf{Obstructions}
\begin{itemize}
    \item[] 2D
    \begin{itemize}
        \item\( p+ip \) SC on $\sigma$ is obstructed by chiral anomaly at $\tau$
    \end{itemize}
\end{itemize}

\textbf{Obstruction-free states}:
\begin{itemize}

\item[] 1D: ($\bzt$) \begin{enumerate}
    \item Majorana chain on $\tau$
\end{enumerate}

\item[] 2D: No obstruction-free states ($\bbz_1$)
\end{itemize}

\textbf{Trivializations}:
\begin{itemize}
    \item Open surface decorations trivializes Majorana chain on $\tau$. Therefore, the 1D classification reduces to $\bbz_1$ (trivial).
\end{itemize}

\textbf{Final classification:}
\begin{itemize}
    \item[] $E_{0,dis}^{1D} = \bbz_1$
    \item[] $E_{0,dis}^{2D} = \bbz_1$
    \item[]  $\mathcal{G}_{0,dis} = E_{0,dis}^{1D} \times E_{0,dis}^{2D} = \bbz_1$
\end{itemize}

\subsubsection*{Disordered Spin-1/2}
\textbf{Block state decorations}:
\begin{itemize}
    \item[] 1D
    \begin{itemize}
        \item $\tau$: \placeholder
    \end{itemize}
    \item[] 2D
    \begin{itemize}
        \item $\sigma$:\( p+ip \) SC
    \end{itemize}
\end{itemize}

\textbf{Obstructions}
\begin{itemize}
    \item[] 2D
    \begin{itemize}
        \item\( p+ip \) SC on $\sigma$ is obstructed by chiral anomaly at $\tau$
    \end{itemize}
\end{itemize}

\textbf{Obstruction-free states}:
\begin{itemize}

\item[] 1D: ($\bzt$) \begin{enumerate}
    \item \placeholder on $\tau$(I)
\end{enumerate} 
\item[] This state is also \textbf{trivialization-free}.

\item[] 2D: No obstruction-free states ($\bbz_1$)
\end{itemize}

\textbf{Final classification:}
\begin{itemize}
    \item[] $E_{1/2,dis}^{1D} = \bzt(I)$
    \item[] $E_{1/2,dis}^{2D} = \bbz_1$
    \item[]  $\mathcal{G}_{1/2,dis} = E_{1/2,dis}^{1D} \times E_{1/2,dis}^{2D} = \bzt(I)$
\end{itemize}

\subsection{$C_{3h}$}
\subsubsection*{Cell Decomposition}
\begin{figure}[!htbp]
    \centering
    \includegraphics[width=0.9\linewidth]{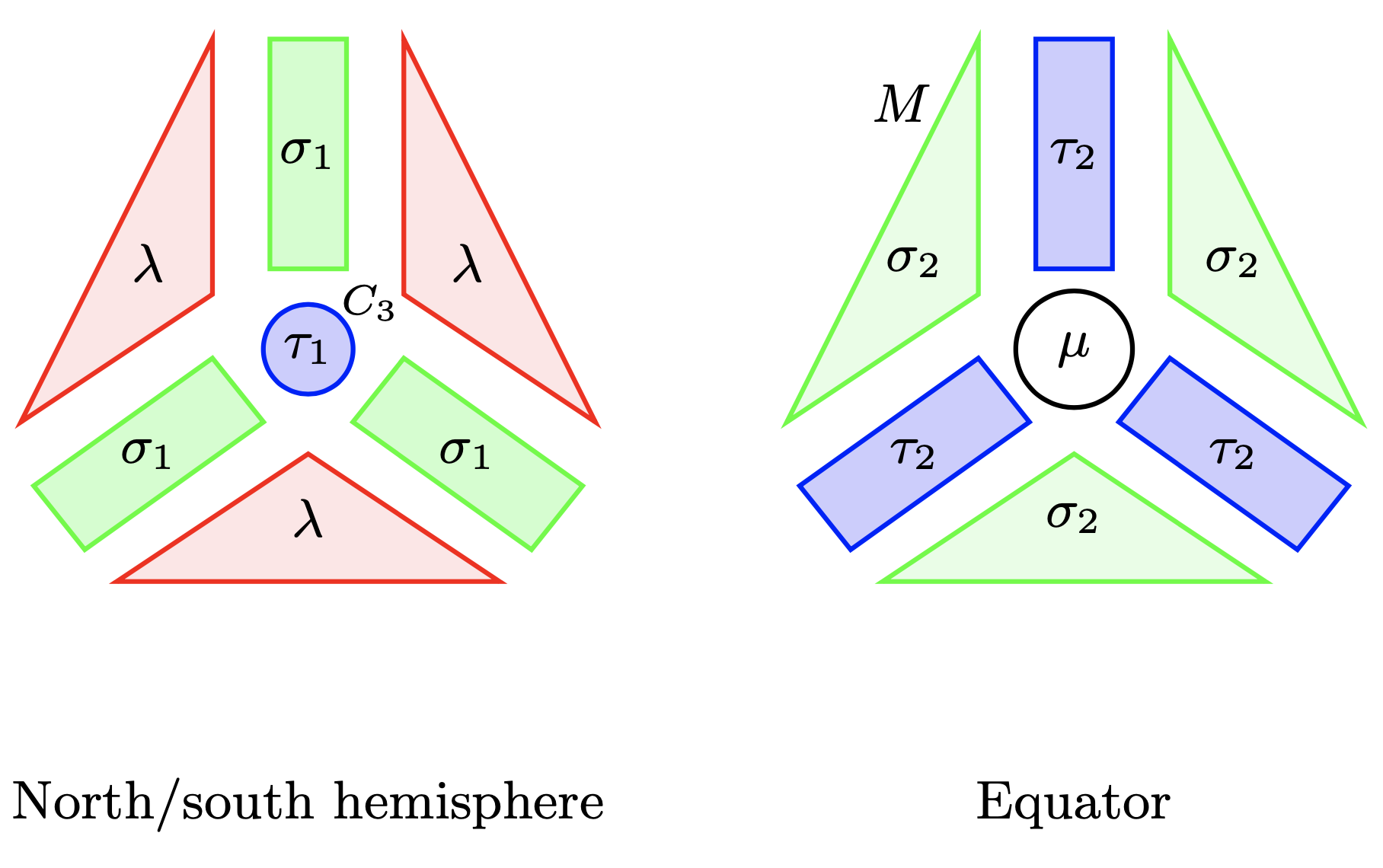}
    \caption{$C_{3h}$ lattice}
\end{figure}
\textbf{Blocks and onsite symmetries}:
\begin{itemize}
    \item 2D: $G_{\sigma_1} = I,\ G_{\sigma_2} = \bzt$
    \item 1D: $G_{\tau_1} = \bbz_3,\ G_{\tau_2} = \bzt$
    \item 0D: $G_{\mu} = \bbz_3\times\bzt$
\end{itemize}
\subsubsection*{Decohered Spinless}
\textbf{Block state decorations}:
\begin{itemize}
    \item[] 0D
    \begin{itemize}
        \item $\mu$: Odd fermion
    \end{itemize}
    \item[] 1D
    \begin{itemize}
        \item $\tau_1$: Majorana chain
        \item $\tau_2$: Majorana chain, $\bzt$ fSPT
    \end{itemize}
    \item[] 2D
    \begin{itemize}
        \item $\sigma_1$:\( p+ip \) SC
        \item $\sigma_2$:\( p+ip \) SC, fLG
    \end{itemize}
\end{itemize}

\textbf{Obstructions}
\begin{itemize}
    \item[] 1D
    \begin{itemize}
        \item Majorana chain on $\tau_1$ is obstructed at $\mu$ by 2-fold rotational symmetry.
        \item Majorana chain on $\tau_2$ is obstructed at $\mu$ (odd number of Majorana modes).
        \item Simultaneous decoration of the above phases is also obstructed.
    \end{itemize}
    \item[] 2D
    \begin{itemize}
        \item\( p+ip \) SC on $\sigma_1$ is obstructed by chiral anomaly at $\tau_1$
        \item\( p+ip \) SC on $\sigma_2$ is incompatible with spinless 3-fold rotational symmetry on equator.
    \end{itemize}
\end{itemize}

\textbf{Obstruction-free states}:
\begin{itemize}
\item[] 0D state ($\bzt$) is obstruction-free(E)

\item[] 1D: ($\bzt$) \begin{enumerate}
    \item $\bzt$ fSPT on $\tau_2$ (E)
\end{enumerate}

\item[] 2D: ($\bbz_4$) \begin{enumerate}
    \item fLG on $\sigma_2$ (E)
\end{enumerate} 
\end{itemize}

\textbf{Trivializations}:
\begin{itemize}
    \item Majorana bubble on $\sigma_1$ $\Rightarrow$ $\bzt$ fSPT on $\tau_2$. Therefore, the 1D classification reduces to $\bbz_1$ (trivial).
    \item Majorana bubble on $\sigma_2$ $\Rightarrow$ Odd fermion on $\mu$. Therefore, the 0D classification reduces to $\bbz_1$ (trivial). 
\end{itemize}

\textbf{Final classification:}
\begin{itemize}
    \item[] $E_{0,dec}^{0D} = \bbz_1$
    \item[] $E_{0,dec}^{1D} = \bbz_1$
    \item[] $E_{0,dec}^{2D} = \bbz_4(E)$
    \item[]  $\mathcal{G}_{0,dec} =  E_{0,dec}^{1D} \times E_{0,dec}^{2D} = \bbz_4(E)$
\end{itemize}

\subsubsection*{Decohered Spin-1/2}
\textbf{Block state decorations}:
\begin{itemize}
    \item[] 0D
    \begin{itemize}
        \item $\mu$: Odd fermion
    \end{itemize}
    \item[] 1D
    \begin{itemize}
        \item $\tau_1$: Majorana chain
        \item $\tau_2$: $\bbz_4^f$ ASPT
    \end{itemize}
    \item[] 2D
    \begin{itemize}
        \item $\sigma_1$:\( p+ip \) SC
        \item $\sigma_2$: No nontrivial block state
    \end{itemize}
\end{itemize}

\textbf{Obstructions}
\begin{itemize}
    \item[] 1D 
    \begin{itemize}
        \item Majorana chain at $\tau_1$ is obstructed at $\mu$ by 2-fold rotational symmetry.
        \item $\bbz_4^f$ ASPT on $\tau_2$ is obstructed at $\mu$ (odd number of edge states)
    \end{itemize}
    \item[] 2D
    \begin{itemize}
        \item\( p+ip \) SC on $\sigma_1$ is obstructed by chiral anomaly at $\tau_1$
    \end{itemize}
\end{itemize}

\textbf{Obstruction-free states}:
\begin{itemize}
\item[] 0D state ($\bzt$) is obstruction-free
\item[] 1D: No obstruction-free states ($\bbz_1$)

\item[] 2D: No obstruction-free states ($\bbz_1$). 
\end{itemize}

\textbf{Trivializations}:
\begin{itemize}
    \item Fermion bubble on $\tau_1$ $\Rightarrow$ Odd fermion on $\mu$. Therefore, the 0D classification reduces to $\bbz_1$ (trivial).
\end{itemize}
\textbf{Final classification:}
\begin{itemize}
    \item[] $E_{1/2,dec}^{0D} = \bbz_1$
    \item[] $E_{1/2,dec}^{1D} = \bbz_1$
    \item[] $E_{1/2,dec}^{2D} = \bbz_1$
    \item[]  $\mathcal{G}_{1/2,dec} = E_{1/2,dec}^{0D} \times E_{1/2,dec}^{1D} \times E_{1/2,dec}^{2D} = \bbz_1$
\end{itemize}

\subsubsection*{Disordered Spinless}
\textbf{Block state decorations}:
\begin{itemize}
    \item[] 1D
    \begin{itemize}
        \item $\tau_1, \tau_2$: Majorana chain
    \end{itemize}
    \item[] 2D
    \begin{itemize}
        \item $\sigma_1$:\( p+ip \) SC
        \item $\sigma_2$:\( p+ip \) SC, fLG
    \end{itemize}
\end{itemize}

\textbf{Obstructions}
\begin{itemize}
    \item[] 1D
    \begin{itemize}
        \item Majorana chain on $\tau_2$ is obstructed at $\mu$ (odd number of Majorana modes)
    \end{itemize}
    \item[] 2D
    \begin{itemize}
        \item\( p+ip \) SC on $\sigma_1$ is obstructed by chiral anomaly at $\tau_1$
        \item\( p+ip \) SC on $\sigma_2$ is incompatible with spinless rotational symmetry on equator.
    \end{itemize}
\end{itemize}

\textbf{Obstruction-free states}:
\begin{itemize}
\item[] 1D ($\bzt$) \begin{enumerate}
    \item Majorana chain on $\tau_1$(I)
\end{enumerate} 

\item[] 2D ($\bzt$) \begin{enumerate}
    \item fLG on $\sigma_2$
\end{enumerate} 
\end{itemize}

\textbf{Trivializations}:
\begin{itemize}
    \item 
\end{itemize}

\textbf{Final classification:}
\begin{itemize}
    \item[] $E_{0,dis}^{1D} = \bzt^2$
    \item[] $E_{0,dis}^{2D} = \bzt$
    \item[] Non-trivial stacking $\Rightarrow$ $\mathcal{G}_{0,dis} = E_{0,dis}^{1D} \times E_{0,dis}^{2D} = \bzt^3$
\end{itemize}

\subsubsection*{Disordered Spin-1/2}
\textbf{Block state decorations}:
\begin{itemize}
    \item[] 1D
    \begin{itemize}
        \item $\tau_1$: Majorana chain
        \item $\tau_2$: \placeholder
    \end{itemize}
    \item[] 2D
    \begin{itemize}
        \item $\sigma_1$:\( p+ip \) SC
        \item $\sigma_2$: 2D $\bbz_4^f$ ASPT
    \end{itemize}
\end{itemize}

\textbf{Obstructions}
\begin{itemize}
    \item[] 1D
    \begin{itemize}
        \item \placeholder on $\tau_2$ is obstructed at $\mu$ (odd number of \placeholdermodes)
    \end{itemize}
    \item[] 2D
    \begin{itemize}
        \item\( p+ip \) SC on $\sigma_1$ is obstructed by chiral anomaly at $\tau_1$
    \end{itemize}
\end{itemize}

\textbf{Obstruction-free states}:
\begin{itemize}
\item[] 1D ($\bzt$) \begin{enumerate}
    \item Majorana chain on $\tau_1$ (I)
\end{enumerate} 

\item[] 2D ($\bzt$) \begin{enumerate}
    \item 2D $\bbz_4^f$ ASPT on $\sigma_2$ (I)
\end{enumerate} 
\end{itemize}

\textbf{Trivializations}:
\begin{itemize}
    \item Majorana bubble on $\sigma_1$ $\Rightarrow$ Majorana chain on $\tau_1$. Therefore, the 1D classification reduces to $\bbz_1$ (trivial).
\end{itemize}
\textbf{Final classification:}
\begin{itemize}
    \item[] $E_{1/2,dis}^{1D} = \bbz_1$
    \item[] $E_{1/2,dis}^{2D} = \bzt(I)$
    \item[]  $\mathcal{G}_{1/2,dis} = E_{1/2,dis}^{1D} \times E_{1/2,dis}^{2D} = \bzt(I)$
\end{itemize}

\subsection{$C_{6h}$}
\subsubsection*{Cell Decomposition}
\begin{figure}[!htbp]
    \centering
    \includegraphics[width=0.9\linewidth]{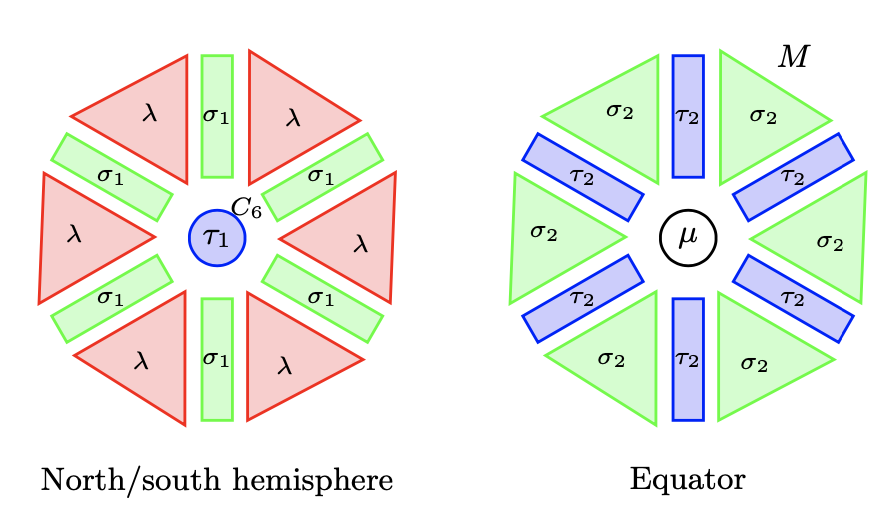}
    \caption{$C_{6h}$ lattice}
\end{figure}
\textbf{Blocks and onsite symmetries}:
\begin{itemize}
    \item $G_{\sigma_1} = I,\ G_{\sigma_2}=\bzt$
    \item $G_{\tau_1} = \bbz_6,\ G_{\tau_2} = \bzt$
    \item $G_{\mu} = \bbz_6\times\bzt$
\end{itemize}
\subsubsection*{Decohered Spinless}
\textbf{Block state decorations}:
\begin{itemize}
    \item[] 0D
    \begin{itemize}
        \item $\mu$: Odd fermion
    \end{itemize}
    \item[] 1D
    \begin{itemize}
        \item $\tau_1$: Majorana chain, $\bbz_6$ fSPT
        \item $\tau_2$: Majorana chain, $\bzt$ fSPT
    \end{itemize}
    \item[] 2D
    \begin{itemize}
        \item $\sigma_1$:\( p+ip \) SC
        \item $\sigma_2$:\( p+ip \) SC, fLG
    \end{itemize}
\end{itemize}

\textbf{Obstructions}
\begin{itemize}
    \item[] 1D
    \begin{itemize}
        \item Majorana chain on $\tau_1$ is obstructed at $\mu$ by 2-fold reflection symmetry.
        \item Majorana chain on $\tau_2$ is obstructed at $\mu$ by 6-fold rotation symmetry.
        \item Simultaneous decoration of the above phases is also obstructed.
    \end{itemize}
    \item[] 2D
    \begin{itemize}
        \item\( p+ip \) SC on $\sigma_1$ is obstructed by chiral anomaly at $\tau_1$
        \item\( p+ip \) SC on $\sigma_2$ is incompatible with spinless rotational symmetry on equator.
    \end{itemize}
\end{itemize}

\textbf{Obstruction-free states}:
\begin{itemize}
\item[] 0D state ($\bzt$) is obstruction-free (E)

\item[] 1D: ($\bzt^2$) \begin{enumerate}
    \item $\bbz_6$ fSPT on $\tau_1$ (I)
    \item $\bzt$ fSPT on $\tau_2$ (I)
\end{enumerate}

\item[] 2D: ($\bbz_4$) \begin{enumerate}
    \item fLG on $\sigma_2$
\end{enumerate} 
\end{itemize}

\textbf{Trivializations}:
\begin{itemize}
    \item Majorana bubble on $\sigma_1$ $\Rightarrow$ Simultaneous decoration of $\bbz_6$ fSPT on $\tau_1$ and $\bzt$ fSPT on $\tau_2$. Therefore, the 1D classification reduces to $\bzt$.
    \item Majorana bubble on $\sigma_2$ $\Rightarrow$ $\bzt$ fSPT on $\tau_2$ and complex fermion on $\mu$. 
\end{itemize}

\textbf{Final classification:}
\begin{itemize}
    \item[] $E_{0,dec}^{0D} = \bzt(E)$
    \item[] $E_{0,dec}^{1D} = \bbz_1$
    \item[] $E_{0,dec}^{2D} = \bbz_4(E)$
    \item[]  $\mathcal{G}_{0,dec} =  E_{0,dec}^{1D} \times E_{0,dec}^{2D} = \bzt\times\bbz_4(E)$
\end{itemize}

\subsubsection*{Decohered Spin-1/2}
\textbf{Block state decorations}:
\begin{itemize}
    \item[] 0D
    \begin{itemize}
        \item $\mu$: Odd fermion
    \end{itemize}
    \item[] 1D
    \begin{itemize}
        \item $\tau_1$: $\bbz_6\rtimes\bzt^f$ ASPT 
        \item $\tau_2$: $\bbz_4^f$ ASPT
    \end{itemize}
    \item[] 2D
    \begin{itemize}
        \item $\sigma_1$:\( p+ip \) SC
        \item $\sigma_2$: No nontrivial block state
    \end{itemize}
\end{itemize}

\textbf{Obstructions}
\begin{itemize}
    \item[] 2D
    \begin{itemize}
        \item\( p+ip \) SC on $\sigma_1$ is obstructed by chiral anomaly at $\tau_1$
    \end{itemize}
\end{itemize}

\textbf{Obstruction-free states}:
\begin{itemize}
\item[] 0D state ($\bzt$) is obstruction-free (I)
\item[] 1D: ($\bzt^2$) \begin{enumerate}
    \item $\bbz_6\rtimes\bzt^f$ ASPT on $\tau_1$ (I)
    \item $\bbz_4^f$ ASPT on $\tau_2$ (I)
\end{enumerate} 

\item[] 2D: No obstruction-free states ($\bbz_1$). 
\end{itemize}
\textbf{Trivializations:}
\begin{itemize}
    \item Fermion bubble on $\tau_1$ $\Rightarrow$ Odd fermion on $\mu$. Therefore, the 0D classification reduces to $\bbz_1$ (trivial).
\end{itemize}

\textbf{Final classification:}
\begin{itemize}
    \item[] $E_{1/2,dec}^{0D} = \bbz_1$
    \item[] $E_{1/2,dec}^{1D} = \bzt^2(I)$
    \item[] $E_{1/2,dec}^{2D} = \bbz_1$
    \item[]  $\mathcal{G}_{1/2,dec} = E_{1/2,dec}^{0D} \times E_{1/2,dec}^{1D} \times E_{1/2,dec}^{2D} = \bzt^2(I)$
\end{itemize}

\subsubsection*{Disordered Spinless}
\textbf{Block state decorations}:
\begin{itemize}
    \item[] 1D
    \begin{itemize}
        \item $\tau_1, \tau_2$: Majorana chain
    \end{itemize}
    \item[] 2D
    \begin{itemize}
        \item $\sigma_1$:\( p+ip \) SC
        \item $\sigma_2$:\( p+ip \) SC, fLG
    \end{itemize}
\end{itemize}

\textbf{Obstructions}
\begin{itemize}
    \item[] 2D
    \begin{itemize}
        \item\( p+ip \) SC on $\sigma_1$ is obstructed by chiral anomaly at $\tau_1$
        \item\( p+ip \) SC on $\sigma_2$ is incompatible with spinless rotational symmetry on equator.
    \end{itemize}
\end{itemize}

\textbf{Obstruction-free states}:
\begin{itemize}
\item[] 1D ($\bzt^2$) \begin{enumerate}
    \item Majorana chain on $\tau_1$ or $\tau_2$ (I)
\end{enumerate} 

\item[] 2D ($\bzt$) \begin{enumerate}
    \item fLG on $\sigma_2$ (E)
\end{enumerate} 
These states are all \textbf{trivialization-free}.
\end{itemize}

\textbf{Final classification:}
\begin{itemize}
    \item[] $E_{0,dis}^{1D} = \bzt^2(I)$
    \item[] $E_{0,dis}^{2D} = \bzt(E)$
    \item[]  $\mathcal{G}_{0,dis} = E_{0,dis}^{1D} \times E_{0,dis}^{2D} = \bzt(E)\times\bzt^2(I)$
\end{itemize}

\subsubsection*{Disordered Spin-1/2}
\textbf{Block state decorations}:
\begin{itemize}
    \item[] 1D
    \begin{itemize}
        \item $\tau_1, \tau_2$: \placeholder
    \end{itemize}
    \item[] 2D
    \begin{itemize}
        \item $\sigma_1$:\( p+ip \) SC
        \item $\sigma_2$: 2D $\bbz_4^f$ ASPT
    \end{itemize}
\end{itemize}

\textbf{Obstructions}
\begin{itemize}
    \item[] 2D
    \begin{itemize}
        \item\( p+ip \) SC on $\sigma_1$ is obstructed by chiral anomaly at $\tau_1$
    \end{itemize}
\end{itemize}

\textbf{Obstruction-free states}:
\begin{itemize}
\item[] 1D ($\bzt^2$) \begin{enumerate}
    \item \placeholder on $\tau_1$ or $\tau_2$ (I)
\end{enumerate} 

\item[] 2D ($\bzt$) \begin{enumerate}
    \item 2D $\bbz_4^f$ ASPT on $\sigma_2$ (I)
\end{enumerate} 
\item[] These states are all \textbf{trivialization-free}.
\end{itemize}

\textbf{Final classification:}
\begin{itemize}
    \item[] $E_{1/2,dis}^{1D} = \bzt^2(I)$
    \item[] $E_{1/2,dis}^{2D} = \bzt(I)$
    \item[]  $\mathcal{G}_{1/2,dis} = E_{1/2,dis}^{1D} \times E_{1/2,dis}^{2D} = \bzt^3(I)$
\end{itemize}

\subsection{$D_6$}
\subsubsection*{Cell Decomposition}
\begin{figure}[!htbp]
    \centering
    \includegraphics[width=0.9\linewidth]{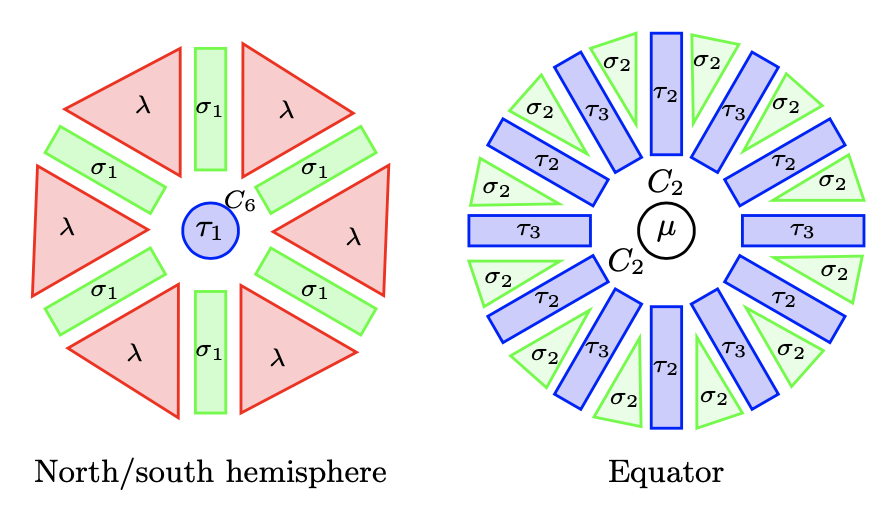}
    \caption{$D_6$ lattice}
\end{figure}
\textbf{Blocks and onsite symmetries}:
\begin{itemize}
    \item 2D: $G_{\sigma_1},\ G_{\sigma_2} = I$
    \item 1D: $G_{\tau_1} = \bbz_6,\ G_{\tau_2},\ G_{\tau_3} = \bzt$
    \item 0D: $G_{\mu} = \bbz_6\rtimes \bzt$
\end{itemize}

\subsubsection*{Decohered Spinless}
\textbf{Block state decorations}:
\begin{itemize}
    \item[] 0D
    \begin{itemize}
        \item $\mu$: Odd fermion
    \end{itemize}
    \item[] 1D
    \begin{itemize}
        \item $\tau_1$: Majorana chain, $\bbz_6$ fSPT 
        \item $\tau_2,\ \tau_3$: Majorana chain, $\bzt$ fSPT
    \end{itemize}
    \item[] 2D
    \begin{itemize}
        \item $\sigma_1,\ \sigma_2$:\( p+ip \) SC
    \end{itemize}
\end{itemize}

\textbf{Obstructions}
\begin{itemize}
    \item[] 1D
    \begin{itemize}
        \item Majorana chain on $\tau_1,\ \tau_2$, or $\tau_3$ is obstructed at $\mu$
    \end{itemize}
    \item[] 2D
    \begin{itemize}
        \item\( p+ip \) SC on $\sigma_1$ is obstructed by chiral anomaly at $\tau_1$
        \item\( p+ip \) SC on $\sigma_2$ is obstructed by chiral anomaly at $\tau_2$
    \end{itemize}
\end{itemize}

\textbf{Obstruction-free states}:
\begin{itemize}
\item[] 0D state ($\bzt$) is obstruction-free (E)

\item[] 1D: ($\bzt^3$) \begin{enumerate}
    \item $\bbz_6$ fSPT on $\tau_1$ (I)
    \item $\bzt$ fSPT on $\tau_2$ or $\tau_3$ (I)
\end{enumerate}

\item[] 2D: No obstruction-free states ($\bbz_1$)
\end{itemize}

\textbf{Trivializations}:
\begin{itemize}
    \item Majorana bubble on $\sigma_1$ $\Rightarrow$ Simultaneous decoration of $\bbz_4$ fSPT on $\tau_1$ and $\bzt$ fSPT on $\tau_2$. Therefore, the 1D classification reduces to $\bzt^2$.
    \item Majorana bubble on $\sigma_2$ $\Rightarrow$ Simultaneous decoration of $\bzt$ fSPT on $\tau_2$ and $\tau_3$. Therefore, the 1D classification further reduces to $\bzt$. 
    \item Chern insulator bubble trivializes odd fermion parity on $\mu$. The 0D classification reduces to $\bbz_1$.
\end{itemize}

\textbf{Final classification:}
\begin{itemize}
    \item[] $E_{0,dec}^{0D} = \bbz_1$
    \item[] $E_{0,dec}^{1D} = \bzt(I)$
    \item[] $E_{0,dec}^{2D} = \bbz_1$
    \item[]  $\mathcal{G}_{0,dec} =  E_{0,dec}^{0D} \times E_{0,dec}^{1D} \times E_{0,dec}^{2D} = \bzt(I)$
\end{itemize}

\subsubsection*{Decohered Spin-1/2}
\textbf{Block state decorations}:
\begin{itemize}
    \item[] 0D
    \begin{itemize}
        \item $\mu$: Odd fermion
    \end{itemize}
    \item[] 1D
    \begin{itemize}
        \item $\tau_1$: $\bbz_6\Rightarrow\bzt^f$ ASPT
        \item $\tau_2,\ \tau_3$: $\bbz_4^f$ ASPT
    \end{itemize}
    \item[] 2D
    \begin{itemize}
        \item $\sigma_1,\sigma_2$:\( p+ip \) SC
    \end{itemize}
\end{itemize}

\textbf{Obstructions}
\begin{itemize}
    \item[] 2D
    \begin{itemize}
        \item\( p+ip \) SC on $\sigma_1$ is obstructed by chiral anomaly at $\tau_1$
        \item\( p+ip \) SC on $\sigma_2$ is obstructed by chiral anomaly at $\tau_2$
    \end{itemize}
\end{itemize}

\textbf{Obstruction-free states}:
\begin{itemize}
\item[] 0D state ($\bzt$) is obstruction-free (I)
\item[] 1D: ($\bzt^3$) \begin{enumerate}
    \item $\bbz_6\Rightarrow\bzt^f$ ASPT on $\tau_1$ (I)
    \item $\bbz_4^f$ ASPT on $\tau_2$ or $\tau_3$ (I)
\end{enumerate} 
\item[] These states are also \textbf{trivialization-free}.

\item[] 2D: No obstruction-free states ($\bbz_1$). 
\end{itemize}

\textbf{Final classification:}
\begin{itemize}
    \item[] $E_{1/2,dec}^{0D} = \bzt(I)$
    \item[] $E_{1/2,dec}^{1D} = \bzt^3(I)$
    \item[] $E_{1/2,dec}^{2D} = \bbz_1$
    \item[]  $\mathcal{G}_{1/2,dec} = E_{1/2,dec}^{0D}\times E_{1/2,dec}^{1D} \times E_{1/2,dec}^{2D} = \bzt^4(I)$
\end{itemize}

\subsubsection*{Disordered Spinless}
\textbf{Block state decorations}:
\begin{itemize}
    \item[] 1D
    \begin{itemize}
        \item $\tau_1,\ \tau_2,\ \tau_3$: Majorana chain
    \end{itemize}
    \item[] 2D
    \begin{itemize}
        \item $\sigma_1,\ \sigma_2$:\( p+ip \) SC
    \end{itemize}
\end{itemize}

\textbf{Obstructions}
\begin{itemize}
    \item[] 2D
    \begin{itemize}
        \item\( p+ip \) SC on $\sigma_1$, $\sigma_2$ is obstructed by chiral anomaly at $\tau_1$, $\tau_2$ respectively
    \end{itemize}
\end{itemize}

\textbf{Obstruction-free states}:
\begin{itemize}
\item[] 1D ($\bzt^3$) \begin{enumerate}
    \item Majorana chain on $\tau_1$, $\tau_2$ or $\tau_3$ (I)
\end{enumerate} 

\item[] 2D: No obstruction-free states ($\bbz_1$)
\end{itemize}

\textbf{Trivializations}:
\begin{itemize}
    \item Open surface decoration $\Rightarrow$ Simultaneous decoration of Majorana chains on $\tau_1,\ \tau_2$, and $\tau_3$. This reduces the 1D classification to $\bzt^2$.
\end{itemize}

\textbf{Final classification:}
\begin{itemize}
    \item[] $E_{0,dis}^{1D} = \bzt^2(I)$
    \item[] $E_{0,dis}^{2D} = \bbz_1$
    \item[]  $\mathcal{G}_{0,dis} = E_{0,dis}^{1D} \times E_{0,dis}^{2D} = \bzt^2(I)$
\end{itemize}

\subsubsection*{Disordered Spin-1/2}
\textbf{Block state decorations}:
\begin{itemize}
    \item[] 1D
    \begin{itemize}
        \item $\tau_1,\ \tau_2,\ \tau_3$: Majorana chain
    \end{itemize}
    \item[] 2D
    \begin{itemize}
        \item $\sigma_1,\ \sigma_2$:\( p+ip \) SC
    \end{itemize}
\end{itemize}

\textbf{Obstructions}
\begin{itemize}
    \item[] 2D
    \begin{itemize}
        \item\( p+ip \) SC on $\sigma_1$, $\sigma_2$ is obstructed by chiral anomaly at $\tau_1$, $\tau_2$ respectively
    \end{itemize}
\end{itemize}

\textbf{Obstruction-free states}:
\begin{itemize}
\item[] 1D ($\bzt^3$) \begin{enumerate}
    \item \placeholder on $\tau_1$, $\tau_2$, or $\tau_3$ (I)
\end{enumerate} 
\item[] These states are also \textbf{trivialization-free}.
\item[] 2D: No obstruction-free states ($\bbz_1$)
\end{itemize}

\textbf{Final classification:}
\begin{itemize}
    \item[] $E_{1/2,dis}^{1D} = \bzt^3(I)$
    \item[] $E_{1/2,dis}^{2D} = \bbz_1$
    \item[]  $\mathcal{G}_{1/2,dis} = E_{1/2,dis}^{1D} \times E_{1/2,dis}^{2D} = \bzt^3(I)$
\end{itemize}

\subsection{$C_{6v}$}
\subsubsection*{Cell Decomposition}
\begin{figure}[!htbp]
    \centering
    \includegraphics[width=0.9\linewidth]{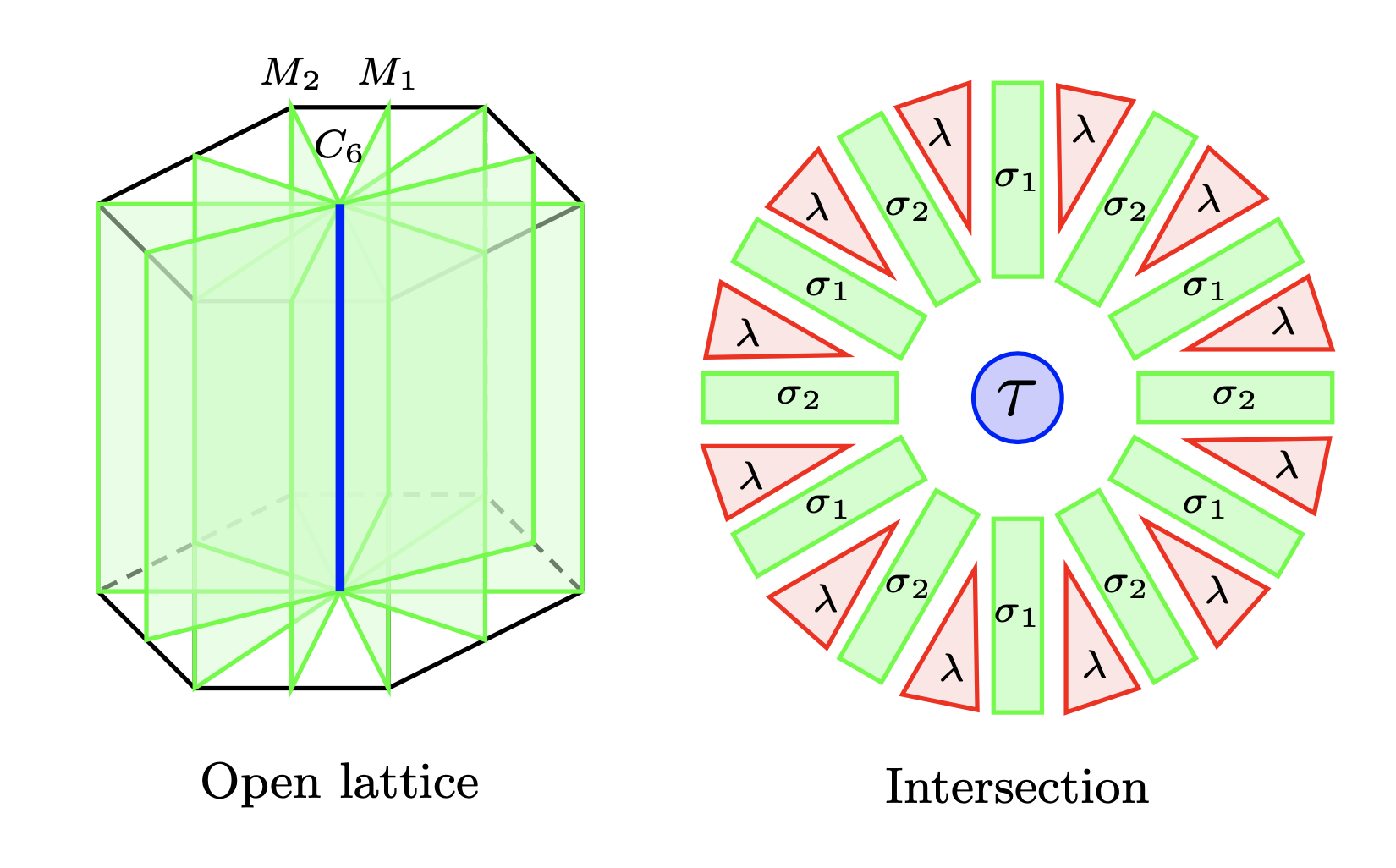}
    \caption{$C_{6v}$ lattice}
\end{figure}
\textbf{Blocks and onsite symmetries}:
\begin{itemize}
    \item $G_{\sigma_1},\ G_{\sigma_2} = \bzt$
    \item $G_{\tau} = \bbz_6\rtimes\bzt$
\end{itemize}

\subsubsection*{Decohered Spinless}
\textbf{Block state decorations}:
\begin{itemize}
    \item[] 1D
    \begin{itemize}
        \item $\tau$: Majorana chain, $\bzt^{M_1}$ fSPT, $\bzt^{M_2}$ fSPT
    \end{itemize}
    \item[] 2D
    \begin{itemize}
        \item $\sigma_1,\ \sigma_2$:\( p+ip \) SC, fLG
    \end{itemize}
\end{itemize}

\textbf{Obstructions}
\begin{itemize}
    \item[] 2D
    \begin{itemize}
        \item\( p+ip \) SC on $\sigma_1$ or $\sigma_2$ is obstructed by chiral anomaly at $\tau$
        \item Decoration of $p\pm ip$-SC with opposite chiralities on $\sigma_1$ and $\sigma_2$ is obstructed at $\tau$ (anomaly indicator $\nu_{M_1}=1/4$) 
        \item fLG on $\sigma_1$ is obstructed at $\tau$ ($\nu_{M_1}=1/2$)
        \item fLG on $\sigma_2$ is obstructed at $\tau$ ($\nu_{M_2}=1/2$)
    \end{itemize}
\end{itemize}

\textbf{Obstruction-free states}:
\begin{itemize}

\item[] 1D: ($\bzt^3$) \begin{enumerate}
    \item Majorana chain on $\tau$ (E)
    \item $\bbz_4 \rtimes \bzt$ fSPT on $\tau$ (E)
\end{enumerate}

\item[] 2D: ($\bzt^2$) \begin{enumerate}
    \item $n=2$ fLG on $\sigma_1$ or $\sigma_2$ (I) 
\end{enumerate}
\end{itemize}

\textbf{Trivializations}:
\begin{itemize}
    \item Majorana bubble on $\sigma_1$ $\Rightarrow$ $\bzt^{M_1}$ fSPT on $\tau$. Therefore, the 1D classification reduces to $\bzt^2$.
    \item Majorana bubble on $\sigma_2$ $\Rightarrow$ $\bzt^{M_2}$ fSPT on $\tau$. Therefore, the 1D classification further reduces to $\bzt$. 
\end{itemize}

\textbf{Final classification:}
\begin{itemize}
    \item[] $E_{0,dec}^{1D} = \bzt(E)$
    \item[] $E_{0,dec}^{2D} = \bzt^2(I)$
    \item[]  $\mathcal{G}_{0,dec} =  E_{0,dec}^{1D} \times E_{0,dec}^{2D} = \bzt(E)\times\bzt^2(I)$
\end{itemize}

\subsubsection*{Decohered Spin-1/2}
\textbf{Block state decorations}:
\begin{itemize}
    \item[] 1D
    \begin{itemize}
        \item $\tau$: $\bbz_4^{f,M_1}$ ASPT, $\bbz_4^{f,M_2}$ ASPT (I)
    \end{itemize}
    \item[] These states are all \textbf{obstruction-free} and \textbf{trivialization-free}
    \item[] 2D
    \begin{itemize}
        \item $\sigma_1,\ \sigma_2$: No nontrivial block state
    \end{itemize}
\end{itemize}

\textbf{Final classification:}
\begin{itemize}
    \item[] $E_{1/2,dec}^{1D} = \bzt^2(I)$
    \item[] $E_{1/2,dec}^{2D} = \bbz_1$
    \item[]  $\mathcal{G}_{1/2,dec} = E_{1/2,dec}^{1D} \times E_{1/2,dec}^{2D} = \bzt^2(I)$
\end{itemize}

\subsubsection*{Disordered Spinless}
\textbf{Block state decorations}:
\begin{itemize}
    \item[] 1D
    \begin{itemize}
        \item $\tau$: Majorana chain
    \end{itemize}
    \item[] 2D
    \begin{itemize}
        \item $\sigma_1,\ \sigma_2$:\( p+ip \) SC, fLG
    \end{itemize}
\end{itemize}

\textbf{Obstructions}
\begin{itemize}
    \item[] 2D
    \begin{itemize}
        \item\( p+ip \) SC on $\sigma_1$ or $\sigma_2$ is obstructed by chiral anomaly at $\tau$
        \item Decoration of $p\pm ip$-SC with opposite chiralities on $\sigma_1$ and $\sigma_2$ is obstructed at $\tau$ (anomaly indicator $\nu_{M_1}=1/4$) 
    \end{itemize}
\end{itemize}

\textbf{Obstruction-free states}:
\begin{itemize}
\item[] 1D ($\bzt$) \begin{enumerate}
    \item Majorana chain on $\tau$ (E)
\end{enumerate} 

\item[] 2D ($\bzt^2$) \begin{enumerate}
    \item fLG on $\sigma_1$ or $\sigma_2$ (I)
\end{enumerate}
\item[] These states are all \textbf{trivialization-free}
\end{itemize}

\textbf{Final classification:}
\begin{itemize}
    \item[] $E_{0,dis}^{1D} = \bzt(E)$
    \item[] $E_{0,dis}^{2D} = \bzt^2(I)$
    \item[]  $\mathcal{G}_{0,dis} = E_{0,dis}^{1D} \times E_{0,dis}^{2D} = \bzt(E)\times\bzt^2(I)$
\end{itemize}

\subsubsection*{Disordered Spin-1/2}
\textbf{Block state decorations}:
\begin{itemize}
    \item[] 1D
    \begin{itemize}
        \item $\tau$: \placeholder (I)
    \end{itemize}
    \item[] 2D
    \begin{itemize}
        \item $\sigma_1,\ \sigma_2$: 2D $\bbz_4^f$ ASPT (I)
    \end{itemize}
    \item[] These states are all \textbf{obstruction-free} and \textbf{trivialization-free}.
\end{itemize}

\textbf{Final classification:}
\begin{itemize}
    \item[] $E_{1/2,dis}^{1D} = \bzt(I)$
    \item[] $E_{1/2,dis}^{2D} = \bzt^2(I)$
    \item[]  $\mathcal{G}_{1/2,dis} = E_{1/2,dis}^{1D} \times E_{1/2,dis}^{2D} = \bzt^3(I)$
\end{itemize}

\subsection{$D_{3h}$}
\subsubsection*{Cell Decomposition}
\begin{figure}[!htbp]
    \centering
    \includegraphics[width=0.9\linewidth]{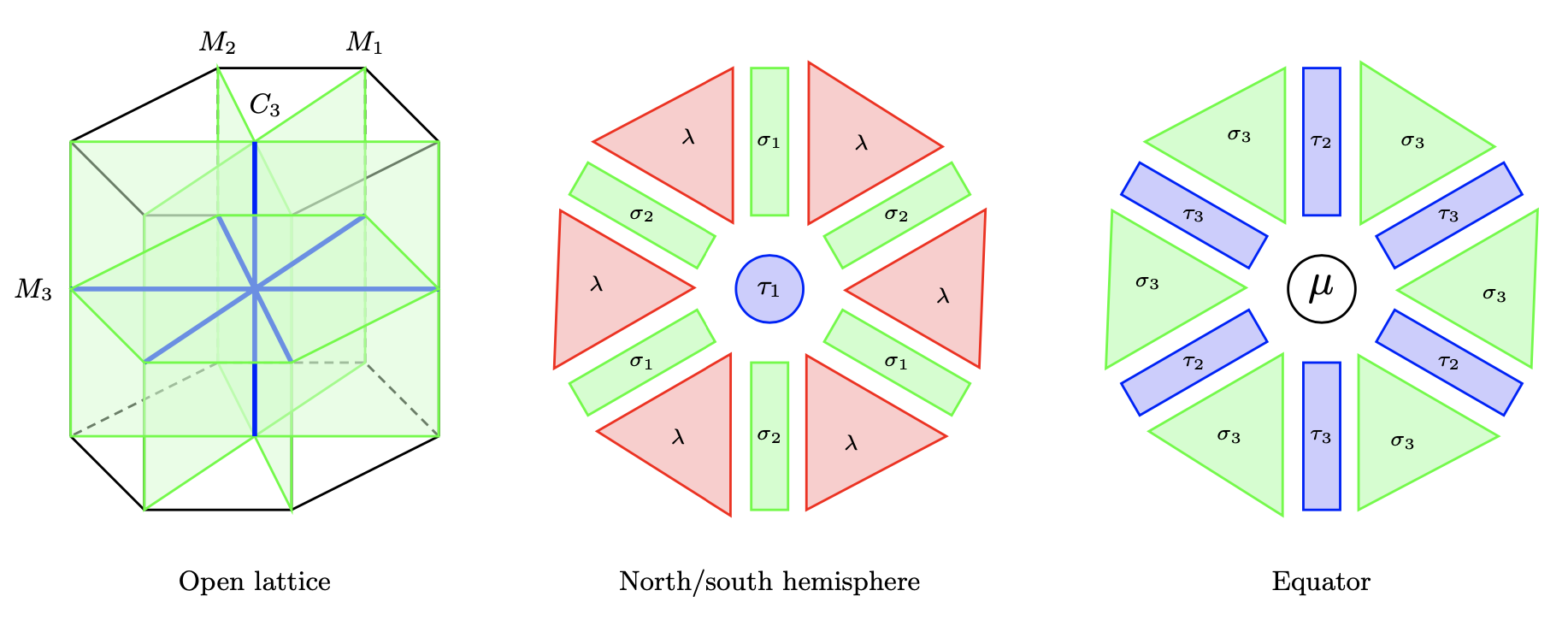}
    \caption{$D_{3h}$ lattice}
\end{figure}
\textbf{Blocks and onsite symmetries}:
\begin{itemize}
    \item 2D: $G_{\sigma_1},\ G_{\sigma_2},\ G_{\sigma_3} = \bzt$
    \item 1D: $G_{\tau_1} = \bbz_3\rtimes\bzt,\ G_{\tau_2},\ G_{\tau_3} = \bzt \times \bzt$
    \item 0D: $G_{\mu} = \bzt \times (\bbz_3\rtimes \bzt)$
\end{itemize}

\subsubsection*{Decohered Spinless}
\textbf{Block state decorations}:
\begin{itemize}
    \item[] 0D
    \begin{itemize}
        \item $\mu$: Odd fermion
    \end{itemize}
    \item[] 1D
    \begin{itemize}
        \item $\tau_1$: Majorana chain, $\bbz_3\rtimes\bzt$ fSPT 
        \item $\tau_2$: Majorana chain, $\bzt^{M_1}$ fSPT, $\bzt^{M_3}$ fSPT
        \item $\tau_3$: Majorana chain, $\bzt^{M_2}$ fSPT, $\bzt^{M_3}$ fSPT
    \end{itemize}
    \item[] 2D
    \begin{itemize}
        \item $\sigma_1,\ \sigma_2,\ \sigma_3$:\( p+ip \) SC, fLG
    \end{itemize}
\end{itemize}

\textbf{Obstructions}
\begin{itemize}
    \item[] 1D
    \begin{itemize}
        \item Majorana chain on $\tau_1$, $\tau_2$, or $\tau_3$ is obstructed at $\mu$. Simultaneous decorations are also obstructed, with the exception of Majorana chains on $\tau_2$ and $\tau_3$.
        \item $\bzt^{M_1}$ fSPT on $\tau_2$ or $\bzt^{M_2}$ fSPT on $\tau_3$ are obstructed at $\mu$, unless simultaneously decorated.
        \item $\bzt^{M_3}$ fSPT on one of $\tau_2$ or $\tau_3$ is obstructed at $\mu$ (odd number of edge modes)
    \end{itemize}
    \item[] 2D
    \begin{itemize}
        \item\( p+ip \) SC on $\sigma_1$ or $\sigma_2$ is obstructed by chiral anomaly at $\tau_1$
        \item\( p+ip \) SC on $\sigma_3$ is obstructed by chiral anomaly at $\tau_3$
        \item Simultaneous decorations are obstructed as well ($\nu_{\bzt}=1/2$ anomaly on 1D blocks).
        \item fLG on $\sigma_1$ is obstructed at $\tau_2$ ($\nu_{M_1}=1/4$)
        \item fLG on $\sigma_2$ is obstructed at $\tau_3$ ($\nu_{M_2}=1/4$)
        \item $n=2$ fLG on one of $\sigma_1$ or $\sigma_2$ is obstructed at $\tau_1$ ($\nu_{M_1}/\nu_{M_2} = 3/2$)
        \item fLG on $\sigma_3$ is obstructed at $\tau_3$ ($\nu_{M_3}=1/2$)
    \end{itemize}
\end{itemize}

\textbf{Obstruction-free states}:
\begin{itemize}
\item[] 0D states are obstruction-free ($\bzt$)
 
\item[] 1D: ($\bzt^4$) \begin{enumerate}
    \item Majorana chains on $\tau_2$ and $\tau_3$ (E)
    \item $\bbz_3\rtimes\bzt$ fSPT on $\tau_1$ (I)
    \item $\bzt^{M_1}$ fSPT on $\tau_2$ and $\bzt^{M_2}$ fSPT on $\tau_3$ (I)
    \item $\bzt^{M_3}$ fSPTs on $\tau_2$ and $\tau_3$ (I)
\end{enumerate}

\item[] 2D: ($\bzt^2$) \begin{enumerate}
    \item $n=2$ fLGs on $\sigma_1$ and $\sigma_2$ (I)
    \item $n=2$ fLG on $\sigma_3$ (I)
\end{enumerate}
\end{itemize}

\textbf{Trivializations}:
\begin{itemize}
    \item Majorana bubble on $\sigma_3$ $\Rightarrow$ $\bzt^{M_1}$ fSPT on $\tau_2$ and $\bzt^{M_2}$ fSPT on $\tau_3$. Therefore, the 1D classification reduces to $\bzt^3$.
    \item Majorana bubbles on $\sigma_1$ and $\sigma_2$ $\Rightarrow$ Simultaneous decoration of $\bbz_3\rtimes\bzt$ fSPT on $\tau_1$ and $\bzt^{M_3}$ fSPTs on $\tau_2$ and $\tau_3$. Therefore, the 1D classification further reduces to $\bzt^2$.
    \item Double Majorana ($\bzt$ fSPT) bubble on $\sigma_1$ $\Rightarrow$ $\bbz_3\rtimes\bzt$ fSPT on $\tau_1$. Therefore, the 1D classification further reduces to $\bzt$. 
    \item Fermion bubble on $\tau_2$ $\Rightarrow$ Odd fermion on $\mu$. Therefore, the 0D classification reduces to $\bbz_1$ (trivial). 
\end{itemize}

\textbf{Final classification:}
\begin{itemize}
    \item[] $E_{0,dec}^{0D} = \bbz_1$
    \item[] $E_{0,dec}^{1D} = \bzt(E)$
    \item[] $E_{0,dec}^{2D} = \bzt^2(I)$
    \item[]  $\mathcal{G}_{0,dec} = E_{0,dec}^{0D} \times E_{0,dec}^{1D} \times E_{0,dec}^{2D} = \bzt(E)\times\bzt^2(I)$
\end{itemize}

\subsubsection*{Decohered Spin-1/2}

\textbf{Block state decorations}:
\begin{itemize}
    \item[] 0D
    \begin{itemize}
        \item $\mu$: Odd fermion
    \end{itemize}
    \item[] 1D
    \begin{itemize}
        \item $\tau_1$: $\bbz_3\rtimes\bzt\rtimes\bzt^f$ ASPT $\cong$ $\bbz_4^f$ ASPT 
        \item $\tau_2$: $\bbz_4^{f,M_1}$ ASPT, $\bbz_4^{f,M_3}$ ASPT
        \item $\tau_3$: $\bbz_4^{f,M_2}$ ASPT, $\bbz_4^{f,M_3}$ ASPT
    \end{itemize}
    \item[] 2D
    \begin{itemize}
        \item $\sigma_1,\ \sigma_2,\ \sigma_3$: No nontrivial block state
    \end{itemize}
\end{itemize}

\textbf{Obstructions}
\begin{itemize}
    \item[] 1D
    \begin{itemize}
        \item $\bbz_4^{f,M_1}$ ASPT on $\tau_2$ or $\bbz_4^{f,M_2}$ ASPT on $\tau_3$ are obstructed at $\mu$, unless simultaneous decorated.
        \item $\bbz_4^{f,M_3}$ ASPT on one of $\tau_2$ or $\tau_3$ is obstructed at $\mu$ (odd number of edge modes)
    \end{itemize}
\end{itemize}

\textbf{Obstruction-free states}:
\begin{itemize}
\item[] 0D states are obstruction-free ($\bzt$)
 
\item[] 1D: ($\bzt^3$) \begin{enumerate}
    \item $\bbz_3\rtimes\bzt\rtimes\bzt^f$ ASPT on $\tau_1$ (I)
    \item $\bbz_4^{f,M_1}$ ASPT on $\tau_2$ and $\bbz_4^{f,M_2}$ ASPT on $\tau_3$ (I)
    \item $\bbz_4^{f,M_3}$ ASPTs on $\tau_2$ and $\tau_3$ (I)
\end{enumerate} 

\end{itemize}

\textbf{Trivializations}:
\begin{itemize}
     \item $\bbz_4$ ASPT bubble on $\sigma_1$ $\Rightarrow$ $\bbz_3\rtimes\bzt\rtimes\bzt^f$ ASPT on $\tau_1$. Therefore, the 1D classification reduces to $\bzt^2$. 
    \item Fermion bubble on $\tau_2$ $\Rightarrow$ Odd fermion on $\mu$. Therefore, the 0D classification reduces to $\bbz_1$ (trivial). 
\end{itemize}

\textbf{Final classification:}
\begin{itemize}
    \item[] $E_{1/2,dec}^{0D} = \bbz_1$
    \item[] $E_{1/2,dec}^{1D} = \bzt^2(I)$
    \item[] $E_{1/2,dec}^{2D} = \bbz_1$
    \item[]  $\mathcal{G}_{1/2,dec} = E_{0,dec}^{0D} \times E_{1/2,dec}^{1D} \times E_{1/2,dec}^{2D} = \bzt^2(I)$
\end{itemize}

\subsubsection*{Disordered Spinless}
\textbf{Block state decorations}:
\begin{itemize}
    \item[] 1D
    \begin{itemize}
        \item $\tau_1,\ \tau_2,\ \tau_3$: Majorana chain
    \end{itemize}
    \item[] 2D
    \begin{itemize}
        \item $\sigma_1,\ \sigma_2,\ \sigma_3$:\( p+ip \) SC, fLG
    \end{itemize}
\end{itemize}

\textbf{Obstructions}
\begin{itemize}
    \item[] 1D 
    \begin{itemize}
        \item Majorana chain on one of $\tau_2$ or $\tau_3$ is obstructed at $\mu$ (odd number of Majorana modes)
    \end{itemize}
    \item[] 2D
    \begin{itemize}
        \item\( p+ip \) SC on $\sigma_1$ or $\sigma_2$ is obstructed by chiral anomaly at $\tau_1$
        \item\( p+ip \) SC on $\sigma_3$ is obstructed by chiral anomaly at $\tau_3$
        \item Simultaneous decorations are obstructed as well ($\nu_{\bzt}=1/2$ anomaly on 1D blocks).
        \item fLG on one of $\sigma_1$ or $\sigma_2$ is obstructed at $\tau_1$
    \end{itemize}
\end{itemize}

\textbf{Obstruction-free states}:
\begin{itemize}
\item[] 1D ($\bzt^2$) \begin{enumerate}
    \item Majorana chain on $\tau_1$ (I)
    \item Majorana chains on $\tau_2$ and $\tau_3$ (E)
\end{enumerate} 

\item[] 2D ($\bzt^2$) \begin{enumerate}
    \item fLG on $\sigma_1$ and $\sigma_2$ (I)
    \item fLG on $\sigma_3$ (I)
\end{enumerate}
\end{itemize}

\textbf{Trivializations}:
\begin{itemize}
    \item Majorana bubble on $\sigma$ $\Rightarrow$ Majorana chain on $\tau_1$. Therefore, the 1D classification reduces to $\bzt$
\end{itemize}

\textbf{Final classification:}
\begin{itemize}
    \item[] $E_{0,dis}^{1D} = \bzt(E)$
    \item[] $E_{0,dis}^{2D} = \bzt^2(I)$
    \item[]  $\mathcal{G}_{0,dis} = E_{0,dis}^{1D} \times E_{0,dis}^{2D} = \bzt(E)\times\bzt^2(I)$
\end{itemize}

\subsubsection*{Disordered Spin-1/2}
\textbf{Block state decorations}:
\begin{itemize}
    \item[] 1D
    \begin{itemize}
        \item $\tau_1,\ \tau_2,\ \tau_3$: \placeholder
    \end{itemize}
    \item[] 2D
    \begin{itemize}
        \item $\sigma_1,\ \sigma_2,\ \sigma_3$: 2D $\bbz_4^f$ ASPT
    \end{itemize}
\end{itemize}

\textbf{Obstructions}
\begin{itemize}
    \item[] 1D 
    \begin{itemize}
        \item \placeholder on one of $\tau_2$ or $\tau_3$ is obstructed at $\mu$ (odd number of \placeholdermodes at $\mu$)
    \end{itemize}
    \item[] 2D
    \begin{itemize}
        \item 2D $\bbz_4^f$ ASPT on one of $\sigma_1$ or $\sigma_2$ is obstructed at $\tau_1$
    \end{itemize}
\end{itemize}

\textbf{Obstruction-free states}:
\begin{itemize}
\item[] 1D ($\bzt^2$) \begin{enumerate}
    \item \placeholder on $\tau_1$ (I)
    \item \placeholders on $\tau_2$ and $\tau_3$ (I) 
\end{enumerate} 

\item[] 2D ($\bzt^2$) \begin{enumerate}
    \item 2D $\bbz_4^f$ ASPT on $\sigma_1$ and $\sigma_2$ (I)
    \item 2D $\bbz_4^f$ ASPT on $\sigma_3$ (I)
\end{enumerate}
\end{itemize}

\textbf{Trivializations}:
\begin{itemize}
    \item \bubbleplaceholder on $\sigma$ $\Rightarrow$ \placeholder on $\tau_1$. Therefore, the 1D classification reduces to $\bzt$
\end{itemize}

\textbf{Final classification:}
\begin{itemize}
    \item[] $E_{1/2,dis}^{1D} = \bzt(I)$
    \item[] $E_{1/2,dis}^{2D} = \bzt^2(I)$
    \item[]  $\mathcal{G}_{1/2,dis} = E_{1/2,dis}^{1D} \times E_{1/2,dis}^{2D} = \bzt^3(I)$
\end{itemize}

\subsection{$D_{6h}$}
\subsubsection*{Cell Decomposition}
\begin{figure}[!htbp]
    \centering
    \includegraphics[width=0.9\linewidth]{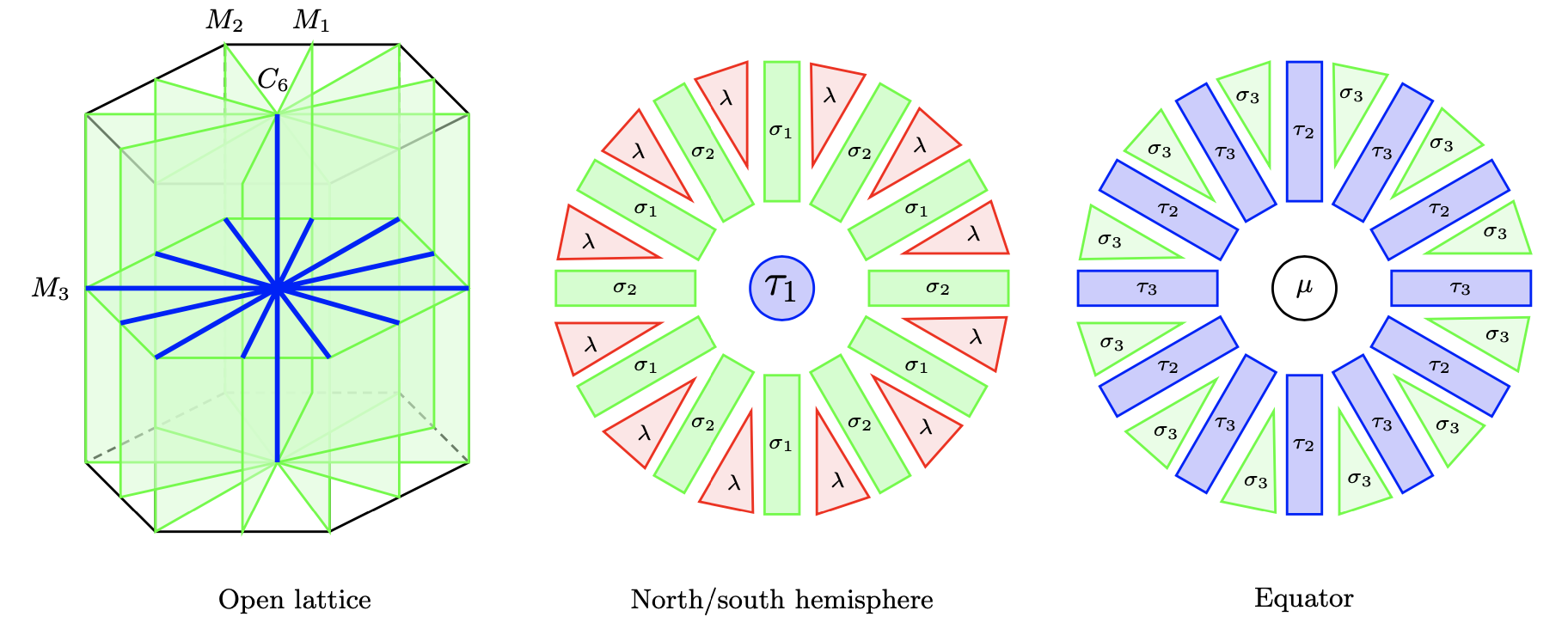}
    \caption{$D_{6h}$ lattice}
\end{figure}
\textbf{Blocks and onsite symmetries}:
\begin{itemize}
    \item $G_{\sigma_1},\ G_{\sigma_2},\ G_{\sigma_3} = \bzt$
    \item $G_{\tau_1} = \bbz_6\rtimes\bzt,\ G_{\tau_2},\ G_{\tau_3} = \bzt\times\bzt$
    \item $G_{\mu} = \bzt \times (\bbz_6\rtimes\bzt)$
\end{itemize}

\subsubsection*{Decohered Spinless}
\textbf{Block state decorations}:
\begin{itemize}
    \item[] 0D
    \begin{itemize}
        \item $\mu$: Odd fermion
    \end{itemize}
    \item[] 1D
    \begin{itemize}
        \item $\tau_1$: Majorana chain, $\bzt^{M_1}$ fSPT, $\bzt^{M_2}$ fSPT
        \item $\tau_2$: Majorana chain, $\bzt^{M_1}$ fSPT, $\bzt^{M_3}$ fSPT
        \item $\tau_3$: Majorana chain, $\bzt^{M_2}$ fSPT, $\bzt^{M_3}$ fSPT
    \end{itemize}
    \item[] 2D
    \begin{itemize}
        \item $\sigma_1,\ \sigma_2,\ \sigma_3$:\( p+ip \) SC, fLG
    \end{itemize}
\end{itemize}

\textbf{Obstructions}
\begin{itemize}
    \item[] 1D
    \begin{itemize}
        \item Majorana chain on $\tau_1$, $\tau_2$, or $\tau_3$ is obstructed at $\mu$
    \end{itemize}
    \item[] 2D
    \begin{itemize}
        \item\( p+ip \) SC on $\sigma_1$, $\sigma_2$, $\sigma_3$ is obstructed by chiral anomaly at $\tau_1$, $\tau_2$, $\tau_3$ respectively
        \item Simultaneous decorations are obstructed as well
        \item fLG on $\sigma_1$ is obstructed at $\tau_1$ ($\nu_{M_1}=1/2$)
        \item fLG on $\sigma_2$ is obstructed at $\tau_2$ ($\nu_{M_2}=1/2$)
        \item fLG on $\sigma_3$ is obstructed at $\tau_3$ ($\nu_{M_3}=1/2$)   
    \end{itemize}
\end{itemize}

\textbf{Obstruction-free states}:
\begin{itemize}
\item[] 0D states are obstruction-free ($\bzt$)
 
\item[] 1D: ($\bzt^6$) \begin{enumerate}
    \item Any combination of fSPTs on $\tau_1$, $\tau_2$, $\tau_3$ (I)
\end{enumerate}

\item[] 2D: ($\bzt^3$) \begin{enumerate}
    \item $n=2$ fLG on $\sigma_1$, $\sigma_2$, or $\sigma_3$ (I)
\end{enumerate}
\end{itemize}

\textbf{Trivializations}:
\begin{itemize}
    \item Majorana bubble on $\sigma_1$ $\Rightarrow$ Simultaneous decoration of $\bzt^{M_1}$ fSPT on $\tau_1$ and $\tau_2$. Therefore, the 1D classification reduces to $\bzt^5$.
    \item Majorana bubble on $\sigma_2$ $\Rightarrow$ Simultaneous decoration of $\bzt^{M_2}$ fSPT on $\tau_1$ and $\tau_3$. Therefore, the 1D classification further reduces to $\bzt^4$.
    \item Majorana bubble on $\sigma_3$ $\Rightarrow$ Simultaneous decoration of $\bzt^{M_3}$ fSPT on $\tau_2$ and $\tau_3$. Therefore, the 1D classification further reduces to $\bzt^3$. 
\end{itemize}

\textbf{Final classification:}
\begin{itemize}
    \item[] $E_{0,dec}^{0D} = \bzt(E)$
    \item[] $E_{0,dec}^{1D} = \bzt^3(I)$
    \item[] $E_{0,dec}^{2D} = \bzt^3(I)$
    \item[]  $\mathcal{G}_{0,dec} = E_{0,dec}^{0D} \times E_{0,dec}^{1D} \times E_{0,dec}^{2D} = \bzt\times\bzt^6(I)$
\end{itemize}

\subsubsection*{Decohered Spin-1/2}
\textbf{Block state decorations}:
\begin{itemize}
    \item[] 0D
    \begin{itemize}
        \item $\mu$: Odd fermion (I)
    \end{itemize}
    \item[] 1D
    \begin{itemize}
        \item $\tau_1$: $\bbz_4^{f,M_1}$ ASPT, $\bbz_4^{f,M_2}$ ASPT (I)
        \item $\tau_2$: $\bbz_4^{f,M_1}$ ASPT, $\bbz_4^{f,M_3}$ ASPT (I)
        \item $\tau_3$: $\bbz_4^{f,M_2}$ ASPT, $\bbz_4^{f,M_3}$ ASPT (I)
    \end{itemize}
    \item[] These states are all \textbf{obstruction-free} and \textbf{trivialization-free}
    \item[] 2D
    \begin{itemize}
        \item $\sigma_1,\ \sigma_2,\ \sigma_3$: No nontrivial block state
    \end{itemize}
\end{itemize}

\textbf{Final classification:}
\begin{itemize}
    \item[] $E_{1/2,dec}^{0D} = \bzt(I)$
    \item[] $E_{1/2,dec}^{1D} = \bzt^6(I)$
    \item[] $E_{1/2,dec}^{2D} = \bbz_1$
    \item[]  $\mathcal{G}_{1/2,dec} = E_{1/2,dec}^{0D} \times E_{1/2,dec}^{1D} \times E_{1/2,dec}^{2D} = \bzt^7(I)$
\end{itemize}

\subsubsection*{Disordered Spinless}
\textbf{Block state decorations}:
\begin{itemize}
    \item[] 1D
    \begin{itemize}
        \item $\tau_1,\ \tau_2,\ \tau_3$: Majorana chain
    \end{itemize}
    \item[] 2D
    \begin{itemize}
        \item $\sigma_1,\ \sigma_2,\ \sigma_3$:\( p+ip \) SC, fLG
    \end{itemize}
\end{itemize}

\textbf{Obstructions}
\begin{itemize}
    \item[] 2D
    \begin{itemize}
        \item\( p+ip \) SC on $\sigma_1$, $\sigma_2$, $\sigma_3$ is obstructed by chiral anomaly at $\tau_1$, $\tau_2$, $\tau_3$ respectively
    \end{itemize}
\end{itemize}

\textbf{Obstruction-free states}:
\begin{itemize}
\item[] 1D ($\bzt^3$) \begin{enumerate}
    \item Majorana chain on $\tau_1$, $\tau_2$, or $\tau_3$ (I) 
\end{enumerate} 

\item[] 2D ($\bzt^3$) \begin{enumerate}
    \item fLG on $\sigma_1$, $\sigma_2$, or $\sigma_3$ (I)
\end{enumerate}
\item[] These states are all \textbf{trivialization-free}
\end{itemize}

\textbf{Final classification:}
\begin{itemize}
    \item[] $E_{0,dis}^{1D} = \bzt^3(I)$
    \item[] $E_{0,dis}^{2D} = \bzt^3(I)$
    \item[]  $\mathcal{G}_{0,dis} = E_{0,dis}^{1D} \times E_{0,dis}^{2D} = \bzt^6(I)$
\end{itemize}

\subsubsection*{Disordered Spin-1/2}
\textbf{Block state decorations}:
\begin{itemize}
    \item[] 1D
    \begin{itemize}
        \item $\tau_1,\ \tau_2,\ \tau_3$: \placeholder (I)
    \end{itemize}
    \item[] 2D
    \begin{itemize}
        \item $\sigma_1,\ \sigma_2,\ \sigma_3$: 2D $\bbz_4^f$ ASPT (I)
    \end{itemize}
    \item[] These states are all \textbf{obstruction-free} and \textbf{trivialization-free}.
\end{itemize}

\textbf{Final classification:}
\begin{itemize}
    \item[] $E_{1/2,dis}^{1D} = \bzt^3(I)$
    \item[] $E_{1/2,dis}^{2D} = \bzt^3(I)$
    \item[]  $\mathcal{G}_{1/2,dis} = E_{1/2,dis}^{1D} \times E_{1/2,dis}^{2D} = \bzt^6(I)$
\end{itemize}

\subsection{$T$}
\subsubsection*{Cell Decomposition}
\begin{figure}[!htbp]
    \centering
    \includegraphics[width=0.9\linewidth]{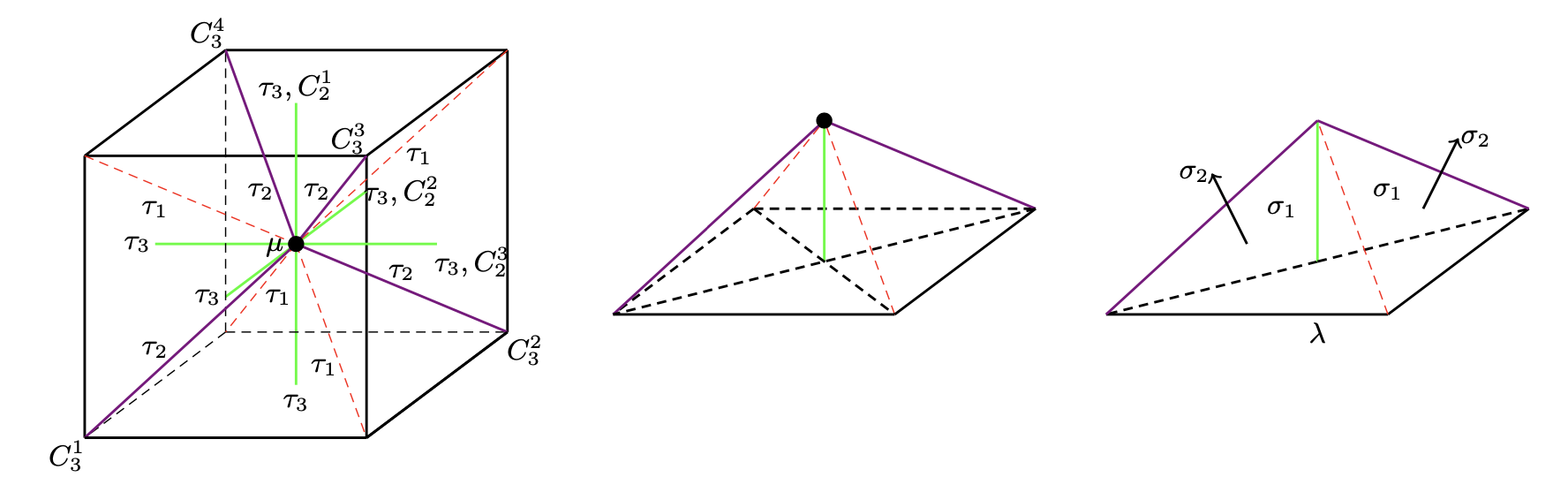}
    \caption{$T$ lattice}
\end{figure}
\textbf{Blocks and onsite symmetries}:
\begin{itemize}
    \item 2D: $G_{\sigma_1},\ G_{\sigma_2} = I$
    \item 1D: $G_{\tau_1},\ G_{\tau_2} = \bbz_3,\ G_{\tau_3} = \bzt$
    \item 0D: $G_{\mu} = A_4$
\end{itemize}
\subsubsection*{Decohered Spinless}
\textbf{Block state decorations}:
\begin{itemize}
    \item[] 0D
    \begin{itemize}
        \item $\mu$: Odd fermion
    \end{itemize}
    \item[] 1D
    \begin{itemize}
        \item $\tau_1,\ \tau_2$: Majorana chain
        \item $\tau_3$: Majorana chain, $\bzt$ fSPT
    \end{itemize}
    \item[] 2D
    \begin{itemize}
        \item $\sigma_1,\ \sigma_2$:\( p+ip \) SC
    \end{itemize}
    
\end{itemize}

\textbf{Obstructions}
\begin{itemize}
    \item[] 2D
    \begin{itemize}
        \item\( p+ip \) SC on $\sigma_1$, $\sigma_2$ is obstructed by chiral anomaly at $\tau_2$, $\tau_1$ respectively
    \end{itemize}
\end{itemize}

\textbf{Obstruction-free states}:
\begin{itemize}
\item[] 0D state ($\bzt$) is obstruction-free(E). 

\item[] 1D: ($\bzt^4$) \begin{enumerate}
    \item Majorana chain on $\tau_1$, $\tau_2$, or $\tau_3$ (I)
    \item $\bzt$ fSPT on $\tau_3$ (I)
\end{enumerate}

\item[] 2D: No obstruction-free states ($\bbz_1$)
\end{itemize}

\textbf{Trivializations}:
\begin{itemize}
    \item Majorana bubble on $\sigma_1$ $\Rightarrow$ Simultaneous decoration of Majorana chain on $\tau_2$ and $\bzt$ fSPT on $\tau_3$. This reduces 1D classification to $\bzt^3$.
    \item Majorana bubble on $\sigma_2$ $\Rightarrow$ Simultaneous decoration of Majorana chains on $\tau_1$ and $\tau_2$. This further reduces 1D classification to $\bzt^2$.
    \item Open surface decoration trivializes Majorana chain on $\tau_3$. This further reduces 1D classification to $\bzt$.
    \item Chern insulator bubble trivializes odd fermion parity on $\mu$. The 0D classification reduces to $\bbz_1$
\end{itemize}

\textbf{Final classification:}
\begin{itemize}
    \item[] $E_{0,dec}^{0D} = \bbz_1$
    \item[] $E_{0,dec}^{1D} = \bzt(I)$
    \item[] $E_{0,dec}^{2D} = \bbz_1$
    \item[]  $\mathcal{G}_{0,dec} = E_{0,dec}^{0D} \times E_{0,dec}^{1D} \times E_{0,dec}^{2D} = \bzt(I)$
\end{itemize}

\subsubsection*{Decohered Spin-1/2}
\textbf{Block state decorations}:
\begin{itemize}
    \item[] 0D
    \begin{itemize}
        \item $\mu$: Odd fermion
    \end{itemize}
    \item[] 1D
    \begin{itemize}
        \item $\tau_1,\ \tau_2$: Majorana chain
        \item $\tau_3$: $\bbz_4^f$ ASPT
    \end{itemize}
    \item[] 2D
    \begin{itemize}
        \item $\sigma_1,\ \sigma_2$:\( p+ip \) SC
    \end{itemize}
    
\end{itemize}

\textbf{Obstructions}
\begin{itemize}
    \item[] 2D
    \begin{itemize}
        \item\( p+ip \) SC on $\sigma_1$, $\sigma_2$ is obstructed by chiral anomaly at $\tau_2$, $\tau_1$ respectively
    \end{itemize}
\end{itemize}

\textbf{Obstruction-free states}:
\begin{itemize}
\item[] 0D state ($\bzt$) is obstruction-free (I). 

\item[] 1D: ($\bzt^3$) \begin{enumerate}
    \item Majorana chain on $\tau_1$ or $\tau_2$ (E) 
    \item $\bbz_4^f$ ASPT on $\tau_3$ (I)
\end{enumerate}

\item[] 2D: No obstruction-free states ($\bbz_1$)
\end{itemize}

\textbf{Trivializations}:
\begin{itemize}
    \item Majorana bubble on $\sigma_1$ $\Rightarrow$ Majorana chain on $\tau_2$. This reduces 1D classification to $\bzt^2$.
    \item Majorana bubbles on $\sigma_1$ and $\sigma_2$ $\Rightarrow$ Majorana chain on $\tau_1$. This further reduces 1D classification to $\bzt$.
\end{itemize}

\textbf{Final classification:}
\begin{itemize}
    \item[] $E_{1/2,dec}^{0D} = \bzt(I)$
    \item[] $E_{1/2,dec}^{1D} = \bzt(I)$
    \item[] $E_{1/2,dec}^{2D} = \bbz_1$
    \item[]  $\mathcal{G}_{1/2,dec} = E_{1/2,dec}^{0D} \times E_{1/2,dec}^{1D} \times E_{1/2,dec}^{2D} = \bzt^2(I)$
\end{itemize}

\subsubsection*{Disordered Spinless}
\textbf{Block state decorations}:
\begin{itemize}
    \item[] 1D
    \begin{itemize}
        \item $\tau_1,\ \tau_2,\ \tau_3$: Majorana chain
    \end{itemize}
    \item[] 2D
    \begin{itemize}
        \item $\sigma_1,\ \sigma_2$:\( p+ip \) SC
    \end{itemize}
    
\end{itemize}

\textbf{Obstructions}
\begin{itemize}
    \item[] 2D
    \begin{itemize}
        \item\( p+ip \) SC on $\sigma_1$, $\sigma_2$ is obstructed by chiral anomaly at $\tau_2$, $\tau_1$ respectively
    \end{itemize}
\end{itemize}

\textbf{Obstruction-free states}:
\begin{itemize}
\item[] 1D: ($\bzt^3$) \begin{enumerate}
    \item Majorana chain on $\tau_1$, $\tau_2$, or $\tau_3$
\end{enumerate}

\item[] 2D: No obstruction-free states ($\bbz_1$)
\end{itemize}

\textbf{Trivializations}:
\begin{itemize}
    \item Majorana bubble on $\sigma_1$ $\Rightarrow$ Majorana chain on $\tau_2$. This reduces 1D classification to $\bzt^2$.
    \item Majorana bubbles on $\sigma_1$ and $\sigma_2$ $\Rightarrow$ Majorana chain on $\tau_1$. This further reduces 1D classification to $\bzt$.
    \item Open surface decoration trivializes Majorana chain on $\tau_3$. This further reduces 1D classification to $\bbz_1$ (trivial).
\end{itemize}

\textbf{Final classification:}
\begin{itemize}
    \item[] $E_{0,dis}^{1D} = \bbz_1$
    \item[] $E_{0,dis}^{2D} = \bbz_1$
    \item[]  $\mathcal{G}_{0,dis} = E_{0,dis}^{1D} \times E_{0,dis}^{2D} = \bbz_1$
\end{itemize}

\subsubsection*{Disordered Spin-1/2}
\textbf{Block state decorations}:
\begin{itemize}
    \item[] 1D
    \begin{itemize}
        \item $\tau_1,\ \tau_2$: Majorana chain
        \item $\tau_3$: \placeholder
    \end{itemize}
    \item[] 2D
    \begin{itemize}
        \item $\sigma_1,\ \sigma_2$:\( p+ip \) SC
    \end{itemize}
    
\end{itemize}

\textbf{Obstructions}
\begin{itemize}
    \item[] 2D
    \begin{itemize}
        \item\( p+ip \) SC on $\sigma_1$, $\sigma_2$ is obstructed by chiral anomaly at $\tau_2$, $\tau_1$ respectively
    \end{itemize}
\end{itemize}

\textbf{Obstruction-free states}:
\begin{itemize}
\item[] 1D: ($\bzt^3$) \begin{enumerate}
    \item Majorana chain on $\tau_1$ or $\tau_2$ (E)
    \item \placeholder on $\tau_3$ (I)
\end{enumerate}

\item[] 2D: No obstruction-free states ($\bbz_1$)
\end{itemize}

\textbf{Trivializations}:
\begin{itemize}
    \item Majorana bubble on $\sigma_1$ $\Rightarrow$ Majorana chain on $\tau_2$. This reduces 1D classification to $\bzt^2$.
    \item Majorana bubbles on $\sigma_1$ and $\sigma_2$ $\Rightarrow$ Majorana chain on $\tau_1$. This further reduces 1D classification to $\bzt$.

\end{itemize}

\textbf{Final classification:}
\begin{itemize}
    \item[] $E_{1/2,dis}^{1D} = \bzt(I)$
    \item[] $E_{1/2,dis}^{2D} = \bbz_1$
    \item[]  $\mathcal{G}_{1/2,dis} = E_{1/2,dis}^{1D} \times E_{1/2,dis}^{2D} = \bzt(I)$
\end{itemize}

\subsection{$T_h$}
\subsubsection*{Cell Decomposition}
\begin{figure}[!htbp]
    \centering
    \includegraphics[width=0.9\linewidth]{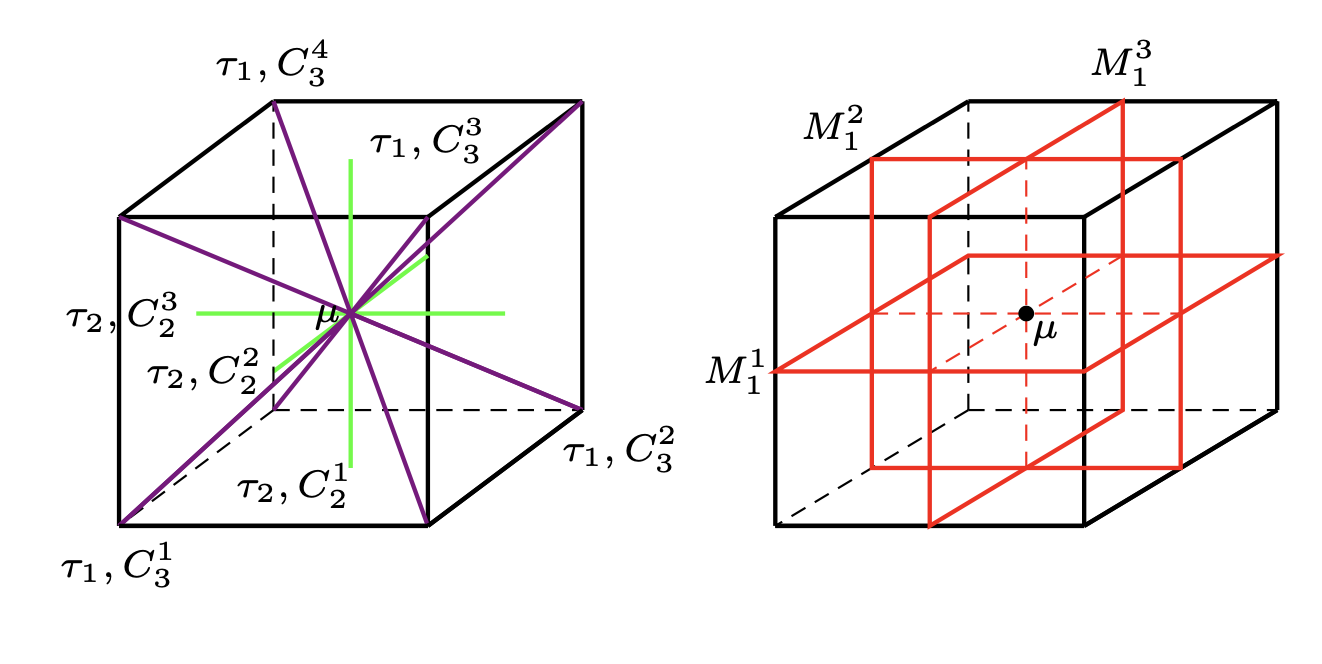}
    \includegraphics[width=0.9\linewidth]{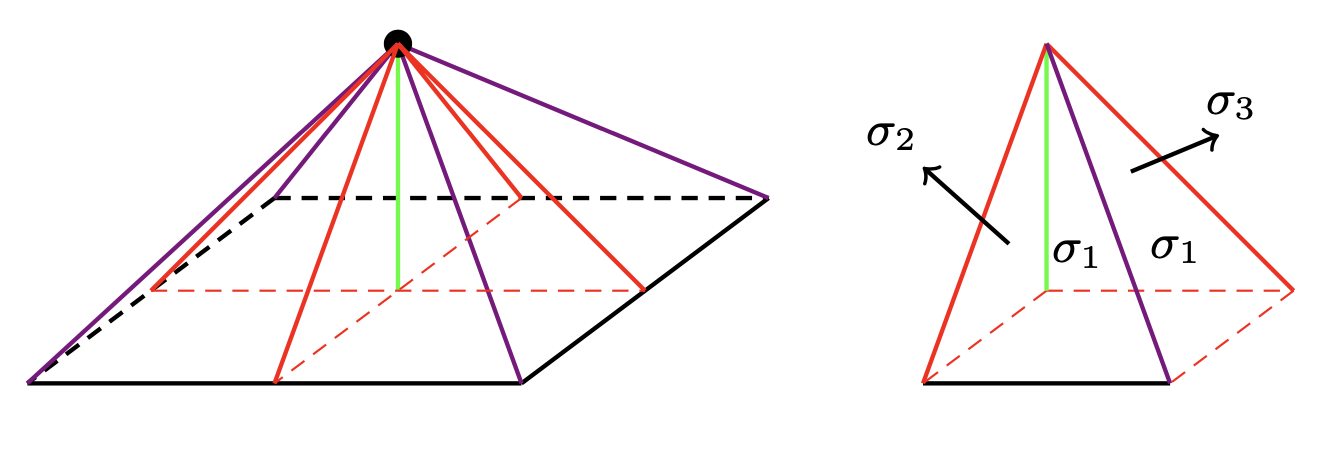}
    \caption{$T_h$ lattice}
\end{figure}
\textbf{Blocks and onsite symmetries}:
\begin{itemize}
    \item 2D: $G_{\sigma_1}=I,\  G_{\sigma_2},\ G_{\sigma_3} = \bzt$
    \item 1D: $G_{\tau_1} = \bbz_3,\ G_{\tau_2} = \bzt \times \bzt,\ G_{\tau_3} = \bzt$ 
    \item 0D: $G_{\mu} = A_4 \times \bzt$
\end{itemize}

\subsubsection*{Decohered Spinless}
\textbf{Block state decorations}:
\begin{itemize}
    \item[] 0D
    \begin{itemize}
        \item $\mu$: Odd fermion
    \end{itemize}
    \item[] 1D
    \begin{itemize}
        \item $\tau_1$: Majorana chain
        \item $\tau_2$: Majorana chain, $\bzt\times\bzt$ fSPT labeled by $(n_{M_1},n_{M_2})$
        \item $\tau_3$: Majorana chain, $\bzt$ fSPT 
    \end{itemize}
    \item[] 2D
    \begin{itemize}
        \item $\sigma_1$:\( p+ip \) SC
        \item $\sigma_2,\ \sigma_3$:\( p+ip \) SC, fLG 
    \end{itemize}    
\end{itemize}

\textbf{Obstructions}
\begin{itemize}
    \item[] 1D
    \begin{itemize}
        \item Majorana chain on $\tau_2$ is obstructed at $\mu$
    \end{itemize}
    \item[] 2D
    \begin{itemize}
        \item\( p+ip \) SC on $\sigma_1$ is obstructed by chiral anomaly at $\tau_1$
        \item If $\sigma_2$ and $\sigma_3$ have non-identical decorations, they are obstructed at $\tau_3$.
        \item $p\pm ip$-SC on $\sigma_2$ and $\sigma_3$ with opposite chiralities are obstructed at $\tau_2$ ($\nu_{M_1}=1/4$)
        \item fLG on $\sigma_2$ and $\sigma_3$ is obstructed at $\tau_2$ ($\nu=1/2$).
    \end{itemize}
\end{itemize}

\textbf{Obstruction-free states}:
\begin{itemize}
\item[] 0D state ($\bzt$) is obstruction-free (E). 

\item[] 1D: ($\bzt^5$) \begin{enumerate}
    \item Majorana chain on $\tau_1$ (E)  
    \item $\bzt \times \bzt$ fSPT on $\tau_2$ (I)
    \item Majorana chain on $\tau_3$ (I)
    \item $\bzt$ fSPT on $\tau_3$ (E)
\end{enumerate}

\item[] 2D ($\bzt$) \begin{enumerate}
    \item Simultaenous decoration of $n=2$ fLG on $\sigma_2$ and $\sigma_3$ (I)
\end{enumerate}
\end{itemize}

\textbf{Trivializations}:
\begin{itemize}
    \item Majorana bubble on $\sigma_1$ $\Rightarrow$ Simultaneous decoration of Majorana chain on $\tau_1$ and $\bzt$ fSPT on $\tau_3$. This reduces 1D classification to $\bzt^4$.
    \item Majorana bubble on $\sigma_2$ $\Rightarrow$ Simultaneous decoration of $(1,0)$ fSPT on $\tau_2$ and Majorana chain on $\tau_3$. This further reduces 1D classification to $\bzt^3$.
    \item Majorana bubble on $\sigma_3$ $\Rightarrow$ Simultaneous decoration of $(0,1)$ fSPT on $\tau_2$ and Majorana chain on $\tau_3$. This further reduces 1D classification to $\bzt^3$.
    \item $\bzt$ fSPT bubble on $\sigma_2$ $\Rightarrow$ $\bzt$ fSPT on $\tau_3$. This further reduces 1D classification to $\bzt$.
\end{itemize}

\textbf{Final classification:}
\begin{itemize}
    \item[] $E_{0,dec}^{0D} = \bzt(E)$
    \item[] $E_{0,dec}^{1D} = \bzt(I)$
    \item[] $E_{0,dec}^{2D} = \bzt(I)$
    \item[]  $\mathcal{G}_{0,dec} = E_{0,dec}^{0D} \times E_{0,dec}^{1D} \times E_{0,dec}^{2D} = \bzt(E)\times\bzt^2(I)$
\end{itemize}

\subsubsection*{Decohered Spin-1/2}
\textbf{Block state decorations}:
\begin{itemize}
    \item[] 0D
    \begin{itemize}
        \item $\mu$: Odd fermion
    \end{itemize}
    \item[] 1D
    \begin{itemize}
        \item $\tau_1$: Majorana chain
        \item $\tau_2$: $\bzt^2\times_{\omega_2^f} \bzt^f$ ASPT labeled by $(n_{1,M_1},n_{1,M_2})$
        \item $\tau_3$: $\bbz_4^f$ ASPT 
    \end{itemize}
    \item[] 2D
    \begin{itemize}
        \item $\sigma_1$:\( p+ip \) SC
        \item $\sigma_2,\ \sigma_3$: No nontrivial block state 
    \end{itemize}    
\end{itemize}

\textbf{Obstructions}
\begin{itemize}    
    \item[] 2D
    \begin{itemize}
        \item\( p+ip \) SC on $\sigma_1$ is obstructed by chiral anomaly at $\tau_1$
    \end{itemize}
\end{itemize}

\textbf{Obstruction-free states}:
\begin{itemize}
\item[] 0D state ($\bzt$) is obstruction-free. (I)

\item[] 1D: ($\bzt^4$) \begin{enumerate}
    \item Majorana chain on $\tau_1$  (E)
    \item $\bzt^2\times_{\omega_2^f} \bzt^f$ ASPT on $\tau_2$ (I)
    \item $\bbz_4^f$ ASPT on $\tau_3$ (I)
\end{enumerate}

\item[] 2D: No obstruction-free states ($\bbz_1$)
\end{itemize}

\textbf{Trivializations}:
\begin{itemize}
    \item Majorana bubble on $\sigma_1$ $\Rightarrow$ Majorana chain on $\tau_1$. This reduces 1D classification to $\bzt^3$.
    \item $\bbz_4^f$ ASPT bubble on $\sigma_2$ $\Rightarrow$ $\bbz_4^f$ ASPT on $\tau_3$. This further reduces 1D classification to $\bzt^2$.
\end{itemize}

\textbf{Final classification:}
\begin{itemize}
    \item[] $E_{1/2,dec}^{0D} = \bzt(I)$
    \item[] $E_{1/2,dec}^{1D} = \bzt^2(I)$
    \item[] $E_{1/2,dec}^{2D} = \bbz_1(E)$
    \item[]  $\mathcal{G}_{1/2,dec} = E_{1/2,dec}^{0D} \times E_{1/2,dec}^{1D} \times E_{1/2,dec}^{2D} = \bzt^3(I)$
\end{itemize}

\subsubsection*{Disordered Spinless}
\textbf{Block state decorations}:
\begin{itemize}
    \item[] 1D
    \begin{itemize}
        \item $\tau_1,\ \tau_2,\ \tau_3$: Majorana chain 
    \end{itemize}
    \item[] 2D
    \begin{itemize}
        \item $\sigma_1$:\( p+ip \) SC
        \item $\sigma_2,\ \sigma_3$:\( p+ip \) SC, fLG 
    \end{itemize}    
\end{itemize}

\textbf{Obstructions}
\begin{itemize}    
    \item[] 2D
    \begin{itemize}
        \item\( p+ip \) SC on $\sigma_1$ is obstructed by chiral anomaly at $\tau_1$
        \item If $\sigma_2$ and $\sigma_3$ have non-identical decorations, they are obstructed at $\tau_3$.
        \item $p\pm ip$-SC on $\sigma_2$ and $\sigma_3$ with opposite chiralities are obstructed at $\tau_2$ ($\nu_{M_1}=1/4$).
    \end{itemize}
\end{itemize}

\textbf{Obstruction-free states}:
\begin{itemize}

\item[] 1D: ($\bzt^3$) \begin{enumerate}
    \item Majorana chain on $\tau_1$ (E)  
    \item Majorana chain on $\tau_2$ or $\tau_3$ (I)
\end{enumerate}

\item[] 2D: ($\bzt$) \begin{enumerate}
    \item Simultaneous decoration of fLG on $\sigma_2$ and $\sigma_3$ (I)
\end{enumerate}
\end{itemize}

\textbf{Trivializations}:
\begin{itemize}
    \item Majorana bubble on $\sigma_1$ $\Rightarrow$ Majorana chain on $\tau_1$. This reduces 1D classification to $\bzt^2$.
    \item Majorana bubble on $\sigma_2$ $\Rightarrow$ Majorana chain on $\tau_3$. This further reduces 1D classification to $\bzt$.
\end{itemize}

\textbf{Final classification:}
\begin{itemize}
    \item[] $E_{0,dis}^{1D} = \bzt(I)$
    \item[] $E_{0,dis}^{2D} = \bzt(I)$
    \item[]  $\mathcal{G}_{0,dis} = E_{0,dis}^{1D} \times E_{0,dis}^{2D} = \bzt^2(I)$
\end{itemize}

\subsubsection*{Disordered Spin-1/2}
\textbf{Block state decorations}:
\begin{itemize}
    \item[] 1D
    \begin{itemize}
        \item $\tau_1$: Majorana chain
        \item $\tau_2,\ \tau_3$: \placeholder 
    \end{itemize}
    \item[] 2D
    \begin{itemize}
        \item $\sigma_1$: $p\pm ip$ SC
        \item $\sigma_2,\ \sigma_3$: 2D $\bbz_4^f$ ASPT 
    \end{itemize}    
\end{itemize}

\textbf{Obstructions}
\begin{itemize}    
    \item[] 2D
    \begin{itemize}
        \item $p\pm ip$ SC on $\sigma_1$ is obstructed by chiral anomaly at $\tau_1$
        \item If $\sigma_2$ and $\sigma_3$ have non-identical decorations, they are obstructed at $\tau_3$.
    \end{itemize}
\end{itemize}

\textbf{Obstruction-free states}:
\begin{itemize}

\item[] 1D: ($\bzt^3$) \begin{enumerate}
    \item Majorana chain on $\tau_1$  (E)
    \item \placeholder on $\tau_2$ or $\tau_3$ (I)
\end{enumerate}

\item[] 2D: ($\bzt$) \begin{enumerate}
    \item Simultaneous decoration of 2D $\bbz_4^f$ ASPT on $\sigma_2$ and $\sigma_3$ (I)
\end{enumerate}
\end{itemize}

\textbf{Trivializations}:
\begin{itemize}
    \item Majorana bubble on $\sigma_1$ $\Rightarrow$ Majorana chain on $\tau_1$. This reduces 1D classification to $\bzt^2$.
    \item \bubbleplaceholder on $\sigma_2$ $\Rightarrow$ \placeholder on $\tau_3$. This further reduces 1D classification to $\bzt$.
\end{itemize}

\textbf{Final classification:}
\begin{itemize}
    \item[] $E_{1/2,dis}^{1D} = \bzt(I)$
    \item[] $E_{1/2,dis}^{2D} = \bzt(I)$
    \item[]  $\mathcal{G}_{1/2,dis} = E_{1/2,dis}^{1D} \times E_{1/2,dis}^{2D} = \bzt^2(I)$
\end{itemize}

\subsection{$T_d$}

\subsubsection*{Cell Decomposition}
\begin{figure}[!htbp]
    \centering
    \includegraphics[width=0.9\linewidth]{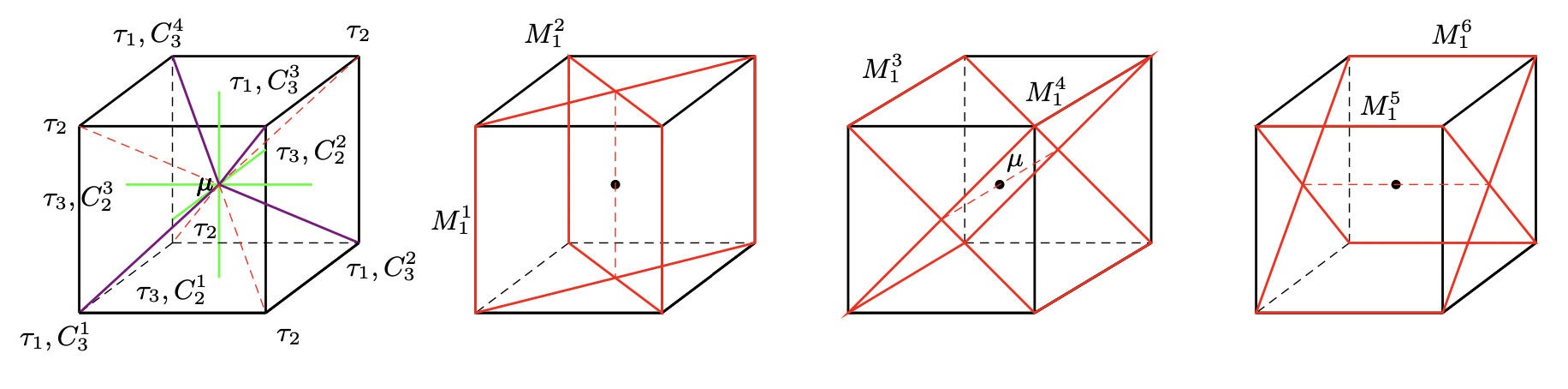}
    \includegraphics[width=0.9\linewidth]{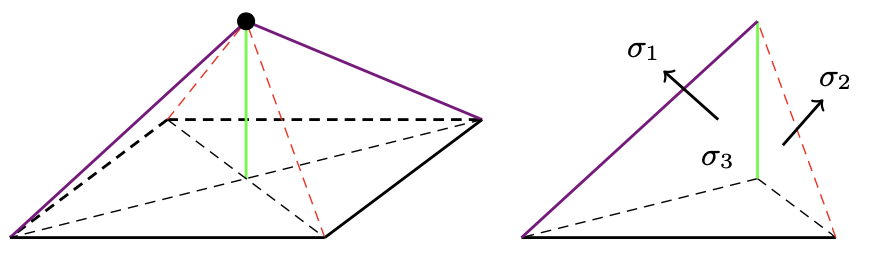}
    \caption{$T_d$ lattice}
\end{figure}
\textbf{Blocks and onsite symmetries}:
\begin{itemize}
    \item 2D: $G_{\sigma_1},\ G_{\sigma_2},\ G_{\sigma_3} = \bzt$
    \item 1D: $G_{\tau_1},\ G_{\tau_2} = \bbz_3\rtimes \bzt,\ G_{\tau_3} = \bzt\times\bzt$
    \item 0D: $G_{\mu} = S_4$
\end{itemize}

\subsubsection*{Decohered Spinless}
\textbf{Block state decorations}:
\begin{itemize}
    \item[] 0D
    \begin{itemize}
        \item $\mu$: Odd fermion
    \end{itemize}
    \item[] 1D
    \begin{itemize}
        \item $\tau_1$,\ $\tau_2$: Majorana chain, $\bbz_3\rtimes\bzt$ fSPT ($\bzt$)
        \item $\tau_3$: Majorana chain, $\bzt\times\bzt$ fSPT ($\bzt^2$)
    \end{itemize}
    \item[] 2D
    \begin{itemize}
        \item $\sigma_1,\ \sigma_2,\ \sigma_3$: $p\pm ip$ SC, fLG 
    \end{itemize}    
\end{itemize}

\textbf{Obstructions}
\begin{itemize}
    \item[] 1D
    \begin{itemize}
        \item Majorana chain on one of $\tau_1$ or $\tau_2$ is obstructed at $\mu$
        \item (1,0)-fSPT or (0,1)-fSPT on $\tau_3$ is obstructed at $\mu$
    \end{itemize}
    \item[] 2D
    \begin{itemize}
        \item If $\sigma_1$,\ $\sigma_2$, and $\sigma_3$ have non-identical decorations, they are obstructed at at least one of $\tau_{1,2,3}$.
        \item Simultaneous decoration of $p\pm ip$ SC on $\sigma_1$, $\sigma_2$, and $\sigma_3$ is obstructed ($\nu_{M_1}=1/4$)
        \item Simultaneous decoration of fLG on $\sigma_1$, $\sigma_2$ and $\sigma_3$ is obstructed ($\nu=1/2$).
    \end{itemize}
\end{itemize}

\textbf{Obstruction-free states}:
\begin{itemize}
\item[] 0D state ($\bzt$) is obstruction-free. 

\item[] 1D: ($\bzt^5$) \begin{enumerate}
    \item Majorana chain on $\tau_1$ and $\tau_2$  (E)
    \item $\bbz_3 \rtimes \bzt$ fSPT on $\tau_1$ (E)
    \item $\bbz_3 \rtimes \bzt$ fSPT on $\tau_2$ (E)
    \item Majorana chain on $\tau_3$ (I)
    \item (1,1) fSPT on $\tau_3$ (I)
\end{enumerate}

\item[] 2D ($\bzt$) \begin{enumerate}
    \item Simultaenous decoration of $n=2$ fLG on $\sigma_1,\ \sigma_2$, and $\sigma_3$ (I)
\end{enumerate}
\end{itemize}

\textbf{Trivializations}:
\begin{itemize}
    \item Majorana bubbles on $\sigma_1$ and $\sigma_2$ $\Rightarrow$ (1,1) fSPT on $\tau_3$.
    \item Majorana bubble on $\sigma_3$ $\Rightarrow$ Majorana chains on $\tau_1$ and $\tau_2$.
    \item $\bzt$ fSPT bubble on $\sigma_1$ $\Rightarrow$ $\bbz_3\rtimes\bzt$ fSPT on $\tau_1$.
    \item $\bzt$ fSPT bubble on $\sigma_2$ $\Rightarrow$ $\bbz_3\rtimes\bzt$ fSPT on $\tau_2$. Ultimately, the 1D classification reduces to $\bzt$
\end{itemize}

\textbf{Final classification:}
\begin{itemize}
    \item[] $E_{0,dec}^{0D} = \bzt(E)$
    \item[] $E_{0,dec}^{1D} = \bzt(I)$
    \item[] $E_{0,dec}^{2D} = \bzt(I)$
    \item[] Non-trivial stacking $\Rightarrow$ $\mathcal{G}_{0,dec} = E_{0,dec}^{0D} \times E_{0,dec}^{1D} \times E_{0,dec}^{2D} = \bzt(I)\times\bbz_4(I)$
\end{itemize}

\subsubsection*{Decohered Spin-1/2}
\textbf{Block state decorations}:
\begin{itemize}
    \item[] 0D
    \begin{itemize}
        \item $\mu$: Odd fermion
    \end{itemize}
    \item[] 1D
    \begin{itemize}
        \item $\tau_1$,\ $\tau_2$: $\bbz_3\rtimes\bzt\rtimes\bzt^f$ ASPT ($\bzt$)
        \item $\tau_3$: $\bzt\times\bzt\rtimes\bzt^f$ ASPT ($\bzt^2$) labeled by $(n_1,n_2)$
    \end{itemize}
    \item[] 2D
    \begin{itemize}
        \item $\sigma_1,\ \sigma_2,\ \sigma_3$: No nontrivial block states. 
    \end{itemize}    
\end{itemize}

\textbf{Obstructions}
\begin{itemize}
    \item[] 1D
    \begin{itemize}
        \item (1,0)-fSPT or (0,1)-fSPT on $\tau_3$ is obstructed at $\mu$
    \end{itemize}

\end{itemize}

\textbf{Obstruction-free states}:
\begin{itemize}
\item[] 0D state ($\bzt$) is obstruction-free (I). 

\item[] 1D: ($\bzt^3$) \begin{enumerate}
    \item $\bbz_3\rtimes\bzt\rtimes\bzt^f$ ASPT on $\tau_1$ or $\tau_2$ (I)
    \item (1,1) ASPT on $\tau_3$ (I)
\end{enumerate}

\end{itemize}

\textbf{Trivializations}:
\begin{itemize}
    \item $\bbz_4^f$ ASPT bubble on $\sigma_1$ $\Rightarrow$ $\bbz_3\rtimes\bzt\rtimes\bzt^f$ ASPT on $\tau_1$.
    \item $\bbz_4^f$ ASPT bubble on $\sigma_2$ $\Rightarrow$ $\bbz_3\rtimes\bzt\rtimes\bzt^f$ ASPT on $\tau_2$. Therefore, the 1D classification further reduces to $\bzt$.
\end{itemize}

\textbf{Final classification:}
\begin{itemize}
    \item[] $E_{1/2,dec}^{0D} = \bzt(I)$
    \item[] $E_{1/2,dec}^{1D} = \bzt(I)$
    \item[] $E_{1/2,dec}^{2D} = \bbz_1$
    \item[]  $\mathcal{G}_{1/2,dec} = E_{1/2,dec}^{0D} \times E_{1/2,dec}^{1D} \times E_{1/2,dec}^{2D} = \bzt^2(I)$
\end{itemize}

\subsubsection*{Disordered Spinless}
\textbf{Block state decorations}:
\begin{itemize}
    \item[] 1D
    \begin{itemize}
        \item $\tau_1,\ \tau_2,\ \tau_3$: Majorana chain 
    \end{itemize}
    \item[] 2D
    \begin{itemize}
        \item $\sigma_1,\ \sigma_2,\ \sigma_3$: $p\pm ip$ SC, fLG 
    \end{itemize}    
\end{itemize}

\textbf{Obstructions}
\begin{itemize}    
    \item[] 2D
    \begin{itemize}
        \item If $\sigma_1$,\ $\sigma_2$, and $\sigma_3$ have non-identical decorations, they are obstructed at at least one of $\tau_{1,2,3}$.
        \item Simultaneous decoration of $p\pm ip$ SC on $\sigma_1$, $\sigma_2$, and $\sigma_3$ is obstructed ($\nu_{M_1}=1/4$)
    \end{itemize}
\end{itemize}

\textbf{Obstruction-free states}:
\begin{itemize}

\item[] 1D: ($\bzt^3$) \begin{enumerate}
    \item Majorana chain on $\tau_1$ or $\tau_2$ (E)
    \item Majorana chain on $\tau_3$ (I)
\end{enumerate}

\item[] 2D: ($\bzt$) \begin{enumerate}
    \item Simultaneous decoration of fLG on $\sigma_1,\ \sigma_2$, and $\sigma_3$ (I).
\end{enumerate}
\end{itemize}

\textbf{Trivializations}:
\begin{itemize}
    \item Majorana bubble on $\sigma_1$ $\Rightarrow$ Majorana chain on $\tau_1$. This reduces 1D classification to $\bzt^2$.
    \item Majorana bubble on $\sigma_2$ $\Rightarrow$ Majorana chain on $\tau_2$. This further reduces 1D classification to $\bzt$.
\end{itemize}

\textbf{Final classification:}
\begin{itemize}
    \item[] $E_{0,dis}^{1D} = \bzt(I)$
    \item[] $E_{0,dis}^{2D} = \bzt(I)$
    \item[]  $\mathcal{G}_{0,dis} = E_{0,dis}^{1D} \times E_{0,dis}^{2D} = \bzt^2(I)$
\end{itemize}

\subsubsection*{Disordered Spin-1/2}
\textbf{Block state decorations}:
\begin{itemize}
    \item[] 1D
    \begin{itemize}
        \item $\tau_1,\ \tau_2,\ \tau_3$: \placeholder 
    \end{itemize}
    \item[] 2D
    \begin{itemize}
        \item $\sigma_1,\ \sigma_2,\ \sigma_3$: 2D $\bbz_4^f$ ASPT
    \end{itemize}    
\end{itemize}

\textbf{Obstructions}
\begin{itemize}    
    \item[] 2D
    \begin{itemize}
        \item If $\sigma_1$,\ $\sigma_2$, and $\sigma_3$ have non-identical decorations, they are obstructed at at least one of $\tau_{1,2,3}$.
    \end{itemize}
\end{itemize}

\textbf{Obstruction-free states}:
\begin{itemize}

\item[] 1D: ($\bzt^3$) \begin{enumerate}
    \item \placeholder on $\tau_1$, $\tau_2$, or $\tau_3$ (I)
\end{enumerate}

\item[] 2D: ($\bzt$) \begin{enumerate}
    \item Simultaneous decoration of 2D $\bbz_4^f$ ASPT on $\sigma_1,\ \sigma_2$, and $\sigma_3$. (I)
\end{enumerate}
\end{itemize}

\textbf{Trivializations}:
\begin{itemize}
    \item \bubbleplaceholder on $\sigma_1$ $\Rightarrow$ \placeholder on $\tau_1$. This reduces 1D classification to $\bzt^2$.
    \item \bubbleplaceholder on $\sigma_2$ $\Rightarrow$ \placeholder on $\tau_2$. This further reduces 1D classification to $\bzt$.
\end{itemize}

\textbf{Final classification:}
\begin{itemize}
    \item[] $E_{1/2,dis}^{1D} = \bzt(I)$
    \item[] $E_{1/2,dis}^{2D} = \bzt(I)$
    \item[]  $\mathcal{G}_{1/2,dis} = E_{1/2,dis}^{1D} \times E_{1/2,dis}^{2D} = \bzt^2(I)$
\end{itemize}

\subsection{$O$}
\subsubsection*{Cell Decomposition}
\begin{figure}[!htbp]
    \centering
    \includegraphics[width=0.9\linewidth]{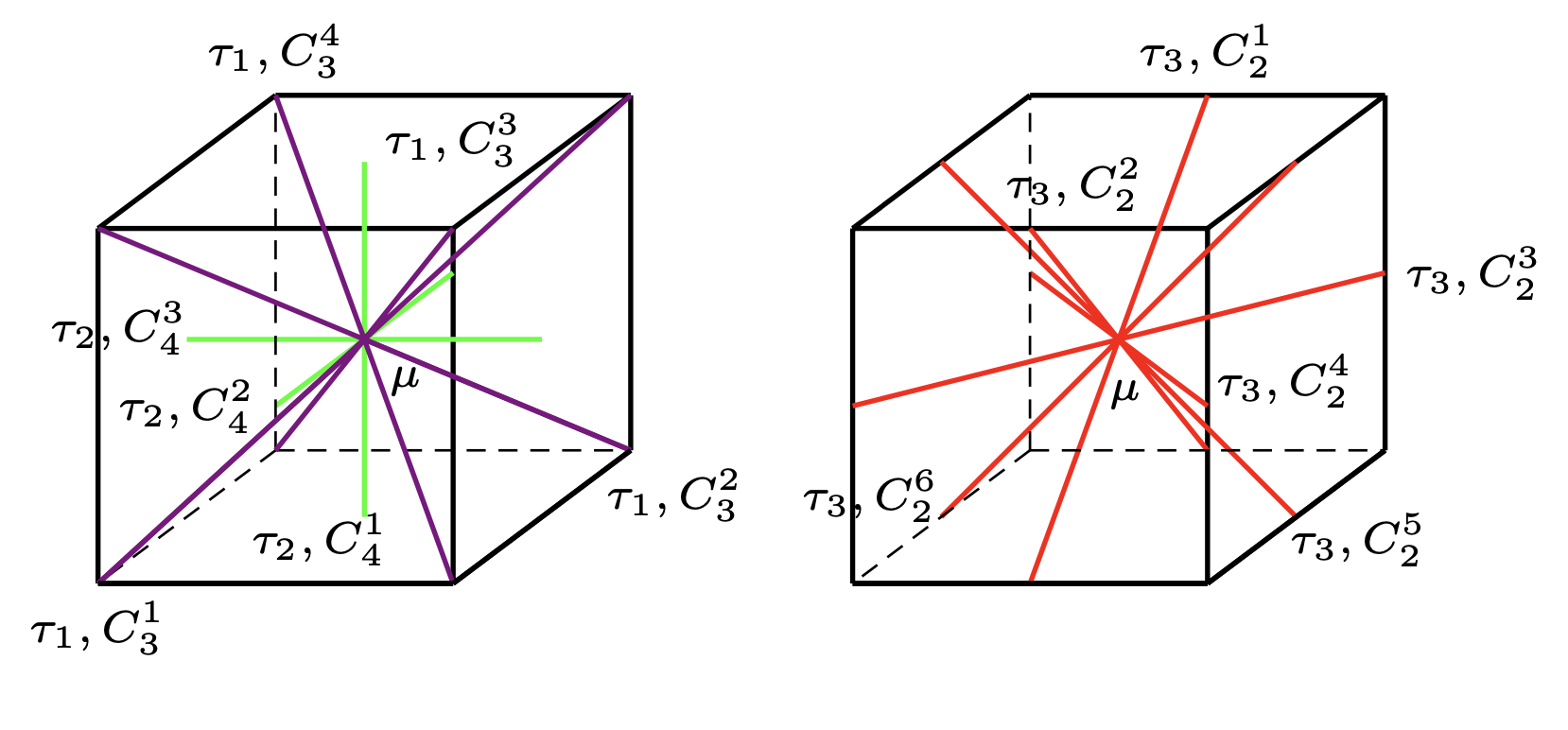}
    \includegraphics[width=0.9\linewidth]{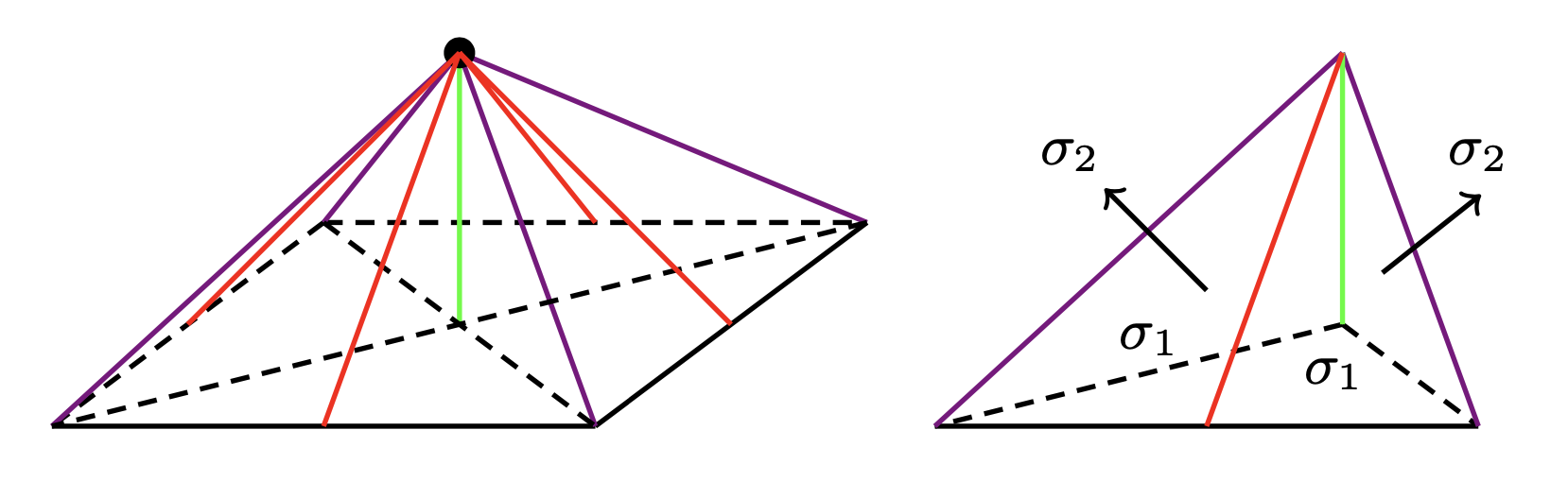}
    \caption{$O$ lattice}
\end{figure}
\textbf{Blocks and onsite symmetries}:
\begin{itemize}
    \item $G_{\sigma_1},\ G_{\sigma_2} = I$
    \item $G_{\tau_1} = \bbz_3,\ G_{\tau_2} = \bbz_4,\ G_{\tau_3} = \bzt$
    \item $G_{\mu} = S_4$
\end{itemize}

\subsubsection*{Decohered Spinless}
\textbf{Block state decorations}:
\begin{itemize}
    \item[] 0D
    \begin{itemize}
        \item $\mu$: Odd fermion
    \end{itemize}
    \item[] 1D
    \begin{itemize}
        \item $\tau_1$: Majorana chain
        \item $\tau_2$: Majorana chain, $\bbz_4$ fSPT
        \item $\tau_3$: Majorana chain, $\bzt$ fSPT
    \end{itemize}
    \item[] 2D
    \begin{itemize}
        \item $\sigma_1,\ \sigma_2$: $p\pm ip$ SC
    \end{itemize}    
\end{itemize}

\textbf{Obstructions}
\begin{itemize}
    \item[] 1D
    \begin{itemize}
        \item Majorana chain on one of $\tau_2$ or $\tau_3$ is obstructed at $\mu$
    \end{itemize}
    \item[] 2D
    \begin{itemize}
        \item $p\pm ip$ SC on $\sigma_1$, $\sigma_2$ is obstructed by chiral anomaly on $\tau_3$, $\tau_2$ respectively
    \end{itemize}
\end{itemize}

\textbf{Obstruction-free states}:
\begin{itemize}
\item[] 0D state ($\bzt$) is obstruction-free (E). 

\item[] 1D: ($\bzt^3$) \begin{enumerate}
    \item Majorana chain on $\tau_1$ (I)
    \item $\bbz_4$ fSPT on $\tau_2$ (I)
    \item $\bzt$ fSPT on $\tau_3$ (I)
\end{enumerate}

\item[] 2D: No obstruction-free states ($\bbz_1$)
\end{itemize}

\textbf{Trivializations}:
\begin{itemize}
    \item Majorana bubble on $\sigma_1$ $\Rightarrow$ Simultaneous decoration of Majorana chain on $\tau_1$ and $\bzt$ fSPT on $\tau_3$. Therefore, the 1D classification reduces to $\bzt^2$.
    \item Majorana bubble on $\sigma_2$ $\Rightarrow$ Simultaneous decoration of Majorana chain on $\tau_1$ and $\bbz_4$ fSPT on $\tau_2$. Therefore, the 1D classification further reduces to $\bzt$.
    \item Chern bubble trivializes odd fermion parity on $\mu$. The 0D classification reduces to $\bbz_1$
\end{itemize}

\textbf{Final classification:}
\begin{itemize}
    \item[] $E_{0,dec}^{0D} = \bbz_1$
    \item[] $E_{0,dec}^{1D} = \bzt(I)$
    \item[] $E_{0,dec}^{2D} = \bbz_1$
    \item[] Non-trivial stacking $\Rightarrow$ $\mathcal{G}_{0,dec} = E_{0,dec}^{0D} \times E_{0,dec}^{1D} \times E_{0,dec}^{2D} = \bzt(I)$
\end{itemize}

\subsubsection*{Decohered Spin-1/2}
\textbf{Block state decorations}:
\begin{itemize}
    \item[] 0D
    \begin{itemize}
        \item $\mu$: Odd fermion
    \end{itemize}
    \item[] 1D
    \begin{itemize}
        \item $\tau_1$: Majorana chain
        \item $\tau_2$: $\bbz_4\rtimes \bzt^f$ ASPT
        \item $\tau_3$: $\bbz_4^f$ ASPT
    \end{itemize}
    \item[] 2D
    \begin{itemize}
        \item $\sigma_1,\ \sigma_2$: $p\pm ip$ SC
    \end{itemize}    
\end{itemize}

\textbf{Obstructions}
\begin{itemize}
    \item[] 2D
    \begin{itemize}
        \item $p\pm ip$ SC on $\sigma_1$, $\sigma_2$ is obstructed by chiral anomaly on $\tau_3$, $\tau_2$ respectively
    \end{itemize}
\end{itemize}

\textbf{Obstruction-free states}:
\begin{itemize}
\item[] 0D state ($\bzt$) is obstruction-free (I). 

\item[] 1D: ($\bzt^3$) \begin{enumerate}
    \item Majorana chain on $\tau_1$ (I) 
    \item $\bbz_4\rtimes \bzt^f$ ASPT on $\tau_2$ (I)
    \item $\bbz_4^f$ ASPT on $\tau_3$ (I)
\end{enumerate}

\item[] 2D: No obstruction-free states ($\bbz_1$)

\end{itemize}

\textbf{Trivializations}:
\begin{itemize}
    \item Majorana bubble on $\sigma_1$ $\Rightarrow$ Majorana chain on $\tau_1$. Therefore, the 1D classification reduces to $\bzt^2$.
\end{itemize}

\textbf{Final classification:}
\begin{itemize}
    \item[] $E_{1/2,dec}^{0D} = \bzt(I)$
    \item[] $E_{1/2,dec}^{1D} = \bzt^2(I)$
    \item[] $E_{1/2,dec}^{2D} = \bbz_1$
    \item[]  $\mathcal{G}_{1/2,dec} = E_{1/2,dec}^{0D} \times E_{1/2,dec}^{1D} \times E_{1/2,dec}^{2D} = \bzt^3(I)$
\end{itemize}

\subsubsection*{Disordered Spinless}
\textbf{Block state decorations}:
\begin{itemize}
    \item[] 1D
    \begin{itemize}
        \item $\tau_1,\ \tau_2,\ \tau_3$: Majorana chain 
    \end{itemize}
    \item[] 2D
    \begin{itemize}
        \item $\sigma_1,\ \sigma_2,\ \sigma_3$: $p\pm ip$ SC 
    \end{itemize}    
\end{itemize}

\textbf{Obstructions}
\begin{itemize}
    \item[] 2D
    \begin{itemize}
        \item $p\pm ip$ SC on $\sigma_1$, $\sigma_2$ is obstructed by chiral anomaly on $\tau_3$, $\tau_2$ respectively
    \end{itemize}
\end{itemize}

\textbf{Obstruction-free states}:
\begin{itemize}
\item[] 1D: ($\bzt^3$) \begin{enumerate}
    \item Majorana chain on $\tau_1$, $\tau_2$, or $\tau_3$ (I)
\end{enumerate}

\item[] 2D: No obstruction-free states ($\bbz_1$)
\end{itemize}

\textbf{Trivializations}:
\begin{itemize}
    \item Majorana bubble on $\sigma_1$ $\Rightarrow$ Majorana chain on $\tau_1$. Therefore, the 1D classification reduces to $\bzt^2$.
    \item Open surface decoration trivializes Majorana chain on $\tau_2$. Therefore, the 1D classification further reduces to $\bzt$.
\end{itemize}

\textbf{Final classification:}
\begin{itemize}
    \item[] $E_{0,dis}^{1D} = \bzt(I)$
    \item[] $E_{0,dis}^{2D} = \bbz_1$
    \item[] $\mathcal{G}_{0,dis} = E_{0,dis}^{1D} \times E_{0,dis}^{2D} = \bzt(I)$
\end{itemize}

\subsubsection*{Disordered Spin-1/2}
\textbf{Block state decorations}:
\begin{itemize}
    \item[] 1D
    \begin{itemize}
        \item $\tau_1$: Majorana chain
        \item $\tau_2,\ \tau_3$: \placeholder 
    \end{itemize}
    \item[] 2D
    \begin{itemize}
        \item $\sigma_1,\ \sigma_2,\ \sigma_3$: $p\pm ip$ SC 
    \end{itemize}    
\end{itemize}

\textbf{Obstructions}
\begin{itemize}
    \item[] 2D
    \begin{itemize}
        \item $p\pm ip$ SC on $\sigma_1$, $\sigma_2$ is obstructed by chiral anomaly on $\tau_3$, $\tau_2$ respectively
    \end{itemize}
\end{itemize}

\textbf{Obstruction-free states}:
\begin{itemize}
\item[] 1D: ($\bzt^3$) \begin{enumerate}
    \item Majorana chain on $\tau_1$  (I)
    \item \placeholder on $\tau_2$ or $\tau_3$ (I)
\end{enumerate}

\item[] 2D: No obstruction-free states ($\bbz_1$)

\end{itemize}

\textbf{Trivializations}:
\begin{itemize}
    \item Majorana bubble on $\sigma_1$ $\Rightarrow$ Majorana chain on $\tau_1$. Therefore, the 1D classification reduces to $\bzt^2$.
\end{itemize}

\textbf{Final classification:}
\begin{itemize}
    \item[] $E_{1/2,dis}^{1D} = \bzt^2(I)$
    \item[] $E_{1/2,dis}^{2D} = \bbz_1$
    \item[]  $\mathcal{G}_{1/2,dis} = E_{1/2,dis}^{1D} \times E_{1/2,dis}^{2D} = \bzt^2(I)$
\end{itemize}

\subsection{$O_h$}
\subsubsection*{Cell Decomposition}
\begin{figure}[!htbp]
    \centering
    \includegraphics[width=0.9\linewidth]{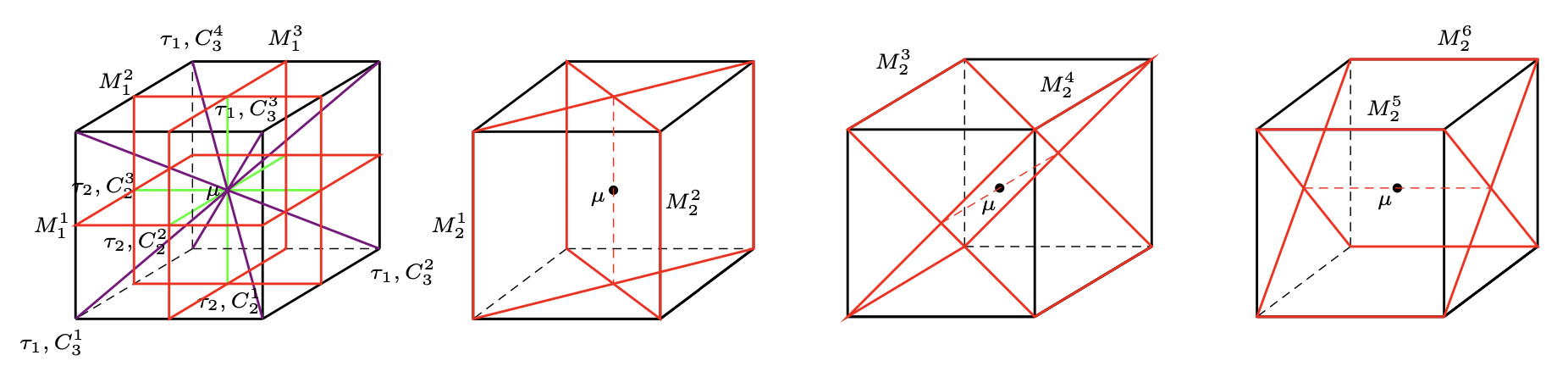}
    \includegraphics[width=0.9\linewidth]{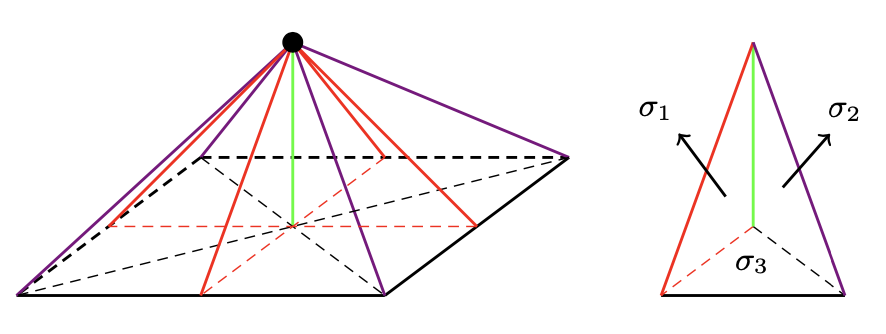}
    \caption{$O_h$ lattice}
\end{figure}
\textbf{Blocks and onsite symmetries}:
\begin{itemize}
    \item $G_{\sigma_1},\ G_{\sigma_2},\ G_{\sigma_3} = \bzt$
    \item $G_{\tau_1} = \bbz_3 \rtimes \bzt,\ G_{\tau_2} = \bbz_4\times \bzt,\ G_{\tau_3} = \bzt\times\bzt$
    \item $G_{\mu} = S_4\times \bzt$
\end{itemize}
\subsubsection*{Decohered Spinless}
\textbf{Block state decorations}:
\begin{itemize}
    \item[] 0D
    \begin{itemize}
        \item $\mu$: Odd fermion
    \end{itemize}
    \item[] 1D
    \begin{itemize}
        \item $\tau_1$: Majorana chain, $\bbz_3\rtimes\bzt$ fSPT ($\bzt$)
        \item $\tau_2$: Majorana chain, $\bbz_4\times\bzt$ fSPT ($\bzt^2$) $\cong$ ($\bbz_4^{n_1}$ fSPT, $\bbz_4^{n_2}$ fSPT)
        \item $\tau_3$: Majorana chain, $\bzt\times\bzt$ fSPT ($\bzt^2$) $\cong$ ($\bzt^{M_1}$ fSPT, $\bzt^{M_2}$ fSPT)
    \end{itemize}
    \item[] 2D
    \begin{itemize}
        \item $\sigma_1,\ \sigma_2,\ \sigma_3$: $p\pm ip$ SC, fLG
    \end{itemize}    
\end{itemize}

\textbf{Obstructions}
\begin{itemize}
    \item[] 1D
    \begin{itemize}
        \item Majorana chain on $\tau_2$ is obstructed at $\mu$ ($M_1$ reflection symmetry) 
        \item Majorana chain on $\tau_3$ is obstructed at $\mu$ (4-fold rotation symmetry about $\tau_2$)
    \end{itemize}
    \item[] 2D
    \begin{itemize}
        \item $p\pm ip$ SC decorations are obstructed by chiral anomaly unless all three $\sigma$ blocks are simultaneously decorated, which in turn is obstructed by $\bzt$ anomaly ($\nu=1/4$).
        \item All decorations with one copy of fLG state are obstructed.
        \item $n=2$ fLG on one of $\sigma_2$ or $\sigma_3$ is obstructed.
    \end{itemize}
\end{itemize}

\textbf{Obstruction-free states}:
\begin{itemize}
\item[] 0D state ($\bzt$) is obstruction-free. 

\item[] 1D: ($\bzt^6$) \begin{enumerate}
    \item Majorana chain on $\tau_1$ (I)
    \item $\bbz_3\rtimes\bzt$ fSPT on $\tau_1$ (E)
    \item $\bbz_4$ fSPT on $\tau_2$ (E)
    \item $\bzt^{M_1}$ fSPT on $\tau_2$ or $\tau_3$ (I)
    \item $\bzt^{M_2}$ fSPT on $\tau_3$ (I)
\end{enumerate}

\item[] 2D: ($\bzt^2$) \begin{enumerate}
    \item $n=2$ fLG on $\sigma_1$ (I)
    \item $n=2$ fLG on $\sigma_2$ and $\sigma_3$ (I)
\end{enumerate} 
\end{itemize}

\textbf{Trivializations}:
\begin{itemize}
    \item Double Majorana ($\bzt$ fSPT) bubble on $\sigma_2$ $\Rightarrow$ $\bbz_3\rtimes\bzt$ fSPT on $\tau_1$. 
    \item Majorana bubble on $\sigma_1$ $\Rightarrow$ Simultaneous decoration of $\bbz_4^{n_1}$ fSPT on $\tau_2$ and $\bzt^{M_2}$ fSPT on $\tau_3$.
    \item Majorana bubble on $\sigma_2$ $\Rightarrow$ Simultaneous decoration of Majorana chain on $\tau_1$ and $\bbz_4^{n_2}$ fSPT on $\tau_2$. 
    \item Majorana bubble on $\sigma_3$ $\Rightarrow$ Simultaneous decoration of Majorana chain on $\tau_1$ and $\bzt^{M_1}$ fSPT on $\tau_3$. Therefore, the 1D classification finally reduces to $\bzt^2$.
\end{itemize}

\textbf{Final classification:}
\begin{itemize}
    \item[] $E_{0,dec}^{0D} = \bzt$
    \item[] $E_{0,dec}^{1D} = \bzt(E)\times\bzt(I)$
    \item[] $E_{0,dec}^{2D} = \bzt^2(I)$
    \item[] Non-trivial stacking $\Rightarrow$ $\mathcal{G}_{0,dec} = E_{0,dec}^{0D} \times E_{0,dec}^{1D} \times E_{0,dec}^{2D} = \bzt^2(E)\times\bzt^3(I)$
\end{itemize}

\subsubsection*{Decohered Spin-1/2}
\textbf{Block state decorations}:
\begin{itemize}
    \item[] 0D
    \begin{itemize}
        \item $\mu$: Odd fermion (I)
    \end{itemize}
    \item[] 1D
    \begin{itemize}
        \item $\tau_1$: $\bbz_3\rtimes\bzt\rtimes\bzt^f$ ASPT ($\bzt$) $\cong$ $\bbz_f^4$ ASPT (I)
        \item $\tau_2$: $\bbz_4\times\bzt\rtimes\bzt^f$ ASPT ($\bzt^2$) $\cong$ ($\bbz_4^{n_1}\rtimes\bzt^f$ ASPT, $\bbz_4^{n_2}\rtimes\bzt^f$ ASPT) (I)
        \item $\tau_3$: $\bzt\times\bzt\rtimes\bzt^f$ ASPT ($\bzt^2$) $\cong$ ($\bbz_4^{f,M_1}$ fSPT, $\bbz_4^{f,M_2}$ ASPT) (I)
    \end{itemize}
    \item[] These states are all \textbf{obstruction-free}.
    \item[] 2D
    \begin{itemize}
        \item $\sigma_1,\ \sigma_2,\ \sigma_3$: No nontrivial block state
    \end{itemize}    
\end{itemize}

\textbf{Trivializations}:
\begin{itemize}
    \item $\bbz_4^f$ ASPT bubble on $\sigma_2$ $\Rightarrow$ $\bbz_4^f$ ASPT on $\tau_1$. Therefore, the 1D classification reduces to $\bzt^4$
\end{itemize}

\textbf{Final classification:}
\begin{itemize}
    \item[] $E_{1/2,dec}^{0D} = \bzt(I)$
    \item[] $E_{1/2,dec}^{1D} = \bzt^4(I)$
    \item[] $E_{1/2,dec}^{2D} = \bbz_1$
    \item[] Non-trivial stacking $\Rightarrow$ $\mathcal{G}_{1/2,dec} = E_{1/2,dec}^{0D} \times E_{1/2,dec}^{1D} \times E_{1/2,dec}^{2D} = \bzt^5(I)$
\end{itemize}

\subsubsection*{Disordered Spinless}
\textbf{Block state decorations}:
\begin{itemize}
    \item[] 1D
    \begin{itemize}
        \item $\tau_1,\ \tau_2,\ \tau_3$: Majorana chain 
    \end{itemize}
    \item[] 2D
    \begin{itemize}
        \item $\sigma_1,\ \sigma_2,\ \sigma_3$: $p\pm ip$ SC, fLG
    \end{itemize}    
\end{itemize}

\textbf{Obstructions}
\begin{itemize}
    \item[] 2D
    \begin{itemize}
        \item\( p+ip \) SC decorations are obstructed by chiral anomaly unless all three $\sigma$ blocks are simultaneously decorated, which in turn is obstructed by $\bzt$ anomaly ($\nu=1/4$).
        \item fLG on one of $\sigma_2$ or $\sigma_3$ is obstructed.
    \end{itemize}
\end{itemize}

\textbf{Obstruction-free states}:
\begin{itemize}
\item[] 1D: ($\bzt^3$) \begin{enumerate}
    \item Majorana chain on $\tau_1$, $\tau_2$, or $\tau_3$ (I)
\end{enumerate}

\item[] 2D: ($\bzt^2$) \begin{enumerate}
    \item fLG on $\sigma_1$ (I)
    \item fLG on $\sigma_2$ and $\sigma_3$ (I)
\end{enumerate}
\end{itemize}

\textbf{Trivializations}:
\begin{itemize}
    \item Majorana bubble on $\sigma_1$ $\Rightarrow$ Majorana chain on $\tau_1$. Therefore, the 1D classification reduces to $\bzt^2$.
\end{itemize}

\textbf{Final classification:}
\begin{itemize}
    \item[] $E_{0,dis}^{1D} = \bzt^2(I)$
    \item[] $E_{0,dis}^{2D} = \bzt^2(I)$
    \item[]  $\mathcal{G}_{0,dis} = E_{0,dis}^{1D} \times E_{0,dis}^{2D} = \bzt^4(I)$
\end{itemize}

\subsubsection*{Disordered Spin-1/2}
\textbf{Block state decorations}:
\begin{itemize}
    \item[] 1D
    \begin{itemize}
        \item $\tau_1,\ \tau_2,\ \tau_3$: \placeholder
    \end{itemize}
    \item[] 2D
    \begin{itemize}
        \item $\sigma_1,\ \sigma_2,\ \sigma_3$: 2D $\bbz_4^f$ ASPT
    \end{itemize}    
\end{itemize}

\textbf{Obstructions}
\begin{itemize}
    \item[] 2D
    \begin{itemize}
        \item 2D $\bbz_4^f$ ASPT on one of $\sigma_2$ or $\sigma_3$ is obstructed.
    \end{itemize}
\end{itemize}

\textbf{Obstruction-free states}:
\begin{itemize}
\item[] 1D: ($\bzt^3$) \begin{enumerate}
    \item \placeholder on $\tau_1$, $\tau_2$, or $\tau_3$ (I)
\end{enumerate}

\item[] 2D: ($\bzt^2$) \begin{enumerate}
    \item 2D $\bbz_4^f$ ASPT on $\sigma_1$ (I)
    \item 2D $\bbz_4^f$ ASPT on $\sigma_2$ and $\sigma_3$ (I)
\end{enumerate}
\end{itemize}

\textbf{Trivializations}:
\begin{itemize}
    \item \bubbleplaceholder on $\sigma_1$ $\Rightarrow$ \placeholder on $\tau_1$. Therefore, the 1D classification reduces to $\bzt^2$.
\end{itemize}

\textbf{Final classification:}
\begin{itemize}
    \item[] $E_{1/2,dis}^{1D} = \bzt^2(I)$
    \item[] $E_{1/2,dis}^{2D} = \bzt^2(I)$
    \item[]  $\mathcal{G}_{1/2,dis} = E_{1/2,dis}^{1D} \times E_{1/2,dis}^{2D} = \bzt^4(I)$
\end{itemize}

\end{document}